\documentclass[pre,twocolumn,twoside,showpacs, amsmath,amssymb, superscriptaddress,floatfix]{revtex4-2}
\usepackage{xcolor}
\usepackage{graphicx}
\usepackage{dcolumn}
\usepackage{bm}
\usepackage{subfig}
\usepackage{tikz}
\usetikzlibrary{calc}
\usepackage{listings}

\newenvironment{keywrds}{%
\small \bfseries  \scshape 

}{%
\vspace{0.1cm}
}

\begin{document}

\author{Alexandre Wagemakers}
\affiliation{Nonlinear Dynamics, Chaos and Complex Systems Group, Departamento de Biolog\'ia y Geolog\'ia, F\'isica Aplicada y Qu\'imica Inorg\'anica, Universidad Rey Juan Carlos,
Tulip\'an s/n, 28933 M\'ostoles, Madrid, Spain}
\email{alexandre.wagemakers@urjc.es}

\title{Basins of Attraction: A Dynamical Zoo}

\date{\today}

\begin{abstract}
Research in multistable systems is a flourishing field with countless examples and applications across scientific disciplines. I present a catalog of multistable dynamical systems covering relevant fields of knowledge. This work is intended to provide a research tool to the community in the form of classified examples with computer codes to reproduce basins of attraction. The companion code to this article can be found at \url{https://github.com/awage/BasinsCollection} or \url{https://doi.org/10.5281/zenodo.15124200}.
\end{abstract}

\maketitle
\tableofcontents

Basins classification glossary:
\begin{itemize}
\item SMB: Smooth basin boundary
\item WD: Wada basins
\item SPF: Sporadically fractal boundary
\item FB: Fractal boundary 
\item SLB: Slim fractal boundary
\item IWB: Intertwined basin boundary
\item IB: Intermingled basin boundary
\item RB: Riddled basin boundary
\item OH: Open Hamiltonian
\item MG: Megastability
\item HA: Hidden Attractors
\item MAP: Discrete map
\item ODE: Ordinary differential equations
\item DDE: Delay differential equations
\end{itemize}

\section{\label{sec:intro}Introduction}

In this article, I wish to offer researchers interested in the multistable phenomenon a library of relevant dynamical systems. This article does not aim at a full review of the field but provides a catalog of computable and reproducible examples in a wide variety of systems and disciplines. Fundamental examples are provided as well as theoretical and applied models from physics, engineering, and natural sciences. The computer code for each example is provided and relies heavily on a specialized dynamical systems library. To help the researcher, I also propose a crude classification of the basins based on the nature of the dynamical system and the properties of its basins. This classification is by no means unique nor exhaustive and serves only as a quick reference index.    

Let us now state a definition of multistability: given a dynamical system with an evolution rule, it is multistable when its state evolves to different classes of behavior depending on the initial condition. The class of behavior should be discernible analytically or numerically with an established criterion. For example, two stable fixed points must be separated enough to be distinguishable by an algorithm. A simple example is the flipping of a coin; the initial kick brings the coin to one face or another depending on the initial impulse. If we consider this system from a deterministic point of view, the outcome is completely defined from the beginning. This example can be extended to dice, roulette, and physical systems that settle in a stable state after a transient period. For these examples, the objects have been designed with multiple outcomes from the beginning. However, multistability can result from side effects of nonlinearities, hysteresis, bifurcations, or a combination of these factors. The outcome of the evolution depends on the initial conditions, but notice that given an external perturbation, the system can eventually switch from one state to another, introducing the idea of control.  

The importance of coexisting final states in dynamical systems is unquestionable from both a theoretical and practical perspective\cite{feudelcomplexmultistable2008}. Knowing the possible outcomes of a system is a necessary exploratory task if we want to know more about its behavior. An intuitive tool for this inquiry is the representation of basins of attraction for each final state as a function of the initial conditions~\cite{daza2024multistability}. This visual representation, often depicted in a two-dimensional projection, provides insight into the types of structures present in phase space at a glance. Nevertheless, it is not the only way to represent or detect multistable behaviors. Quantitative indicators are also available~\cite{menck_how_2013, daza2022classifying}. 

I provide the necessary computer code to reproduce the presented results using an algorithm based on recurrences~\cite{datseris2022effortless}. The strength of the algorithm is its flexibility and adaptability to a wide variety of systems. Briefly put, the algorithm tracks the passage of a trajectory on a portion of phase space using a finite grid. If the trajectory recurs along the same path repeatedly, we consider that the dynamics have reached an attractor. The algorithm counts the number of consecutive steps of the integrator along an already visited region. When the counter reaches an arbitrary threshold, the program stores the attractor location in memory. A similar process occurs to detect initial conditions that have reached an registered attractor to estimate the basins. The important metaparameters are discussed briefly in the numerical methods section to adapt the algorithm to each particular case. All the systems in this paper have been freshly simulated with an implementation of the technique. 

The range of systems is restricted to ordinary differential equations and discrete maps due to the numerical limitations of the tools used. As an exception, I present two of multistable systems in delay differential equations at the end of the article. The extent of systems described in the literature is substantial~\cite{pisarchik2022multistability}. An exhaustive review is not the intention here, and I introduce only a small sample of this corpus. 

There are two independent classifications proposed. First, the sections group the systems into foundational models, theoretical examples, open Hamiltonian systems, life sciences, economics, physics, and finally engineering. Some systems can obviously straddle different categories, and the choice to fit the system in one category or another is not easy. The second classification is included at the beginning of each section and marks some important features of the basins, such as the nature of the boundaries and the type of dynamical systems. The acronyms are listed at the beginning of the paper. 

For each example, a brief description and motivation of the model is presented, along with the equations to simulate the system, the parameters and initial conditions needed to reproduce the basins of attraction. There is only one emblematic basin represented for each publication, although most systems exhibit a flourishing variety of basins. 

Finally, I encourage the reader to download and simulate some of the examples, even changing some parameters. The code is archieved in a github repository under a MIT license: \url{https://github.com/awage/BasinsCollection}. It is very rewarding to generate our own set of basins. For most of the systems, less than a few minutes of computation is enough to obtain a decent resolution of the basins. 

Please take a walk through the zoo and admire this fantastic fauna.

\section{\label{sec:num_met}Numerical Methods}

Suppose we have a physical system with accessible state; let us evolve the dynamics after setting the initial condition. If the state of the system, or part of the system, is observable, we can attempt to identify the long-term behavior by detecting recurrent events. For example, a fixed point is recurrent in the sense that the state eventually remains in a tiny bounded region. For a periodic orbit, the state will come back arbitrarily close to a previous point of the past trajectory. The estimation of the basins and detection of the attractors are based on this observation. 

As mentioned in the introduction, the core of the numerical treatment is a mapper pairing an initial condition with an attractor or a final state~\cite{datseris2022effortless}. The algorithm needs as an input a numerical solver that evolves the trajectory step by step and a grid over a bounded region of the phase space. This grid can be regularly spaced or even irregular. In some situations, the trajectories of several attractors can be very close to the origin, and it is convenient to have more detail in certain regions. The grid is only for tracking purposes and does not necessarily coincide with the set of initial conditions of the basins of interest. The algorithm keeps track of the passage of the orbit through the defined cells of the reference grid. When a cell has already been visited, a counter starts and increases if the next cell has also been visited previously. If the next cell is freshly visited, the counter resets. Eventually, the counter reaches the threshold \texttt{consecutive\_recurrences}, and we decide that we have found a new attractor. A different process starts to ensure we have all the available pieces of the attractor and mark them suitably. At the end of this step, the cell containing the initial condition is marked as being part of the basin of this attractor. Once we have at least one attractor detected, when an initial condition converges to the cells marked as belonging to the attractor, the algorithm waits a number of consecutive steps equal to \texttt{consecutive\_attractor\_steps} before declaring that the initial conditions belong to the basin of this attractor. With these minimal requirements, the algorithm works out of the box in many cases. However, some systems require adjustments due to the specifics of the attractors or due to the peculiar geometry of the phase space.

The numerical integration of the differential equations is achieved with Verner's ``Most Efficient'' 9/8 Runge-Kutta solver with error tolerances provided by the keywords. The time step is chosen automatically by the algorithm except when specified in the example.

For some examples presented in this collection, this algorithm is unsuited since there is no attractor. See, for example, the Hénon-Heiles Hamiltonian or the open billiards examples. These cases consist of trajectories diverging to infinity through a specific path. A custom computer code must be tailored to detect the path followed by the trajectory. We either use a specialized code for the detection of the exits or, for the dynamical billiards, a library for the setup of the geometry and the detection of the passage through an open exit~\cite{datseris2017dynamicalbilliards}. The numerical integration of the Hamiltonian dynamics is also performed with the Verner algorithm. The dissipation issue for these systems is not a problem since trajectories will escape quickly.

\section{Foundational Models}

This section covers historically important models that have been studied extensively in the literature. Some of these models remain up to date and are paradigms in the field. Other examples are lesser known or newer, but they still have their place in this section.

\subsection{The Forced Damped Pendulum} 
\begin{keywrds}FB, WD, ODE\end{keywrds}

It is hard to avoid the pendulum in physics, as it represents the fundamental idea of periodic motion. In this particular version, there is a twist, as we add two ingredients: a dissipative term and a periodic external forcing. The loss of energy is compensated by the external driving. The equation of motion is:  
\begin{equation}\label{eq:forced_pend}
\ddot{x} + d\dot{x} + \sin x = F\cos( \omega t)
\end{equation}
where $d$ is the strength of the damping, $F$ is the amplitude of the driving, and $\omega$ is its angular frequency. Such a simple model contains a wealth of behaviors~\cite{hubbard1999forced} and often serves as a guinea pig for researchers in dynamical systems. By changing the three parameters $d$, $F$, and $\omega$, we can obtain multistable behaviors, making the basins interesting. The key is to keep the dissipation $d$ low enough to allow multiple coexisting solutions. As $d$ increases, the system is driven to a single stable orbit. 

The preparation of the model requires some care if we want to detect the attractors correctly. The phase $x$ can increase unbounded, so we must restrict its motion to the interval $[-\pi, \pi[$. To do this, we implement a special procedure named callback to detect when the variable $x$ steps out of the interval. If it happens, the numerical solver stops, rescales the variable $x$ modulus $2\pi$ within the interval, and resumes the integration. Another interesting numerical trick is to transform the continuous system into a discrete system using a stroboscopic map. The solver integrates the trajectory over a period of $2\pi/\omega$. Such reduction is permissible since we have an external periodic forcing. These standard numerical procedures will be used often in this article. 

\begin{figure}
\begin{center}
\includegraphics[width=\columnwidth]{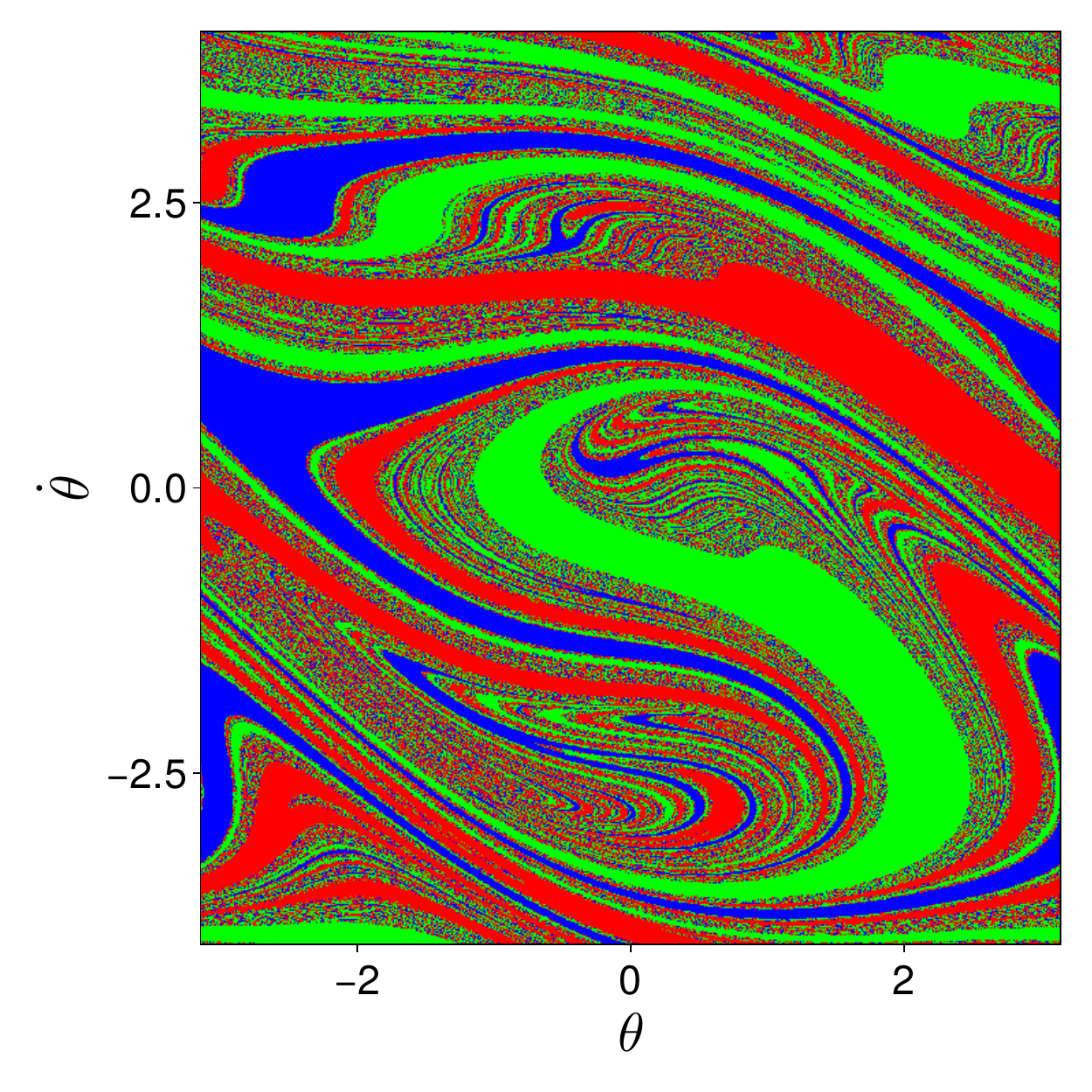}
\end{center}
\caption{\label{fig:fig_forced_pend}Basins of attraction of the forced damped pendulum in Eq. \ref{eq:forced_pend} with $d=0.2$, $F=1.66$, and $\omega = 1$ with a resolution of 1200$\times$1200. These basins have the Wada property.}
\end{figure}

Figure \ref{fig:fig_forced_pend} represents the basins of the pendulum  appearing in numerous publications dealing with fractal structures~\cite{nusse1996basins,aguirre2009fractal}. Not only is the boundary between the three basins fractal, but it also possesses the Wada property~\cite{wagemakers2020detect}. This unique boundary can separate three or more basins, a singular property that is not uncommon in fractal basins as it will appear in several examples along the article. 

Noteworthily the forced pendulum can also exhibit more stable states and a riddled basin~\cite{alexander_riddled_1992} for the parameters $d = 0.2$, $F = 1.36$, and $\omega = 0.5$. More spectacular basins with multiple attractors can be found by varying these parameters.

\subsection{A Map for Understanding the Unpredictability}
\begin{keywrds}FB, MAP\end{keywrds}

\begin{figure}
\begin{center}
\includegraphics[width=\columnwidth]{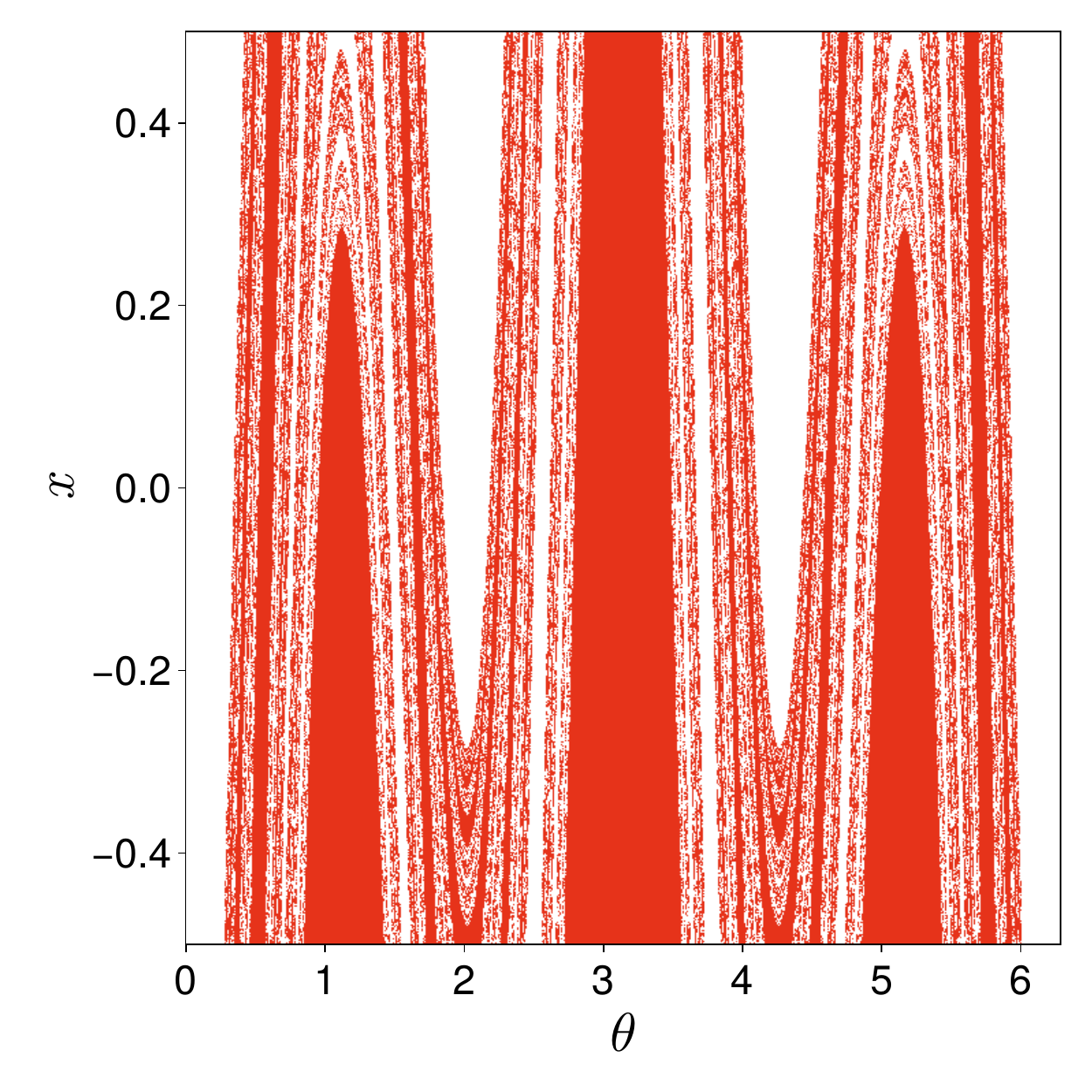}
\end{center}
\caption{\label{fig:grebogi_map}Basins of attraction of the map \ref{eq:grebogi_map} for the parameters $J_0 = 0.3$, $a = 1.32$, and $b = 0.9$. A fractal boundary appears between the basins of the two fixed point attractors.} 
\end{figure}

In an important paper, Grebogi et al.~\cite{grebogi1983final} defined the fundamental notion of final state unpredictability. This seminal paper introduces a numerical technique for computing the fractal dimension of the boundary between basins. For this measurement, the algorithm sample initial conditions within a ball of radius $\varepsilon$. If the ball contains initial conditions leading to different basins, it is tagged as uncertain. Balls with initial conditions that lead to only one basin are called certain. The ratio $f(\varepsilon)$ between the number of uncertain and total number of balls is measured for a range of ball sizes $\varepsilon$. The ratio is expected to scale with a power related to the dimension of the boundary. A linear fit in logarithmic scale of $f(\varepsilon)$ versus $\varepsilon$ recovers the uncertainty exponent $\alpha = D - d$, where $D$ is the dimension of phase space and $d$ the fractal dimension of the boundary. Part of the article focuses on a map with two attractors: 
\begin{align}\label{eq:grebogi_map}
\begin{split}
    \theta_{n+1} &= \theta_n + a\sin(2\theta_n) - b\sin(4\theta_n) - x_n\sin(\theta_n)\text{mod}(2\pi),\\
x_{n+1} &= -J_0 \cos(\theta_n).
\end{split}
\end{align}
For the parameters $J_0 = 0.3$, $a = 1.32$, and $b = 0.9$, we obtain a fractal boundary separating the two fixed points at $(\theta, x) = (0, -J_0)$ and $(\pi, J_0)$, as seen in Fig. \ref{fig:grebogi_map}. The authors measured a fractal dimension of $d = 1.8$ for this boundary. This data can be useful for calibrating an algorithm for the measure of the fractal dimension.

Computationally, this map does not require any particular care. Basins are obtained swiftly with high resolution.

\subsection{A Simple Map for the Wada Property}
\begin{keywrds}FB, WD, MAP\end{keywrds}
\begin{figure}
\begin{center}
\includegraphics[width=\columnwidth]{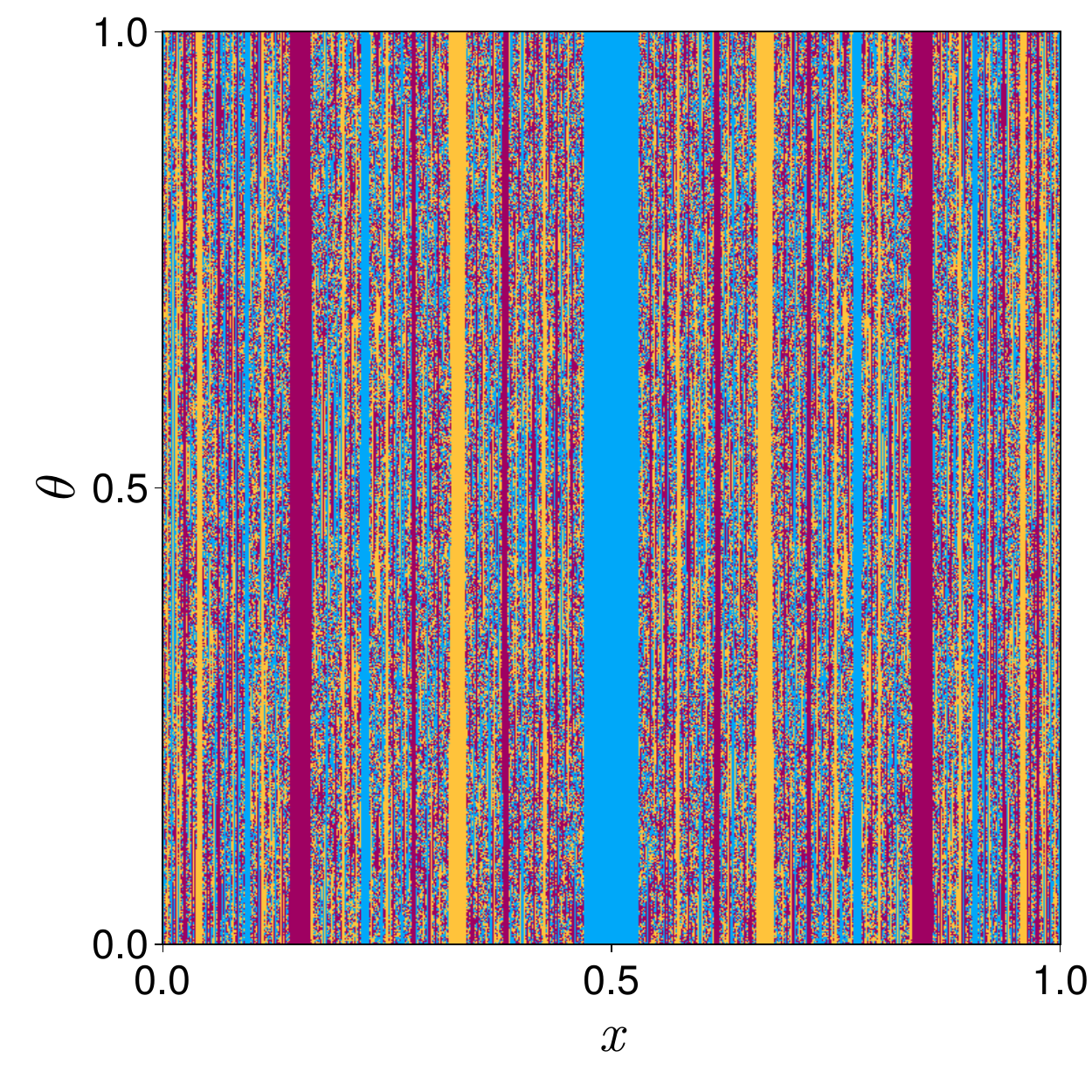}
\end{center}
\caption{\label{fig:feudel_wada}Basins of attraction of the map \ref{eq:feudel_wada} for the parameters $r = 3.833$ and $a = 0.0015$. This discrete map has been designed to produce boundaries with the Wada property.} 
\end{figure}

The following simple dynamical system~\cite{Feudel_1998_bif} has been designed such that the basins have the Wada property: 
\begin{align}\label{eq:feudel_wada}
\begin{split}
    x_{n+1} &= M(x_n) + a\cos(2\pi\theta_n),\\
\theta_{n+1} &= \theta_n + \omega \text{ mod}(1), 
\end{split}
\end{align}
The function $M$ represents the logistic application $f(x) = r x (1-x)$ iterated three times; that is, $M(x) = f(f(f(x)))$. Notice the weak periodic driving on the variable $x$; this is a key ingredient in the dynamics. For the parameters $\omega = (\sqrt{5} - 1)/2$, $r=3.833$, and $a = 0.0015$, the boundary separating the basins of attraction of the three attractors is Wada, as shown in Fig.~\ref{fig:feudel_wada}. The creation of the Wada boundary is carefully demonstrated in the article~\cite{Feudel_1998_bif}, and it is the result of the shape of the function $M$ in combination with the periodic forcing. To verify the Wada property, one of the methods consists of following the unstable manifold of a fixed point and checking that this manifold crosses every basin~\cite{wagemakers2020detect}. The authors use a similar technique by checking the preimages of special intervals containing parts of the boundary. If the preimages of these intervals contain all three basins, then inductive reasoning suffices to assert the Wada property of the boundary. 

This map can also be used to study basin bifurcations, which are dramatic events occurring in the basins when a parameter changes. In this example, the change is caused by the intersection of critical curves with the boundary of a basin, creating islands of one basin within another. See also Sec.~\ref{basin_mira} and reference \cite{mira1994basin} for further explanations of the phenomenon.

\subsection{The Kicked Double Rotor}
\begin{keywrds}FB, IWB, RB, MAP\end{keywrds}

\begin{figure*}
\begin{center}
    \subfloat[]{\includegraphics[width=\columnwidth]{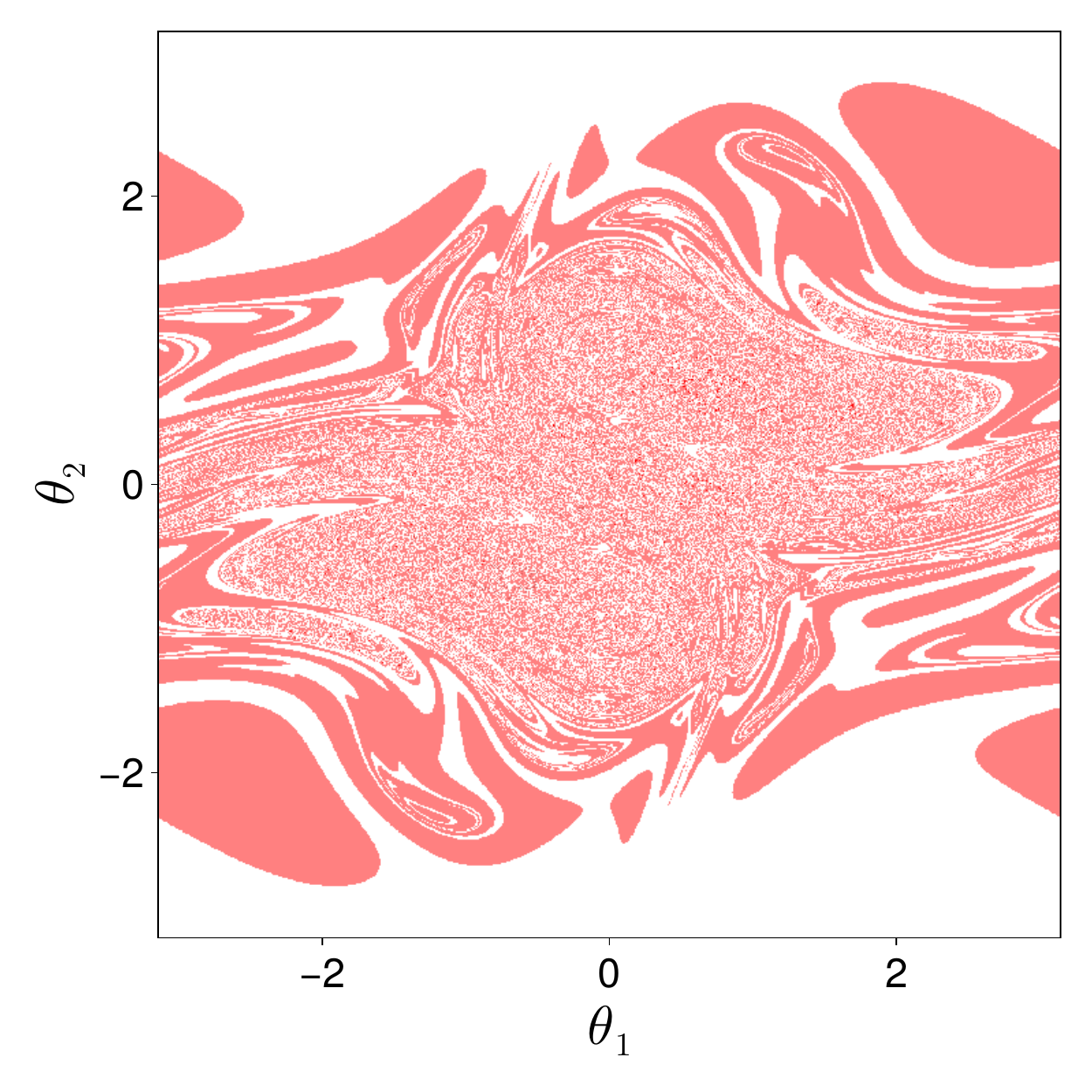}}
\subfloat[]{\includegraphics[width=\columnwidth]{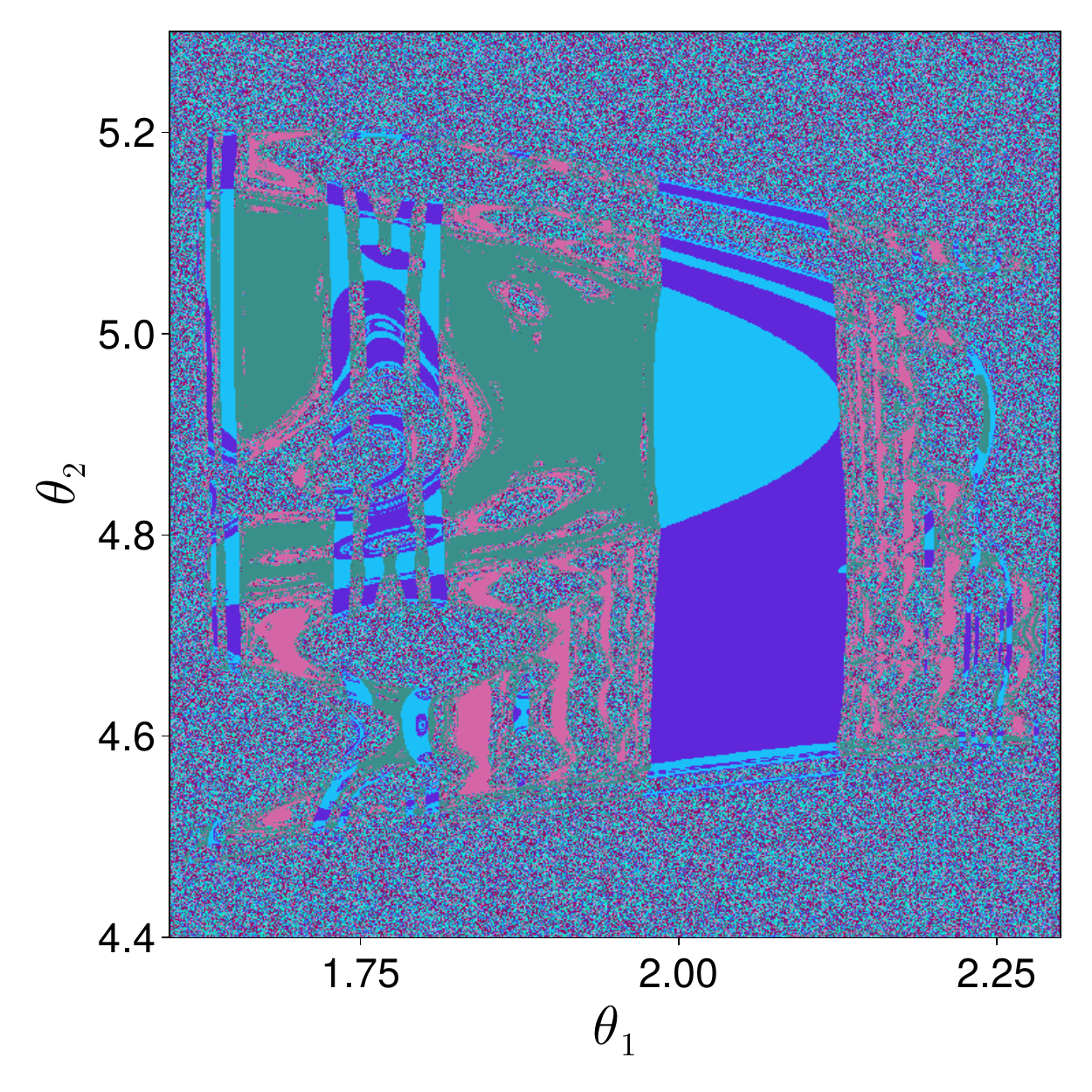}}
\end{center}
    \caption{\label{fig:kicked_rotor}Basins of attraction of the kicked double rotor for the model in Eqs.~\ref{eq:4d_kicked_rotor_1} in (a). The basins are represented for the projection on the $\theta_1$-$\theta_2$ plane with $\dot \theta_1 = 0$ and $\dot \theta_2 = 0$, $\nu = 0.1$, and $f_0 = 0.1$. In (b), a similar system is represented with a different set of parameters: $\nu = 0.2$ and $f_0 = 6.5$. The equations are described in \cite{Feudel_1998}. The algorithm has found 19 basins in this region of phase space. For this figure, the initial conditions are $\dot \theta_1 = 9$ and $\dot \theta_2 = 2.5$.}
\end{figure*}

The kicked double rotor is a model of a thin rod attached to a second mobile rod with two small masses at its ends. The second rod can spin freely at its center, and the first rod can rotate around its free end. When it is not spinning, it resembles a $\bot$ shape. Of course, the rods are massless, frictionless, and free from gravitational forces. In these conditions, the dynamics can be described by a four-dimensional equation representing the evolution of the angles $\theta_1$ and $\theta_2$ formed by the rods. One end of the spinning rod receives a periodic kick always in the same direction and with constant strength. The evolution of the rotor can be integrated from one kick to the next, resulting in a discrete dynamical system that provides the angles only at the time of the kicks. The expression of the discrete map is~\cite{Grebogi_1987}: 
\begin{align}\label{eq:4d_kicked_rotor_1}
\begin{split}
\left(\begin{matrix}\theta^1_{n+1}\\ \theta^2_{n+1} \end{matrix}\right) &= M(T) 
\left(\begin{array}{c}\dot{\theta}^1_{n}\\ \dot{\theta}^2_{n} \end{array}\right) + 
\left(\begin{array}{c}\theta^1_{n}\\ \theta^2_{n} \end{array}\right) \textrm{ mod}(2\pi) \\
\left(\begin{array}{c}\dot{\theta}^1_{n+1}\\ \dot{\theta}^2_{n+1} \end{array}\right) &= L(T)
\left(\begin{array}{c}\dot{\theta}^1_{n}\\ \dot{\theta}^2_{n} \end{array}\right) + 
\left(\begin{array}{c}\dfrac{l_1 f_0}{I_1} \sin\theta_{n+1}^1 \\ \dfrac{l_2 f_0}{I_2} \sin\theta_{n+1}^2 \end{array}\right) \\
\end{split}
\end{align} 
The default parameters are: $T=1$, $m_1 = m_2 = 1$, $\nu_1 = \nu_2 = \nu = 0.1$, $l_1 = 1/\sqrt{2}$, $l_2 = 1$, $I_1 = (m_1 + m_2)l_1^2$, $I_2 = m_2 l_2^2$, and $f_0 = 0.1$. The matrices $L(T)$ and $M(T)$ are expressed as: 
\begin{align*}
L(T) &= W_1 e^{-s_1} + W_2 e^{-s_2},\\
M(T) &= \dfrac{W_1}{s_1} e^{-s_1} + \dfrac{W_2}{s_2} e^{-s_2},\\
s_1 &= \nu (3+\sqrt{5})/2,\\ 
s_2 &= \nu (3-\sqrt{5})/2.\\ 
\end{align*} 
with $W_i = u_i^T u_i$, where the row vector $u_i$ is given by
\begin{equation*}
u_i = \frac{1}{\sqrt{\nu^2+ (\nu- s_i)^2}} 
\left(\begin{array}{cc}(\nu - s_i) &, \nu \end{array}\right)  
\end{equation*} 
This model has been studied in~\cite{Grebogi_1987} to show the existence of fractal basins of attraction, as shown in Fig.~\ref{fig:kicked_rotor} (a). These basins have different fractal dimensions depending on the studied region. The authors showed that every fractal region has subregions where the boundary is smooth. This property is known as intertwined basins. In~\cite{Feudel_1998}, the same model has been studied for a smaller parameter $\nu$, revealing the appearance of a much larger number of coexisting attractors. The parameter $\nu$ controls the friction of the rod. When this parameter tends to zero, the system returns to a conservative equation, and infinitely many orbits are present. As we approach the conservative case, more stable orbits appear. Additionally, the basins are clearly fractal, as shown in Fig.~\ref{fig:kicked_rotor} (b). The equations for this second model are slightly different and have not been included in the text, as the parameters do not correspond.

\subsection{Riddled Basins}
\begin{keywrds}RB, ODE\end{keywrds}

\begin{figure}
\begin{center}
\includegraphics[width=\columnwidth]{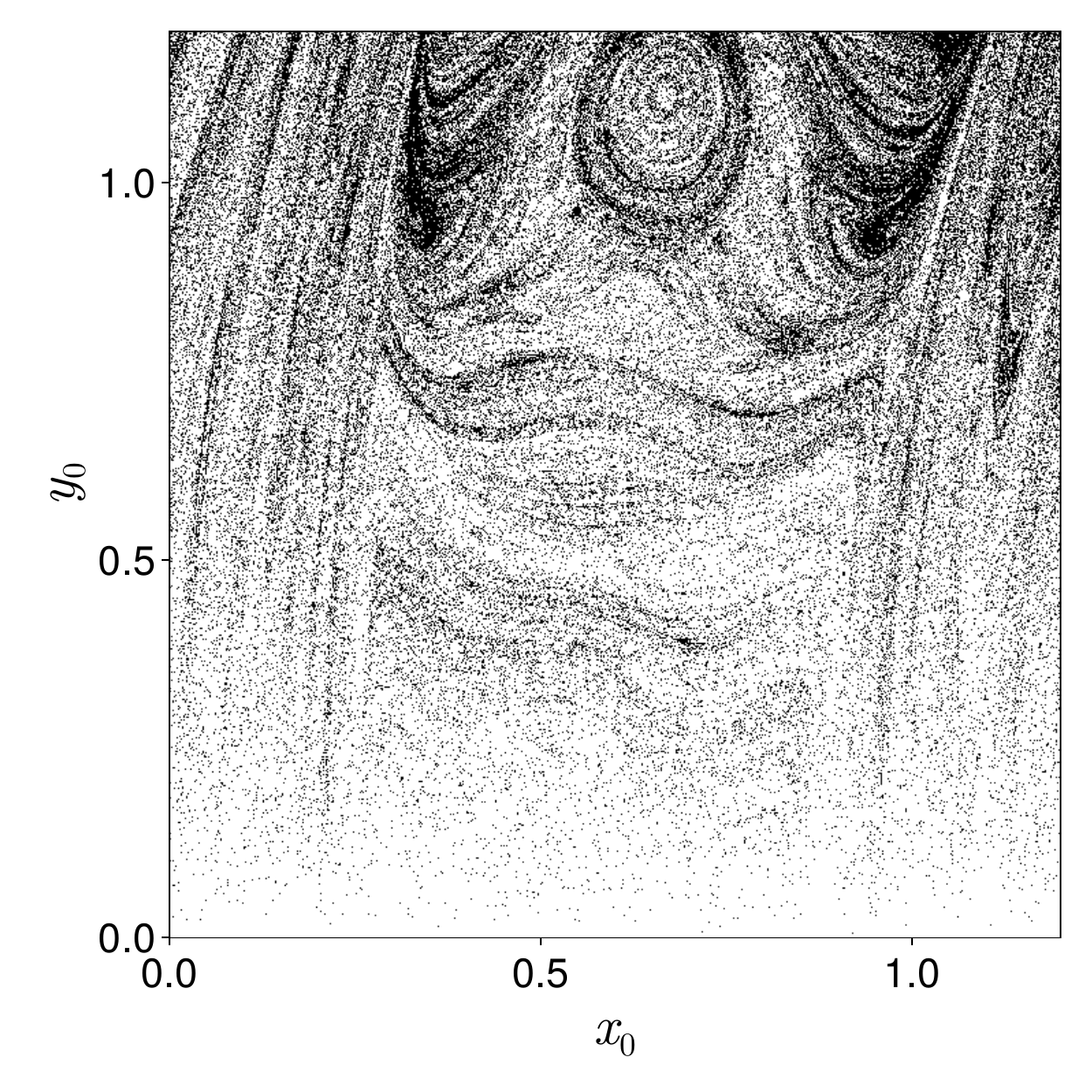}
\end{center}
\caption{\label{fig:riddled}Basins of attraction of a particle in a potential subject to forcing and friction as described in Eqs.~\ref{eq:riddled}. The black dots represent the initial conditions converging to a chaotic attractor, while the white points represent those diverging. The parameters are $\gamma = 0.05$, $\bar x = 1.9$, $f_0 = 2.3$, and $\omega = 3.5$.} 
\end{figure}

Basins of attraction can display a fractal boundary between different basins. But what happens when the boundary fills the phase space? In riddled basins, for each initial condition of the basins, we can always find a neighborhood with an initial condition belonging to another basin. In a celebrated paper, Ott et al.~\cite{Ott_1994} study the motion of a point particle subject to friction and sinusoidal forcing in a two-dimensional potential given by \( V(x,y) = (1-x^2)^2 + (x + \bar x) y^2 \). The equations of motion in the two-dimensional plane are:
\begin{align}\label{eq:riddled}
\begin{split}
\frac{d^2x}{dt^2} & + \gamma \frac{dx}{dt} - 4x(1-x^2) + y^2 = f_0 \sin(\omega t) \\
\frac{d^2y}{dt^2} & + \gamma \frac{dy}{dt} + 2y(x+\bar x) = 0
\end{split}
\end{align}
Due to the symmetry of the potential, a chaotic attractor exists in the invariant hyperplane \( \frac{dy}{dt}=0 \) for an appropriate choice of parameters \( \gamma \), \( \bar x \), \( f_0 \), and \( \omega \). The initial conditions on the plane \( x_0 \), \( y_0 \) for \( \dot x_0 = 0 \) and \( \dot y_0 = 0 \) either converge to the chaotic attractor on the invariant plane or diverge to \( \pm \infty \). There is a natural measure zero set of orbits on the attractor for which the perturbations from the plane grow exponentially. The resulting structure of this basin of attraction is riddled, as shown in Fig.~\ref{fig:riddled}, although there are visible clusters of points. The density of black points is not uniform across the basin.

The computation of this basin is demanding because of the long transients before the trajectory settles into one of the final states. The dynamics can be followed on the projected plane \( x,y \). Moreover, since we have periodic forcing, the stroboscopic map with a period of \( 2\pi/\omega \) can be defined to follow the orbit. Despite these simplifications, computing the basins in Fig.~\ref{fig:riddled} can take several hours.

A similar version of this system has been published in \cite{sommerer1995end, sommerer1996intermingled} in an attempt to demonstrate that deterministic systems and predictability can be orthogonal concepts, see Sec.~\ref{sec:Intermingled_sommerer}.

\subsection{Intermingled Basins in the Triangular Map}
\begin{keywrds}IB, MAP\end{keywrds}
\begin{figure}
\begin{center}
\includegraphics[width=\columnwidth]{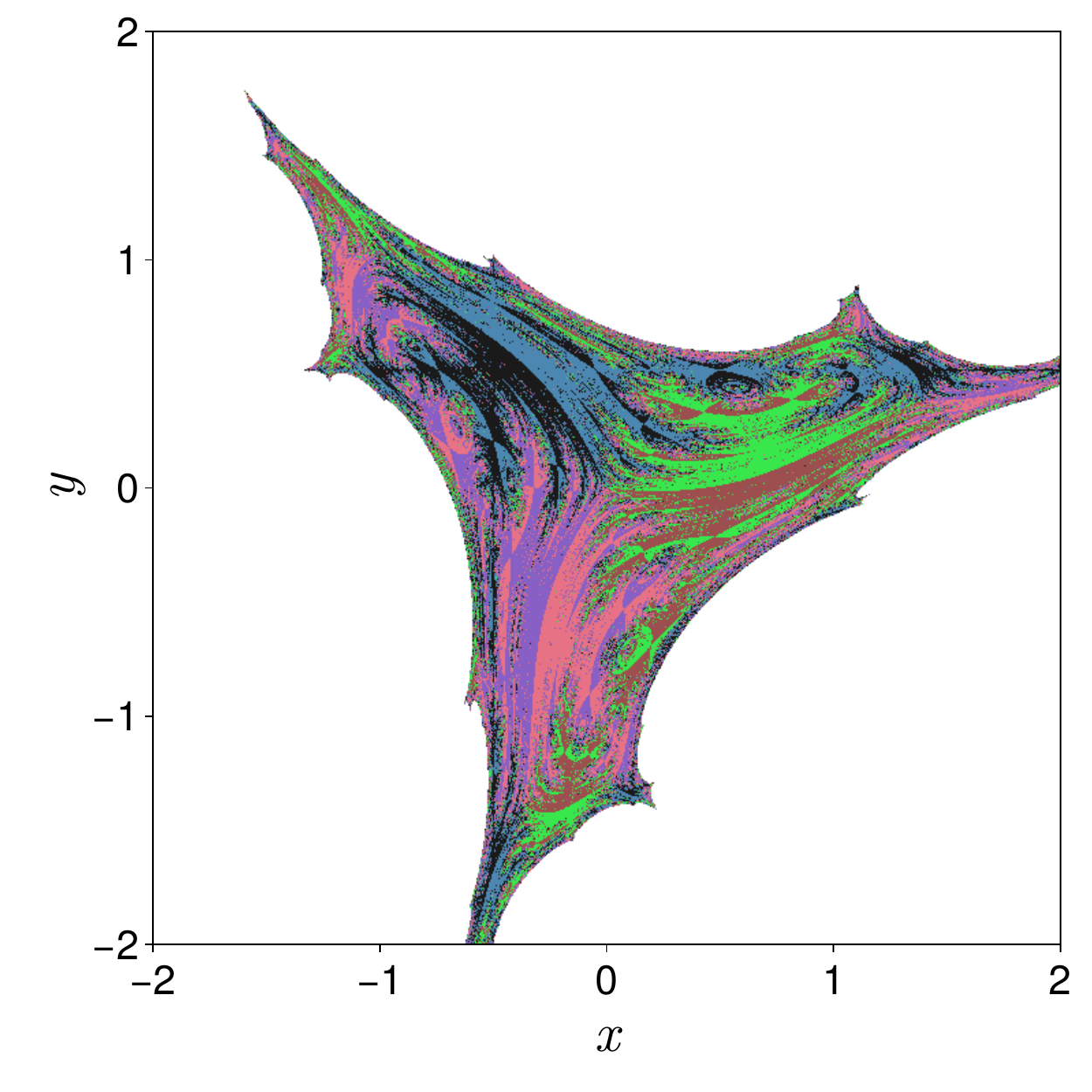}
\end{center}
    \caption{\label{fig:triang_img}Basins of attraction of the triangular map in Eq.~\ref{eq:triang_img}. The 6 basins have an intermingled boundary, where each basins of attraction is riddled with another basin.}
\end{figure}

The paper~\cite{alexander1996intermingled} provides a rigorous mathematical verification of the presence of intermingled basins for the triangle map, a family of quadratic maps of the plane. Using a combination of analytical proofs and numerical verification to show that for a specific parameter value, the map has three attractors with intermingled basins. 

Consider a riddled basin where every point within it contains segments of another basin positioned arbitrarily close. If the second basin is also riddled by the first, we refer to the basins as intermingled. This argument can be extended to any number of basins. In a more formal definition, for every open set S, if the basin of attraction of one of the attractors intersects S in a set of positive Lebesgue measure, then so do the other basins.

For the first versions of the map, the study identified three attractors with intermingled basins. Extending the analysis to the second iterate of the map revealed six attractors for which the same phenomenon occurs for a particular value of the parameter. The map in its complex form is: 
\begin{equation}\label{eq:triang_img}
    z_{n+1} = F(z) = z^2 - (1 + i \lambda)\bar z
\end{equation}
The authors proved the intermingling of the basins of $F$ and $F^2$, the composition of the function $F$ with itself. The basins showed in Fig.~\ref{fig:triang_img} have been computed with $F^2$ and have 6 attractors plus diverging trajectories outside the triangular shaped bass outside the triangular shaped basins.

\subsection{\label{sec:mag_pend}The Magnetic Pendulum: Double Transient System} 
\begin{keywrds}SLB, ODE\end{keywrds}
\begin{figure}
\begin{center}
\includegraphics[width=\columnwidth]{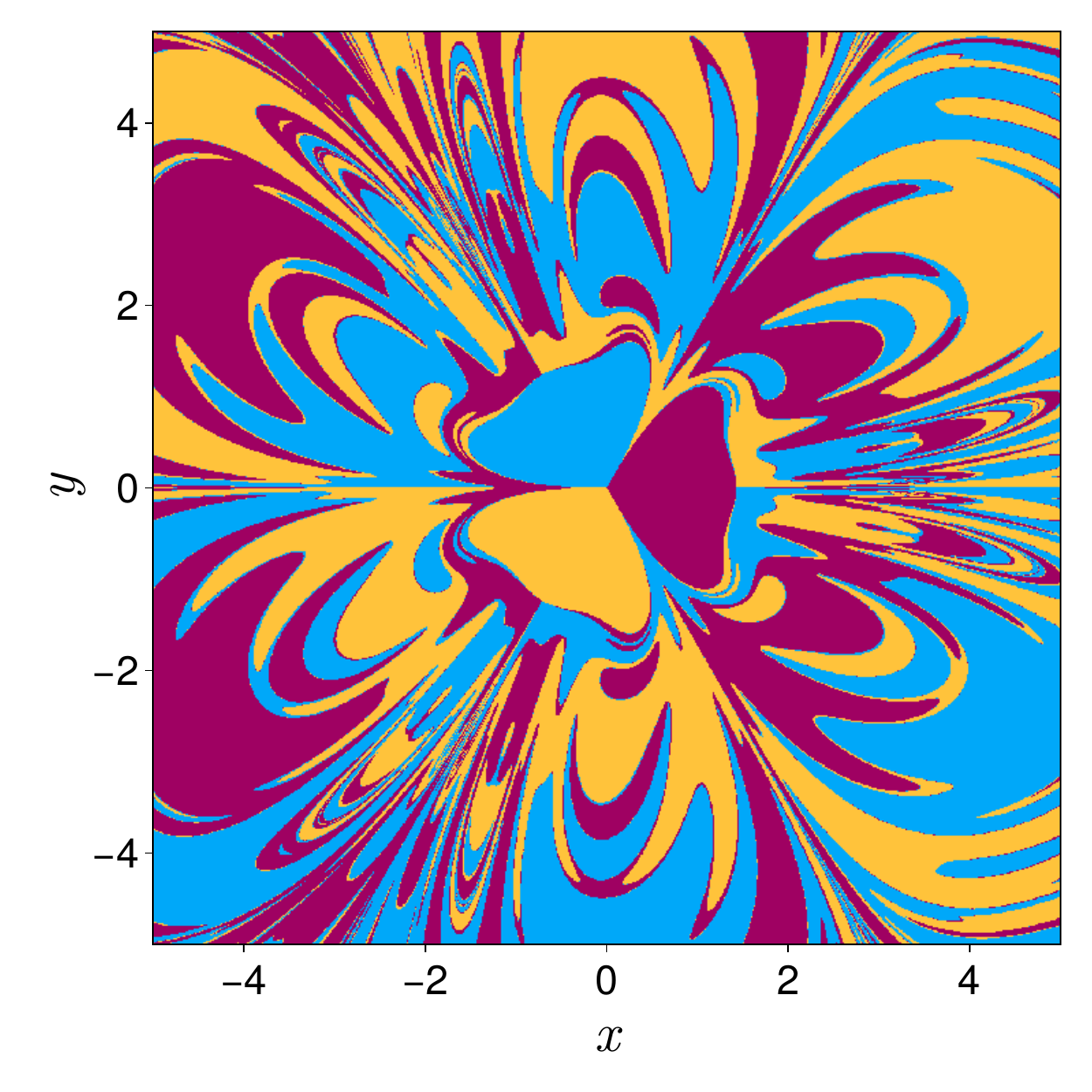}
\end{center}
\caption{\label{fig:mag_pend}Basins of attraction of the magnetic pendulum with three magnets. Although the pattern seems fractal, it is a deceiving impression. When the boundaries are studied more closely, the fractal pattern disappears. The boundary is not self-similar at all scales. Parameters for this plot are $\alpha = 0.2$, $\omega = 0.5$, $d = 0.3$, $N=3$.}
\end{figure}

This is a lovely example of how a toy made of a rod and three magnets can give rise to interesting behavior. The paper~\cite{Motter_2013} studies in detail the fine structure of the basins of this mechanical system exhibiting a double transient: one transient from the initial condition to the final state and another transient from fractal to non-fractal in the basin. If we look at the basin closely enough, the fractal pattern breaks. This is unexpected, as we tend to imagine the patterns continuing at all scales. The model for the magnetic pendulum is the following: 
\begin{align}\label{eq:pendulum}
\begin{split}
\ddot{\mathbf{v}} &= -\omega ^2\mathbf{v} - \alpha \dot{\mathbf{v}} - \sum_{i=1}^N \frac{ \mathbf{v} - \mathbf{v}_i}{D_i^3},\\
\mathbf{v} &= (x,y) \\
D_i &= \sqrt{(x-x_i)^2  + (y-y_i)^2 + d^2}
\end{split}
\end{align} 
The vectors $\mathbf{v}_i$ represent the positions of the magnets on the plane, and $\mathbf{v}$ is the position of the tip of the rod swinging above the plane. Figure~\ref{fig:mag_pend} represents the basins for three magnets spaced equidistantly on the unit circle with coordinates $(1,0), (-0.5,\sqrt{3}/2), (-0.5, -\sqrt{3}/2)$. The number of magnets and positions can be chosen arbitrarily. Despite the pattern of the basins looking fractal, the boundary is smooth at very small scales. The authors in~\cite{Motter_2013} named this behavior doubly transient chaos since the chaotic saddle is itself a transient phenomenon. An extension of this work can be found in Sec.~\ref{slim_fractal}.

\subsection{Sporadical Fractals} 
\begin{keywrds}SPF, MAP\end{keywrds}
\begin{figure}
\begin{center}
\includegraphics[width=\columnwidth]{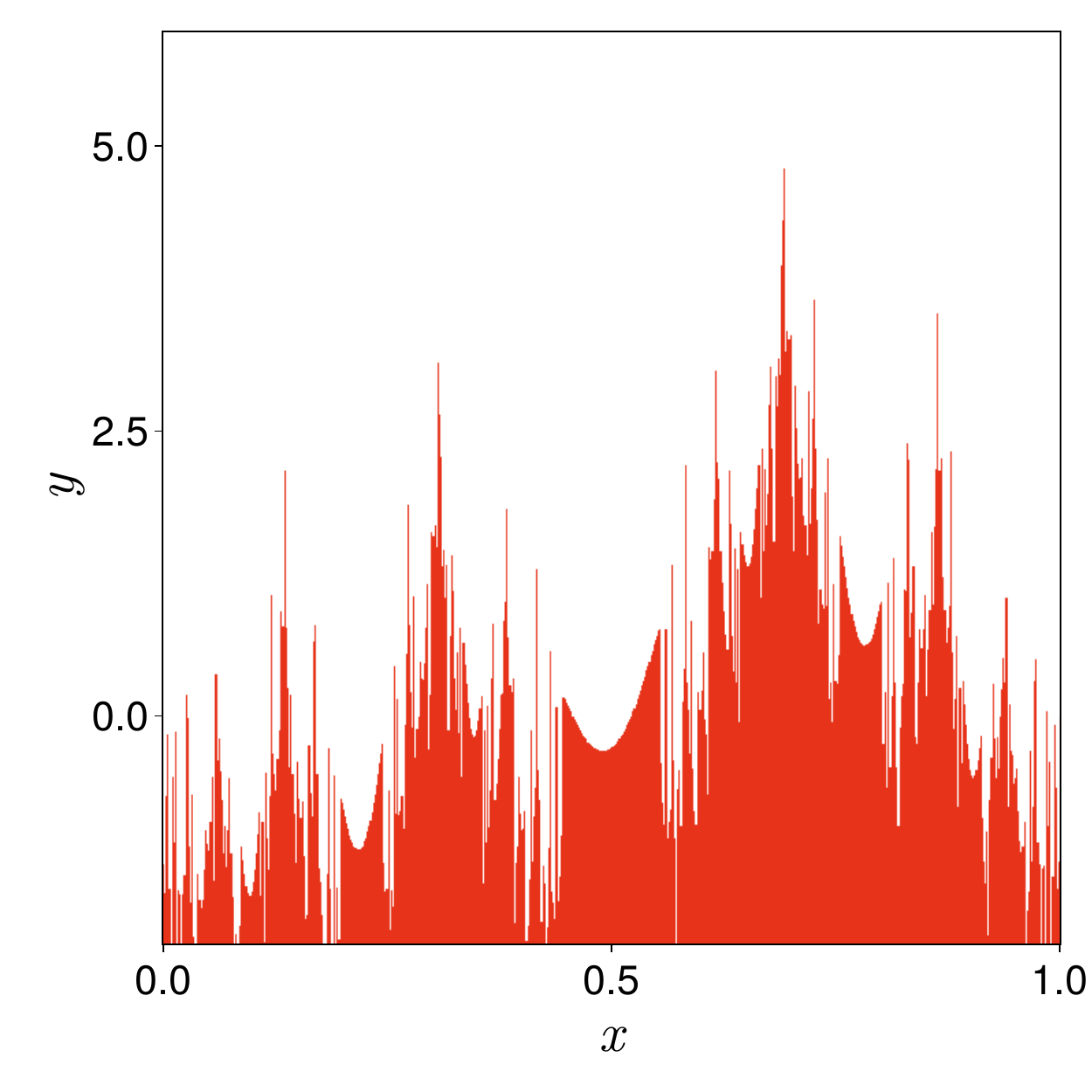}
\end{center}
\caption{\label{fig:sporadical}Basin of attraction of the system in Eqs.~\ref{eq:sporadical}. The boundary between basins displays a mix of fractal and smooth differentiable curves, a specific property of sporadical fractals. The figure has been computed with the parameter $\lambda = 1.1$.} 
\end{figure}

Usually, when we speak about fractal boundaries, we mentally picture complex structures everywhere. At the other end of the spectrum, smooth boundaries are differentiable everywhere. Sporadically fractal curves lie in between. These sporadically fractal boundaries are smooth and differentiable almost everywhere but have specific points where they exhibit fractal characteristics. In~\cite{Hunt_1999}, the authors demonstrate this concept using a two-dimensional map example, revealing that the boundaries can be characterized by a continuous function with regions of non-differentiability associated with a fractal dimension less than one. The result in a fractal dimension of the boundary itself between one and two.

The following model is a construction to illustrate this property: 
\begin{align}\label{eq:sporadical}
\begin{split}
x_{n+1} &= M(x_n)\\
y_{n+1} &= \lambda y_n + \sin(2\pi x_n) 
\end{split}
\end{align}
where $M$ is a continuous piecewise function: 
\begin{equation}
M(x) = \left\{\begin{array}{ll} 
9x/(4-5x) & \textrm{if } x \leq 0 \\
9/4x & \textrm{if } 0 < x \leq 4/9 \\
81/4(x-x^2)-4 & \textrm{if } 4/9 < x \leq 5/9 \\
9/4(1-x) & \textrm{if } x > 5/9 
\end{array}\right.
\end{equation}
This two-dimensional map does not have attractors; the trajectories diverge to $\pm \infty$. In this case, we assign one basin to $+\infty$ and the other to $-\infty$. For the computation of these basins, the algorithm checks at each step if the orbit passes a certain threshold and then returns the basin associated with the initial condition.

The result shown in Fig.~\ref{fig:sporadical} displays a mix of fractal and smooth boundaries specific to sporadical fractals.

\subsection{Hidden Attractor in the Chua Circuit}
\begin{keywrds}HA, ODE\end{keywrds}

\begin{figure}
\begin{center}
\includegraphics[width=\columnwidth]{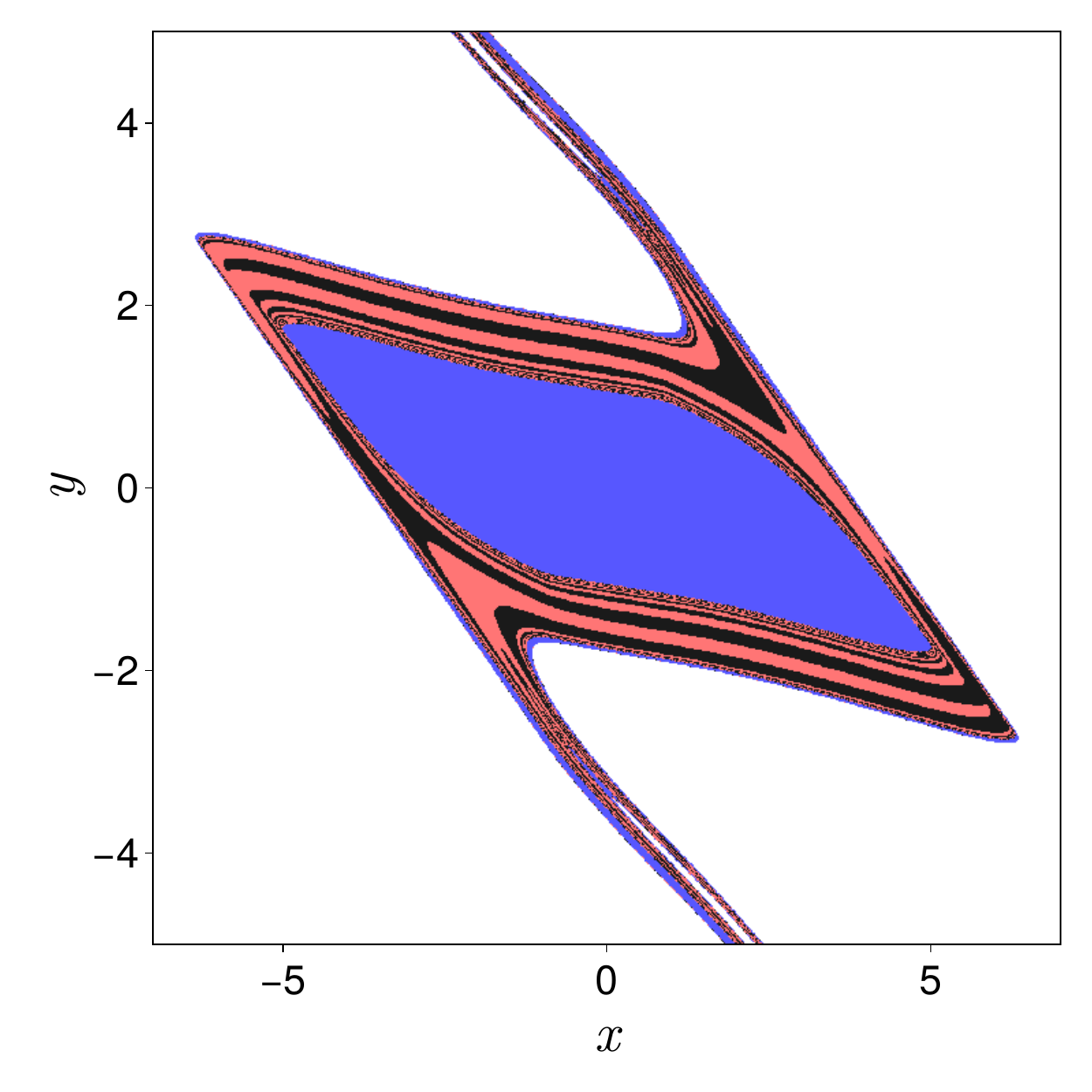}
\end{center}
    \caption{\label{fig:hidden_chua}Basins of attraction of Chua's oscillator with two hidden attractors and a stable fixed point. The parameters are $\alpha = 8.4562218418$, $\beta = 12.0732335925$, $\gamma = 0.0051631393$, $m_0 = -0.1767573476$, and $m_1 = -1.1467573476$, with the digits specified in~\cite{Leonov_2011}.}
\end{figure}

An attractor is called a self-excited attractor if its basin of attraction intersects with any open neighborhood of an unstable fixed point; otherwise, it is called a hidden attractor.

This definition of a hidden attractor tells us that its basins do not intersect with any unstable fixed points. Fixed points, both stable and unstable, are easy to locate in dynamical systems and may lead us to find the self-excited attractors. In other situations, brute-force basin exploration is needed to find the hidden attractors, as is the case for the Chua circuit model in~\cite{Leonov_2011}. The authors propose a custom analytical-numerical algorithm to localize hidden attractors, which involves a sequence of transformations applied to Chua's system and the use of harmonic linearization techniques to find periodic solutions. Here, we reproduce the basins using our numerical method with a purely numerical exploration of the phase space.

The equations of the Chua oscillator are:
\begin{align}\label{eq:hidden_chua}
\begin{split}
\dot{x} & = \alpha(y - x - f(x)), \\
\dot{y} & = x - y + z, \\
\dot{z} & = -(\beta y + \gamma z),
\end{split}
\end{align}
where $x, y, z$ are the state variables representing voltage and current in the circuit, and the nonlinear function is $f(x) = m_1 x + (m_0 - m_1)\frac{1}{2}( |x + 1| - |x - 1| )$.

The basins of attraction for this system are shown in Fig.~\ref{fig:hidden_chua} for initial conditions on the $x-y$ plane, with $z= 0$. The values of the parameters are listed in the figure caption. The inner island represents the basin of a locally stable fixed point at the origin. The whirling basins around the central island are the basins of the two hidden attractors. For initial conditions outside this region, the trajectory diverges.

\subsection{Basins with Tentacles}
\begin{keywrds} ODE \end{keywrds}

\begin{figure}
\begin{center}
\includegraphics[width=\columnwidth]{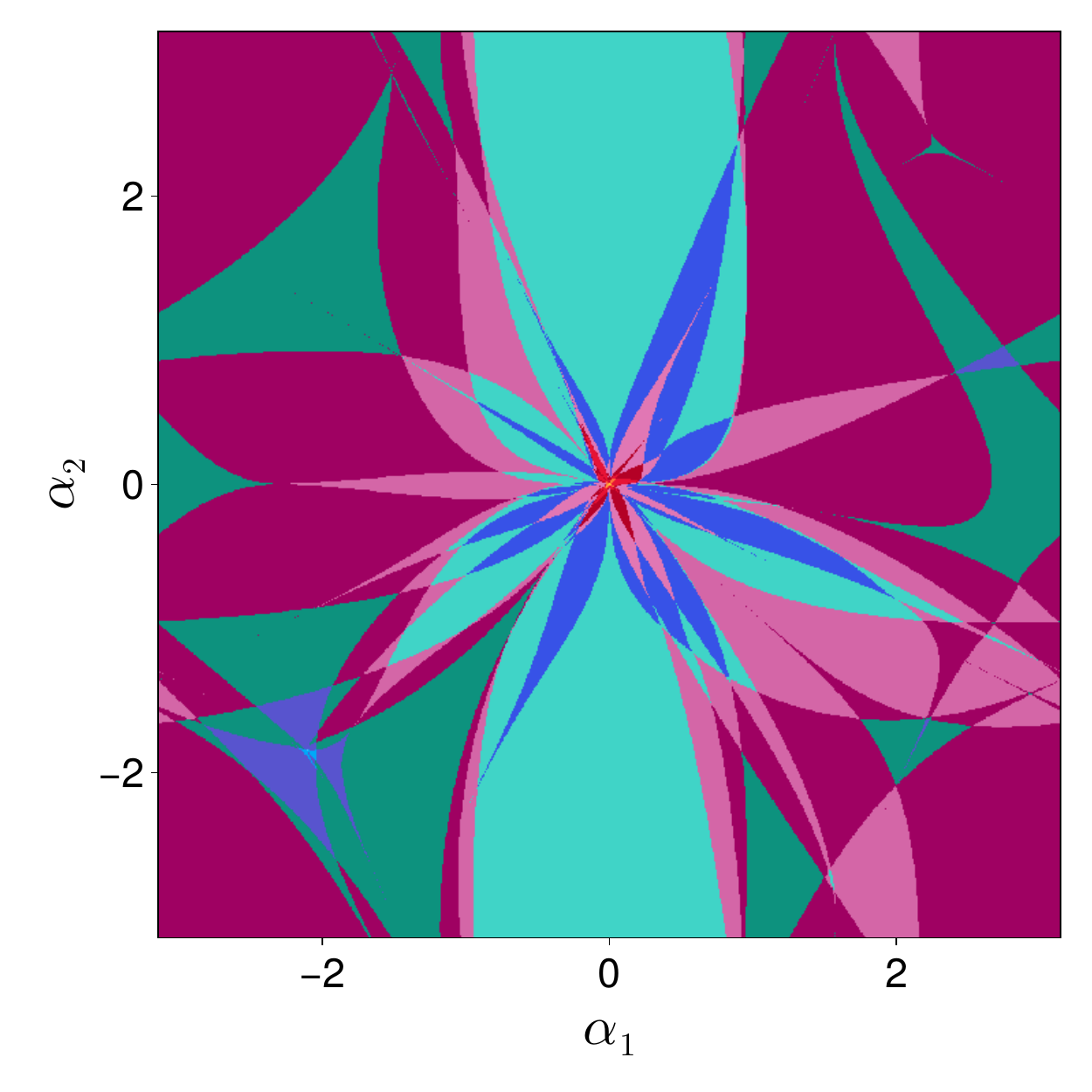}
\end{center}
    \caption{\label{fig:kur_tent}Basins of attraction of Eqs.~\ref{eq:kur_tent}. These basins are a slice of a plane spanned by two n-dimensional vectors $P_1$ and $P_2$ with random components. The axes $\alpha_1$ and $\alpha_2$ are scalars multiplying the vectors, and a constant vector $\theta_{10}$ is added. This represents the 10-twisted state at the origin of the plot.}
\end{figure}

Basins of high-dimensional systems are technically difficult to investigate. The curse of high dimension strikes when we want to learn more about the structures of basins in nonlinear dynamical systems with many variables.

The study~\cite{Zhang_2021} explores the geometry of basins of attraction in high dimensions using a model of identical Kuramoto oscillators arranged in a ring. The attractors for this configuration, known as $q$-twisted states, are easy to identify since they have a formal definition. With a sampling of the phase space, the authors investigate the distribution and relative sizes of the $q$-twisted basins. The findings reveal that their sizes are proportional to $ e^{-kq^2} $, contradicting previous results suggesting an exponential decay with $|q|$. The basins exhibit an octopus-like structure, where the majority of their volume is found in the tentacles rather than the central head. High-dimensional basins often have complex geometries that are not well characterized by simpler shapes, such as hypercubes. However, even basins in the form of hypercubes in high dimensions can be difficult to characterize~\cite{martiniani2023you}.

The dynamics of each Kuramoto oscillator in this work are described by the equation:  
\begin{equation}\label{eq:kur_tent}
   \dot{\theta}_i = \sin(\theta_{i+1} - \theta_i) + \sin(\theta_{i-1} - \theta_i)
\end{equation}
for $ i = 1, \ldots, n $, where $ \theta_i$ represents the phase of oscillator $ i $ at time $ t $. The system assumes periodic boundary conditions with $ \theta_{n+1}(t) = \theta_1(t) \mod 2\pi $.

The $q$-twisted states are characterized by the oscillator phases $\theta_i = 2\pi i q/n + C$. With a simple linear fit of $\theta_i$ as a function of $i$, we can detect the winding number $q$ of the system in this state. Stability analysis shows that $q$-states occur only for $|q| < n/4$. The basins in Fig.~\ref{fig:kur_tent} represent a slice of the subspace spanned by two n-dimensional vectors $P_1$ and $P_2$ with random components containing an equal number of zeros and ones in each vector. The initial conditions in the slice are given by: $\vec v_i = \vec \theta_0 + \alpha_1 P_1 + \alpha_2 P_2$ with $\vec \theta_q = \{\theta_i = i 2\pi q/n \textrm{ for } i \in [1,n]\}$. $\alpha_i \in (-\pi,\pi]$ are two scalars spanning the axes of Fig.~\ref{fig:kur_tent}.

The choice of $\theta_q$ is important since the basins will show the structure around the chosen $q$-twisted state. In our simulations, we have chosen $q=10$.

There are several open questions regarding the structure of these basins. In the classification of the basins I left only {\bf ODE} since there is no systematic study of these boundaries. The picture show a smooth boundary, but the nature of high-dimensional systems is often deceiving. Intuitions about the topology of space break down beyond 4 dimensions, and the available tools are not practical. Specific sampling techniques are necessary to obtain a clearer picture of phase space~\cite{martiniani2023you}.

\subsection{Megastability in Basins}
\begin{keywrds}HA, MG, ODE\end{keywrds}
\begin{figure}
\centering
\includegraphics[width=\columnwidth]{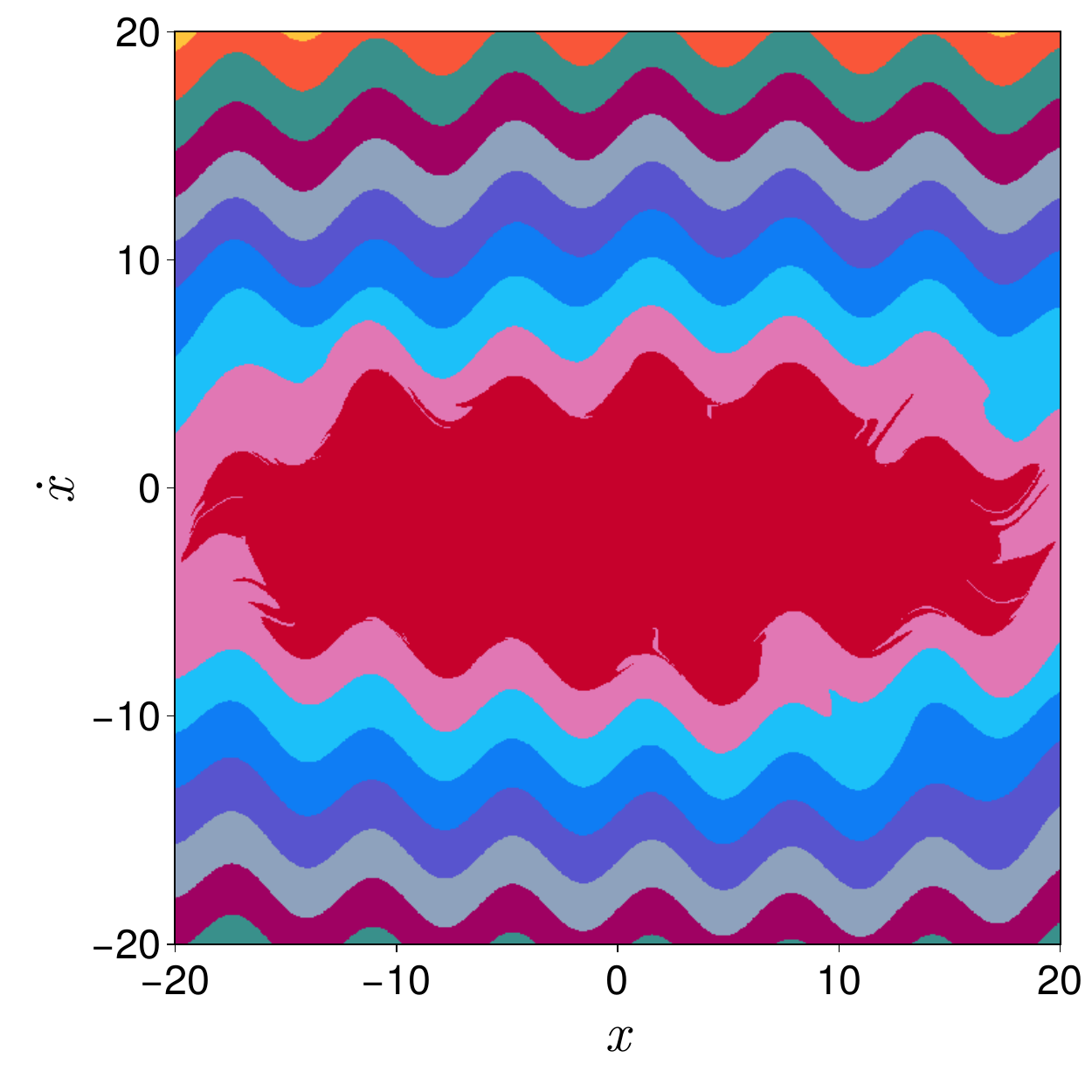}
\caption{\label{fig:megastability}Basins of attraction of~\ref{eq:megastability}. The characteristics of these basins include the onion-like structure of the basins. The attractors are nested and can be of different natures: strange attractors, tori, etc.}
\end{figure}

It is possible to have an infinite collection of nested attractors whose basins form an onion shape. This phenomenon has been termed megastability. 

A clear example of megastability appears in Ref.~\cite{sprott2017megastability}, where a periodically forced oscillator with spatially periodic damping exhibits this property. The construction of the model starts with a second-order differential equation including a term depending on a periodic function. The authors introduced the term $\dot{x} \cos(x)$, creating an infinite number of nested stable limit cycles. Periodic forcing is also present to maintain the trajectories, as the damping tends to drive each trajectory to the origin. 

The model can be stated as:
\begin{align}\label{eq:megastability}
\begin{split}
\ddot{x} + (0.33)^2 x - \dot{x} \cos x - \sin(0.73t)  = 0
\end{split}
\end{align}
The plot in Fig.~\ref{fig:megastability} shows this nested structure of basins. Most of the attractors are hidden and include limit cycles, attracting tori, and strange attractors. Regarding the computational aspect of these basins, the detection grid for tracking the attractors should be especially large. This is the only precaution to take. 

The structure of the differential equation generating megastability has been elucidated in~\cite{lopez2025megastable}, where a general framework for analyzing and generating megastable oscillators has been proposed.

\subsection{Matryoshka Multistability}
\begin{keywrds}MG, FB, ODE\end{keywrds}

\begin{figure}
\begin{center}
\includegraphics[width=\columnwidth]{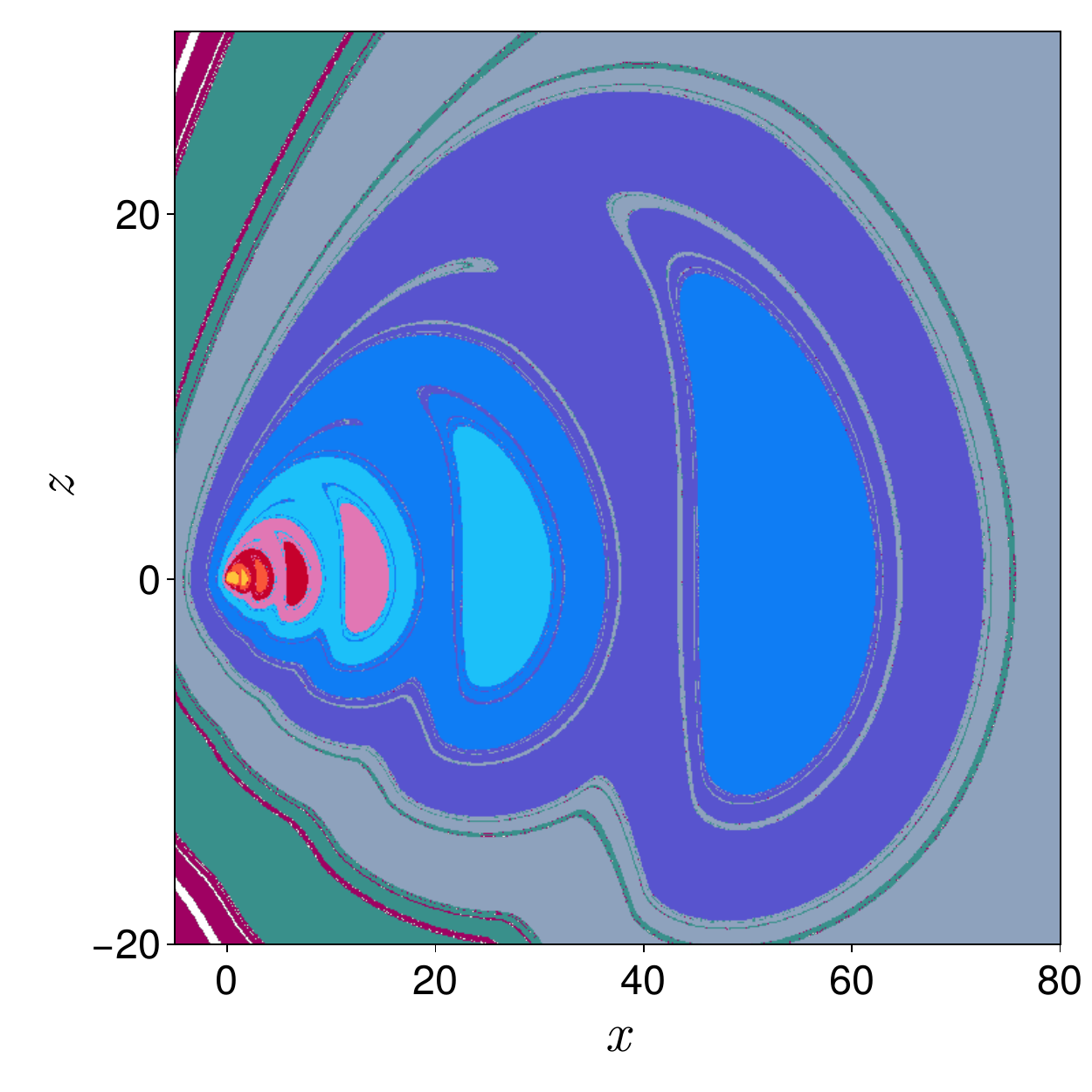}
\end{center}
\caption{\label{fig:matryoshka}Basins of attraction for Eqs.~\ref{eq:matryoshka}. The basins of attraction are self-similar at every scale, displaying a special kind of fractality called Matryoshka stability. Parameters for this simulation are $a = 1.9$, $b = -1.8$, $c = 3.9$, and $\varepsilon = 0.01$.}
\end{figure}

Megastability involves an infinite number of nested basins of attraction, but there exists a basin with a finite volume around the origin. This concept has been extended to attractors nested at every scale, extending toward infinity. The article~\cite{karimov2024matryoshka} introduces the concept of ``matryoshka multistability,'' where an infinite number of self-similar nested attractors coexist in phase space. The authors use two well-known chaotic systems (Chua's circuit and Sprott Case J) as a testbed by modifying their equations to include a self-similar nonlinear function. The algorithmic construction of this function is at the core of the work. The simulations demonstrate the existence of attractors precisely self-similar and nested within each other, akin to matryoshka dolls. The Lyapunov spectra for the attractors are found to be similar across different scales. We must note, however, that there is a lower arbitrary bound limited by machine precision for these nested attractors. This critique also holds for the megastable attractors, as the portion of the explored phase space is also finite. 

Here we reproduce the result for the Sprott Case J system:
\begin{align}\label{eq:matryoshka}
\begin{split}
\dot{x} &= a z  \\
\dot{y} &= b y + z \\
\dot{z} &= -x + y + c~ \mathcal{F}_R(y)
\end{split}
\end{align}
The function $\mathcal{F}_R$ is a Lipschitz continuous function showing a self-similar pattern. The function is evaluated using an algorithmic function:

\begin{lstlisting}
function Fr(y) 
    d = 1.0; m = 1.0; 
    P = 1.23; R = 2.0
    ay = abs(y)
    while true
        if ay < d 
            d = d/R; m = m/R
        elseif ay > 2*d
            d = d*R; m = m*R
        else 
            break
        end
    end
    epsilon = 0.01
    if d > epsilon
        if ay < P*d 
            b = -m*(R-P*R+1)/(R*(P-1))
            k = -m/(R*d*(1-P))
        else
            b = -m*(R-P*R+1)/(P-R)
            k = m*(-R^2+R+1)/(R*d*(P-R))
        end
        return k*ay + b
    else
        return 0 
    end
end
\end{lstlisting}

This function is responsible for the nested basins of attraction visible in Fig.~\ref{fig:matryoshka}. The parameters for this plot are $a = 1.9$, $b = -1.8$, and $c = 3.9$. The parameters of the function are specified in the code listing. Notice the parameter \texttt{epsilon} bounding the lower scale of the attractor, defining the smallest attractor possible. The accumulation of trajectories of nested attractors near the origin complicates detection with the recurrence algorithm. The solution is to use a logarithmically spaced detection grid, where the resolution of the cells is much higher near the origin.

\section{Theoretical Models}

Some models in this section are inspired by physical considerations, such as the Duffing oscillator or the Lorenz discrete reduced model. However, the relation to the physical object is somewhat secondary, and the systems have been studied for their dynamical properties. Some of the models presented are motivated solely by mathematical arguments to explore a particular dynamical phenomenon. The heterogeneity of the models grouped in this section reflects the richness of the dynamical systems found in the literature.

\subsection{The Duffing Oscillator}
\begin{keywrds}FB, WD, ODE\end{keywrds}
\begin{figure}
\begin{center}
\includegraphics[width=\columnwidth]{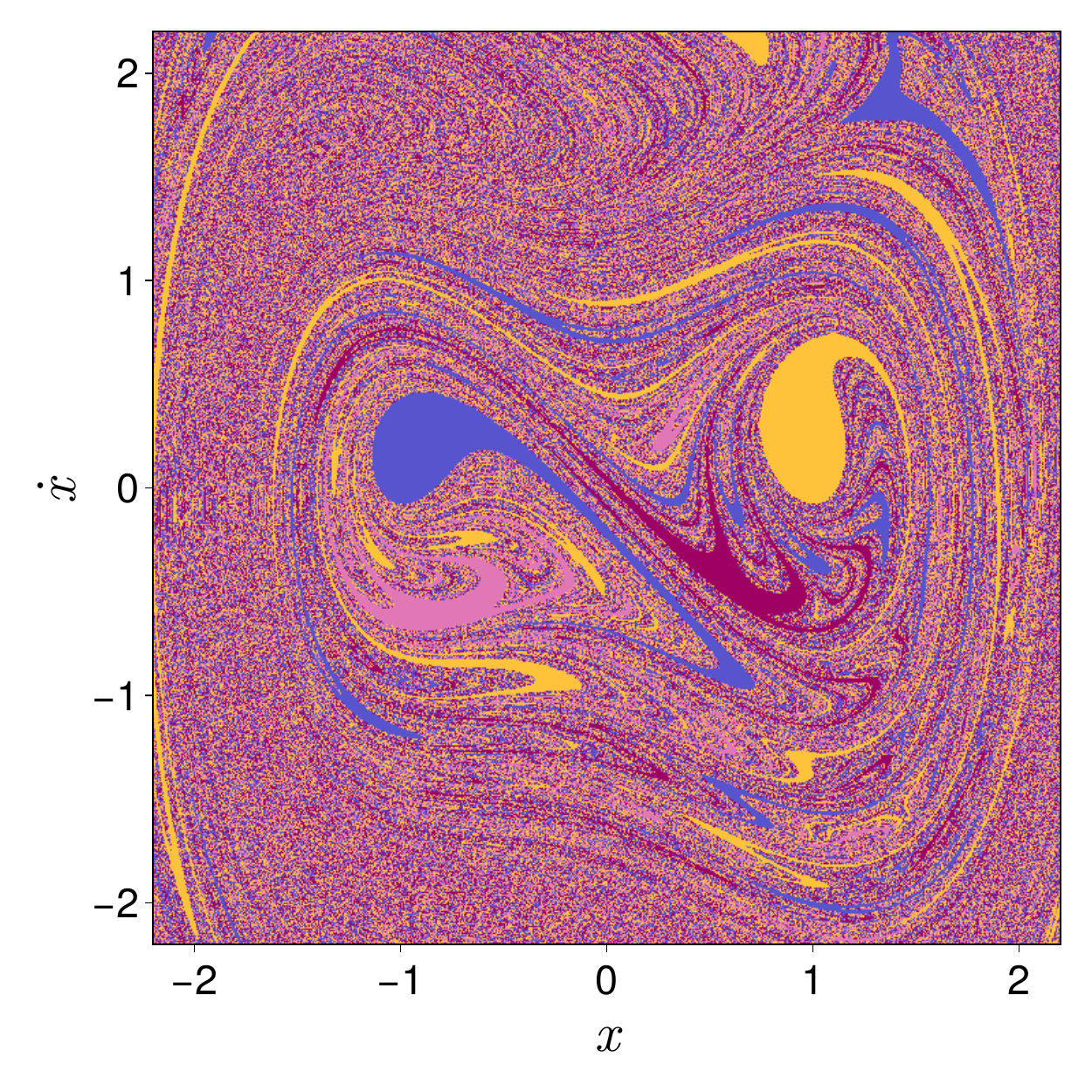}
\end{center}
\caption{\label{fig:duffing}Basins of attraction of the Duffing oscillator described in Eq.~\ref{eq:duffing} for the parameters $d = 0.05$, $F = 0.098$, and $\omega = 1.15$. This figure depicts four basins represented with fractal boundaries exhibiting the Wada property.}
\end{figure}

The Duffing oscillator is a fundamental model for studying nonlinear dynamical systems. It can be conceptualized as a model of the motion of a periodically forced unit mass particle along an axis within a double-well potential. This oscillator has been studied extensively in various contexts and is a paradigm of dissipative nonlinear oscillators\cite{kovacic_duffing_2011}. One of the formulations of the model is: 
\begin{equation}\label{eq:duffing}
\ddot{x} + d\dot{x} + x - x^3 = F\cos(\omega t)
\end{equation}
We consider three relevant parameters: the dissipation $d$, the forcing amplitude $F$, and the angular frequency of the forcing $\omega$. Notice the nonlinear potential $V(x) = -\frac{1}{2}(x^2 - \frac{1}{2}x^4)$ in the shape of a double well, which differentiates this oscillator from the forced damped pendulum described in Eq.~\ref{eq:forced_pend}. The amount of literature on the Duffing oscillator is vast and cannot be reviewed here. We will only present an interesting example of four coexisting attractors in the phase space, as shown in Fig.~\ref{fig:duffing}. Numerically, this oscillator does not require much care. A useful simplification is to transform the continuous system into a discrete application using a stroboscopic map. The numerical solver integrates the trajectory automatically over one period $2\pi/\omega$ and returns only the coordinates at this precise time. The mapper detects the attractors more efficiently. The basins in Fig.~\ref{fig:duffing} also exhibit the Wada properties.

\subsection{The Hénon Map}
\begin{keywrds}FB, MAP\end{keywrds}
\begin{figure}
\begin{center}
\includegraphics[width=\columnwidth]{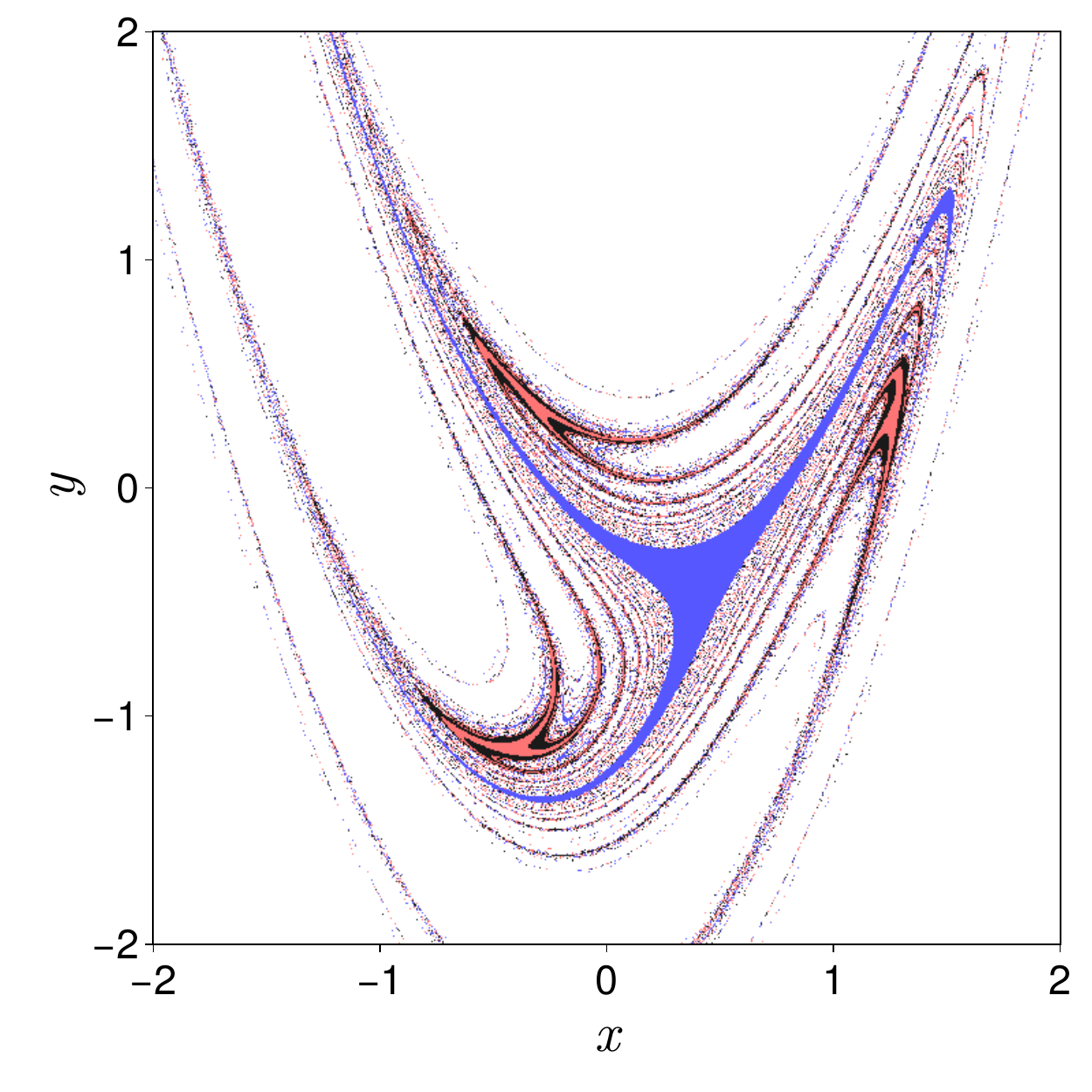}
\end{center}
\caption{\label{fig:henon_map}Basins of attraction of the Hénon map $\mu = 1.08$, $J = 0.9$ for the Eqs.~\ref{eq:henon_map}. For the chosen parameters, several attractors are present: a period 1, period 3, and period 9 attractor.}
\end{figure}

Michel Hénon introduced what is now known as the Hénon map as a simplified version of the Poincaré section of the Lorenz model. It is now one of the landmarks in the study of dynamical systems and chaos due to its simplicity and yet remarkable variety of behaviors. Bifurcations, fractal boundaries, strange attractors, chaos: this system has it all. Countless works in dynamical systems use this system as a guinea pig to perform numerical experiments. Its equations are \cite{henon1976two}:
\begin{align}\label{eq:henon_map} 
\begin{split}
x_{n+1} &= 1 - \mu x_n^2 + y_n\\
y_{n+1} &= -J x_n
\end{split}
\end{align}
For a given choice of parameters, multistable states exist in the phase space with fractal boundaries. In this system, trajectories diverging to infinity are also considered a state. One of the basins represents these trajectories diverging to infinity. The basins presented in Fig.~\ref{fig:henon_map} have orbits of period 1, 3, and 9. In another setting, the map is expressed in a slightly different form: $x_{n+1} = A - x_n^2 - J y_n$, $y_{n+1} = x_n$. For the parameters $A = 2.12467$, $J = 0.3$, we can find a chaotic attractor separated from the attractor to infinity by a fractal boundary~\cite{grebogi1987basin}. This system has been studied in all sorts of possible scenarios and plays a central role in the theory of dynamical systems.

Numerically, this system is very easy to compute for all ranges of parameters, which is one of the main reasons for its popularity in the nonlinear dynamics community.

\subsection{The Newton Algorithm for Root Finding}
\begin{keywrds}FB, WD, MAP\end{keywrds}

\begin{figure}
\begin{center}
\includegraphics[width=\columnwidth]{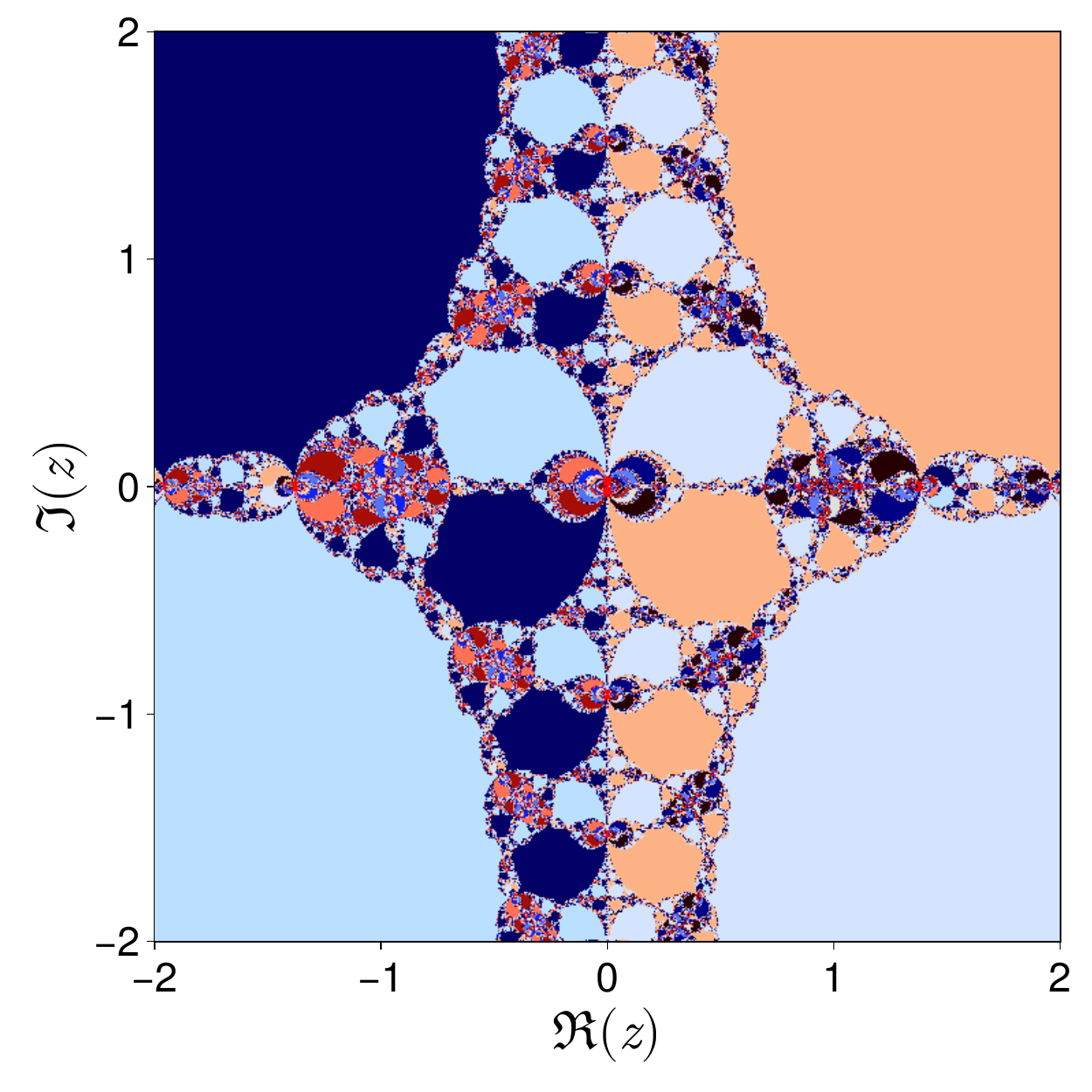}
\end{center}
\caption{\label{fig:newton}Basins of the roots of the nonlinear function $f(x) = x + x^2\sin(2/x)$ when the Newton algorithm is iterated. Ninety-six basins are present in this region of the complex plane. This explosion in the number of possible roots is due to the $1/x$ term in the function $f$. There are however large domains leading to a single root.} 
\end{figure}

Finding the root of nonlinear functions is fundamental in many fields of engineering and theoretical science. Iterative methods such as the Newton algorithm are easy to implement and very efficient for locating the roots accurately. The functions can have multiple zeros, and the final state of the orbit will depend on the initial condition. The Newton algorithm can be summarized by a simple dynamical process: 
\begin{equation}\label{eq:newton}
z_{n+1} = z_n - \frac{f(z_n)}{f'(z_n)}
\end{equation}
where $f$ is a nonlinear function. When $z_n$ is extended to the complex plane, we obtain a mapping from the complex plane to itself. The basins of the roots can be visualized in this two-dimensional space. The initial condition will lead to one of the roots unless it belongs to the invariant set defined by the boundary, the Julia set. The dynamical properties of this discrete application have been studied in detail both theoretically and numerically~\cite{curry1983iteration,varona2002graphic}. More recently, basin entropy has been used to characterize the basins of such a family of iterated methods~\cite{jo2024annealing}. Fractal boundaries are evident when we apply the algorithm to the nonlinear function. In Fig.~\ref{fig:newton}, we have represented the basins of the function $f(x) = x + \sin(2/x) x^2$. For this set of initial conditions, the algorithm has found 96 basins and displays a very intricate structure. This method is of interest if we want to set a particular number of basins. A simple polynomial with $n$ roots will suffice.

\subsection{\label{basin_mira}Basin Bifurcations}
\begin{keywrds} FB, MAP\end{keywrds} 
\begin{figure}
\begin{center}
\includegraphics[width=\columnwidth]{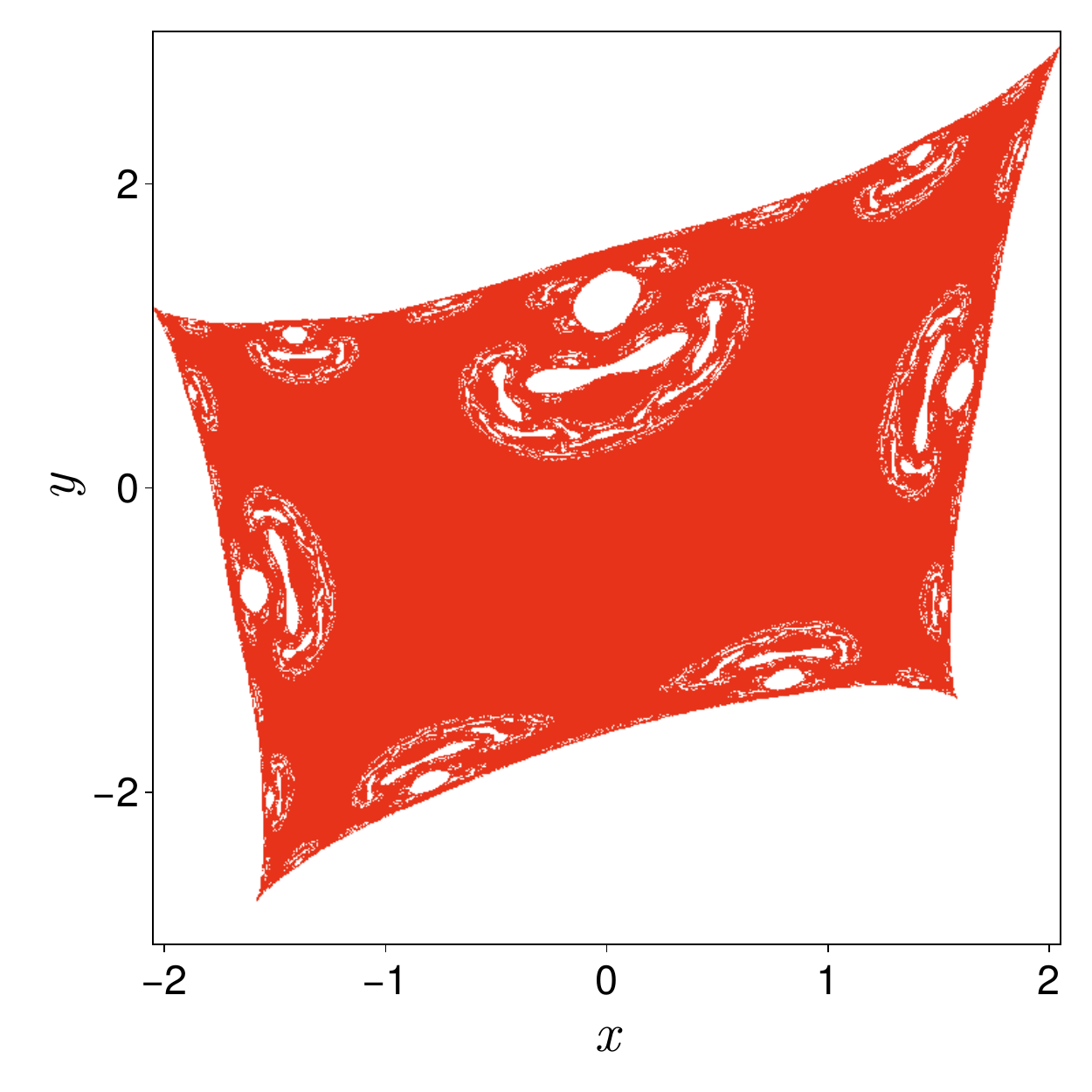}
\end{center}
\caption{\label{fig:map_mira}Basins of attraction of the quadratic map in Eq.~\ref{eq:map_mira}. The diamond-shaped basins contain inner islands of the other basin after the occurrence of a basin bifurcation.}
\end{figure}

A well-known method of analyzing dynamical systems consists of plotting the evolution of the attractors as a parameter changes, a technique known as a bifurcation diagram. Attractors are not the only structures in phase space affected by parameter evolution; basins can also evolve. In~\cite{mira1994basin}, the authors study the transformations of basins as a function of a parameter and detect the appearance of new structures when a threshold is crossed. Three transitions are detailed: connected to disconnected basins, simply connected to multiply connected, and smooth to fractal boundary.

The transformations, called basin bifurcations, happen when the critical curve of an unstable fixed point becomes tangent to the boundary of a basin. The critical curves are the images of the set of points where the Jacobian vanishes. Beyond this collision, inward islands of a basin appear within the connected component of the first basin, as shown in the basins of Fig.~\ref{fig:map_mira}. The authors establish the conditions for this bifurcation and apply the analysis to a two-dimensional map in Eqs.~\ref{eq:map_mira}.

Basins metamorphoses have been studied earlier in~\cite{grebogi1987basin} with the Hénon map, detailing dramatic changes in the basins as a function of a parameter change. In~\cite{wagemakers2023bif}, some of the possible transformations of the basins are analyzed with basin entropy allowing for numerical detection in the parameter space. This provides numerical hints for the appearance of this transformation with the change of the system parameters.  

The model used in the paper is this quadratic two-dimensional map:
\begin{align}\label{eq:map_mira}
\begin{split}
x_{n+1} &= a x_n + y_n\\
y_{n+1} &= x_n^2 + b
\end{split}
\end{align}
with the parameters $a = -0.42$ and $b = -1.32$, we get the basins of Fig.~\ref{fig:map_mira} where the critical curve of the map has crossed the connected basins in diamond shape, generating the fractal islands.

\subsection{Eruptions in Non-Invertible Maps}
\begin{keywrds} RB, MAP\end{keywrds}
\begin{figure}
\begin{center}
\includegraphics[width=\columnwidth]{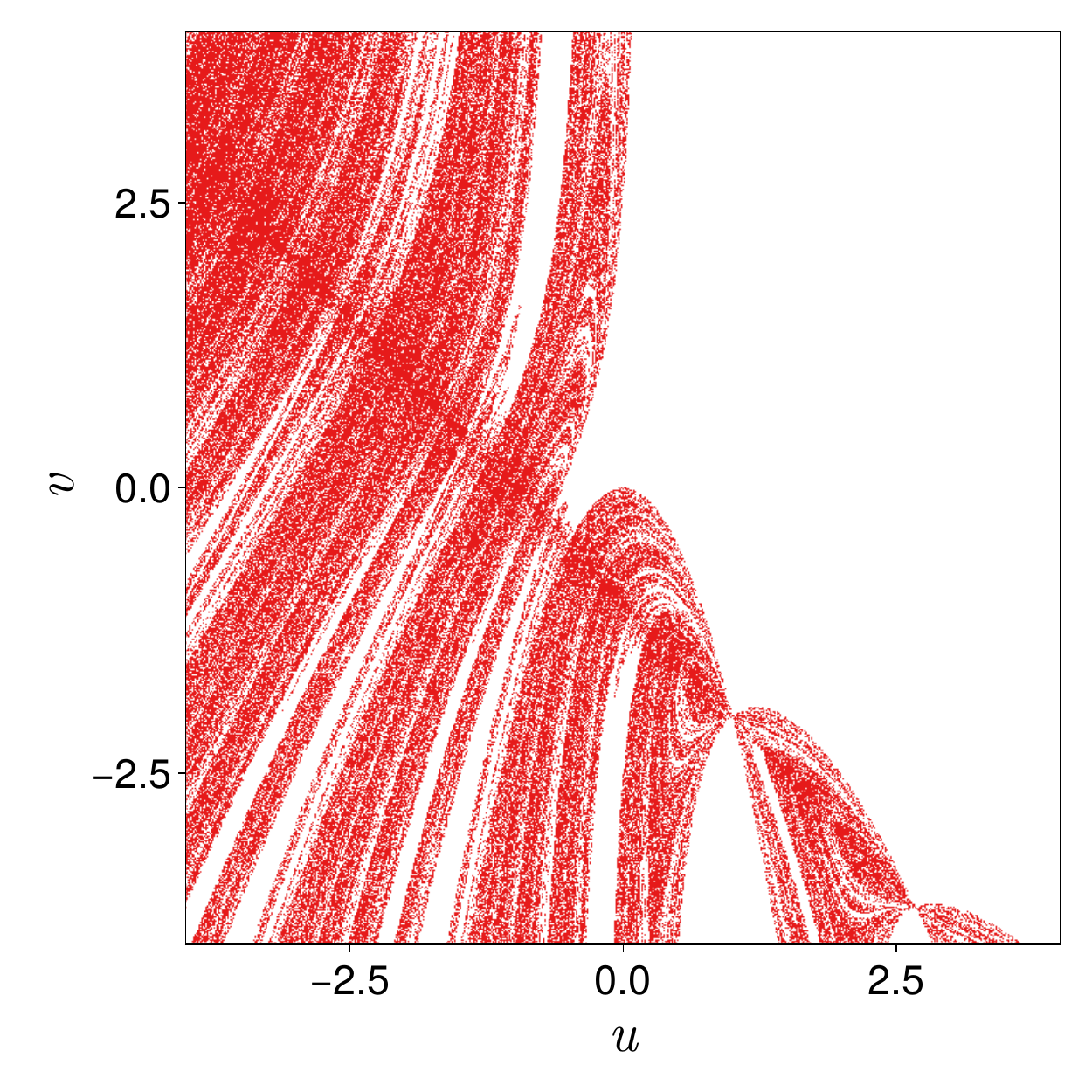}
\end{center}
\caption{\label{fig:eruptions}Basins of attraction of the Bairstow map in Eqs.~\ref{eq:eruptions} for the parameter $a = 0.8$. The two attractors in this figure are a stable fixed point and a chaotic attractor on a diagonal line, which is the diffuse line visible in the plot.}
\end{figure}

In a special case, when two stable fixed points collide and vanish at a singularity, the bifurcation gives birth to an infinite number of unstable periodic and aperiodic orbits. The authors in~\cite{billings1996noninvertible} coined this sudden change ``eruption''.

In the proposed system, the eruption affects the structure of the boundary in several ways. For example, the number of attractors decreases by two, leaving a fractal boundary generated by the preimages of a singular curve. The two possible final states after the bifurcation are a stable fixed point and chaotic trajectories on the invariant line. 

One of the models used in the publication is the Bairstow map:
\begin{align}\label{eq:eruptions}
\begin{split}
    u_{n+1} &= (u_n^3 + u_n (v_n-a+1) + a)/(2 u_n^2 + v_n)\\
    v_{n+1} &= (v_n (u_n^2 + a - 1) + 2 a u_n)/(2 u_n^2 + v_n)
\end{split}
\end{align}
The basins in Fig.~\ref{fig:eruptions} have been computed with the parameter $a=0.8$. The largest connected basin corresponds to the fixed point at $(1, a)$. A diagonal line with the equation $v = -1 - u$ crossing the basins is visible. The initial conditions starting on the red basin converge to this invariant line, where the unstable periodic orbits reside. The dynamics on the invariant line reduce to the Newton root-finding algorithm without any stable root. 

The basins have been computed using an escape basin mechanism: if the trajectory has not converged to the stable fixed point after 200 iterations, it surrely has converged to the invariant line.

\subsection{Coupled Logistic Equations with Quasiperiodic Parametric Modulation}
\begin{keywrds}FB, MAP\end{keywrds}
\begin{figure}
\begin{center}
\includegraphics[width=\columnwidth]{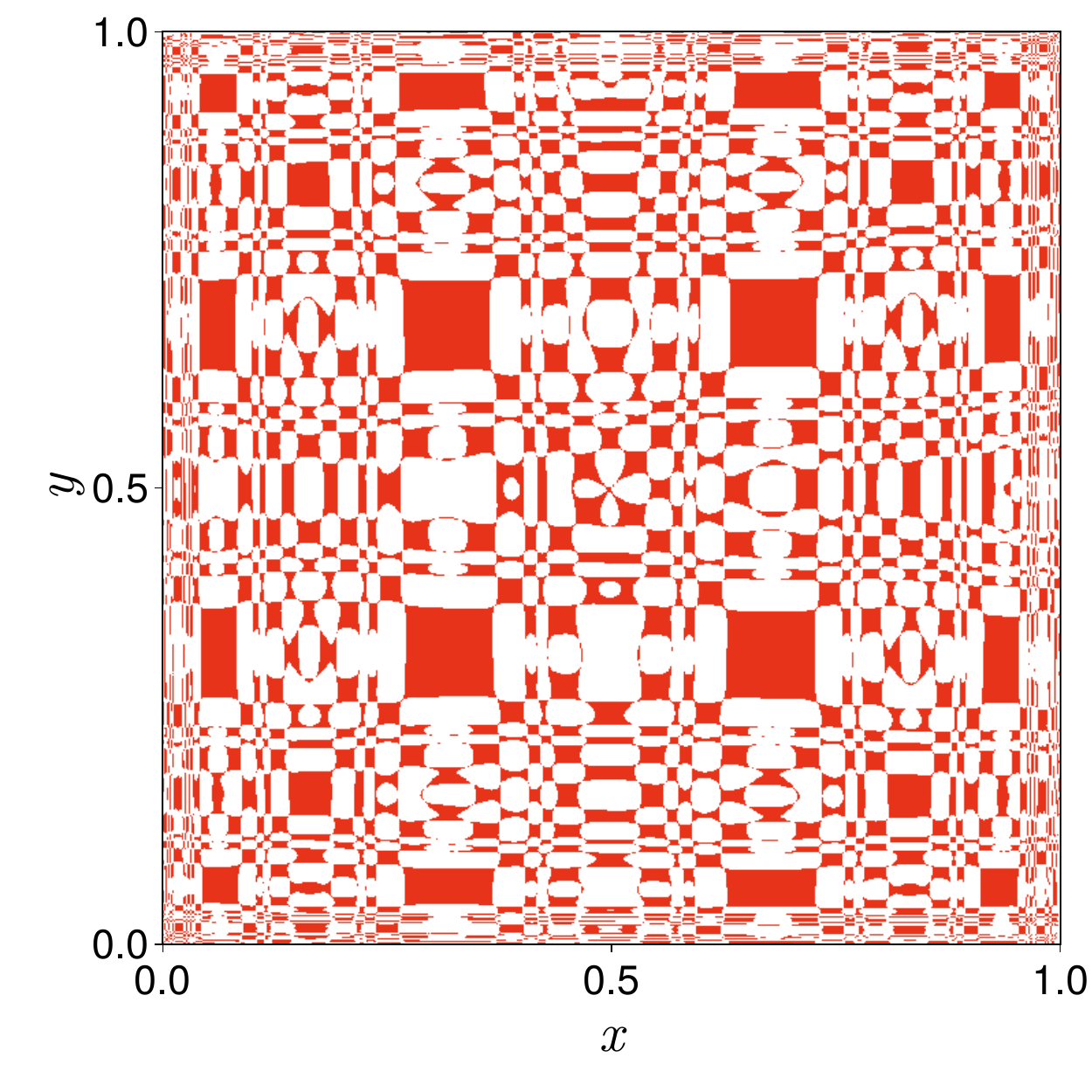}
\end{center}
    \caption{\label{fig:shrimali_map}Basins of attraction of two coupled identical logistic maps with quasiperiodic forcing described in Eqs.~\ref{eq:shrimali_map}. Parameters for this figure are $\alpha = 3.25$, $\beta = 0.01$, $\varepsilon = 0.5$, $\tau = (\sqrt{5}-1)/2$.} 
\end{figure}

The article~\cite{shrimali2005basin} investigates the effects of quasiperiodic forcing on a system of coupled identical logistic maps, focusing on the emergence of bistability and multistability. The system can exhibit a variety of dynamic regimes, including quasiperiodic behaviors, chaotic dynamics, and strange nonchaotic attractors. 

However, the most notable result is the identification of abrupt changes in basin sizes at the onset of bistability. The authors propose a power-law growth of one of the basin volumes as a parameter is varied. This increase in the size of the basins has been studied in a follow-up article~\cite{shrimali2008nature} in more detail to emphasize this power-law dependence.  

Structures in the basins are formed through basin bifurcations as critical points interact with basin boundaries. This interaction leads to the creation of holes and islands within a previously connected component of a basin. The model studied in the article consists of two symmetrically coupled identical logistic maps subjected to a common quasiperiodic force. The equations governing the dynamics of this coupled system are as follows: 
\begin{align}\label{eq:shrimali_map} 
\begin{split} 
x_{n+1} &= \alpha \left(1 + \varepsilon \cos(2\pi \theta_n)\right)x_n(1 - x_n) + \beta(y_n - x_n)\\ 
y_{n+1} &= \alpha\left(1 + \varepsilon \cos(2\pi \theta_n)\right)y_n(1 - y_n) + \beta(x_n - y_n)\\ 
\theta_{n+1} &= \theta_n + \tau \mod 1 
\end{split} 
\end{align} 
The two identical maps can synchronize depending on the parameters and initial conditions. Figure~\ref{fig:shrimali_map} represents the synchronization basins of the two maps for the parameters $\alpha = 3.25$, $\beta = 0.01$, $\varepsilon = 0.5$, and $\tau = (\sqrt{5} - 1)/2$. The structures in the basins are caused by the basin bifurcation mechanism explained in Sec.~\ref{basin_mira}. For the computation of the basins, the state has been tracked on the $x,y$ plane with the initial condition $\theta_0 =  0$ for all trajectories.

\subsection{\label{sec:vcsd}VCSD Chen System} 
\begin{keywrds}FB, MAP\end{keywrds}
\begin{figure} 
\begin{center} 
\includegraphics[width=\columnwidth]{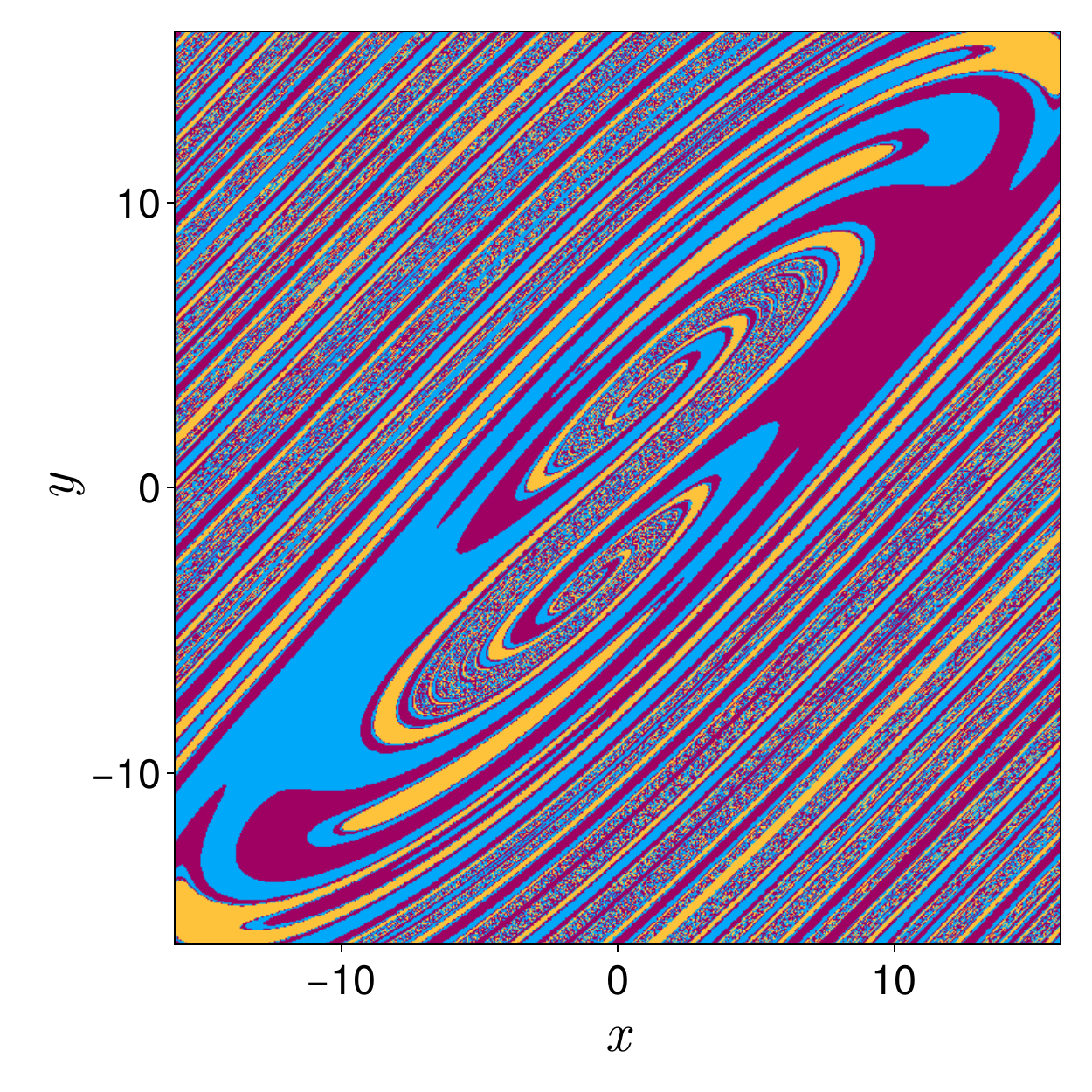} 
\end{center} 
\caption{\label{fig:vcsd_chen}Basins of attraction of the discretized Chen system with the VCSD technique. The system described in Eqs.~\ref{eq:vcsd_chen} has been simulated with the parameters $a=40$, $b=3$, $c=28$, $h=0.01$, and $S=0.77$.} 
\end{figure} 

The discretization of a dynamical system introduces all sorts of behaviors that would otherwise be absent. The numerical integration of a differential equation is essentially a careful discretization taking care of possible unwanted side effects.

In~\cite{ostrovskii2022inducing}, the authors embrace the discretization of a continuous ODE to design multistable systems. The researchers used a numerical integration method with variable symmetry, focusing on the well-known Chen system as a case study. They formulated a two-stage algorithm consisting of the combination of a semi-implicit and a semi-explicit Euler integration method. The method induces multistability within a specific range of a parameter controlling the integration scheme. They identified up to six coexisting attractors in the discretized Chen system. 

The Variable Symmetry Composition Diagonal (VSCD) method applied to the continuous Chen system results in a finite-difference discrete model. The original continuous Chen system is described by the following set of ordinary differential equations:
\begin{align*}
\begin{split}
\dot{x} &= a(y - x), \\
\dot{y} &= (c - a)x - xz + cy, \\
\dot{z} &= xy - bz,
\end{split}
\end{align*}
where $a = 40$, $b = 3$, and $c = 28$. After applying the VSCD method, the equations for the discrete system can be expressed as:
\begin{align} \label{eq:vcsd_chen}
\begin{split}
    x_{n+s} &= (x_n + h_1 a y_n)/(1 + h_1 a), \\
    y_{n+s} &= (y_n + h_1 ((c - a) x_{n+s} - x_{n+s} z_n))/(1- h_1 c)\\
    z_{n+s} &= (z_n + h_1 x_{n+s}y_{n+s})/(1 + h_1 b)\\
    z_{n+1} &= z_{n+s} + h_2(x_{n+s} y_{n+s} - b z_{n+s})\\
    y_{n+1} &= y_{n+s} + h_2((c - a) x_{n+s} - x_{n+s} z_{n+1} + cy_{n+s} )\\
    x_{n+1} &= x_{n+s} + h_2a(y_{n+1} - x_{n+s}) 
\end{split}
\end{align}
where $h_1 = s~h$ and $h_2 = (1 - s)h$, with $s$ being the symmetry coefficient and $h$ the time step size. Fig~\ref{fig:vcsd_chen} represents the slice $x,y$ of the phase space with the choice $z_0=20$ for all initial conditions in the picture. The parameters are $h = 0.01$ and $s = 0.77$ for this computation. 

The first three equations are not dynamical variables and are just function of the state $x_n, y_n, z_n$. The phase space is three dimensional. The basins have a symmetric structure with an interesting spiraling shape.

\subsection{Carpet Oscillator}
\begin{keywrds}MG,  HA, ODE\end{keywrds}
\begin{figure}
\begin{center}
\includegraphics[width=\columnwidth]{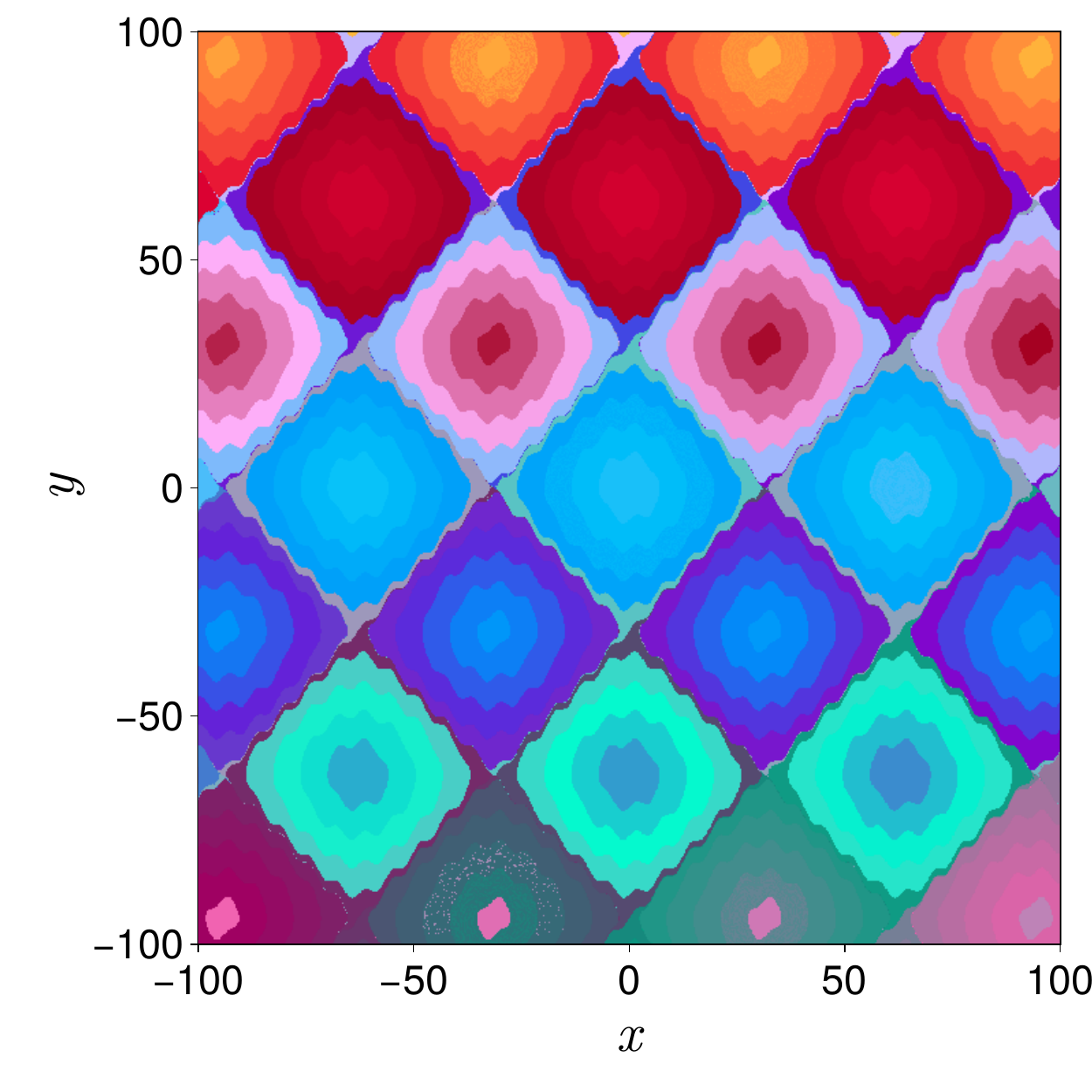}
\end{center}
\caption{\label{fig:carpet}Basins of attraction of the carpet oscillator in Eqs.~\ref{eq:carpet} resemble an ancient Persian carpet. The equations have been designed to produce a symmetric and ongoing repeating pattern of basins.}
\end{figure}

Sometimes the nonlinear system is analyzed for its basins due to interesting properties or aspects of these basins. This is the case for the ``carpet oscillator'' discussed in article~\cite{tang2018carpet}. The system exhibits an infinite number of coexisting limit cycles with a periodic spatial repetition of their basins in phase space. The idea behind it is to modify an existing simple nonlinear oscillator by adding a periodic function to all variables. The results illustrate the presence of infinite attractors, with approximately one-third being self-excited and two-thirds being hidden. The layout of the basins produces a fascinating picture due to the symmetry and the variety of attractors. One version of the system is given by:  
\begin{align}\label{eq:carpet}
\begin{split}
   \dot{x} &= \sin(0.1y)  \\
   \dot{y} &= -\sin(0.1x) + \sin(0.1y)\cos(x)
\end{split}
\end{align}
The basins in Fig.~\ref{fig:carpet} truly resemble a Persian carpet with repeating cells. Each of these cells consists of nested basins of attraction in an onion-shaped structure. This onion-like structure is related to the concepts of megastability and matryoshka basins. The attractors are grouped into cells with nested limit cycles. Some of the basins do not intersect with an unstable fixed point, which is a sign of a hidden attractor inside the basin.

\subsection{Basins of Chimera States}
\begin{keywrds}SB, ODE\end{keywrds}

\begin{figure}
\begin{center}
\includegraphics[width=\columnwidth]{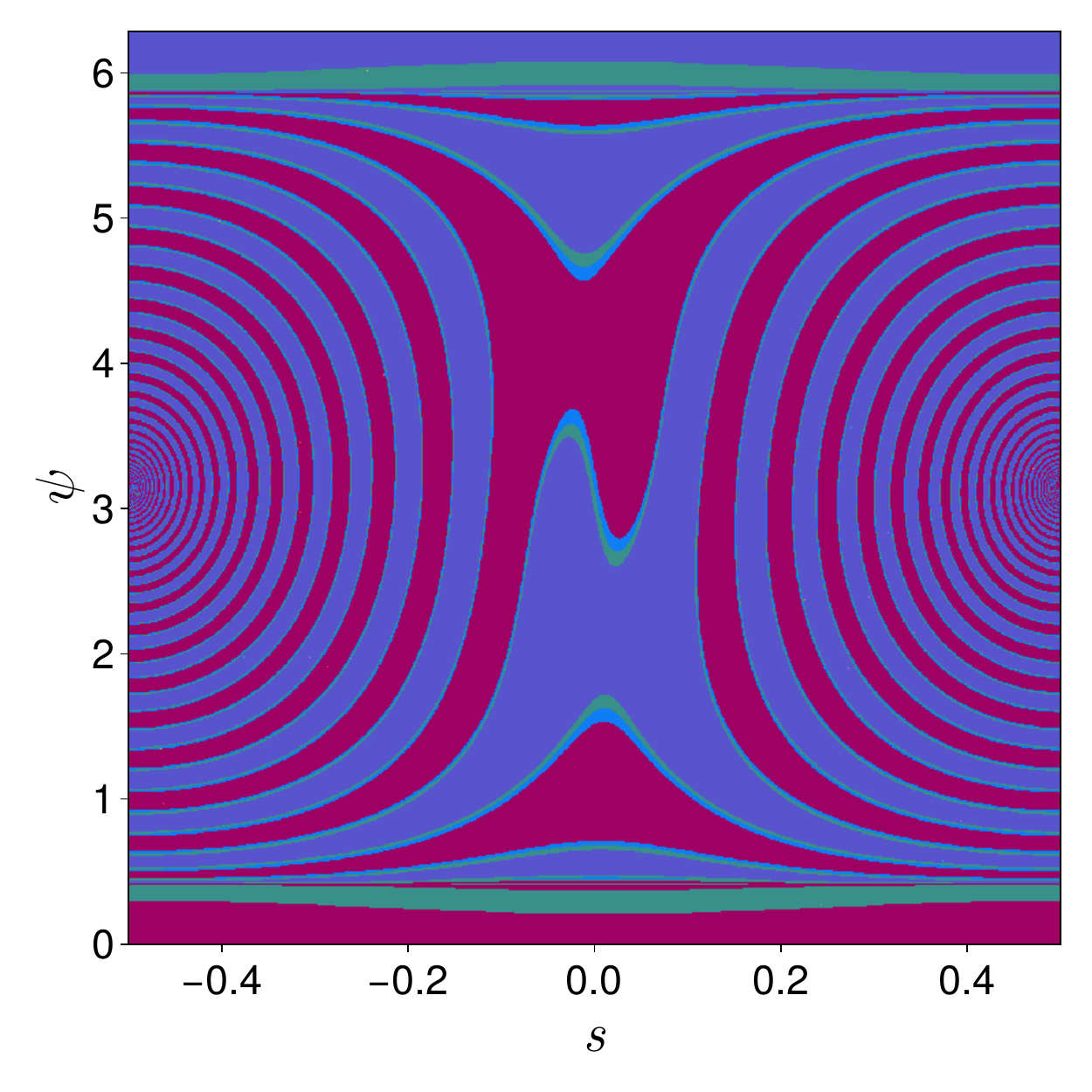}
\end{center}
\caption{\label{fig:chimeras}Basins of attraction for two interacting populations of oscillators. The reduced model of the network in Eqs.~\ref{eq:chimeras} allows tracking the state of the complete network. The three possible states are two chimera states and one stable synchronized state.}
\end{figure}

The article~\cite{martens2016basins} analyzes the basins of attraction for chimera states in a system comprised of two populations of coupled oscillators, interacting with each other through coupling. The model consists of two ensembles of Kuramoto-Sakaguchi oscillators with all-to-all coupling. The oscillators are separated into two equal sets, but the internal coupling of these sets is stronger than the coupling between the sets. This difference allows for a reduction of the model using the Ott-Antonsen ansatz technique~\cite{ott2008low}.  

Once the low-dimensional model has been obtained, the authors continue the study of the synchronized and chimera states as a function of the initial states of the populations. The achievement of the article is to demonstrate that the global final state of the network can be predicted using the information from the initial state of the reduced model. They also propose statistics on the probability of obtaining a chimera state and control strategies to switch the state of the network.  

The dynamics of the system are described by the following equations, which capture the evolution of the phase $\psi$ and the order parameters $\rho_i$ for two populations. The relevant equations are as follows:
\begin{align}\label{eq:chimeras}
\begin{split}
    \frac{d\rho_1}{dt} &= \frac{1-\rho_1^2}{2}(\mu\rho_1\sin(\beta) + \nu\rho_2\sin(\beta-\psi))  \\
    \frac{d\rho_2}{dt} &= \frac{1-\rho_2^2}{2}(\mu\rho_2\sin(\beta) + \nu\rho_1\sin(\beta+\psi))  \\
    \frac{d\psi}{dt} &=  \frac{1+\rho_2^2}{2\rho_2}(\mu\rho_2\cos(\beta) + 
    \nu\rho_1\cos(\beta+\psi)) \\ 
    &- \frac{1+\rho_1^2}{2\rho_1}(\mu\rho_1\cos(\beta) + \nu\rho_2\cos(\beta-\psi))  
\end{split}
\end{align}
The parameters for the basins in Fig.~\ref{fig:chimeras} are $A = 0.1$, $\beta = 0.025$, $\mu = (A+1)/2$; $\nu = 1 - \mu$. The twisted basins represent the two stable chimera states, while the green basin represents the synchronized state. To reflect the symmetries of the phase space, the following variable change is proposed: $s =(\rho_1 + \rho_2)/2$, and $d =(\rho_1 - \rho_2)/2$. Attractors are detected in the $s,d,\psi$ space. Fig.~\ref{fig:chimeras} is a slice of the phase space with the $s,\psi$ variables and $d = 0.56625$ for all initial conditions. For the computations, it is necessary to wrap the phase $\psi$ within the bounds $[0,\pi]$. 

\subsection{Riddled Basins in Coupled Quadratic Map}
\begin{keywrds}RB, MAP\end{keywrds}

\begin{figure}
\begin{center}
\includegraphics[width=\columnwidth]{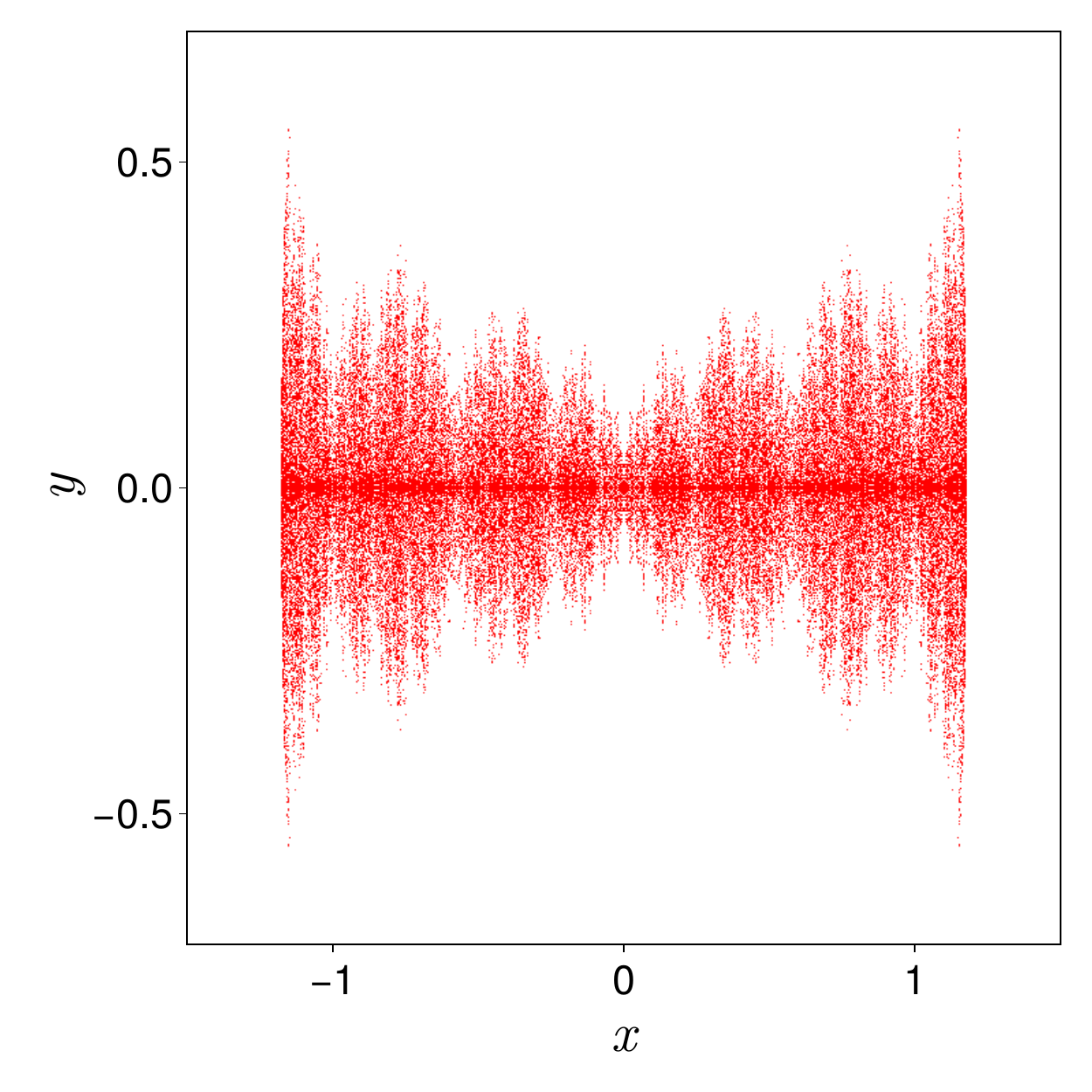}
\end{center}
\caption{\label{fig:map_ashwin}Basins of attraction of the map in Eq.~\ref{eq:map_ashwin}. For the parameters $\nu = 1.28$, $\alpha = 0.7$, and $\varepsilon = 0.5$, there is a Milnor attractor whose basin is riddled with the attractor at infinity.}
\end{figure}

The article~\cite{ashwin1996attractor} starts with the following research problem: suppose a dynamical system possesses an invariant submanifold, and the restriction of the system to this submanifold has a chaotic attractor $A$. Under which conditions is $A$ an attractor for the original system, and in what sense?

The authors tackle this question by examining the spectrum of normal Lyapunov exponents close to the invariant manifold. This approach allows for a classification of the role of the attractor in the submanifold as a function of the values of the exponents. Specifically, the paper identifies conditions for an attractor to transition to a chaotic saddle, characterized by a locally riddled basin or becoming normally repelling. The results reveal how these transitions can be robust under typical parameter changes and present emergent behaviors in numerical simulations. Besides the analytical proofs and definitions, the authors also provide numerical examples to illustrate the results: 
\begin{align}\label{eq:map_ashwin}
\begin{split}
    x_{n+1} &= \frac{3 \sqrt{3}}{2} x_n (x_n^2-1) + \varepsilon x_n y_n^2\\
    y_{n+1} &= \nu \exp(-\alpha x_n^2) y_n + y_n^3
\end{split}
\end{align}
with $\nu = 1.28$, $\alpha = 0.7$, and $\varepsilon = 0.5$. The maximum Lyapunov exponent is bounded by $\log(|\nu|)$, and we can classify the behaviors of the attractor $A$ as a function of this parameter. The attractor $A$ is embedded in the submanifold $y=0$. In the example shown in Fig.~\ref{fig:map_ashwin}, the value of $\nu$ corresponds to a Milnor attractor with a riddled basin boundary. The other attractor is at infinity in this case, corresponding to the white basin. The authors demonstrate how the nature of the basins changes following the theoretical classification as the parameter $\nu$ is varied.

\subsection{Threshold-Linear Networks with Multistable Patterns}
\begin{keywrds} FB, ODE\end{keywrds}

\begin{figure}
\begin{center}
\subfloat[]{\includegraphics[width=\columnwidth]{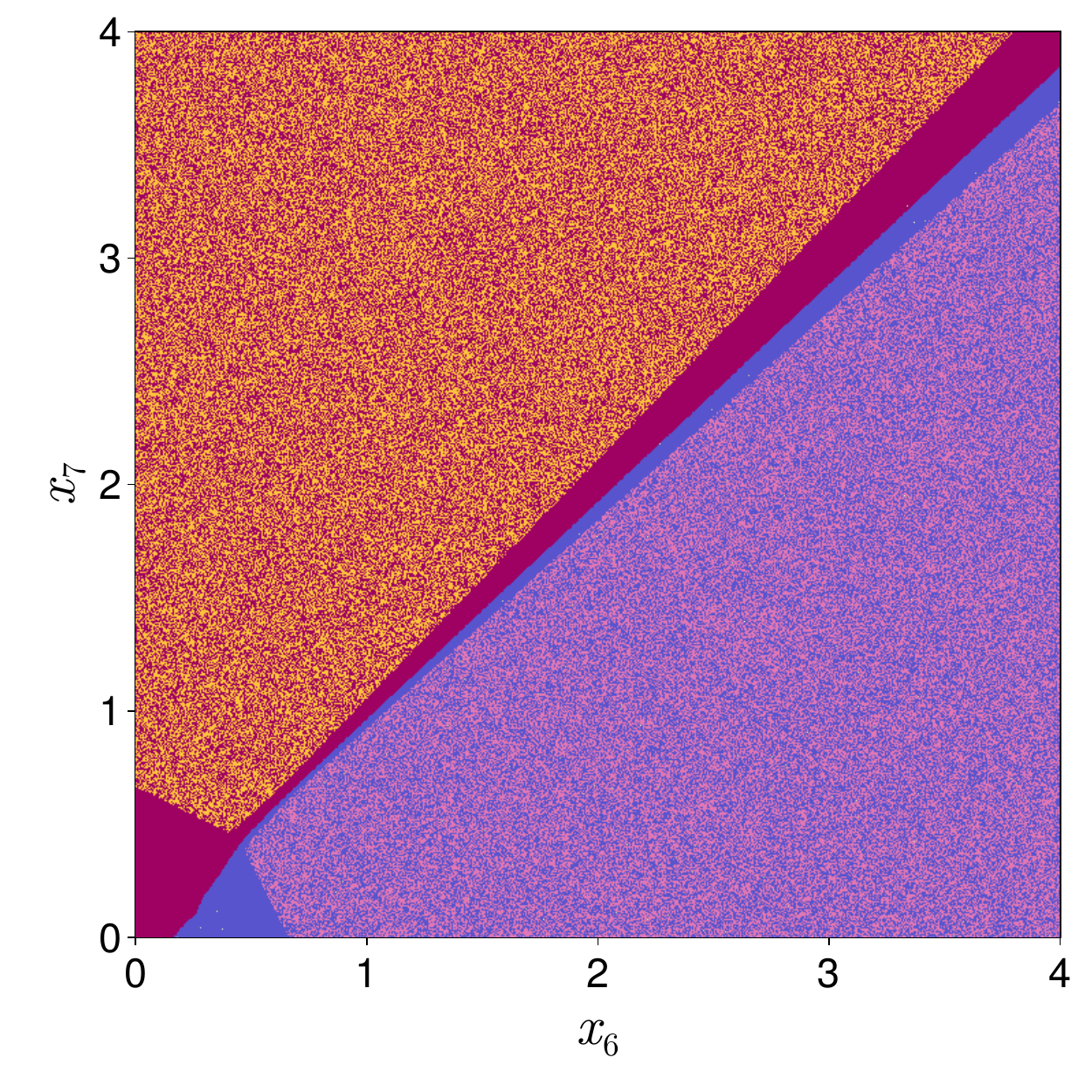}}

\subfloat[]{\begin{tikzpicture}[
    > = stealth, 
    node distance = 2cm, 
    every node/.style = {circle, draw, fill=gray!20, minimum size=5mm, font=\sffamily},
    every edge/.style = {draw, thick, ->} 
]
\node[fill=red!70] (1) {1};
\node[fill=blue!70, right of=1] (2) {2};
\node[fill=green!70, below of=2] (3) {3};
\node[fill=brown!70, below of=1] (4) {4};
\node[fill=gray!70, below of=1, xshift=1cm, yshift=1cm] (5) {5};
\node[fill=orange!70, right of=3] (6) {6};
\node[fill=orange!70, left of=4] (7) {7};
\node[fill=purple!70, left of=1] (8) {8};
\node[fill=pink!70, above of=1] (9) {9};

\path[->] (1) edge (2);
\path[->] (5) edge (1);
\path[->] (1) edge (8);
\path[->] (1) edge (4);
\path[->] (1) edge (9);
\path[->] (3) edge (2);
\path[->] (2) edge (6);
\path[->] (6) edge (3);
\path[->] (3) edge (4);
\path[->] (5) edge (3);
\path[->] (5) edge (6);
\path[->] (4) edge (5);
\path[->] (4) edge (8);
\path[->] (8) edge (4);
\path[->] (2) edge (5);
\path[->] (7) edge (8);
\path[->] (7) edge (1);
\path[->] (8) edge (9);
\path[->] (9) edge (8);
\path[->] (8) edge (1);
\path[->] (8) edge (7);
\path[->] (9) edge (1);
\path[->] (9) edge (2);
\end{tikzpicture}}
\end{center}
    \caption{\label{fig:tlnm}(a) Basins of attraction of the network described in Eq.~\ref{eq:tlnm}. The basins represent a projection of the phase space for the variables $x_6$ and $x_7$. (b) The graph represents the topology of the network for the multistable dynamics depicted in (a).}
\end{figure}
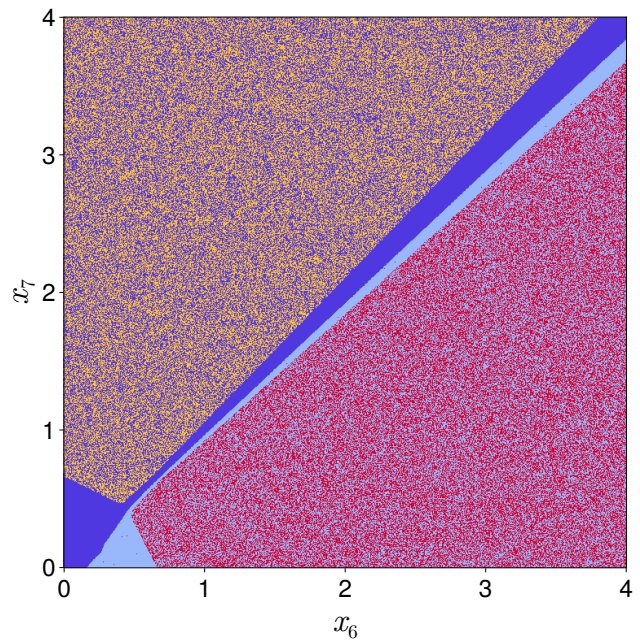
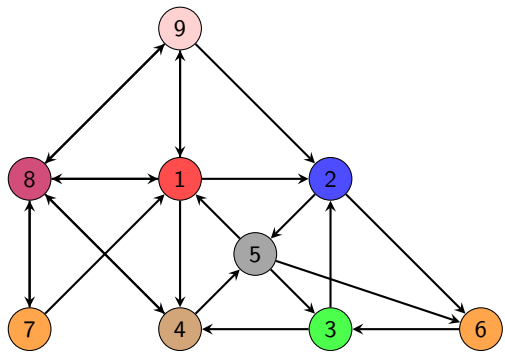

The dynamics of interacting agents have been a subject of sustained interest with a long scientific history. Network science has developed rapidly since the early 2000s and is now a well-studied subject. The agents can be of very different natures, but the important aspect is the exchange of information between the nodes hosting the agents. In the following, the agents are inspired by neuronal activity in living beings nervous systems. These are best known as neural networks. 

Research on neural networks has usually focused on fixed-point dynamics, as exemplified by the Hopfield model. However, other dynamical behaviors are possible, as shown in the study~\cite{Parmelee_2022}. This article investigates the dynamics of a class of threshold-linear networks (TLNs) and identifies conditions leading to the {\em absence} of steady states. Threshold-linear networks have been used in computational neuroscience as a model of recurrent networks. The study focuses on a special class called combinatorial threshold-linear networks, which are mathematically and numerically tractable.  

The authors define CTLNs through directed graphs and categorize stable and unstable fixed points based on network connectivity. The main point of the article is to adjust the parameters of the network to avoid stable fixed points. A variety of oscillations and dynamical behaviors can be observed after destabilization. 

The dynamics of a competitive threshold-linear network are described by the equations:
\begin{equation}\label{eq:tlnm}
   \frac{dx_i}{dt} = -x_i + [\sum_{j=1}^{n} W_{ij} x_j + b_i]^+, \quad i = 1, \ldots, n
\end{equation}
where $[y]^+ = \max\{y, 0\}$ represents the threshold nonlinearity, $W$ is a weighted adjacency matrix, and $b$ is the external input vector. 

In Fig.~\ref{fig:tlnm}, we represent in (a) the basins of attraction of the network sketched in (b). The matrix $W$ is constructed as follows: 
\begin{equation}
W_{ij} = \left\{ \begin{array}{ll}
0 & \textrm{ if } i = j\\
-1 + \varepsilon  & \textrm{ if } i \leftarrow  j\\
-1 - \delta  & \textrm{ if } i \nleftarrow  j
\end{array}\right.
\end{equation}
with $\varepsilon = 0.25$ and $\delta = 0.5$ to guarantee competitive behavior between the dynamics of the nodes. The column vector $b$ has uniform values $b_i = 1$, and the initial conditions are zero for every node except for the two chosen nodes 6 and 7. The basins in Fig.~\ref{fig:tlnm} are fractalized, but a smooth boundary also exists. The basins of this system have not been studied in detail; the authors in~\cite{Parmelee_2022} have reported multistable behavior but have not further pursued its study. The structure of these basins is an open question. In general, there are very few studies on the boundaries of neural networks, although it may be of interest, for example, in the learning process of deep-learning networks~\cite{sohldickstein2024boundary}.

\subsection{Li-Sprott System}
\begin{keywrds}FB, ODE\end{keywrds}

\begin{figure}
\begin{center}
\includegraphics[width=\columnwidth]{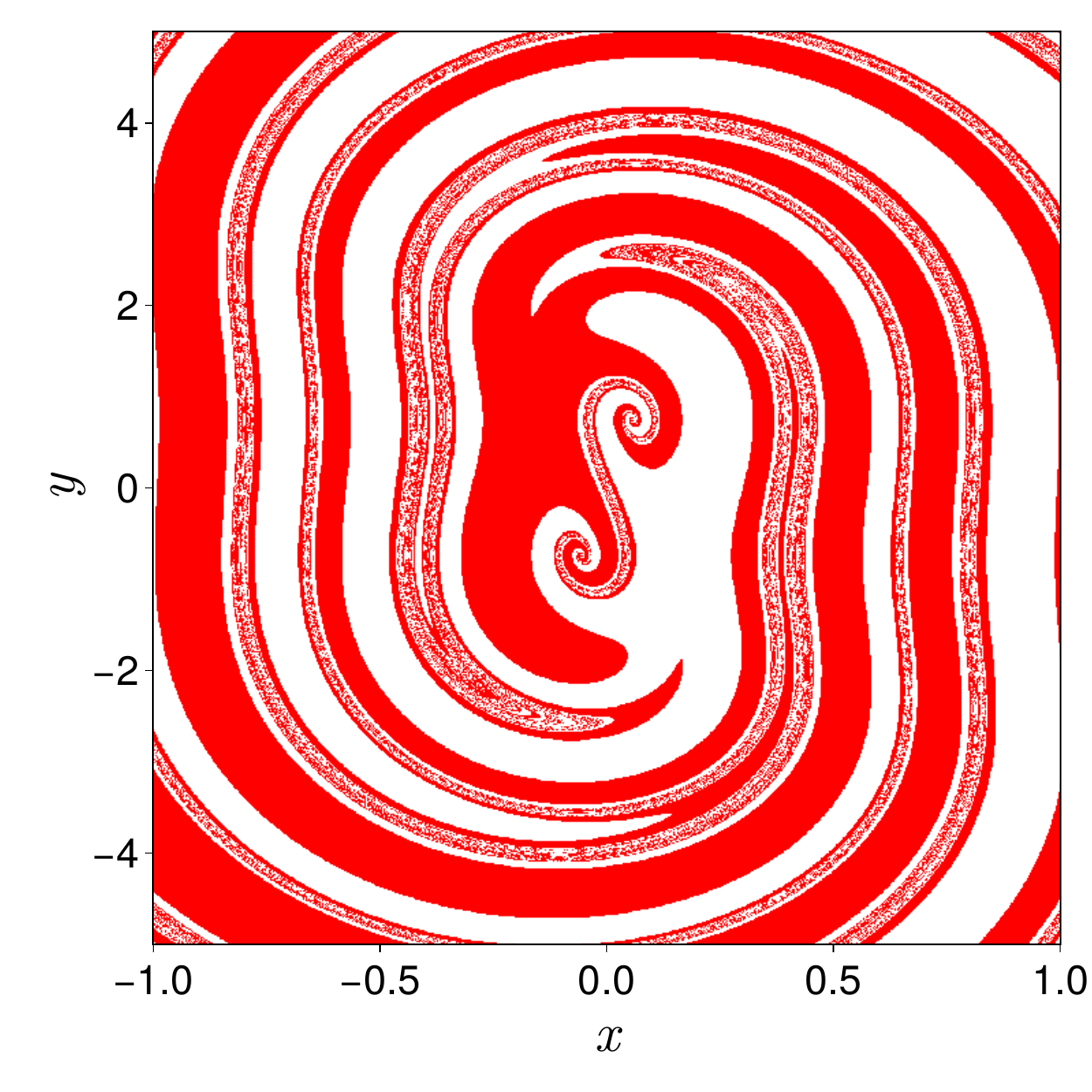}
\end{center}
\caption{\label{fig:li_sprott}Basins of attraction of the system described in Eqs.~\ref{eq:li_sprott}. The two basins are spiraling around each other. Parameters are $a = 13$, $b = 0.55$.}
\end{figure}

In Ref.~\cite{Li_2016}, the authors control the amplitude of multistable chaotic systems. They introduce a chaotic system with a line of equilibria to illustrate how multistability can complicate amplitude control. A scaling parameter is introduced for controlling the amplitude of oscillations. However, the same parameter can cause basins to expand or shrink, leading to unpredictable state switching in the case of an external perturbation. The proposed system is: 
\begin{align}\label{eq:li_sprott}
\begin{split}
\dot{x} & = y + yz, \\
\dot{y} & = yz - axz, \\
\dot{z} & = bz^2 - y^2,
\end{split}
\end{align}
The basins shown in Fig.~\ref{fig:li_sprott} are computed for the initial conditions $(x_0,y_0, -1)$. This is an interesting example of two symmetric basins in a coupled ODE system. In the article, an additional parameter $m$ is introduced in the last equation to scale the amplitude for control. Here, we exclude the control parameter and focus on the basins of the dynamical system only.

\subsection{Dissipative Standard Nontwist Map}
\begin{keywrds}FB, MAP\end{keywrds}
\begin{figure}
\begin{center}
\includegraphics[width=\columnwidth]{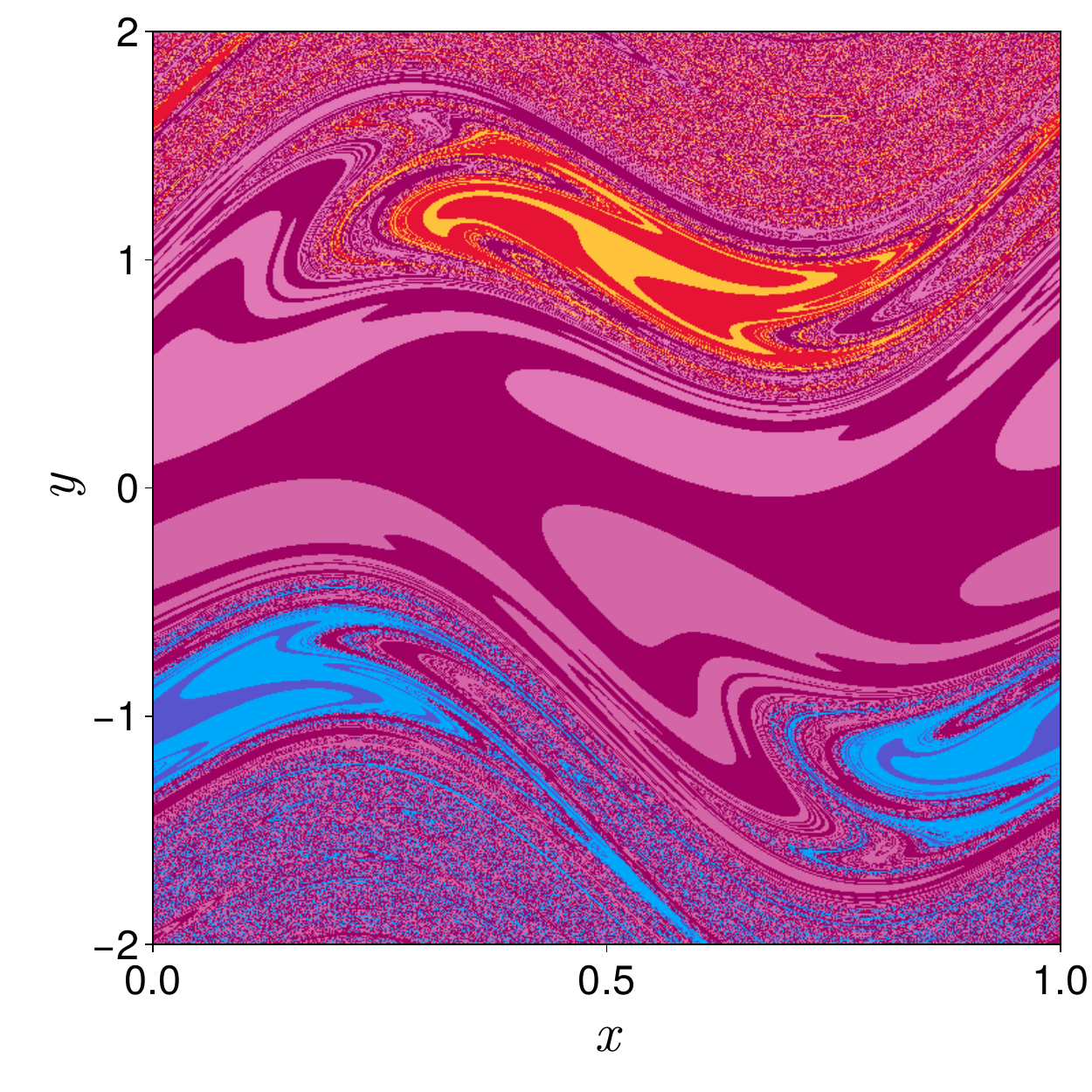}
\end{center}
\caption{\label{fig:dsnm}Basins of attraction of the dissipative standard nontwist map in Eqs.~\ref{eq:dsnm} for the parameters $a = 0.55$, $b = 0.45$, and $\gamma = 0.1$. The basins possess the partially Wada property.}
\end{figure}

The study~\cite{baroni2023chaotic} investigates a dissipative version of the standard nontwist map, focusing on the dynamics associated with chaotic saddles and the interior crises they generate.

The findings show the creation of two types of chaotic saddles: a global chaotic saddle and a local chaotic saddle. The presence of these saddles significantly increases transient times. The realization of the interior crises, both local and global, induces sudden changes in the chaotic attractor size and causes intermittent behavior. The crises are triggered by parameter changes in the system.

The map discussed in the article is the dissipative standard nontwist map (DSNM). It is defined by the following equations:
\begin{align}\label{eq:dsnm}
\begin{split}
y_{n+1} &= (1 - \gamma) y_n - b \sin(2 \pi x_n) \\
x_{n+1} &= x_n + a \left(1 - y_{n+1}^2\right)
\end{split}
\end{align}
where $\gamma = 0.1$ is a dissipation parameter, $b$ controls the nonlinear perturbation, and $a$ is related to the unperturbed rotation number profile. The map features interesting multistable behavior, as shown in Fig.~\ref{fig:dsnm}. However, the intermittent behavior studied in the article occurs in another parameter regime. Crises can be created at will in a certain parameter regime, which is also an attractive feature of the map.

\subsection{Wada in a Cubic Map}
\begin{keywrds}WD, MAP\end{keywrds}
\begin{figure}
\begin{center}
\includegraphics[width=\columnwidth]{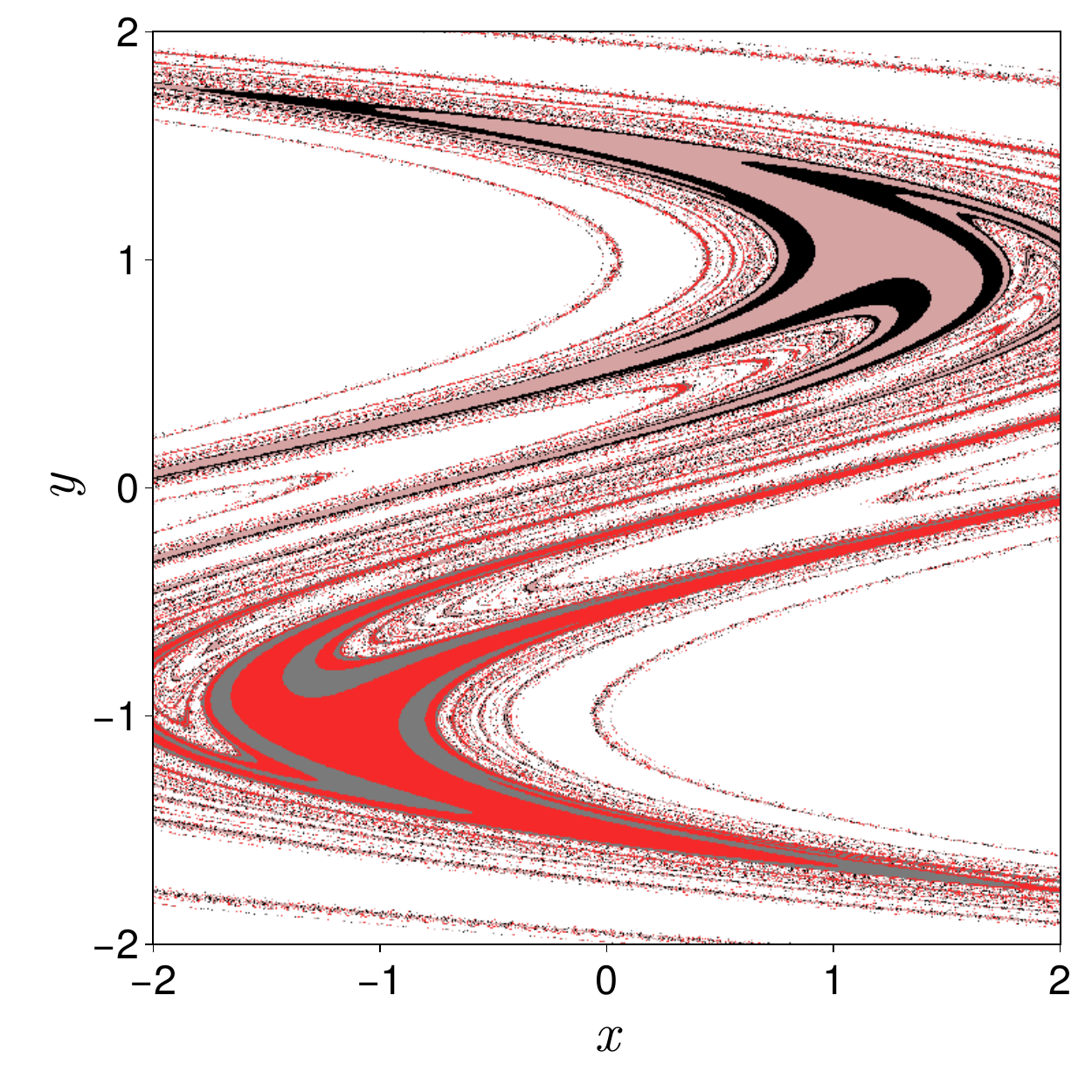}
\end{center}
    \caption{\label{fig:cbic_map}Basins of attraction in a two-dimensional cubic map akin to the Hénon map (Eqs.~\ref{eq:cbic_map}). The parameters of the map are $\mu = 2.9$, $J = 0.66$. The basins boundary has the partially Wada property.}
\end{figure}

The article~\cite{Zhang_2013} focuses on a special phenomenon coined Wada bifurcations. The work also studies the emergence of partially Wada basin boundaries within a two-dimensional cubic map.

The authors manage to isolate a basin cell in the phase space to show the existence of Wada and partially Wada boundaries. A basin cell is a trapping region formed from stable and unstable manifold branches of an accessible periodic orbit on the boundary. A trapping region can be included in a basin of attraction~\cite{aguirre2009fractal}.

When an unstable manifold branch of the scaffolding formed by the pieces of stable and unstable manifolds of the periodic orbits crosses all the basins, then the boundary has the Wada property.

Using this tool, the study shows that basin cell erosion can produce significant changes in basin structures, transitioning from Wada to partially Wada basin boundaries. A new basin can appear inside the basin cell with a smooth boundary. This transition is called a Wada bifurcation.

The article also discusses another interesting example of basin transformation called the Wada metamorphosis. This happens when the basin cell evolves after a parameter change due to the appearance of a new accessible periodic orbit. The new basin cells also have an unstable manifold branch crossing all the basins. In this case, we have a Wada basin boundary metamorphosis.

The equations of the studied map are:
\begin{align}\label{eq:cbic_map}
\begin{split}
x_{n+1} &= y_n\\
y_{n+1} &= \mu y_n - y_n^3 - J x_n
\end{split}
\end{align}
$J$ is a Jacobian parameter constrained within the interval $0 \leq J \leq 1$. The basins represented in Fig.~\ref{fig:cbic_map} have been computed for the parameters $\mu = 2.9$, $J = 0.66$. The authors have identified the presence of the partial Wada property in these basins. Notice the similarity of the equations with the Hénon map.

\subsection{\label{sec:kapitaniak}Riddled Basins in 2D Map} 
\begin{keywrds}RB, MAP\end{keywrds}

\begin{figure}
\begin{center}
\includegraphics[width=\columnwidth]{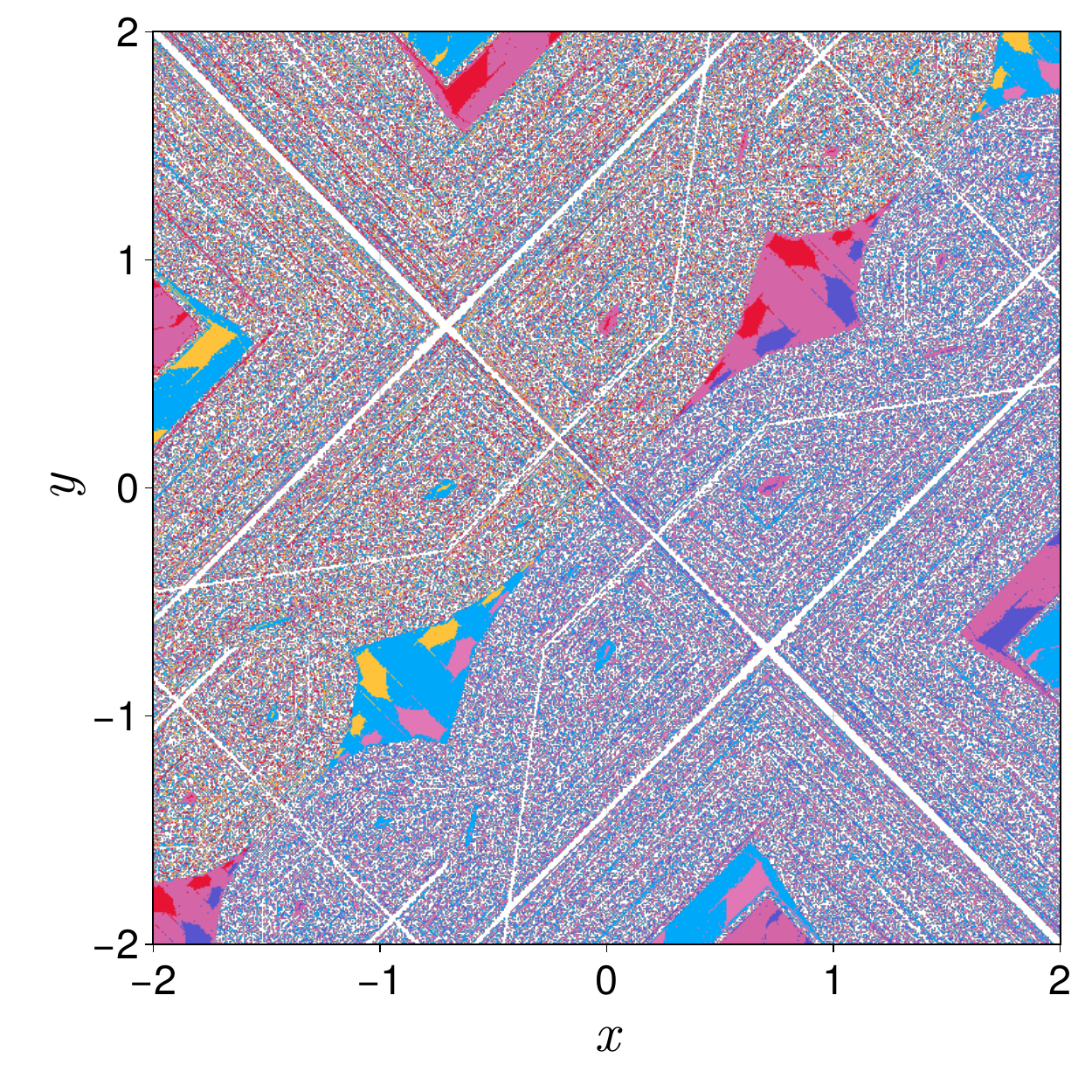}
\end{center}
\caption{\label{fig:riddled_2dmap}Basins of attraction of a two-dimensional discrete map to illustrate the bifurcation between local and global riddling using Eqs.~\ref{eq:riddled_2dmap}.}
\end{figure}

The objective of the article~\cite{Kapitaniak_1998} is to provide sufficient conditions for the existence of locally and globally riddled basins in discrete dynamical systems and to describe bifurcations resulting in this transition. 

The study describes the locally riddled basin as a situation where a trajectory leaves a neighborhood $U$ of the invariant manifold and eventually returns to the attractor. This is a kind of bursting mechanism. In globally riddled basins, the trajectory leaves the neighborhood and goes to another attractor or to infinity. 

The transition from locally to globally riddled basins depends on the appearance or change in stability of an attractor. Local riddling evolves into global riddling, as the trajectories leaving the neighborhood $U$ of the attractor do not return and converge to a different attractor with their basins riddled. Another mechanism is the breaking of the boundary due to a boundary crisis, resulting in basins that are also riddled. 

The map used to illustrate the bifurcation is as follows:
\begin{align}\label{eq:riddled_2dmap}
\begin{split}
   x_{n+1} = f(x_n) + d_1 (y_n - x_n) \\
   y_{n+1} = f(y_n) + d_2 (x_n - y_n)
\end{split}
\end{align}
with:
\[
f(x) = 
\begin{cases} 
l x & \text{if } |x| < \frac{1}{l} \\
    px + \text{sign}(x) (1-p/l) & \text{if } |x| \geq \frac{1}{l}
\end{cases}
\]
In~\cite{Kapitaniak_1998}, the basins are restricted to a small portion of the phase space. In Fig.~\ref{fig:riddled_2dmap}, we show a larger area to appreciate the riddling of the basins for the parameters $l = \sqrt{2}$, $p = -\sqrt{2}$, and $d_1 = d_2 = -0.935$. The islands visible along the diagonal exhibit the local riddled property, although they do not appear to the naked eye.

\subsection{Lorenz Discrete Reduced Model}
\begin{keywrds}FB, MAP\end{keywrds}

\begin{figure}
\begin{center}
\includegraphics[width=\columnwidth]{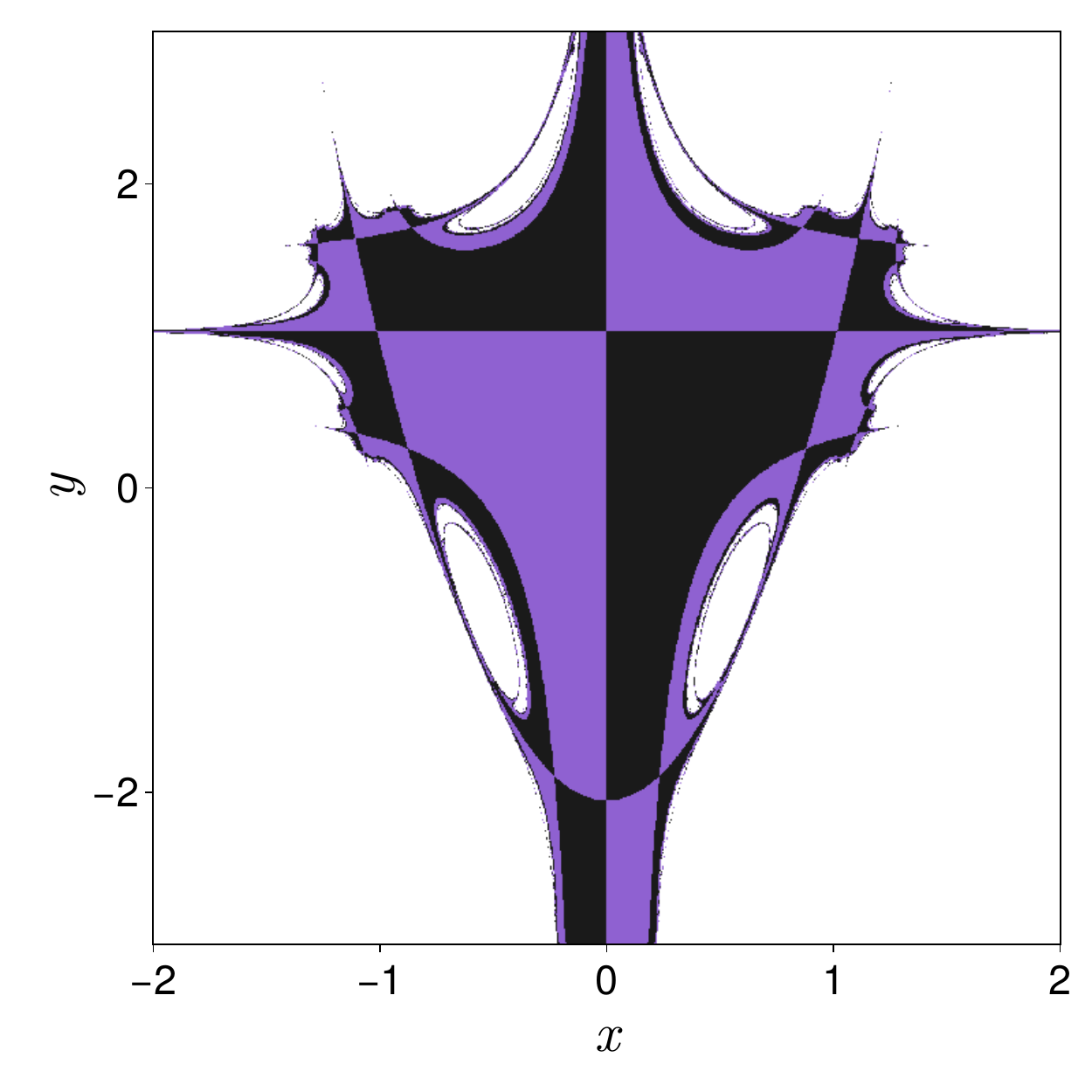}
\end{center}
\caption{\label{fig:lorenz_comp}Basins of attraction of a simplified and discretized Lorenz model described in Eqs.~\ref{eq:lorenz_comp}.}
\end{figure}

E. Lorenz introduced numerous versions of his renowned model throughout the years. In~\cite{lorenz1989computational}, he explores the shift from accurate to inadequate and potentially chaotic approximations, causing computational instability when applying discretization schemes with progressively larger time steps. 

The article examines the limit of discretizing a continuous flow using the time step as a control parameter. As it turns out, it can also be used as a means to create chaotic systems and to study them for their own sake.

The discretized equations of the simplified model are: 
\begin{align}\label{eq:lorenz_comp}
\begin{split}
    x_{n+1} &= (1 + a\tau)x_n - \tau x_n y_n\\
    y_{n+1} &= (1-\tau )y_n + \tau x_n^2
\end{split}
\end{align}
where the parameter $\tau$ represents the time step of the discretization. The parameters for the basins in Fig.~\ref{fig:lorenz_comp} are $\tau = 1.5$ and $a = 0.36$. For these parameters, there are three basins corresponding to two stable symmetric attractors and diverging trajectories to infinity. 

The idea of obtaining new dynamics by discretizing a continuous system also appears in Sec.~\ref{sec:vcsd}, where a numerical integration scheme is taken as a new dynamical system. This highlights an interesting problem related to numerical integrators. We must keep in mind the limitations of numerical solvers for ordinary differential equations. At the core of the solver lies a discretization process. The safeguards in place make them safe to use, but Lorenz work serves as a reminder of the limitations of these methods.

\subsection{Coupled Logistic Map}
\begin{keywrds} RB, IB, MAP \end{keywrds}
\begin{figure}
\begin{center}
\includegraphics[width=\columnwidth]{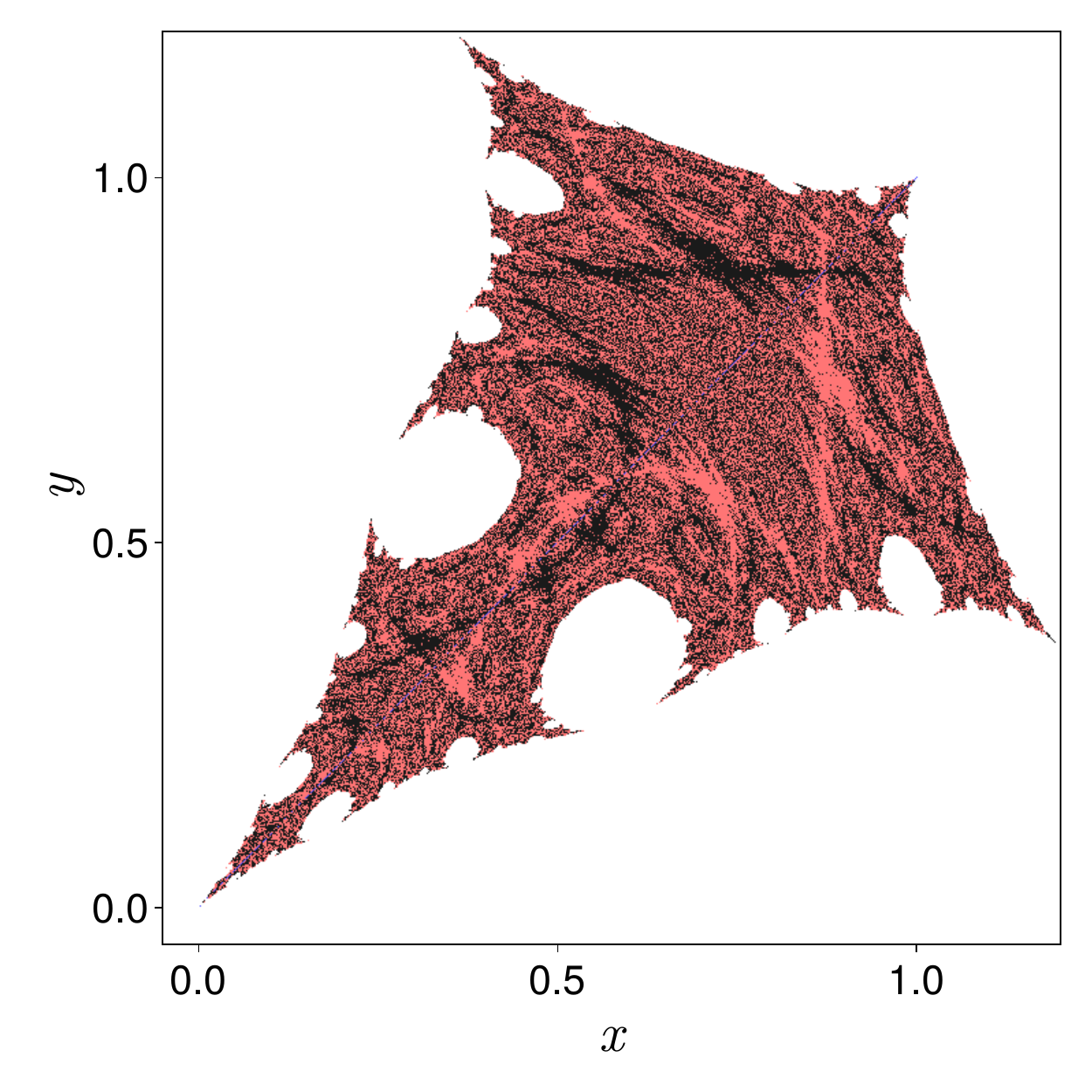}
\end{center}
\caption{\label{fig:coupled_logistics}Basins of attraction of two coupled logistic equations in Eqs.~\ref{eq:coupled_logistics}. The basins of the two chaotic attractors are intermingled due to the presence of a chaotic saddle on the diagonal.}
\end{figure}

The paper~\cite{Maistrenko_1998} investigates the emergence of riddled basins of attraction in a system of two symmetrically coupled logistic maps.

The transverse instability of an attractor can lead to phenomena such as on-off intermittency, as the trajectory leaves the chaotic set before returning after a given time if there is a reinjection mechanism, see also Sec.~\ref{sec:kapitaniak}. This instability will depend on the transverse Lyapunov exponent of the chaotic set. If there is no reinjection of the trajectory to the chaotic set, a transient appears, and all trajectories eventually leave the attractor.

In conditions where the transverse Lyapunov exponent is small or negative, a riddled basin may appear since the chaotic set is attracting, but some embedded unstable periodic orbits are transversally unstable. This leads to a weak attraction in the sense of Milnor, and the corresponding basins are riddled. The article sets the condition for the appearance of riddled basins in a system of two coupled logistic maps defined as:
\begin{align}\label{eq:coupled_logistics}
\begin{split}
    x_{n+1} &= a x_n(1-x_n) + \varepsilon(y_n-x_n) \\
    y_{n+1} &= a y_n(1-y_n) + \varepsilon(x_n-y_n),
\end{split}
\end{align}
with parameters $a = 3.6$ and $\varepsilon = -1$. For this setup, the dynamical system has two symmetric chaotic attractors and also an attractor to infinity. The basins in Fig.~\ref{fig:coupled_logistics} of the chaotic attractors are intermingled. The synchronized states on the diagonal line form a chaotic saddle, causing the initial conditions to end up in one of the two symmetric attractors. The authors computed a small negative transverse Lyapunov exponent for these parameters.

\subsection{4D Memristive Sprott B Oscillator}
\begin{keywrds} FB, ODE \end{keywrds}
\begin{figure}
\begin{center}
\includegraphics[width=\columnwidth]{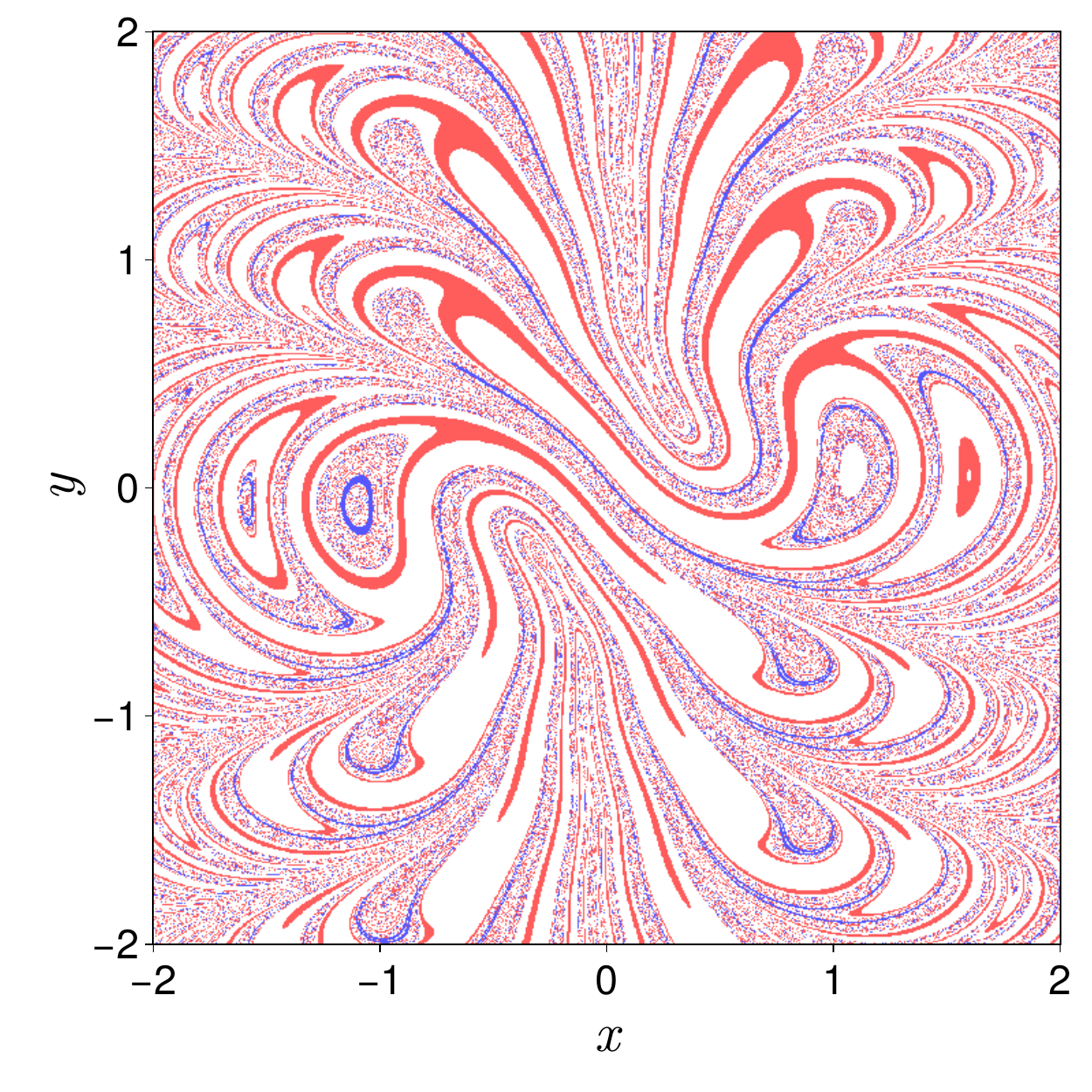}
\end{center}
    \caption{\label{fig:4d_memristive}Basins of attraction of the 4D chaotic Sprott B system formulated in Eqs.~\ref{eq:4d_memristive}. There are two periodic and one chaotic interlaced attractor. This figure shows a projection of the entire phase space onto the $x,y$ plane with the initial conditions $z(0) = 0$ and $u(0) = 0$.}
\end{figure}

A memristive differential equation includes a resistive element whose value depends on the history of the state variables. It can be a function of a delayed variable and time. In the paper~\cite{Ramamoorthy_2022}, the authors introduce a memristive system derived from the Sprott B model by incorporating the memristor element. In this case, the element is modeled by a nonlinear function of one of the variables of the system. The connection with the physical system is tenuous, but the resulting dynamical effects are interesting.

The study focuses on amplitude control through parameter changes and symmetry breaking of the attractors with the introduction of a constant bias. Interesting features of the basins are reported, such as symmetric and entangled attractors. 

The equations of the system are given by:
\begin{align}\label{eq:4d_memristive}
\begin{split}
\dot{x} &= ryz + g, \\
\dot{y} &= x - y, \\
\dot{z} &= 1 - m W(u)xy, \\
\dot{u} &= axy - u.
\end{split}
\end{align}
with the function $W$ defined as:
\begin{equation}
W(u) = \alpha + \gamma |u| + \beta u^2, 
\end{equation}
Parameters are $\alpha = 1$, $\beta = 0.05$, $\gamma = 0.5$, $g = 0.03$, $r = 5.8$, and $m = 11$. For this set of parameters, there are three coexisting attractors: two periodic and one chaotic, interlaced in the phase space. Figure~\ref{fig:4d_memristive} shows a slice of the phase space in the $x,y$ plane with the initial conditions $z(0) = 0$ and $u(0) = 0$.

\subsection{Parametrically Forced Pendulum}
\begin{keywrds} FB, ODE \end{keywrds}
\begin{figure}
\begin{center}
\includegraphics[width=\columnwidth]{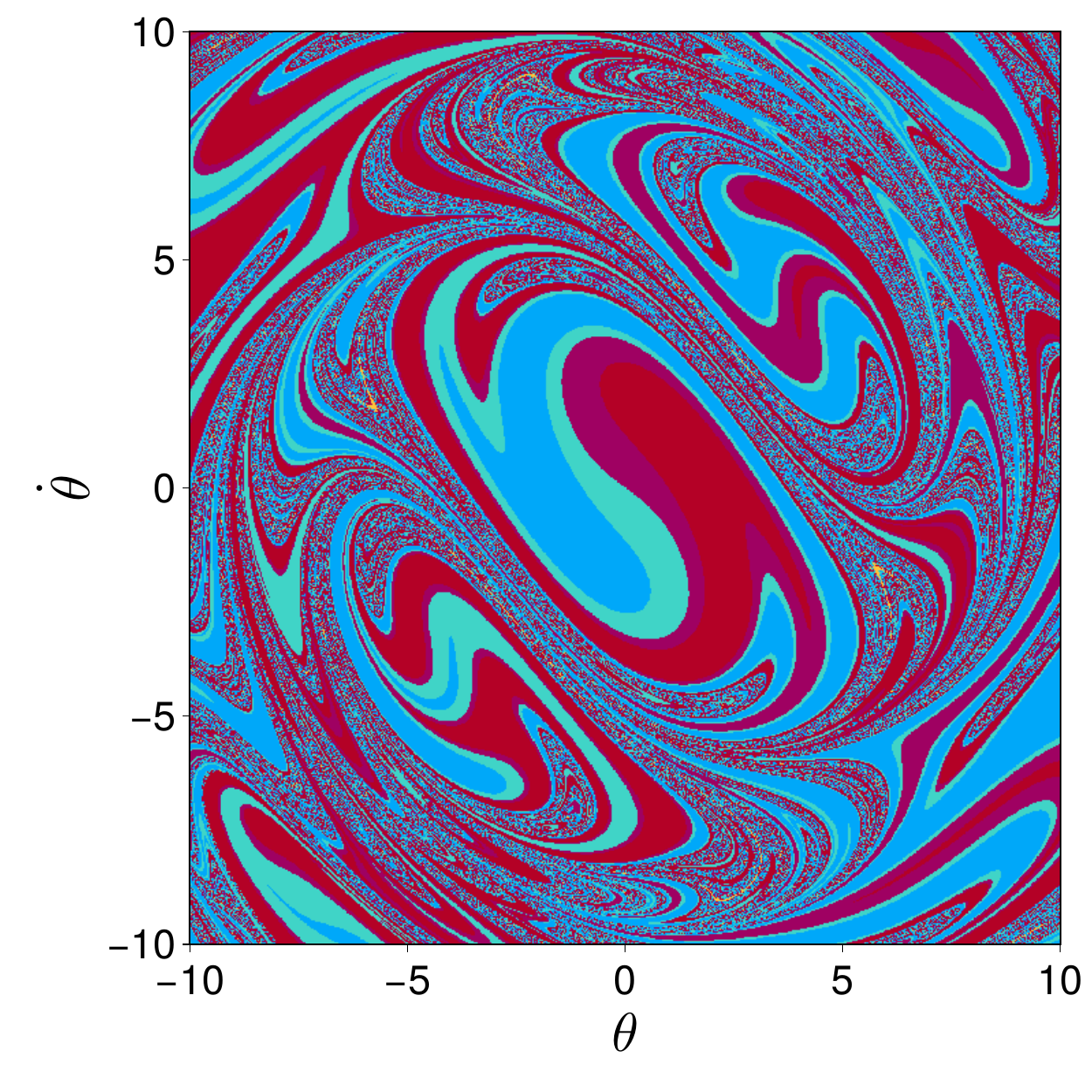}
\end{center}
\caption{\label{fig:parametric_forcing}Basins of attraction of a parametrically forced pendulum described by Eqs.~\ref{eq:parametric_forcing}. The parameters for this figure are: $A_1 = 0$, $A_2 = 4.1$, $\omega = 1.5$, $\mu = 9.81$; $L = m = c = 1$, $b = 0.2$. Note that the coefficient $A_1$ is set to zero, and one of the forcing terms disappears, leaving only the parametric forcing.}
\end{figure}

The study~\cite{klokov2011parametrically} conducts a comprehensive bifurcation analysis of parametrically excited pendulum systems. The model is inspired by a pendulum with a periodically vibrating suspension point in two orthogonal directions. A continuation method called complete bifurcation groups allows for the detection of stable and unstable periodic regimes. Poincaré mappings complete the analysis of the dynamical systems. The analysis reveals various bifurcations and numerous rare attractors, both regular and chaotic. The attractors are considered rare in the sense that they appear only in a narrow range of parameters. There is no restriction on the size of the basins in the phase space.

The equation of the pendulum is: 
\begin{align}\label{eq:parametric_forcing}
\begin{split}
    mL^2 \ddot{\theta} &+ b \dot{\theta} + c \theta + mL(\mu - A_2 \omega^2 \cos(\omega t)) \sin(\theta) \\
    &+ mLA_1 \omega^2 \sin(\omega t) \cos(\theta) = 0
\end{split}
\end{align}
$m$ is the mass of the pendulum, $L$ is the length of the pendulum, $b$ is the linear damping coefficient, $c$ is the linear stiffness coefficient, $\mu$ is the gravitational constant, and finally, $A_1$ and $A_2$ are the amplitudes of the horizontal and vertical oscillations of the suspension point. The basins shown in Fig.~\ref{fig:parametric_forcing} have been obtained with the parameters $A_1 = 0$, $A_2 = 4.1$, $\omega = 1.5$, $\mu = 9.81$; $L = m = c = 1$, $b = 0.2$. Since we are dealing with a periodically forced system, we define a stroboscopic map to improve computational efficiency. While the model is inspired by a physical system, the primary considerations in the article are the exploration of the dynamics. Nevertheless, the authors raise an interesting point by showing that a pendulum with a vibrating suspension point can behave erratically and change its behavior if the parameters or the initial conditions change slightly.

\subsection{The Bogdanov Map}
\begin{keywrds}FB, MAP\end{keywrds}
\begin{figure}
\begin{center}
\includegraphics[width=\columnwidth]{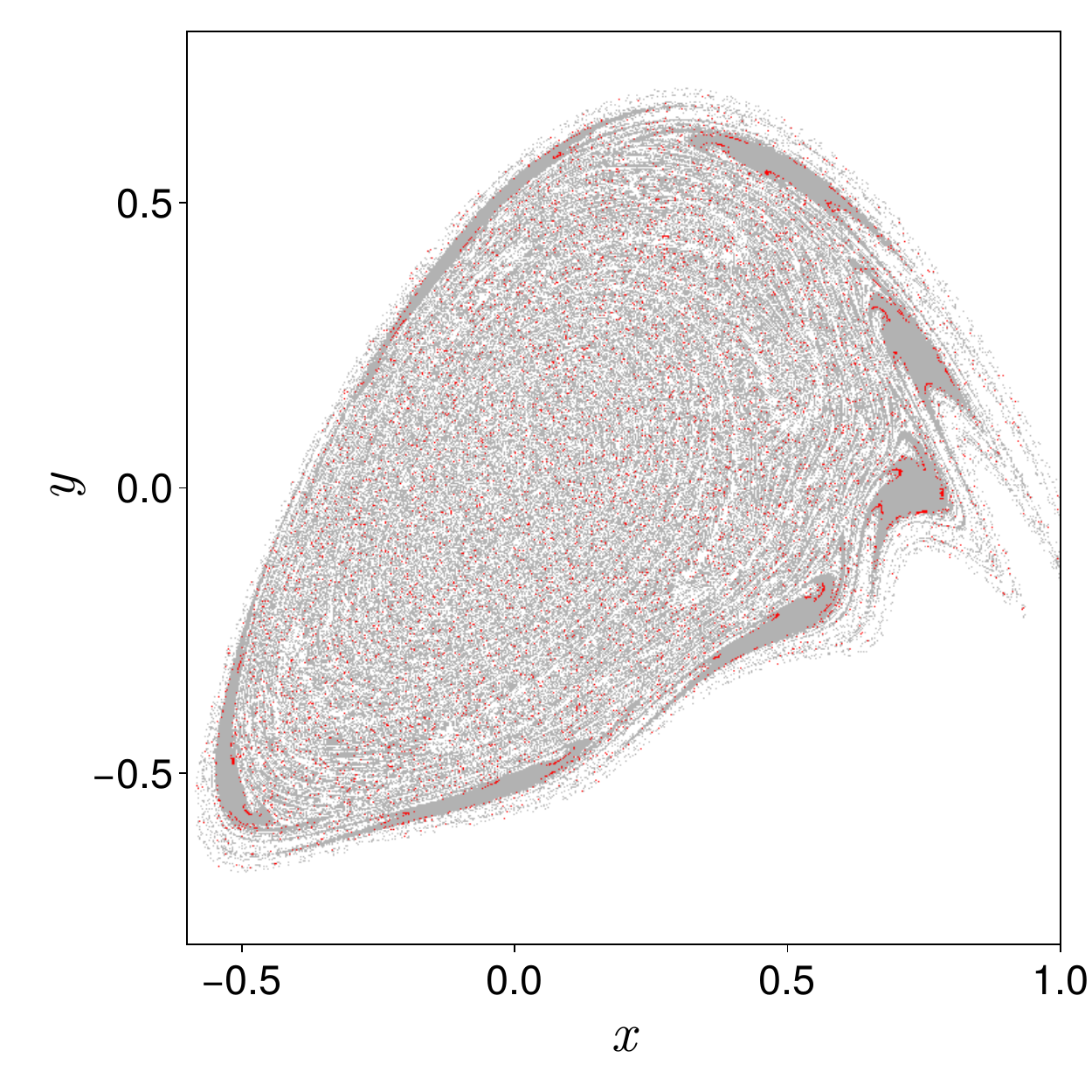}
\end{center}
\caption{\label{fig:bogdanov}Basins of attraction of the Bogdanov map described by Eqs.~\ref{eq:bogdanov} for the parameters $\mu = -0.1$, $k = 1.2$, and $\varepsilon = 0.0125$. The white basin corresponds to diverging trajectories; red indicates a period-36 attractor, and gray represents a period-7 orbit.}
\end{figure}

The article~\cite{arrowsmith1993bogdanov} is a comprehensive study of the dynamics of the Bogdanov map. The theoretical analysis starts with the discretization of the Bogdanov vector field to obtain a two-dimensional quadratic map. One of the parameters controls the dissipation of the system, providing both the conservative and dissipative versions of the map.

The inquiry exhibits the complex dynamics of the map, including Hopf bifurcations and mode locking illustrated by the formation of Arnold tongues. The interaction between the invariant circle and periodic points yields quasi-periodic and chaotic behavior. The authors find a rich structure of periodic orbits associated with saddle-node and Hopf bifurcations, as well as the emergence of chaotic dynamics through the creation of homoclinic tangles. This article serves as a good example of a thorough analysis of dynamical systems using the relevant tools: bifurcation theory, phase plane analysis, Arnold tongues, and so on.

The equations of the quadratic map are:
\begin{align}\label{eq:bogdanov}
\begin{split}
y_{n+1} &= y_n + \varepsilon y_n + k x_n (x_n-1) + \mu x_n y_n\\
x_{n+1} &= x_n + y_{n+1}
\end{split}
\end{align}
Figure~\ref{fig:bogdanov} shows the basins of attraction of the map for the parameters $\mu = -0.1$, $k = 1.2$, and $\varepsilon = 0.0125$. The basins contain two periodic attractors and one diverging trajectory (in white). The basin of one of the attractors is barely visible but is presented in red, corresponding to a period-36 attractor.

\subsection{Coupled Lorenz Model}
\begin{keywrds}RB, ODE\end{keywrds}
\begin{figure}
\begin{center}
\includegraphics[width=\columnwidth]{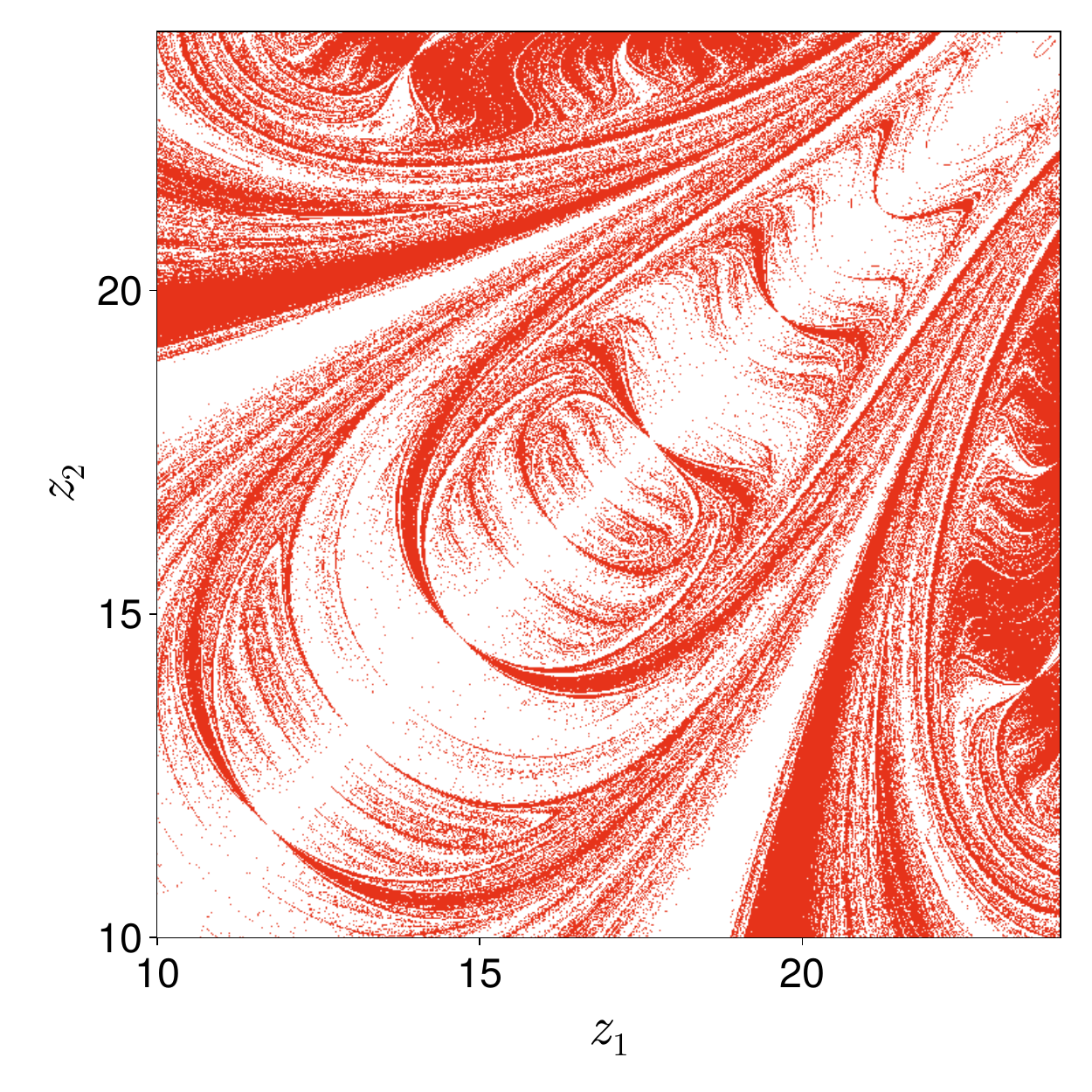}
\end{center}
\caption{\label{fig:coupled_lorenz}Basins of attraction of two identical coupled Lorenz models described in Eqs.~\ref{eq:coupled_lorenz}. The two symmetric attractors correspond to synchronized oscillators for the parameters $\alpha = 10$, $\beta = 24.76$, $\gamma = \frac{8}{3}$, and $\varepsilon = 1.1$.}
\end{figure}

Wontchui et al. in~\cite{Wontchui_2017} coupled two identical chaotic Lorenz oscillators and set the parameters close to a subcritical Hopf bifurcation. A single oscillator, in the regime near the bifurcation, is already multistable with four attractors. When the coupling is enabled, the system can converge to a synchronized, antisynchronized, or desynchronized state. Additionally, riddled basins appear from small to moderate coupling strength. 

The coupled Lorenz oscillators are described by:
\begin{align}\label{eq:coupled_lorenz}
\begin{split}
\dot{x}_1 &= \alpha(y_1 - x_1) \\
\dot{y}_1 &= \beta x_1 - y_1 - x_1 z_1 \\
\dot{z}_1 &= -\gamma z_1 + x_1 y_1 + \varepsilon(z_2 - z_1) \\
\dot{x}_2 &= \alpha(y_2 - x_2) \\
\dot{y}_2 &= \beta x_2 - y_2 - x_2 z_2 \\
\dot{z}_2 &= -\gamma z_2 + x_2 y_2 + \varepsilon(z_1 - z_2).
\end{split}
\end{align}
with $\alpha = 10$, $\beta = 24.76$, $\gamma = \frac{8}{3}$, and $\varepsilon = 1.1$ for Fig.~\ref{fig:coupled_lorenz}. Two symmetric synchronized attractors have been detected in this projection on the $z_1, z_2$ plane with the initial conditions for the other variables $x_1 = y_1 = x_2 = y_2 = 1$. The two systems are either synchronized on the same attractor or antisynchronized. In the latter case, each subsystem is oscillating on a symmetric oscillator. The basin boundary between the two stable states is riddled.

\subsection{Thomas Cyclical Oscillator}
\begin{keywrds} SMB, ODE \end{keywrds}
\begin{figure}
\begin{center}
\includegraphics[width=\columnwidth]{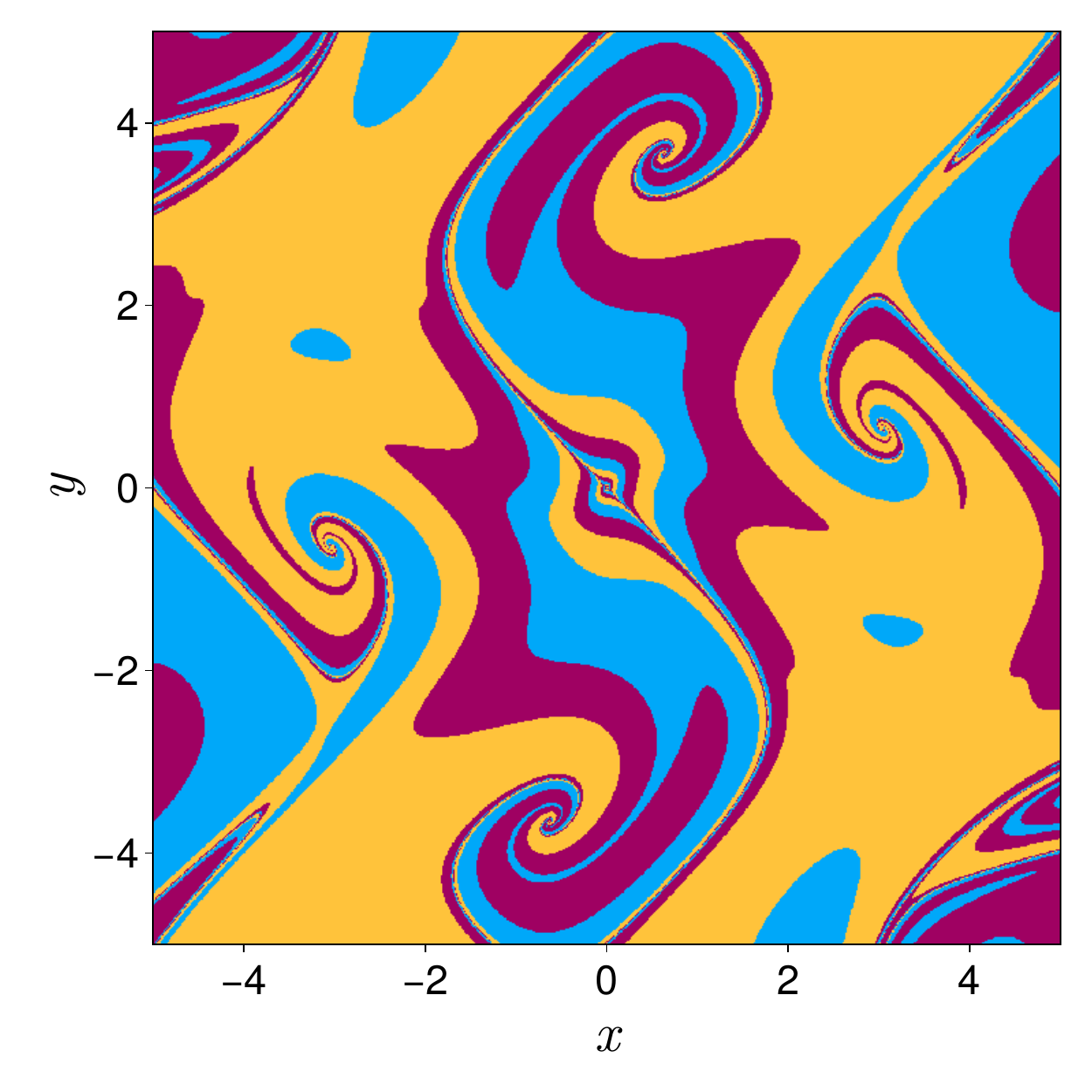}
\end{center}
    \caption{\label{fig:thomas}Basins of attraction of the Thomas cyclical oscillator in Eqs.~\ref{eq:thomas}. The basins correspond to three interlaced periodic orbits. The basins have been computed for $b = 0.1665$.}
\end{figure}

Thomas' cyclical oscillator is a symmetric system originally proposed by René Thomas~\cite{thomas1999deterministic}: 
\begin{align}\label{eq:thomas}
\begin{split}
\dot{x} &= \sin(y) - bx\\
\dot{y} &= \sin(z) - by\\
\dot{z} &= \sin(x) - bz
\end{split}
\end{align}
It has a simple form cyclically symmetric in the $x$, $y$, and $z$ variables and can be viewed as the trajectory of a frictionally damped particle moving in a 3D lattice of forces. For the parameter $b = 0.1665$, the system exhibits interesting multistable behavior reported in~\cite{datseris2022effortless}. The original publication mentioned the existence of multiple stable states but does not delve into the structure of the basins. In Fig.~\ref{fig:thomas}, we can observe a projection on the plane defined by $x$ and $y$ with the initial condition $z_0 = 0$. There are three interlaced stable periodic orbits. This symmetry results from the invariance under cyclic permutation of the variables $x$, $y$, and $z$.

\subsection{Slim Fractals\label{slim_fractal}}
\begin{keywrds}SLB, ODE\end{keywrds}
\begin{figure}
\begin{center}
\includegraphics[width=\columnwidth]{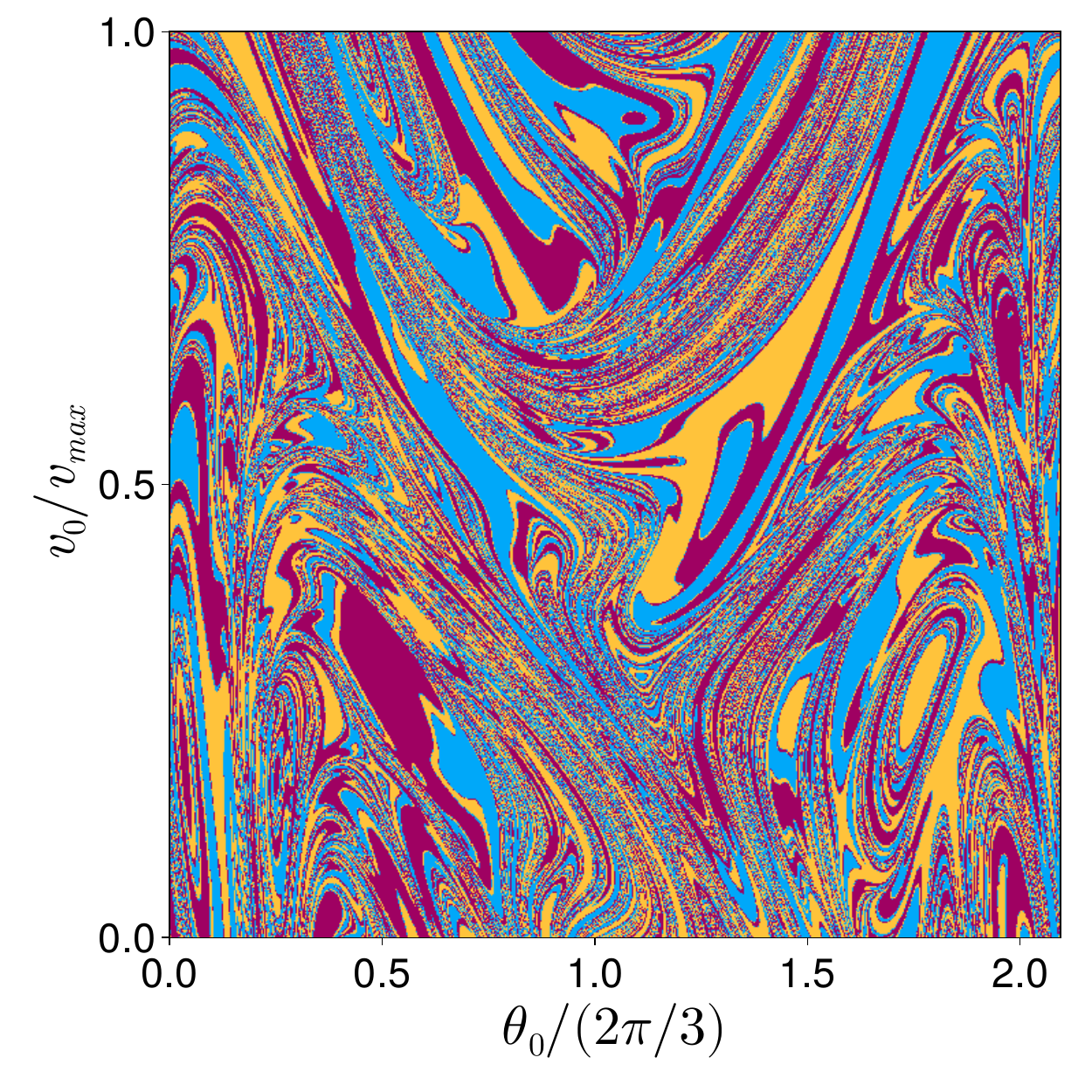}
\end{center}
\caption{\label{fig:slim_fractal}Basin of attraction of the roulette system in Eqs.~\ref{eq:slim_fractal} with the potential function $U(r,\theta) = -r^3\cos(3\theta) + \dfrac{3}{4} r^4$. Although the basins appear fractal, the boundary is smooth at a small scale (below a $10^{-16}$ resolution).} 
\end{figure}

The magnetic pendulum in Sec.~\ref{sec:mag_pend} undergoes a double transient in time and space. The orbits settle to a fixed point after a transient, and the boundary seems fractal unless observed closely enough. In~\cite{Chen_2017}, the researchers study the geometric properties of the basin boundaries in undriven dissipative systems exhibiting doubly transient chaos. Specifically, they aim to determine whether these boundaries are true or slim fractals. To quantify the notion of slim fractal, they introduce a new measure called the ``equivalent dimension'' to capture the sensitive dependence on initial conditions across different scales.

The authors propose a roulette-like dissipative system where a unit mass particle evolves in a carefully designed energy potential. The general equations of motion are: 
\begin{align}\label{eq:slim_fractal}
\begin{split}
\ddot x + \mu \dot x &= -\frac{\partial U}{\partial x}\\
\ddot y + \mu \dot y &= -\frac{\partial U}{\partial y}
\end{split}
\end{align}
Among the proposed potentials, we reproduce the basins generated by $ U(r,\theta) = -r^3\cos(3\theta) + \dfrac{3}{4} r^4 $. This is a three-well potential with a soft barrier between the minima. Using high-precision numerical simulations, the authors demonstrate that the basins generated by this potential are smooth at a very small scale. The basins in Fig.~\ref{fig:slim_fractal} appear fractal at first glance, pushing the authors to propose a novel method for measuring the fractal dimension and defining these objects as slim fractals. The order of magnitude at which the fractal pattern breaks is $10^{-16}$, which is far below the usual concept of ``small'' perturbation. Thus, in practical situations, the unpredictability will be determined by the patterns at larger scales of the boundary. It remains a measurable effect in the phase space of a general class of dynamical systems, namely undriven dissipative systems.

\subsection{Piecewise Smooth Dynamical System}
    \begin{keywrds} FB, MAP \end{keywrds}
\begin{figure}
\begin{center}
\includegraphics[width=\columnwidth]{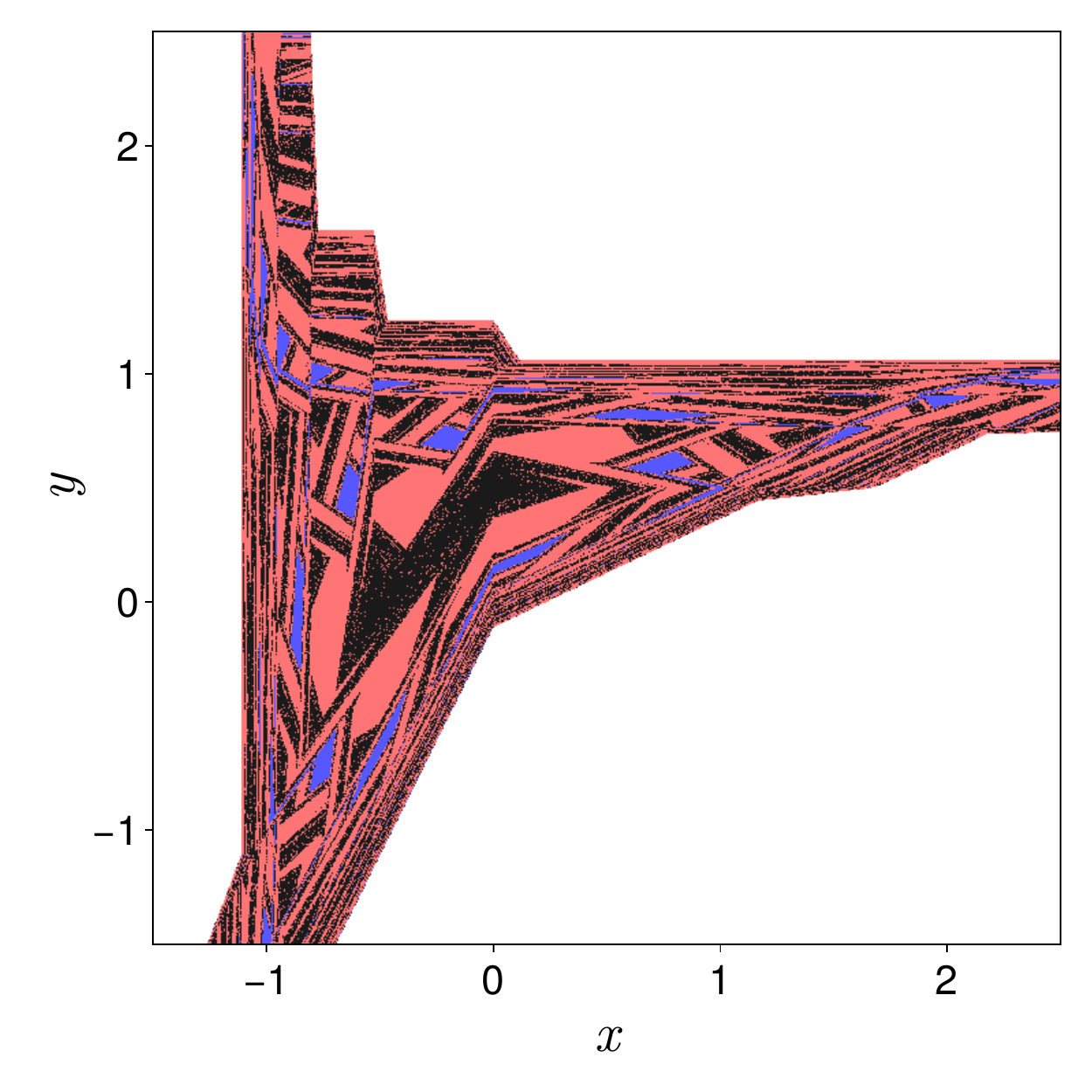}
\end{center}
    \caption{\label{fig:pcw_lai}Basins of attraction of~\ref{eq:pcw_lai}, a discrete piecewise dynamical system. The basins correspond to three periodic attractors and trajectories diverging (in white).}
\end{figure}

The paper~\cite{do2008multistability} investigates multistability in piecewise smooth dynamical systems, focusing on the phenomenon of arithmetically period-adding bifurcations. As the system approaches the weakly dissipative regime, it can exhibit periodic attractors through a series of saddle-node bifurcations. The newly created attractors have periods following an arithmetic sequence, a feature notably absent in smooth dynamical systems. The authors present a detailed study of the interwoven structure of the basins of attraction, highlighting the differences from the basins observed in smooth systems.

The discrete dynamical system is expressed as:
\begin{equation}\label{eq:pcw_lai}
X_{n+1} = F(X_n) = 
\begin{cases} 
f_0(X_n) & \text{if } X_n \in S_0 \\ 
f_1(X_n) & \text{if } X_n \in S_1 
\end{cases}
\end{equation}
where $X_n = (x_n, y_n) \in \mathbb{R}^2$, and the regions $S_0=\{x \leq 0, y\in \mathbb{R}\}$ and $S_1 = \{x >0, y\in\mathbb{R}\}$ define the domains of the two piecewise smooth functions. The specific affine subsystems can be described as:
\begin{align}
f_0(X_n) = 
\begin{pmatrix} 
a & 1 \\ 
b & 0 
\end{pmatrix}
\begin{pmatrix} 
x_n \\ 
y_n 
\end{pmatrix} + 
\begin{pmatrix} 
\mu \\ 
0 
\end{pmatrix}, 
\end{align}
\begin{align}
f_1(X_n) = 
\begin{pmatrix} 
c & 1 \\ 
d & 0 
\end{pmatrix}
\begin{pmatrix} 
x_n \\ 
y_n 
\end{pmatrix} + 
\begin{pmatrix} 
\mu \\ 
0 
\end{pmatrix}.
\end{align}
The basins in Fig.~\ref{fig:pcw_lai} have been computed with the parameters $a = -2$, $b = -0.95$, $d = b$, and $\mu = 1$. There are three periodic attractors of periods 2, 5, and 8, along with trajectories approaching infinity. The article is a good example of the analysis of a system near the conservative regime, and the authors prove the existence and stability of periodic orbits using symbolic representation and classical stability analysis. 

\section{Open Systems Examples}

Open systems are characterized by transient dynamics before the trajectory diverges. In this section, we present a few open Hamiltonian systems, which follow a conservation principle. This class of dynamical systems is of central importance in physics. The studies include models of chaotic scattering of trajectories where a particle interacts with an invariant structure known as a chaotic saddle before escaping to infinity. The scattering can occur in smooth potentials or through impacts on hard walls, as happens in billiards.

Open Hamiltonian systems may seem boring at first sight; we set up an initial condition in a bounded region, and the particle eventually diverges to infinity. However, if the original bounded region has several escape paths, we can track the trajectory and associate an exit with an initial condition, forming what we call escape basins. Under certain conditions, a fractal structure appears on the boundary between the basins of possible escapes. The reader will find examples of conservative dynamical systems in other sections (see Sec.~\ref{sec:cold_atoms} and Sec.~\ref{sec:binary_BH}).

\subsection{The Hénon–Heiles Open Hamiltonian}
\begin{keywrds}FB, WD, OH\end{keywrds}
\begin{figure}
\begin{center}
\includegraphics[width=\columnwidth]{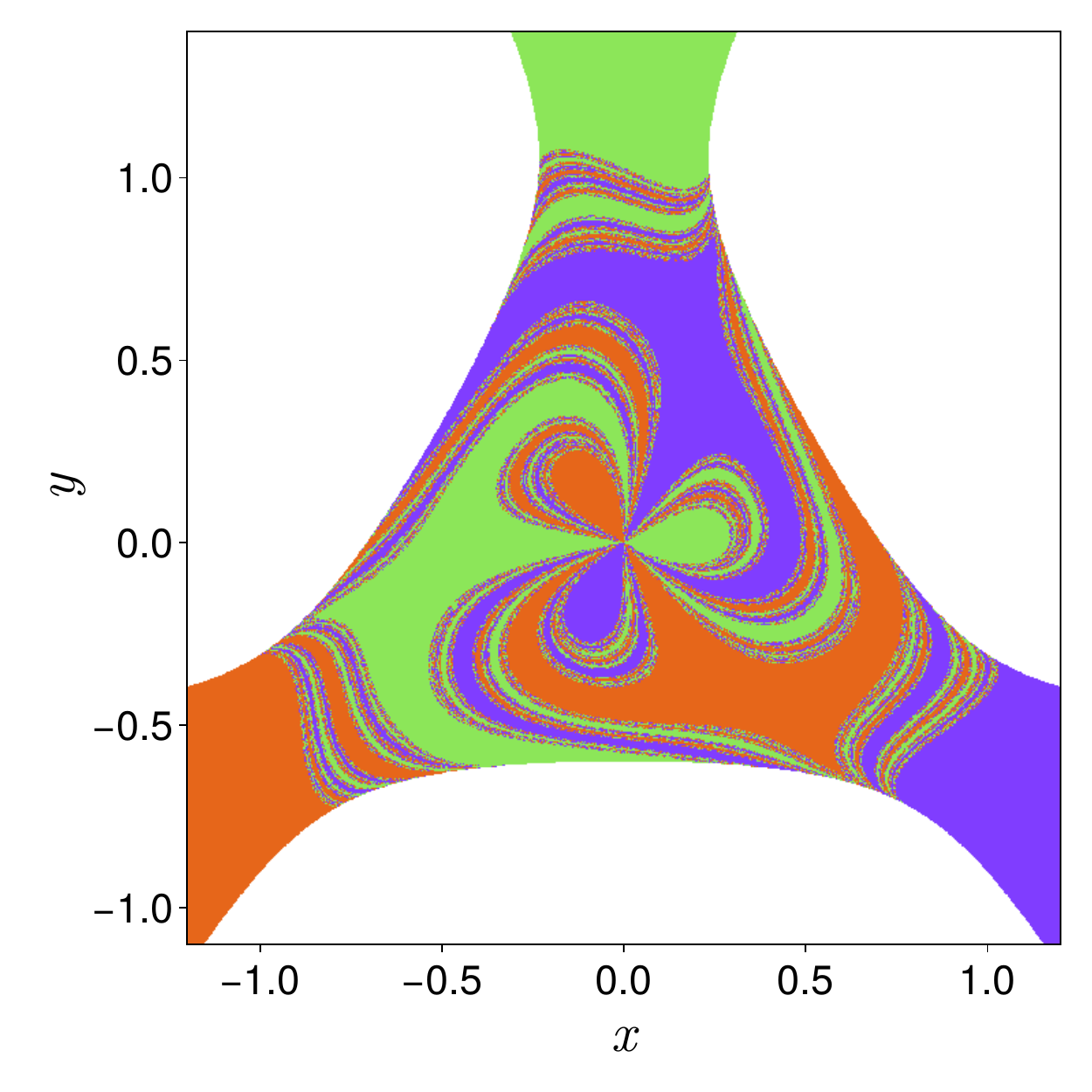}
\end{center}
    \caption{\label{fig:henonheiles}Exit basins of the Hénon–Heiles Hamiltonian system in the $x-y$ space (Eqs.~\ref{eq:hh_pot}). There are three exits reachable from the center of the potential. The energy is set to $E = 0.25$.}
\end{figure}

The Hénon–Heiles potential originally modeled the time evolution of celestial bodies within a two-dimensional galactic potential. The energy of the evolving mass is conserved over time and obeys Newton's laws of dynamics. For energy values below a certain threshold, the motion of the particle is bounded in a region close to the origin. However, above a critical level, the mass can escape to infinity through a limited number of exits. A particle arriving from outside in a straight line to the region near the origin will interact with the potential and eventually exit through one of the openings. Since the motion in the region of interest is chaotic, the particle is scattered after some chaotic transient. This is one of the reasons for its popularity in modeling chaotic scattering processes~\cite{seoane2012new}.

The Hamiltonian energy of the potential is described as:  
\begin{equation} \label{eq:hh_pot}
V(x,y) = \dfrac{1}{2} (x^2 + y^2) + \dfrac{1}{2}(x^2 y - \dfrac{1}{3} y^3)
\end{equation}
The equations of motion for the particle are: 
\begin{align}
\begin{split}
\dot x &= \frac{\partial H}{\partial p_x} = p_x, \\
\dot y &= \frac{\partial H}{\partial p_y} = p_y, \\
\dot p_x &= -\frac{\partial H}{\partial x} = - x - 2xy,\\ 
\dot p_y &= -\frac{\partial H}{\partial y} = - y - (x^2 - y^2). 
\end{split}
\end{align}
where $H$ is the Hamiltonian energy: $H = 0.5(p_x^2 + p_y^2) + V(x,y)$. There are no attractors in this system, so the algorithm described in Sec.~\ref{sec:num_met} is ineffective. Instead, we can associate an exit with specific initial conditions. The set of initial conditions leading through the same exit is called an escape basin. The nature of the escape basins has been studied in~\cite{aguirre_wada_2001}, and an example is reproduced in Fig.~\ref{fig:henonheiles} for an energy $E=0.25$. Since the system is four-dimensional, the initial conditions are established using the tangential shooting method to determine the initial momenta $p_x^0$ and $p_y^0$. Given $x_0$, $y_0$, and $E$, we choose zero radial velocity and positive angular velocity. This translates into the Cartesian coordinates:
\begin{align}
\begin{split}
p_x^0 &= \dot x_0 = -\frac{y_0}{\sqrt{x_0^2 + y_0^2}}\sqrt{2(E - V(x_0,y_0))}\\
p_y^0 &= \dot y_0 = -\frac{x_0}{\sqrt{x_0^2 + y_0^2}}\sqrt{2(E - V(x_0,y_0))}
\end{split}
\end{align}
Once the distance to the origin of the trajectory has reached a certain threshold, the solver stops. We deduce from the position which exit the particle has passed through. The result shown in Fig.~\ref{fig:henonheiles} has the Wada property, as the boundary between the three exits is unique. These rich dynamical properties have generated an endless list of publications in the study of nonlinear dynamical systems and their applications.

\subsection{Open Sinai Billiards}
\begin{keywrds}WD, MAP\end{keywrds}

\begin{figure}
\begin{center}
\subfloat[]{\includegraphics[width=\columnwidth]{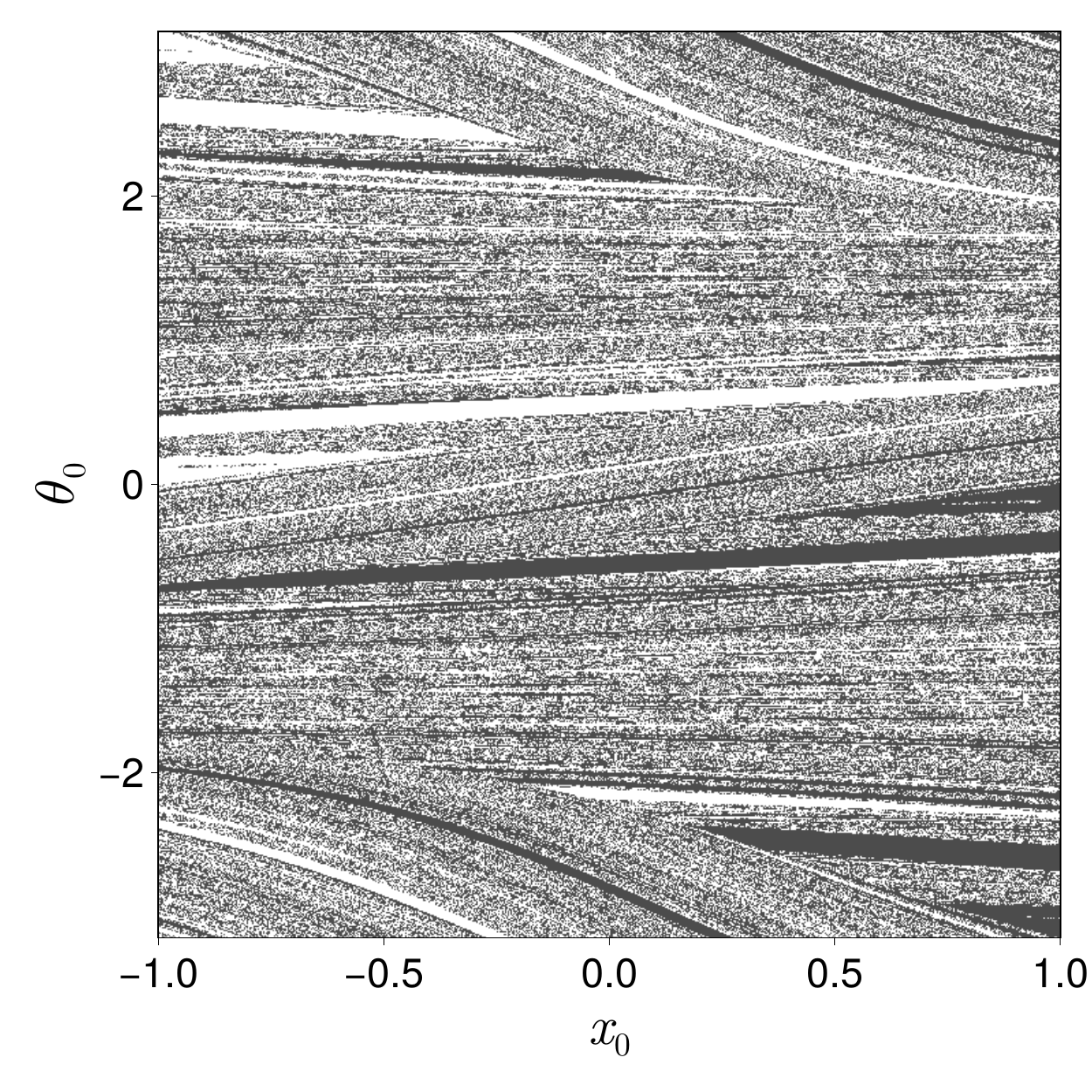}}\\
\subfloat[]{\begin{tikzpicture}

    \def\L{6}     
    \def\a{1}     
    \def\b{2}     
    \def\r{1.3}   

    
    \draw[thick] (0,0) -- (0,\L) -- (\L,\L) -- (\L,0);   
    \draw[thick] (0,0) -- (\a,0);
    \draw[thick] (\a + 1,0) -- (\a + 1 + \b,0);
    \draw[thick] (\a + 1 + \b + 1,0) -- (\L,0);

    \draw[thick] (\L/2,\L/2) circle (\r);

    \draw[<->] (0,\L+0.5) -- (\L,\L+0.5) node[midway, above] {$L$};
    \draw[<->] (\L+0.5,0) -- (\L+0.5,\L) node[midway, right] {$L$};

    \draw[<->] (0,-0.5) -- (\a,-0.5) node[midway, below] {$a$};
    \draw[<->] (\a + 1,-0.5) -- (\a+ 1 + \b,-0.5) node[midway, below] {$b$};
    \draw[<->] (\a+\b + 2,-0.5) -- (\L,-0.5) node[midway, below] {$a$};

    \node at (\a + 0.5,0) [below] {A};
    \node at (\a+\b + 1.5,0) [below] {B};

    \draw[->] (\L/2,\L/2) -- (\L/2+\r, \L/2) node[midway, above] {$r$};


    \draw[dashed] (\L/2 - \r*0.707, 3*\L/4) -- (\L/2+\r*0.707, 3*\L/4);
    
    \coordinate (P) at ({\L/2 - \r*0.707}, {3*\L/4});

    \draw[thick] (P) -- ++(0,1.2);

    \draw[thick, ->] (P) -- ++(-1,0.5);

    \draw[thick, -> ] ($(P)+(0,0.5)$) arc (90:155:0.5);
    \node at ($(P)+(-0.4,0.7)$) {$\theta_0$};

    \node at ($(P)+(0,-0.3)$) {$x_0$};

    \fill (P) circle (2pt);
\end{tikzpicture}}
\end{center}
\caption{\label{fig:open_sinai}(a) Escape basins of the billiard described in the schematics (b). Dimensions are: $a = 0.1$, $b = 0.2$, $L = 4.0$, $r = 1$. The size of each exit is $\Delta = 0.8$.}
\end{figure}
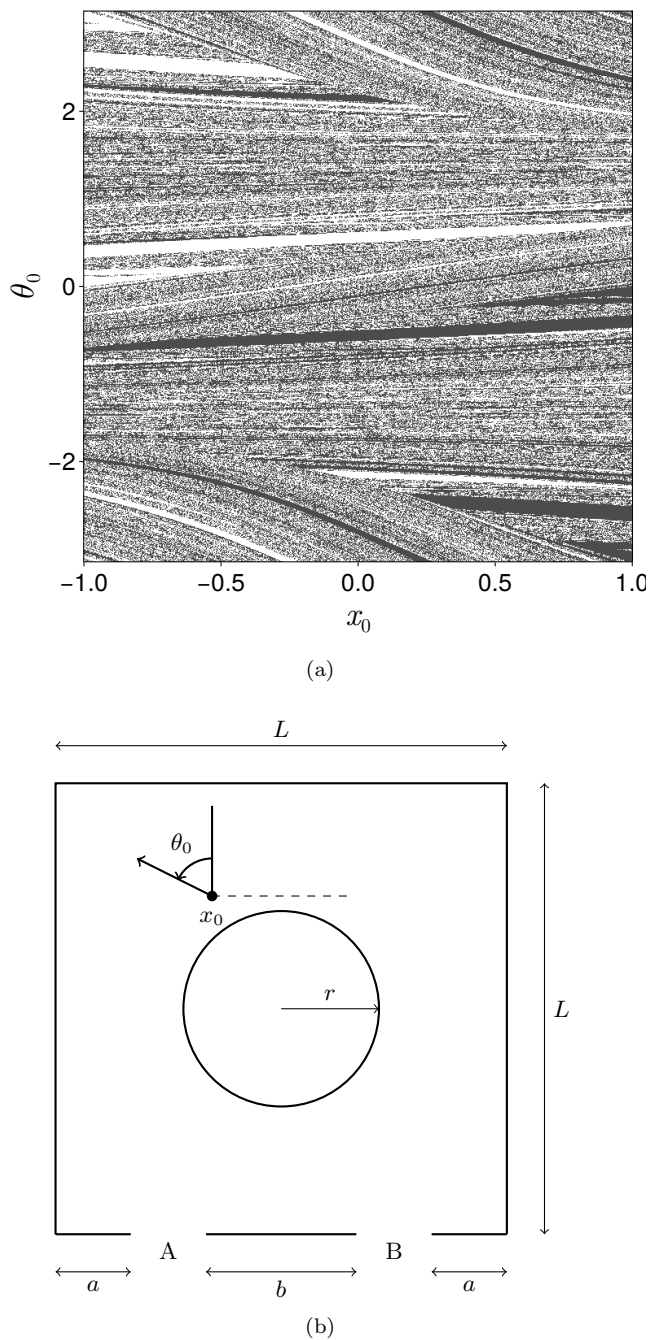

An alternative to open Hamiltonian potentials is open billiards, where a massless particle bounces off on hard walls in a region until it escapes from the initial area. The billiard presented in~\cite{Bleher_1988} consists of a square box with a hard disk at its center. If a particle is launched in the box, it collides with the disk and the walls until the simulation is stopped. It is well known that, when launched appropriately, the particle will visit all areas of the interior and will never reach a stable periodic orbit.

In~\cite{Bleher_1988}, the authors allow the particle to escape through two possible holes in one wall of the box, as shown in Fig.~\ref{fig:open_sinai} (b). Particles launched with unit velocity from a coordinate $x_0$ along the dashed line, with an angle $\theta_0$ measured counterclockwise from the vertical, will exit through hole A or B in a fractal pattern, as indicated in Fig.~\ref{fig:open_sinai} (a). The escape basins are indeed fractal, as measured in the article. The dashed line has a length of 2 and is situated at $y=1.25$ above the center of the circle. 

To simulate this system, a specialized library~\cite{datseris2017dynamicalbilliards} implemented in the Julia programming language simulates the trajectory of the particle in this environment. The trajectory is evolved until a crossing through one of the holes is detected. At this point, we store the exit associated with the initial condition.

The authors of~\cite{Bleher_1988} conjectured that such fractal patterns are common in Hamiltonian systems with multiple exits. This has been confirmed in many subsequent publications.

\subsection{Wada Boundaries in Chaotic Scattering}
\begin{keywrds}WD, MAP\end{keywrds}
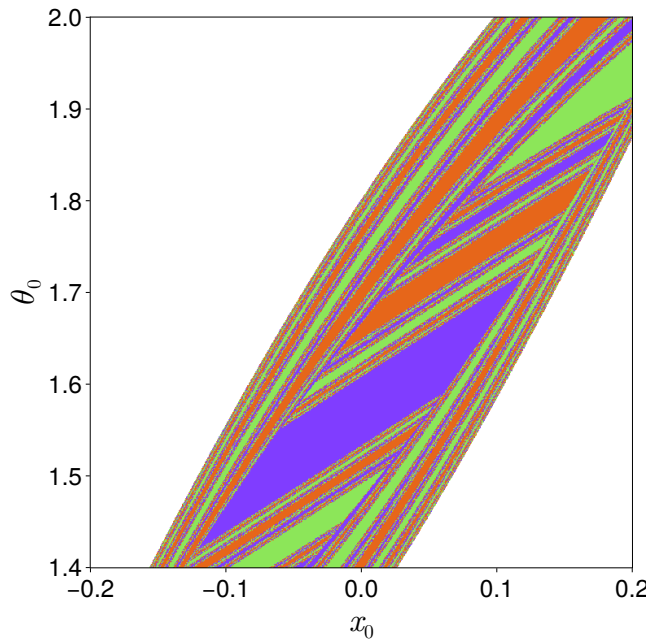
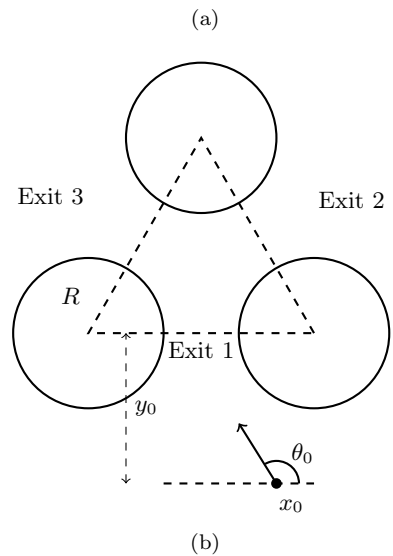
\begin{figure}
\begin{center}
\subfloat[]{\includegraphics[width=\columnwidth]{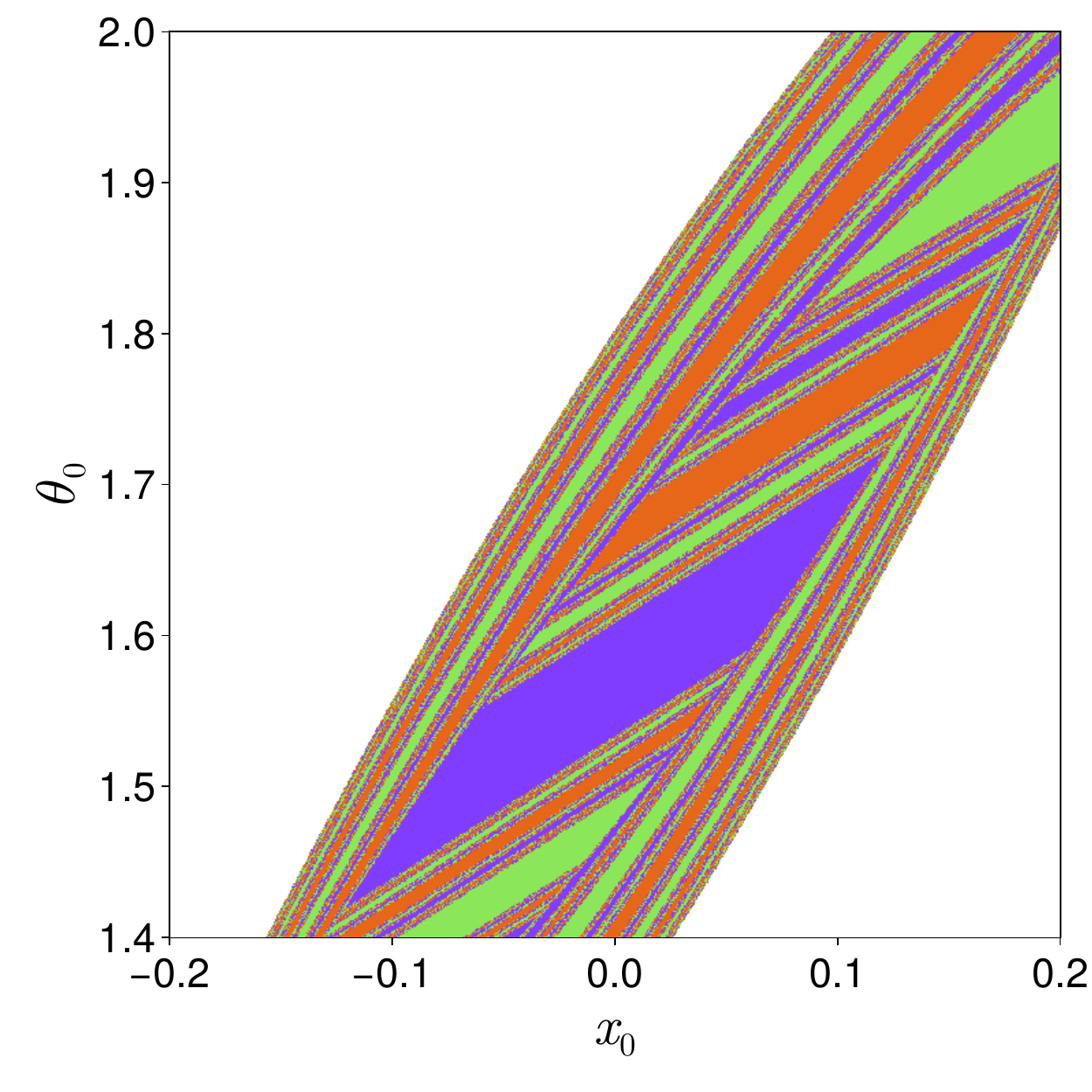}}\\
\subfloat[]{\begin{tikzpicture}

    \def\R{1.}
    \def\w{3}
    
    \coordinate (C1) at (0, 0);             
    \coordinate (C2) at (\w, 0);            
    \coordinate (C3) at (\w/2, {sqrt(3)*\w/2});  

    \draw[thick] (C1) circle (\R); 
    \draw[thick] (C2) circle (\R);
    \draw[thick] (C3) circle (\R);

    \draw[thick, dashed] (C2) -- (C3) -- (C1) -- cycle;
 
    \node at ($(C1)+(0,0.5)$) [left] {$R$};

    \node[below] at ($(C1)+(1.5,0)$) {Exit 1};
    \node[below] at ($(C2)+(0.5,2)$) {Exit 2};
    \node[above] at ($(C3)+(-2,-1.)$)  {Exit 3};


    \coordinate (P) at (2.5, -2);
    \draw[thick, ->] (P) -- ++(-0.5,0.8);
    \draw[thick] ($(P)+(0.3,0)$) arc (0:120:0.3);
    \node at ($(P)+(0.35,0.4)$) {$\theta_0$};
    \node at ($(P)+(0.2,-0.3)$) {$x_0$};
    \fill (P) circle (2pt);
    \draw[thick, dashed] (1, -2) -- (3, -2) ;  
    \draw[<->,dashed] (.5,-2) -- (.5,0) node[midway, right] {$y_0$};
\end{tikzpicture}}
\end{center}
\caption{\label{fig:poon_wada}(a) Escape basins of the billiard described in the schematics (b). The boundary is fractal and has the Wada property. Parameters are $y_0 = -0.5$.}
\end{figure}

In~\cite{poon1996wada}, the research focuses on the existence of the Wada property in the exit basins of open Hamiltonian systems with multiple exit modes. The authors simulated a chaotic billiard system composed of three circular disks arranged in an equilateral triangle, as shown in Fig.~\ref{fig:poon_wada} (b). They analyzed the exit modes based on initial conditions defined by the position and direction of incoming particles. Using graphical inspection methods, they established the presence of the Wada property.

The study found that the basin boundaries in this chaotic scattering system are fractal and Wada. The setup in Fig.~\ref{fig:poon_wada} (b) is slightly different from the original publication, as the three disks are tilted slightly with respect to the launching segment at the bottom. Nevertheless, the results are unaltered, and the Wada property is conserved. In Fig.~\ref{fig:poon_wada} there are four colors since some initial conditions may never lead to the interior of the three disks and may diverge directly to $\infty$. The centers of the three disks are separated by a unit distance and the radius is set to $R = (1 - w)/2$, where $w = 0.1$ is the minimal distance between two disks (the exit).

This is a classical example of chaotic scattering of a particle bouncing on hard disks before escaping in an unpredictable direction.

\subsection{Basins in the Limit of Small Exits}
\begin{keywrds}IB, MAP\end{keywrds}
\begin{figure}
\begin{center}
\subfloat[]{\includegraphics[width=\columnwidth]{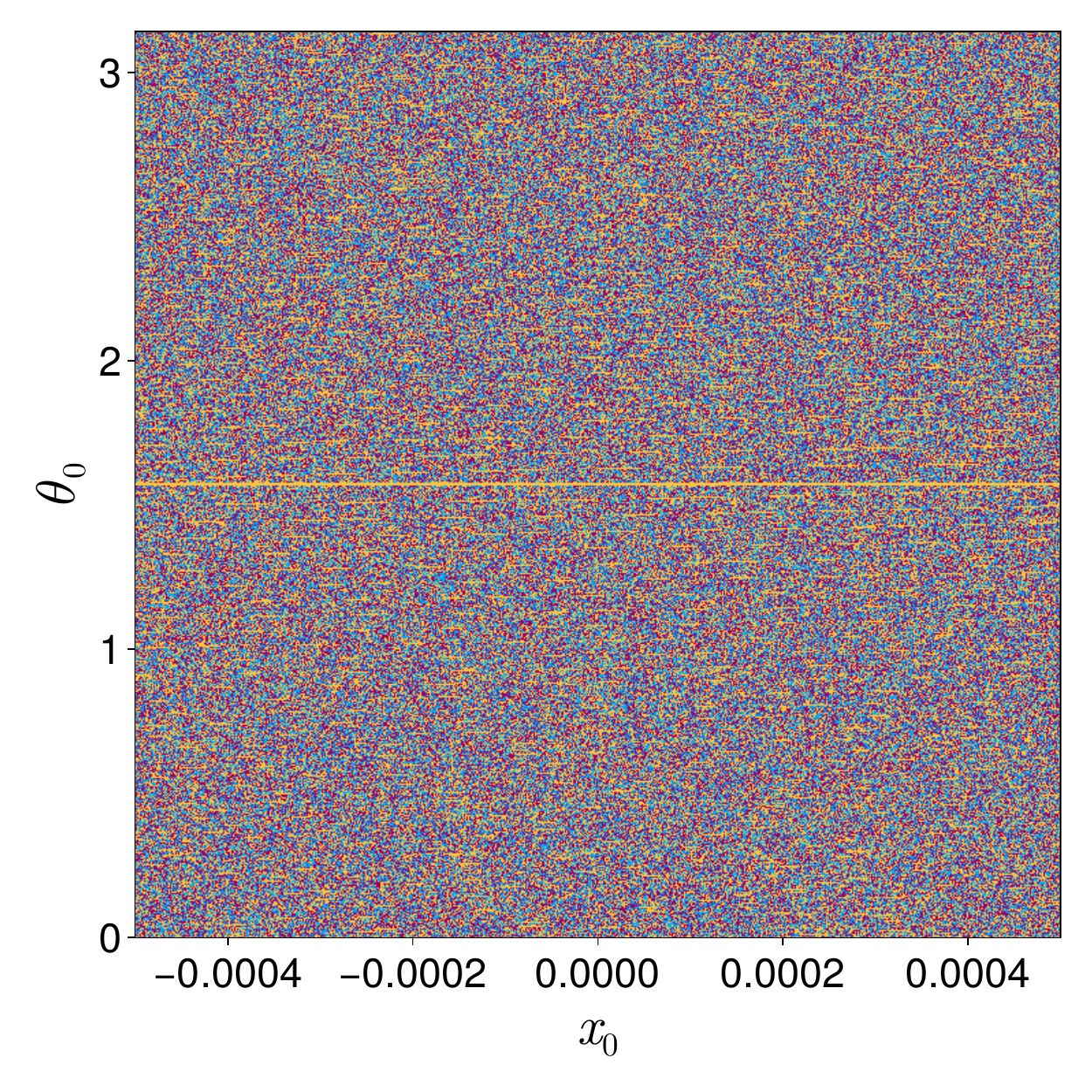}}\\

\subfloat[]{\begin{tikzpicture}

    \def\R{1.}
    \def\w{3}
    
    \coordinate (C1) at (0, 0);             
    \coordinate (C2) at (\w, 0);            
    \coordinate (C3) at (\w/2, {sqrt(3)*\w/2});  

    \draw[thick] (C1) circle (\R); 
    \draw[thick] (C2) circle (\R);
    \draw[thick] (C3) circle (\R);

    \draw[thick, dashed] (C2) -- (C3) -- (C1) ;
 
    \node at ($(C1)+(0,0.5)$) [left] {$R$};

    \draw[thick] ($(C1) + ({\R*0.5}, {\R*sqrt(3)/2})$) -- ($(C3) - ({\R*0.5}, {\R*sqrt(3)/2})$)  ;
    \node at ($(C1) + ({1.5*0.5}, {1.5*sqrt(3)/2})$) [left] {$w$};
    \node[below] at ($(C1)+(1.5,-1)$) {Exit 1};
    \node[below] at ($(C2)+(0.5,2)$) {Exit 2};
    \node[above] at ($(C3)+(-2,-1.)$)  {Exit 3};

    \draw[->, thick, dashed] (-0.5, 0) -- (4.5, 0) node[right] {$x$};  
    \draw[->, thick, dashed] (1.5, -1) -- (1.5, 4) node[above] {$y$};  

    \coordinate (P) at (1.5, 0);
    \draw[thick, ->] (P) -- ++(-0.5,0.8);
    \draw[thick] ($(P)+(0.3,0)$) arc (0:120:0.3);
    \node at ($(P)+(0.35,0.4)$) {$\theta_0$};
    \node at ($(P)+(0.2,-0.3)$) {$x_0$};
    \fill (P) circle (2pt);
\end{tikzpicture}}
\end{center}
\caption{\label{fig:small_limit}(a) Escape basins of the billiard described in the schematics (b). When the size of the exits is reduced, the transient inside the scattering region lasts longer. The consequence is complete uncertainty regarding the initial conditions in the basins. Parameters are $R = 1$, $w = 0.001$.}
\end{figure}
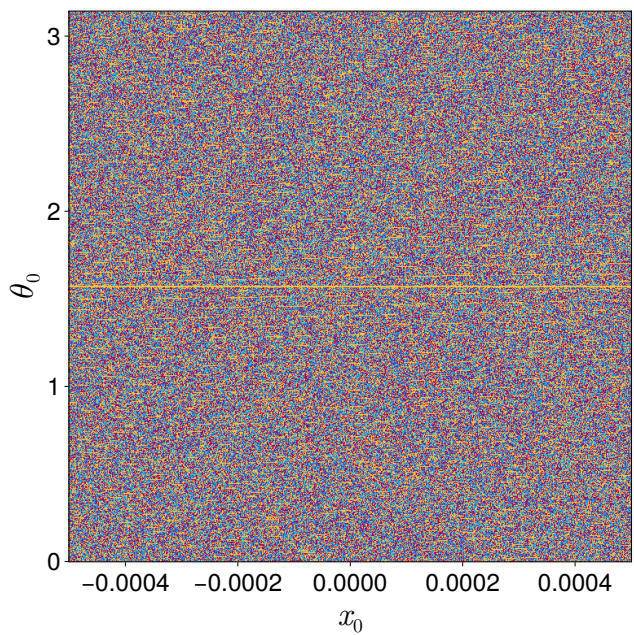
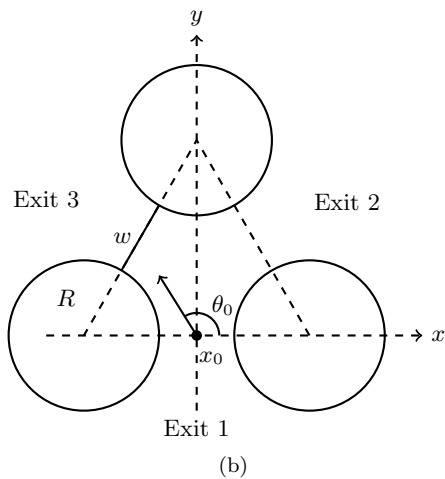

The objective of the paper~\cite{Aguirre_2003} is to analyze the structure of the exit basins, focusing on what happens when the size of the exits decreases and tends to zero. In this limit, the invariant sets tend to fill up the whole phase space, leading to boundary filling the entire basins. The implications are a complete unpredictability of the outcomes.

The hyperbolic system, a simple two-dimensional billiard consisting of three hard disks, has a width parameter $w$ determining the size of the exits. As the size of the exits decreases, the chaotic saddle and its stable and unstable manifolds tend to fill up the entire phase space. This leads to a total fractalization of the basins, where any information about the possible exit given an initial condition is lost.

The exit basins are computed by launching particles from the segment between two disks (see Fig.~\ref{fig:small_limit} (b)) at the coordinate $x_0$ and angle $\theta_0$. The simulation stops when the particle hits one of the three possible exit limits. The escape basins in Fig. \ref{fig:small_limit} (a) are completely fractalized for an exit width of $w = 0.001$.

\section{Examples in Life Science and Economic Science}

The appearance of coexisting stable solutions in a biological mechanism is not surprising. The response to an input stimulus may depend on the internal state of the system. In economic models, the focus is on the equilibrium of the different agents or entities, which depends on the initial state. Both fields have a long tradition of using mathematical modeling of these processes. The dynamical systems presented are a small peak at the rich literature on the subject.

\subsection{Multistability in the Cournot Game}
\begin{keywrds} SMB, MAP\end{keywrds}
\begin{figure}
\begin{center}
    \subfloat[]{\includegraphics[width=\columnwidth]{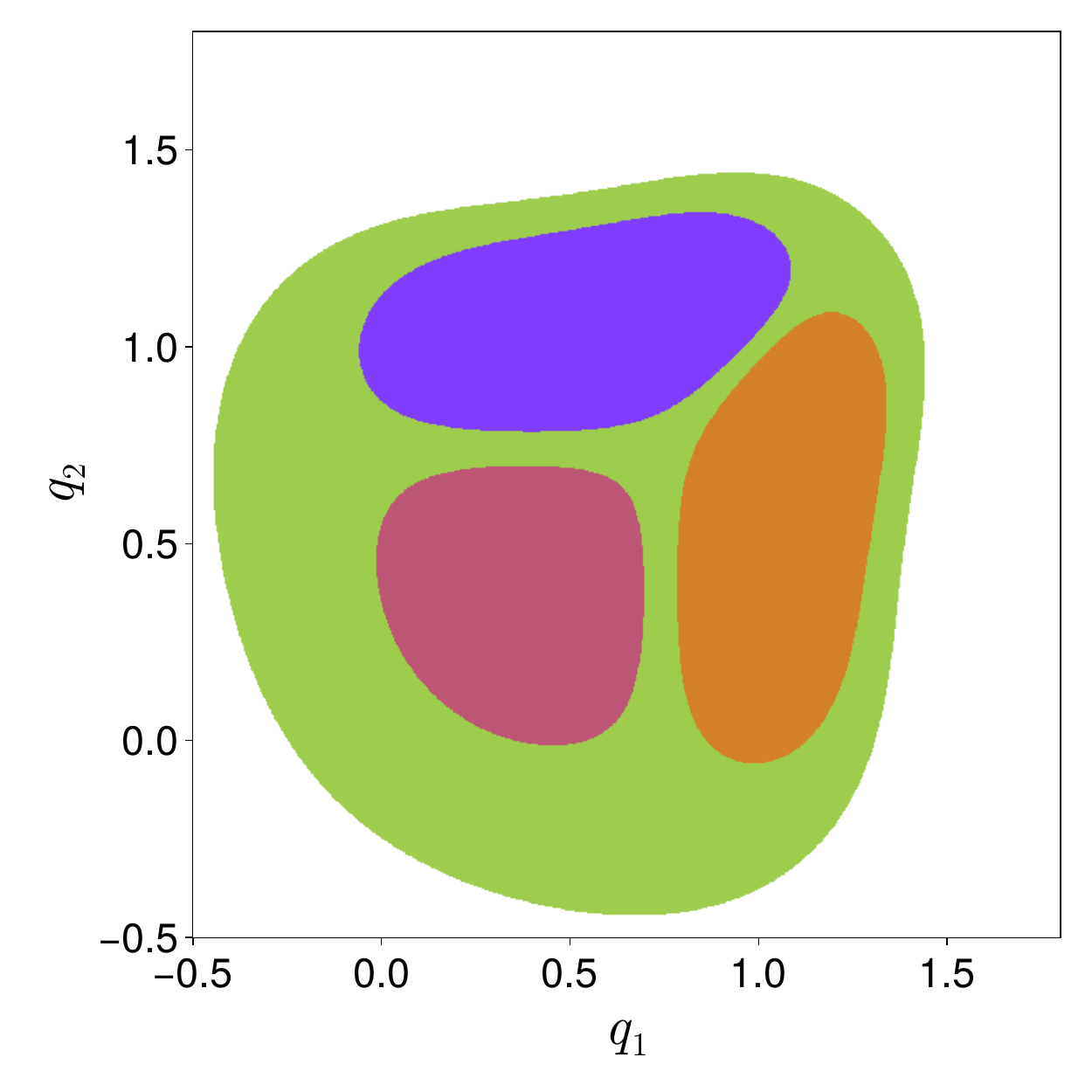}}\\
    \subfloat[]{\includegraphics[width=\columnwidth]{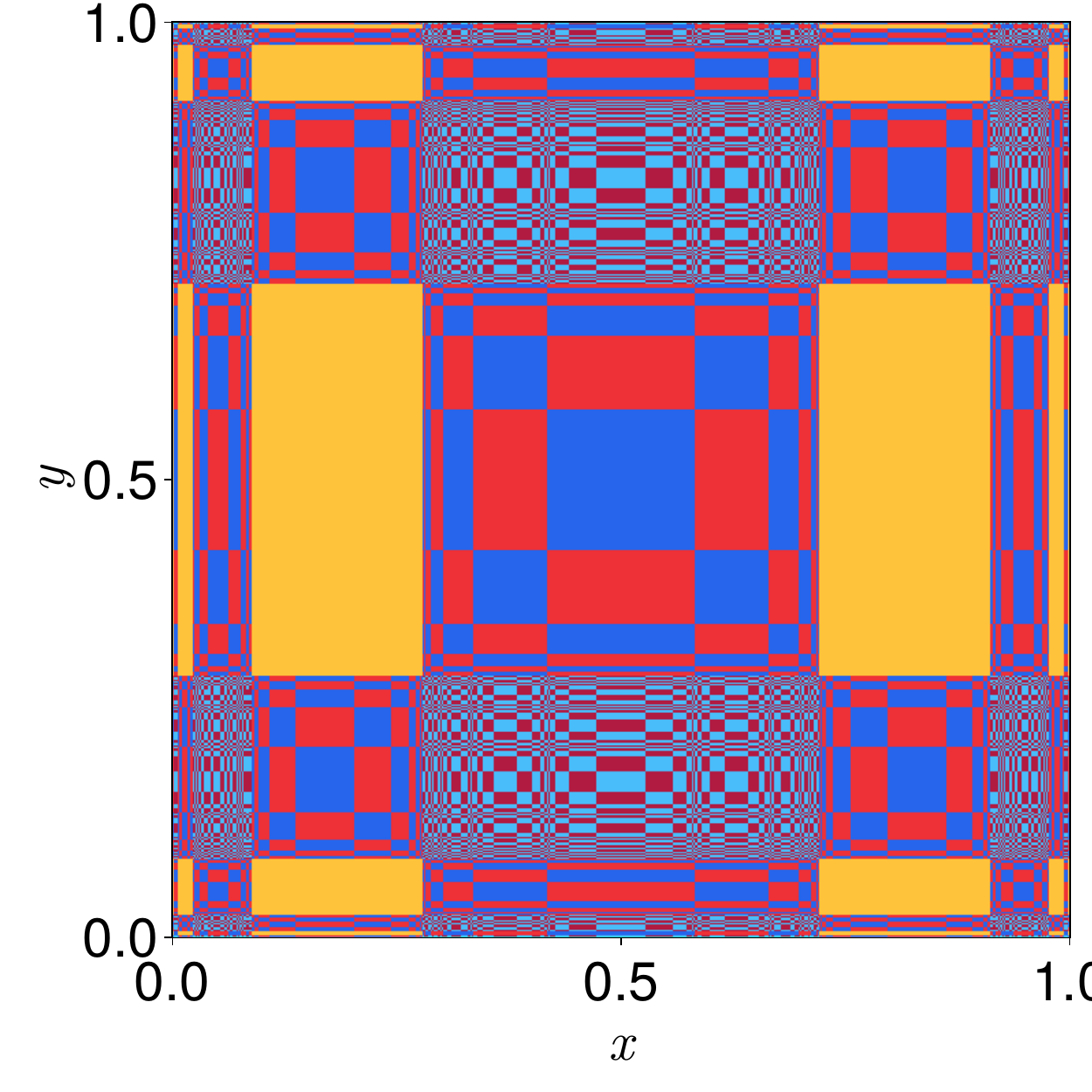}}
\end{center}
    \caption{\label{fig:nash_eq} (a) Basins of attraction of coexisting Nash equilibria in the context of the three-player Cournot game (Eqs.~\ref{eq:nash_eq}). Three identical systems represent the output of competing firms. In (b), the two-player Cournot model described by Eqs.~\ref{eq:cournot2d} is presented. The boundary is fractal, and five coexisting attractors have been found. Parameters are $\mu_1 = 3.53$ and $\mu_2 = 3.55$.}
\end{figure}

The article~\cite{agiza1999multistability} analyzes the dynamics of a Cournot oligopoly model with three competing firms, focusing on the stability of Nash equilibria. In this oligopoly model, a few firms compete and try to outperform each other. Each firm formulates a model of what the competitors might do and then acts based on this model to maximize expected profit. If the firms do not deviate from their strategy, the system settles in a Cournot-Nash equilibrium.

The study explores both the symmetric case of identical firms and the more general case of heterogeneous players. The model is formulated as a discrete dynamical system represented by a three-dimensional non-invertible map. Up to four coexisting Nash equilibria were identified, with several coexisting stable equilibria. This serves as an interesting example of how a simple economic model can display multiple stable solutions.

The simplest model in the article is a discrete nonlinear map with a symmetric structure: 
\begin{align}\label{eq:nash_eq}
\begin{split}
q_{1,n+1} &= (1 - \lambda_1) q_{1,n} + \lambda_1 \mu_1 (q_{2,n} (1-q_{2,n}) + q_{3,n} (1-q_{3,n}))\\
q_{2,n+1} &= (1 - \lambda_2) q_{2,n} + \lambda_2 \mu_2 (q_{3,n} (1-q_{3,n}) + q_{1,n} (1-q_{1,n}))\\
q_{3,n+1} &= (1 - \lambda_3) q_{3,n} + \lambda_3 \mu_3 (q_{1,n} (1-q_{1,n}) + q_{2,n} (1-q_{2,n}))
\end{split}
\end{align}
The authors studied the case where $\lambda_1 = \lambda_2 = \lambda_3 = 0.5$ and $\mu_1 = \mu_2 = \mu_3 = 1.95$. Variables $q_k$ represent the output quantity of each firm. The basins in Fig.~\ref{fig:nash_eq} (a) have been computed on the $q_1, q_2$ plane by setting $q_3 = 1 - 1/(2\mu)$. The choice of the grid for the recurrence detection requires careful consideration, as the trajectories of different attractors come close to each other and may intersect in one cell. A very fine grid is needed. The obtained basin boundaries are smooth and symmetric due to the symmetry of the original model.

In this section, we also present a related and simpler model of the Cournot game authored by Bischi et al.~\cite{bischi2000multistability}, which involves only two dimensions. The Cournot game is restricted to two agents and takes the form of two standard logistic maps with a twist: 
\begin{align}\label{eq:cournot2d}
\begin{split}
    x_{n+1} &=  \mu_1 (1 - y_n) y_n\\ 
    y_{n+1} &=  \mu_2 (1 - x_n) x_n\\ 
\end{split}
\end{align}
The variables $x$ and $y$ depend solely on the other variable. The competition settles into multiple equilibria depending on the parameters. A slight asymmetry is introduced in the parameters of the model, and the results are shown in Fig.~\ref{fig:nash_eq} (b). Five coexisting attractors are displayed for the parameters $\mu_1 = 3.53$ and $\mu_2 = 3.55$. The article~\cite{bischi2000multistability} focuses on proving analytically the dynamical properties of the system described by Eq.~\ref{eq:cournot2d}.

\subsection{Economic Geographic Model}
\begin{keywrds}FB, MAP\end{keywrds}
\begin{figure}
\begin{center}
\includegraphics[width=\columnwidth]{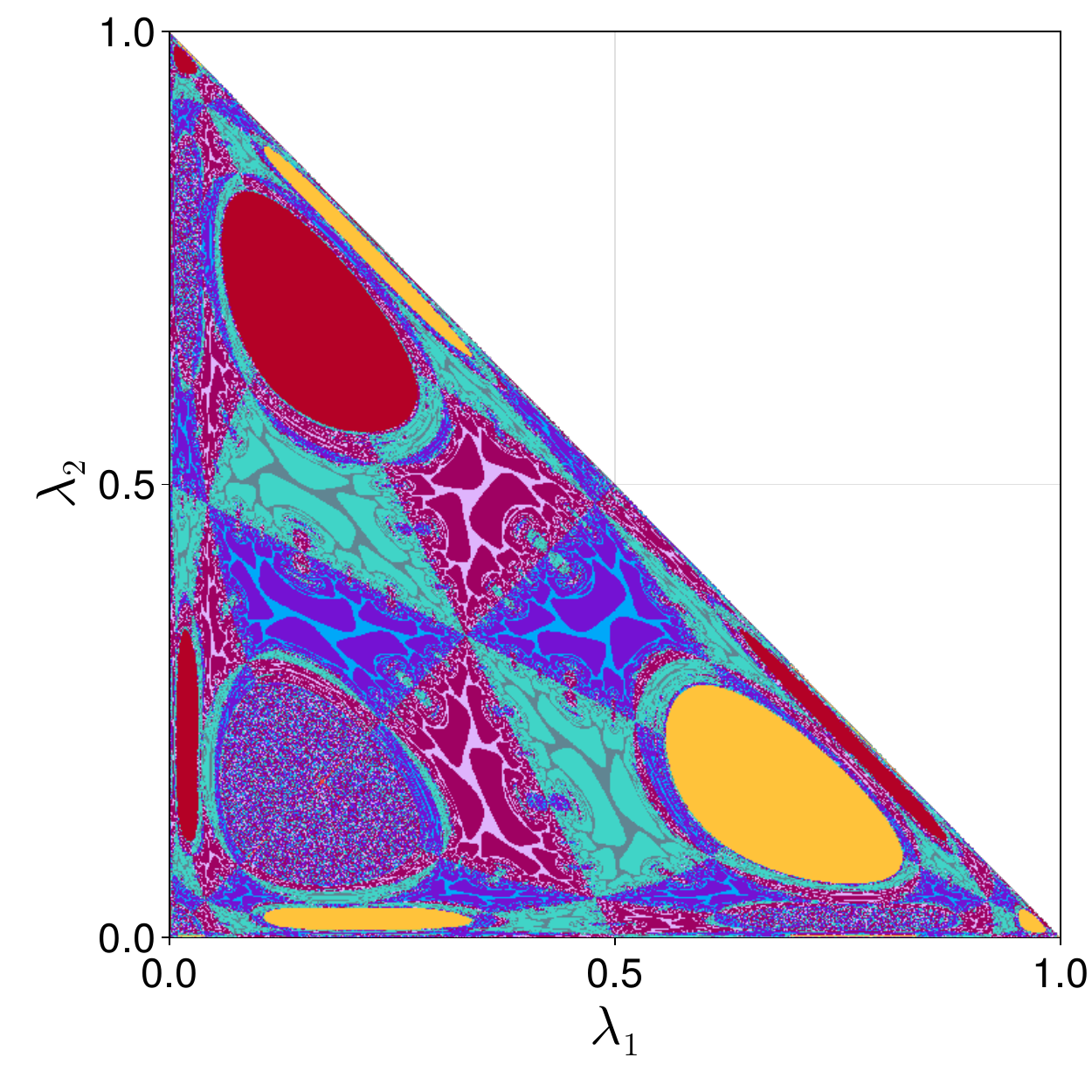}
\end{center}
\caption{\label{fig:econo_geo}Basins of attraction of an economic geographic model described in Eqs. ~\ref{eq:econo_geo}. The basins have been computed with the parameters $\mu = 0.4$, $\gamma = 5.0$, and $\sigma = 5.0$.}
\end{figure}

The study~\cite{commendatore2015typical} investigates the global dynamics of a modified economic geography model. The interplay between agglomeration and dispersion forces influences the spatial distribution of industrial activity across three identical regions. A two-dimensional discrete map models these interactions, revealing an unpredictable economic agglomeration. The initial conditions represent historical factors and determine the long-term economic distributions across regions.

The model used in the article is fairly complex. It involves, first, the definition of some constants: $\phi = 0.085$, $\mu = 0.4$, $\gamma = 5.0$, and $\sigma = 5.0$. The dynamical variables $\lambda_{1,n}$, $\lambda_{2,n}$, and $\lambda_{3,n} = 1 - (\lambda_{1,n} + \lambda_{2,n})$ will serve to compute two other parameters $M_1$ and $M_2$. First, we need the following definitions:
\begin{align*}
\begin{split}
\Delta_1 &= \lambda_1 + \Phi(1 - \lambda_1) \\
\Delta_2 &= \lambda_2 + \Phi(1 - \lambda_2) \\
\Delta_3 &= 1 - (\lambda_1 + \lambda_2)(1 - \Phi) \\
C_1 &= \frac{\sigma - \mu}{3(\sigma - \mu \lambda_1 \left( \frac{1}{\Delta_1} - \frac{\Phi}{\Delta_3} \right))} \\
C_2 &= \frac{\mu \Phi \lambda_1}{\sigma - \mu \lambda_1 \left( \frac{1}{\Delta_1} - \frac{\Phi}{\Delta_3} \right)} \\
C_3 &= \frac{\sigma - \mu}{3(\sigma - \mu \lambda_2 \left( \frac{1}{\Delta_2} - \frac{\Phi}{\Delta_3} \right))} \\
C_4 &= \frac{\mu \Phi \lambda_2}{\sigma - \mu \lambda_2 \left( \frac{1}{\Delta_2} - \frac{\Phi}{\Delta_3} \right)} \\
\end{split}
\end{align*}
The following linear system is solved to obtain the values of $s_1$, $s_2$, and $s_3$:
\begin{equation}
\begin{bmatrix}
1 & -C_2 \left( \frac{1}{\Delta_2} - \frac{1}{\Delta_3} \right) & 0 \\
-C_4 \left( \frac{1}{\Delta_1} - \frac{1}{\Delta_3} \right) & 1 & 0 \\
1 & 1 & 1
\end{bmatrix} 
\begin{bmatrix}
s_1 \\ s_2 \\ s_3
\end{bmatrix} 
=
\begin{bmatrix}
C_1 + \frac{C_2}{\Delta_3} \\
C_3 + \frac{C_4}{\Delta_3} \\
1
\end{bmatrix} 
\end{equation}
The next step involves the computation of intermediate variables $D$, $K_1$, and $K_2$:
\begin{equation}
D = \lambda_1 \Delta_1^{\frac{\mu}{\sigma-1}} \left( \frac{s_1}{\Delta_1} + \Phi \left( \frac{s_2}{\Delta_2} + \frac{s_3}{\Delta_3} \right) \right) + \lambda_2 \Delta_2^{\frac{\mu}{\sigma-1}} \left( \frac{s_2}{\Delta_2} + \Phi \left( \frac{s_1}{\Delta_1} + \frac{s_3}{\Delta_3} \right) \right) + \lambda_3 \Delta_3^{\frac{\mu}{\sigma-1}} \left( \frac{s_3}{\Delta_3} + \Phi \left( \frac{s_1}{\Delta_1} + \frac{s_2}{\Delta_2} \right) \right) 
\end{equation}

\begin{align}
\begin{split}
K_1 &= \frac{\Delta_1^{\frac{\mu}{\sigma-1}} \left( \frac{s_1}{\Delta_1} + \Phi \left( \frac{s_2}{\Delta_2} + \frac{s_3}{\Delta_3} \right) \right)}{D} \\
K_2 &= \frac{\Delta_2^{\frac{\mu}{\sigma-1}} \left( \frac{s_2}{\Delta_2} + \Phi \left( \frac{s_1}{\Delta_1} + \frac{s_3}{\Delta_3} \right) \right)}{D} \\
M_1 &= \lambda_1(1 + \gamma(K_1 - 1)) \\
M_2 &= \lambda_2(1 + \gamma(K_2 - 1)) 
\end{split}
\end{align}

With these values $M_1$ and $M_2$, we will evaluate a function $f(x,y)$: 
\begin{equation}
f(x,y) = \left\{
\begin{array}{l}
0 \textrm{ if } x \leq 0\\
x \textrm{ if } x > 0 \textrm{ and } y > 0 \textrm{ and } x + y < 1\\  
x/(x+y) \textrm{ if } x > 0 \textrm{ and } y > 0 \textrm{ and } x + y \geq 1\\  
x/(1 - y) \textrm{ if } x > 0 \textrm{ and } y \leq 0 \textrm{ and } x + y < 1\\  
1 \textrm{ if } x > 0 \textrm{ and } y \leq 0 \textrm{ and } x + y \geq 1
\end{array}
\right.
\end{equation}
Finally, with this function, we get the next iterates of $\lambda_1$ and $\lambda_2$:
\begin{align}\label{eq:econo_geo}
\begin{split}
\lambda_{1,n+1} &= f(M_1,M_2)\\
\lambda_{2,n+1} &= f(M_2,M_1)
\end{split}
\end{align}
The variables $\lambda_i$ correspond to the geographic distribution of entrepreneurs in each of the three regions. The model takes into account considerations of prices, demand, entrepreneur concentration, and migration. 

All the previous operations are necessary to compute the next iterate. Nevertheless, the computation of the basins takes only a few minutes for a $1200 \times 1200$ grid. The initial conditions for the basins in Fig.~\ref{fig:econo_geo} have been computed with the parameters $\mu = 0.4$, $\gamma = 5.0$, $\sigma = 5.0$, and $\phi = 0.085$. The basins are fractalized with nine stable attractors detected. The region where $\lambda_1 + \lambda_2 > 1$ is forbidden and appears as a divergent attractor. The basins show a fascinating and intricate structure.

\subsection{Rock-Paper-Scissors Cyclic Competition Model}
\begin{keywrds} SMB, MAP \end{keywrds}
\begin{figure}
\begin{center}
\includegraphics[width=\columnwidth]{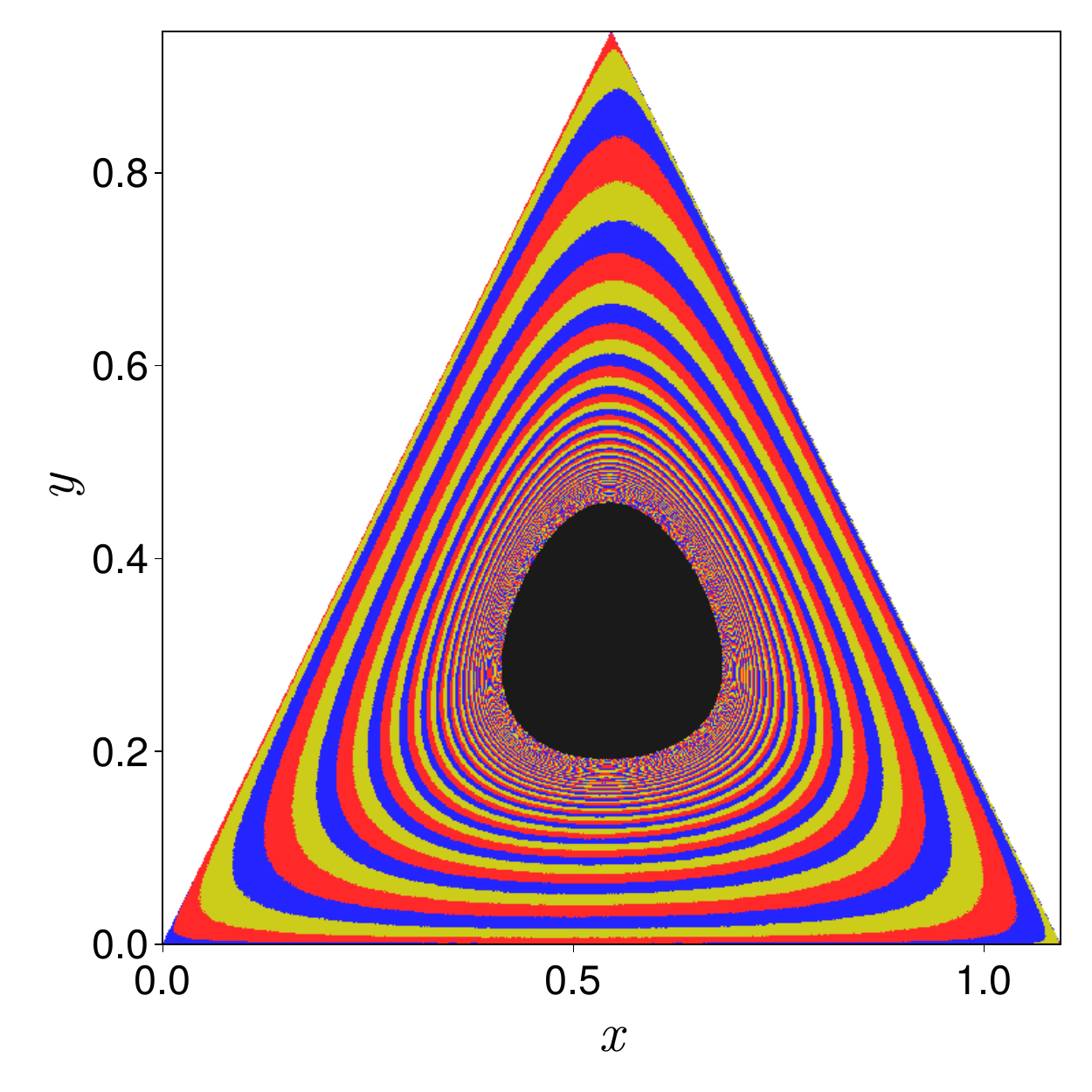}
\end{center}
\caption{\label{fig:cyclic_competition}Basins of attraction of a model ecosystem with cyclic competition among three species, where species $a$ predates species $b$, species $b$ predates species $c$, and species $c$ predates species $a$, as shown in Eqs.~\ref{eq:cyclic_competition}.}
\end{figure}

Ref.~\cite{Park_2018} explores multistability in a cyclic competition model that simulates biodiversity in ecosystems. Specifically, the research question is how the nature of intraspecific competition affects the dynamics of coexistence and extinction among three competing species.

The model incorporates density-dependent intraspecific competition described by logistic growth. The multistability arises at moderately strong levels of intraspecific competition, where the system can exhibit either coexistence or extinction depending on initial species densities. 

The dynamics of species $a$, $b$, and $c$ can be represented by the following set of rate equations:
\begin{align}\label{eq:cyclic_competition}
\begin{split}
    \frac{da}{dt} &= a \left( \mu(1 - \rho) - \sigma c - r \frac{a^2(1-a)}{2} \right), \\
    \frac{db}{dt} &= b \left( \mu(1 - \rho) - \sigma a - r \frac{b^2(1-b)}{2} \right), \\
    \frac{dc}{dt} &= c \left( \mu(1 - \rho) - \sigma b - r \frac{c^2(1-c)}{2} \right),
\end{split}
\end{align}
where $\rho = a + b + c$ represents the total density of the three species. The initial conditions for the model are taken on the surface $a + b + c = \sqrt{2/r}$. We first define two orthogonal vectors on this surface as $\vec u_1 = [-1, 1, 0]/\sqrt{2}$ and $\vec u_2 = [-1, -1, 2]/\sqrt{6}$. The initial condition vector $\vec u = [a, b, c]$ is computed as $\vec u = x \vec u_1 + y \vec u_2 + [\sqrt{r/2}, 0, 0]$. This vector defines an initial condition on the defined plane with the additional boundaries: $u_x > 0$ and $u_y > 0$. 

The predation among the three species is circular, with species $a$ predating species $b$, species $b$ predating species $c$, and species $c$ predating species $a$. For the parameters chosen, there are four stable states: three of them indicate the dominance of one predator with the extinction of the other two. The last stable equilibrium corresponds to equal densities of the three species. The results in Fig.~\ref{fig:cyclic_competition} for the parameters $\sigma = 1$, $\mu = 1$, and $r = 3.35$ show an accumulation of smooth basin boundaries near the central basin (the stable state with equal densities of the predators).

\subsection{Discrete Predator-Prey System}
\begin{keywrds} FB, MAP \end{keywrds}
\begin{figure}
\begin{center}
\includegraphics[width=\columnwidth]{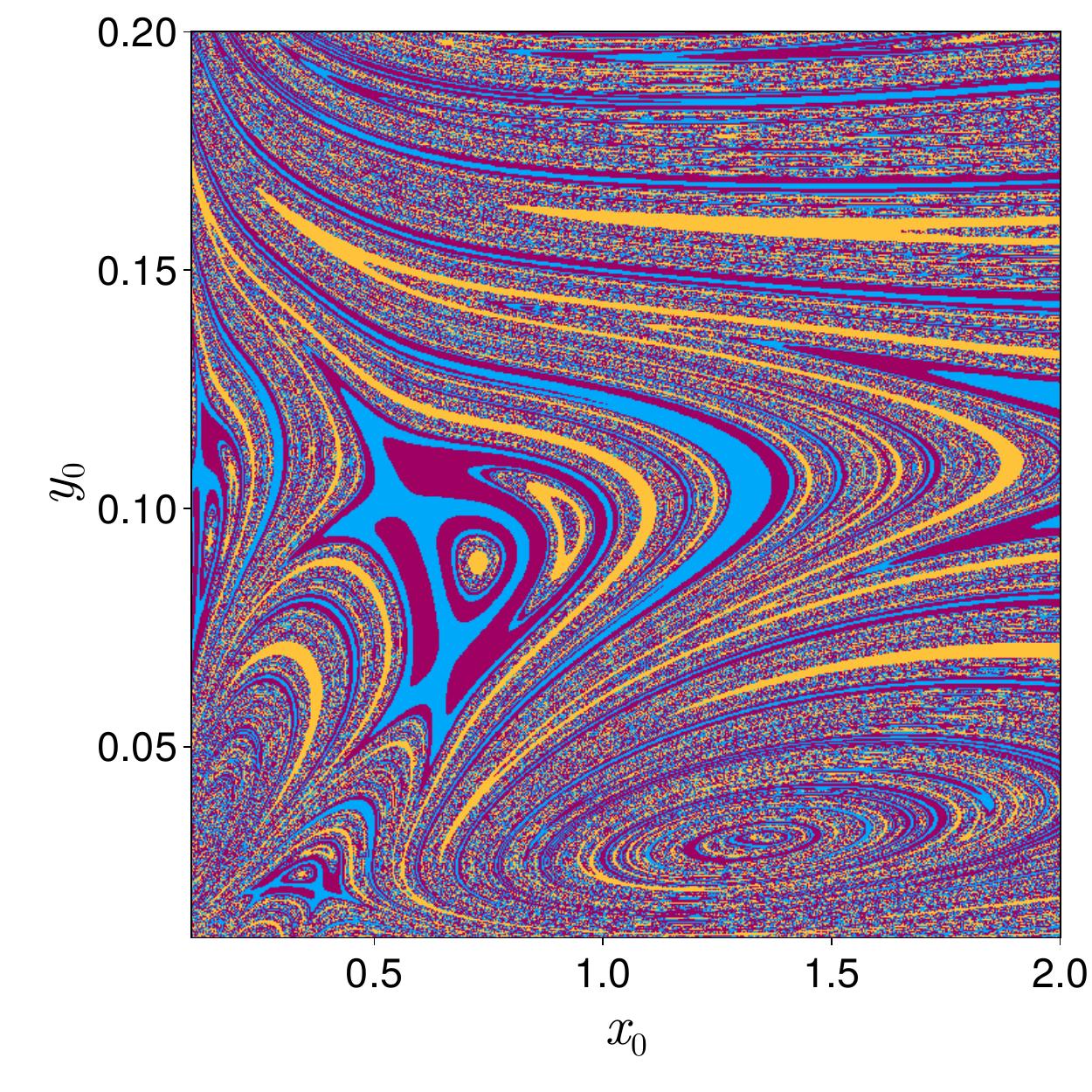}
\end{center}
\caption{\label{fig:pred_prey}Basins of attraction of a discrete predator-prey model described in Eqs.~\ref{eq:pred_prey}.}
\end{figure}

The work~\cite{Garai_2022} incorporates the effects of prey refuge and fear of predation into a predator-prey system. The authors proceed with a thorough numerical exploration of the model using the relevant tools of nonlinear dynamics.

The findings reveal the existence of various structured patterns, including Arnold tongues and shrimp-shaped structures, within the dynamics of the predator-prey model. The model is:
\begin{align}\label{eq:pred_prey}
\begin{split}
   x_{n+1} &= x_n \exp\left(R m +  \frac{R(1 - m)}{1 + k y_n} - D_1 - P x_n \dots\right. \\
   & \left.\dots - \frac{A(1 - m) y_n}{B + (1 - m)x_n}\right)\\
   y_{n+1} &= y_n \exp\left( \frac{c A (1 - m) x_n}{B + (1 - m)x_n} - D_2 \right)
\end{split}
\end{align}
$x_n$ and $y_n$ are the prey and predator population densities. Important parameters of the model are: $P = 0.1$ (the rate of intra-species competition among the prey), $m = 0.104$ (the constant proportion of prey that can find refuge from predators, $0 \leq m < 1$), and $k = 7.935$ (indicating the strength of predator-induced fear affecting the prey). The other parameter values are set to $R = 3.2$, $D_1 = 0.3$, $A = 2$, $B = 5$, $c = 0.9$, and $D_2 = 0.1$. 

The computation of the basins using recurrences is slightly more complex due to the presence of the exponential functions in the model. The trajectories of the attractors tend to accumulate close to zero, making it difficult to distinguish between two attractors. The solution to this problem involves using an irregular grid with a logarithmic size of boxes along one or both axes. The basins shown in Fig.~\ref{fig:pred_prey} are fascinating, featuring self-repeating structures and fractal boundaries.

\subsection{Multispecies Competition} 
\begin{keywrds} FB, ODE\end{keywrds}
\begin{figure}
\begin{center}
\includegraphics[width=\columnwidth]{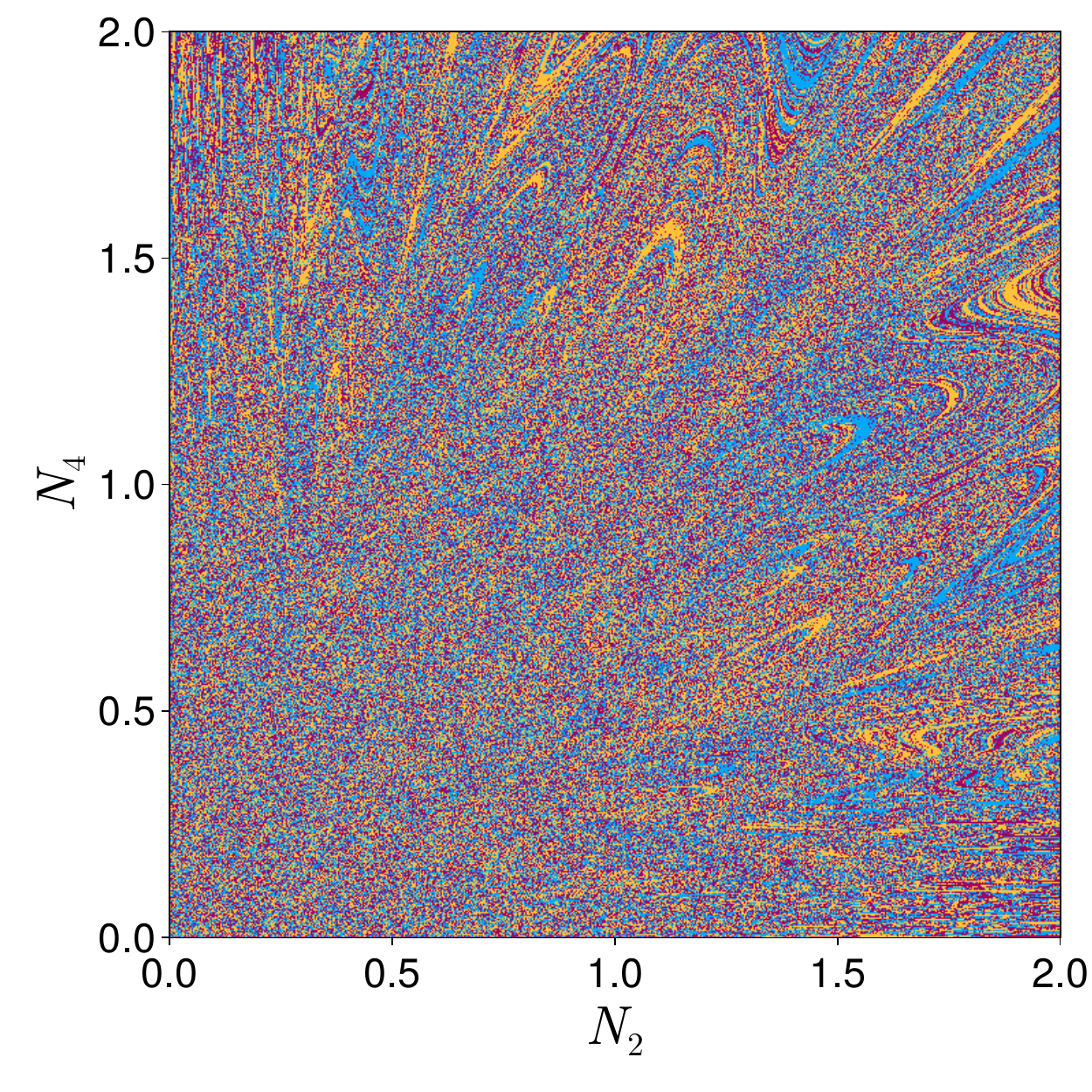}
\end{center}
\caption{\label{fig:multi_comp}Basin of attraction of the system in Eqs.~\ref{eq:multi_comp}. The figure represents the different equilibria of eight different species competing for three resources. The plot is a slice of the phase space with the initial abundances of species 2 and 4. The parameters and settings of the initial conditions are detailed in the text.} 
\end{figure}

In a scenario with different species competing for resources, the final abundance of individuals of a given species may depend significantly on the initial count of each population. In~\cite{Huisman_2001}, the authors set up a competition model commonly applied in phytoplankton and plant ecology, where multiple species compete for three resources. They analyze the dynamics of species abundance and resource availability, studying the behavior by varying the initial conditions to explore the basins for each possible outcome of the competition.

The competition model discussed in the article by Huisman and Weissing describes the dynamics of eight species competing for three abiotic resources. The equations governing this model are as follows:
\begin{align}\label{eq:multi_comp}
\begin{split}
   \frac{dN_i}{dt} &= N_i \left( \mu_i (R_1, R_2, R_3) - m_i \right)\\
   \frac{dR_j}{dt} &= D(S_j - R_j) - \sum_{i=1}^{n} c_{ji} \mu_i(R_1, R_2, R_3) N_i
\end{split}
\end{align}
where $N_i$ is the abundance of species $i$, $m_i$ is the specific mortality rate, and $R_i$ is the availability of resource $i$. $D$ is the resource turnover rate, $S_i$ is the supply of resource $i$, and $c_{ji}$ is the content of resource $j$ in species $i$.
The specific growth rates $\mu_i$ are determined by the most limiting resource: 
\begin{equation}
   \mu_i(R_1, R_2, R_3) = \min\left( \frac{r_i R_1}{K_{1i} + R_1}, \frac{r_i R_2}{K_{2i} + R_2}, \frac{r_i R_3}{K_{3i} + R_3} \right)
\end{equation}
where $r_i$ is the maximum specific growth rate of species $i$, and $K_{ji}$ is the half-saturation constant for resource $j$ for species $i$. The maximum specific growth rate $r_i = 1/d$ for all species, the specific mortality rate $m_i = 0.25/d$, the resource turnover rate $D= 0.25/d$, and the supply of each resource $S_j = 10$ mmol L$^{-1}$ for all resources. The value of $d = 1.$ is used for the simulations. The half-saturation constants for resources $K_{ji}$ between species and resources are given by the following matrix:
\[
K = \begin{bmatrix}
0.2 & 0.05 & 0.50 & 0.05 & 0.50 & 0.03 & 0.51 & 0.51\\
0.15 & 0.06 & 0.05 & 0.50 & 0.30 & 0.18 & 0.04 & 0.31\\
0.15 & 0.50 & 0.30 & 0.06 & 0.05 & 0.18 & 0.31 & 0.04
\end{bmatrix}
\]
The resource contents for each species are represented by the matrix:
\[
C = \begin{bmatrix}
0.2   & 0.10 & 0.10 & 0.10 & 0.10 & 0.22 & 0.10 & 0.10\\
0.10 & 0.20 & 0.10 & 0.10 & 0.20 & 0.10 & 0.22 & 0.10\\
0.10 & 0.10 & 0.20 & 0.20 & 0.10 & 0.10 & 0.10 & 0.22
\end{bmatrix}
\]
In the basins shown in Fig.~\ref{fig:multi_comp}, the choice for the initial conditions differs from the usual approach to computing basins. First, we must integrate a trajectory with the chosen initial conditions $N_2$ and $N_4$, setting $N_i(0) = 0.1$ for $i=1,3$ and $R_i(0)= 10$ until we reach a stable state. The initial abundances of species 6, 7, and 8 are $N_i(0) = 0$ until $t=1000$, when they invade the arena with $N_i(1000) = 0.1$. Then the algorithm searches for the new stable equilibrium of the species. The result is shown in Fig.~\ref{fig:multi_comp} for the initial abundances of species 2 and 4. The basins are clearly fractal and possibly intermingled, but this property has not been studied numerically yet. The simulation of these basins is very demanding and may last several days depending on the resolution of the basins.

\subsection{The Ricker-Gatto Model}
\begin{keywrds}RB, MAP\end{keywrds}
\begin{figure}
\begin{center}
\includegraphics[width=\columnwidth]{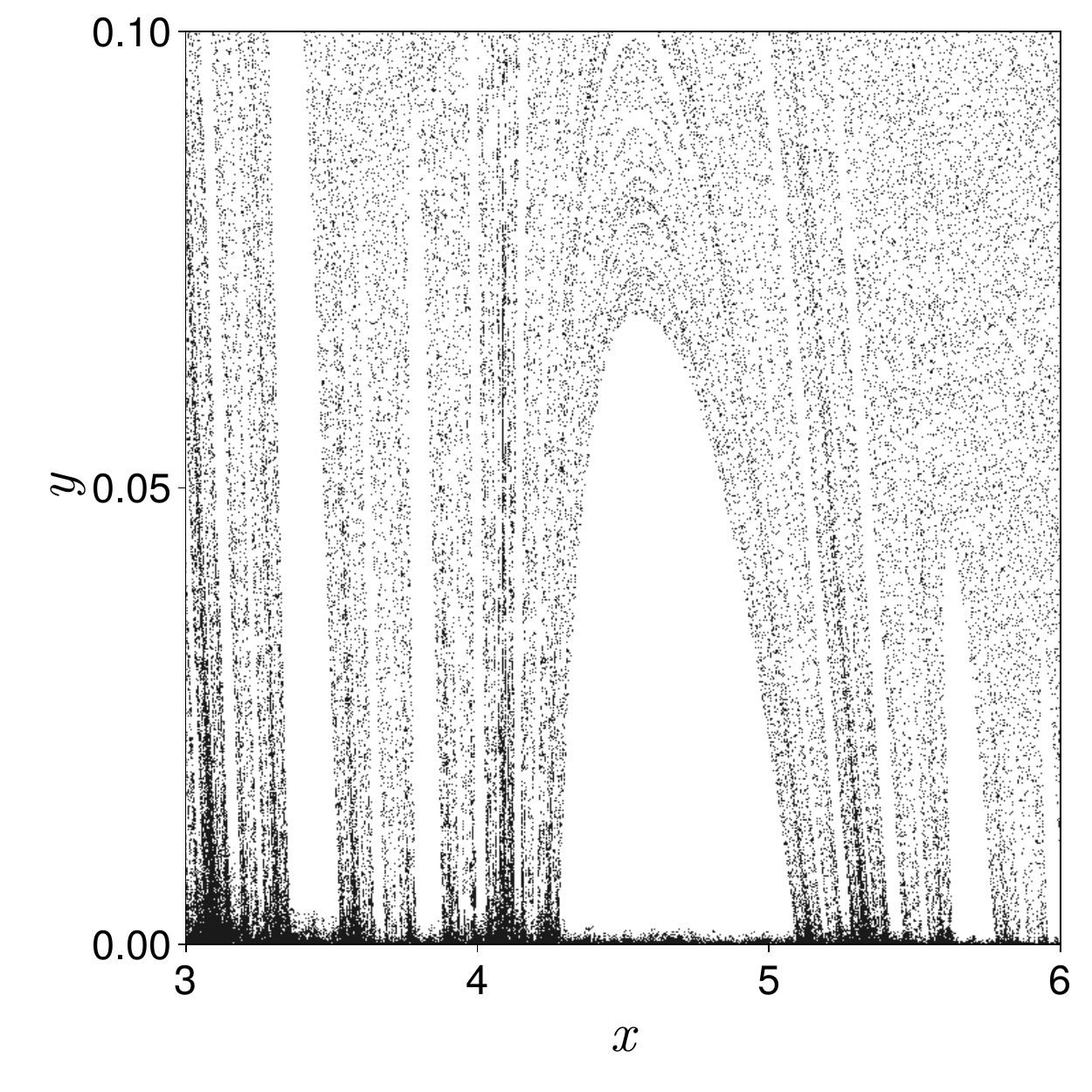}
\end{center}
\caption{\label{fig:ricker_gatto}Basins of attraction of the Ricker-Gatto species competition in Eqs.~\ref{eq:ricker_gatto}. The basin of one of the attractors is riddled with the other. Parameters are $r_1 = 22$,  $s_1 = 0.007815$, $r_2 = 2.95$ and $s_2 = 0.5$.}
\end{figure}

In Ref.~\cite{cazelles2001dynamics} Cazelles investigates several ecological models of interacting populations with riddled basins of attraction. The article study the dynamics of the two models with quantitative tools and study their basins in details. The second model proposed in the article, the Ricker-Gatto model, considers the survival of adults post-reproduction and the effects of competition on recruitment. 

Both models exhibit riddled basins of attraction, where even minor changes in initial conditions lead to drastically different long-term behaviors. In the Ricker-Gatto model, the coexistence attractor is periodic, and the extinction of the $y$ species corresponds to a chaotic attractor. 

The Ricker-Gatto model is:
\begin{align}\label{eq:ricker_gatto}
    \begin{split}
    x_{n+1} &= x_n(r_1e^{(-x_n - y_n)} + s_1)\\
    y_{n+1} &= y_n(r_2e^{(-x_n - y_n)} + s_2)
    \end{split}
\end{align}
Figure~\ref{fig:ricker_gatto} represents the riddled basins between a periodic and chaotic attractor for the parameters: $r_1 = 22$,  $s_1 = 0.007815$, $r_2 = 2.95$ and $s_2 = 0.5$. 

\subsection{Network of Coupled Neurons with Random Weights}
\begin{keywrds}SMB, ODE\end{keywrds}
\begin{figure}
\begin{center}
\includegraphics[width=\columnwidth]{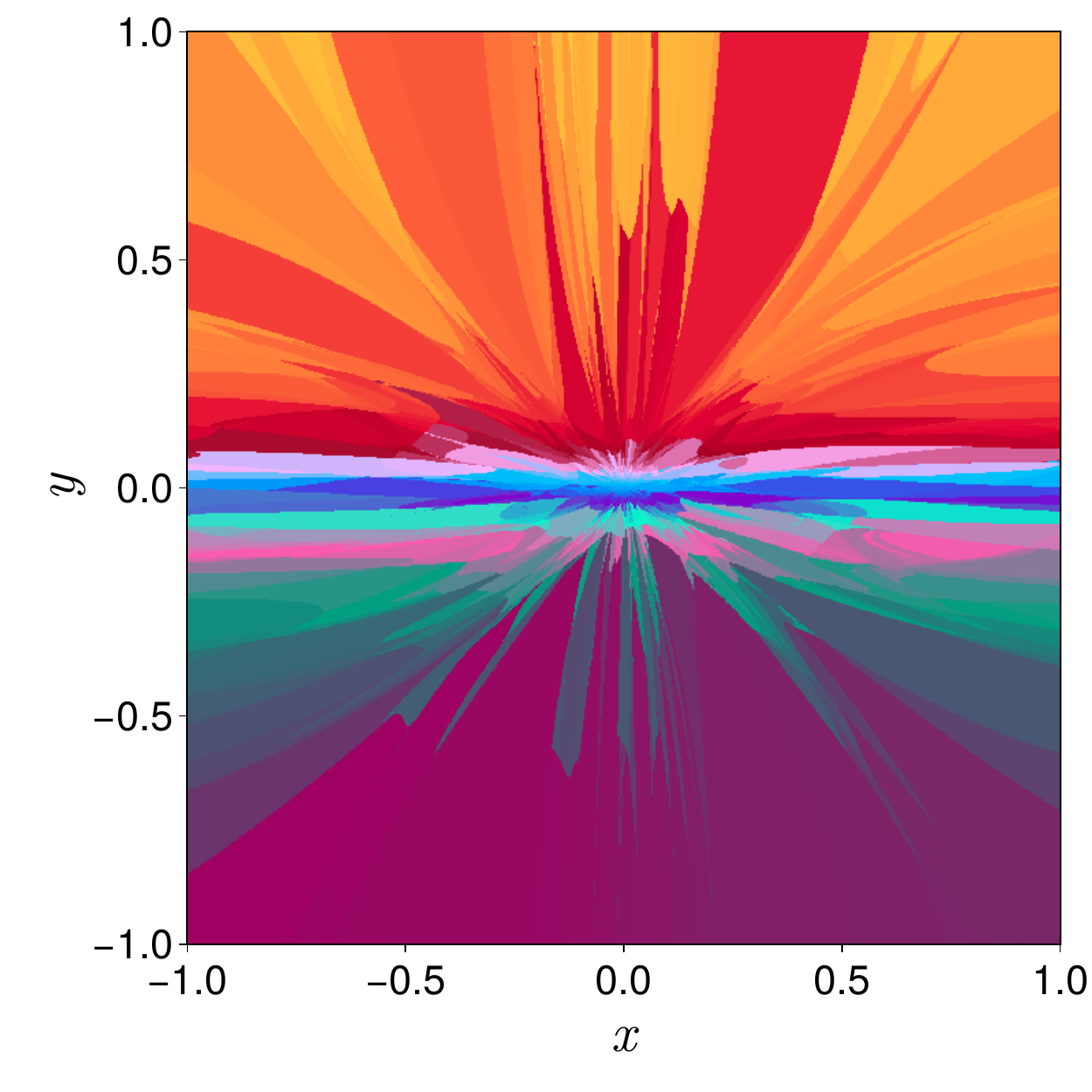}
\end{center}
\caption{\label{fig:stern_net}Basins of attraction of the neuron network in Eqs.~\ref{eq:stern_net}. There are 2654 stable fixed points in this figure. Because colors are cycling they seem less fewer in this plot. Parameters are $N=100$ units, $g=2.5$, $s = 4.5$. }
\end{figure}

Pattern activity in the central nervous system is a mesmerizing subject. Very early in the 20th century, models of electrical activity of neuron cells appeared, followed by studies on the collective activities of the connected cells. In this spirit, the study by Stern et al. \cite{stern2014dynamics} explores the space of final states of a neural network consisting of connected bistable units. The model incorporates strong intra-cluster interactions and random inter-cluster connections, reflecting the long-range interactions. The authors analyze the different dynamical regimes with simulations and a mean-field approach, averaging the activity of the long-range interaction. The strong self-interactions within the network render individual units bistable, while the random inter-unit connections tend to promote chaotic activity. By changing the intra- and inter-coupling parameters, the network transitions between decaying activity, persistent chaos, and long-lived chaotic activity that ultimately converges to multiple fixed points. The model interpolates between chaotic networks and models with a large number of stable fixed points.

One of the keys to the study is the randomly distributed inter-cluster weights. Random matrix theory gives insights into the distribution of the eigenvalue spectra, connecting directly to the stability of the states in the large network limit. The study also identifies long-lived chaotic transients and measures their lifetime. 

The model for the neuron network is:
\begin{equation}\label{eq:stern_net}
   \frac{dx_i}{dt} = -x_i + s \tanh(x_i) + g \sum_{j \neq i} J_{ij} \tanh(x_j),
\end{equation}
where $s$ represents the strength of the self-coupling, $g$ is the strength of the inter-unit coupling, and $J_{ij}$ describes the connections between the units, which are drawn from a Gaussian distribution with zero mean and variance $1/N$. The basins in Fig.~\ref{fig:stern_net} contain 2654 different stable fixed points for a slice on the initial state of the first two neurons. The rest of the initial conditions of the 98 units have been set to zero. Although there is a large number of fixed points, a structure in the phase space can be appreciated. The parameters $g$ and $s$ set the dynamical regime of the network. For $g\geq 2$ and $s \geq 2$ trajectories follow a chaotic transients before stabilizing into a fixed point. The structure of basins for these networks is an open problem. 

\subsection{Adaptive Synapse-Based Neuron}
\begin{keywrds}FB, ODE\end{keywrds}
\begin{figure}
\begin{center}
\includegraphics[width=\columnwidth]{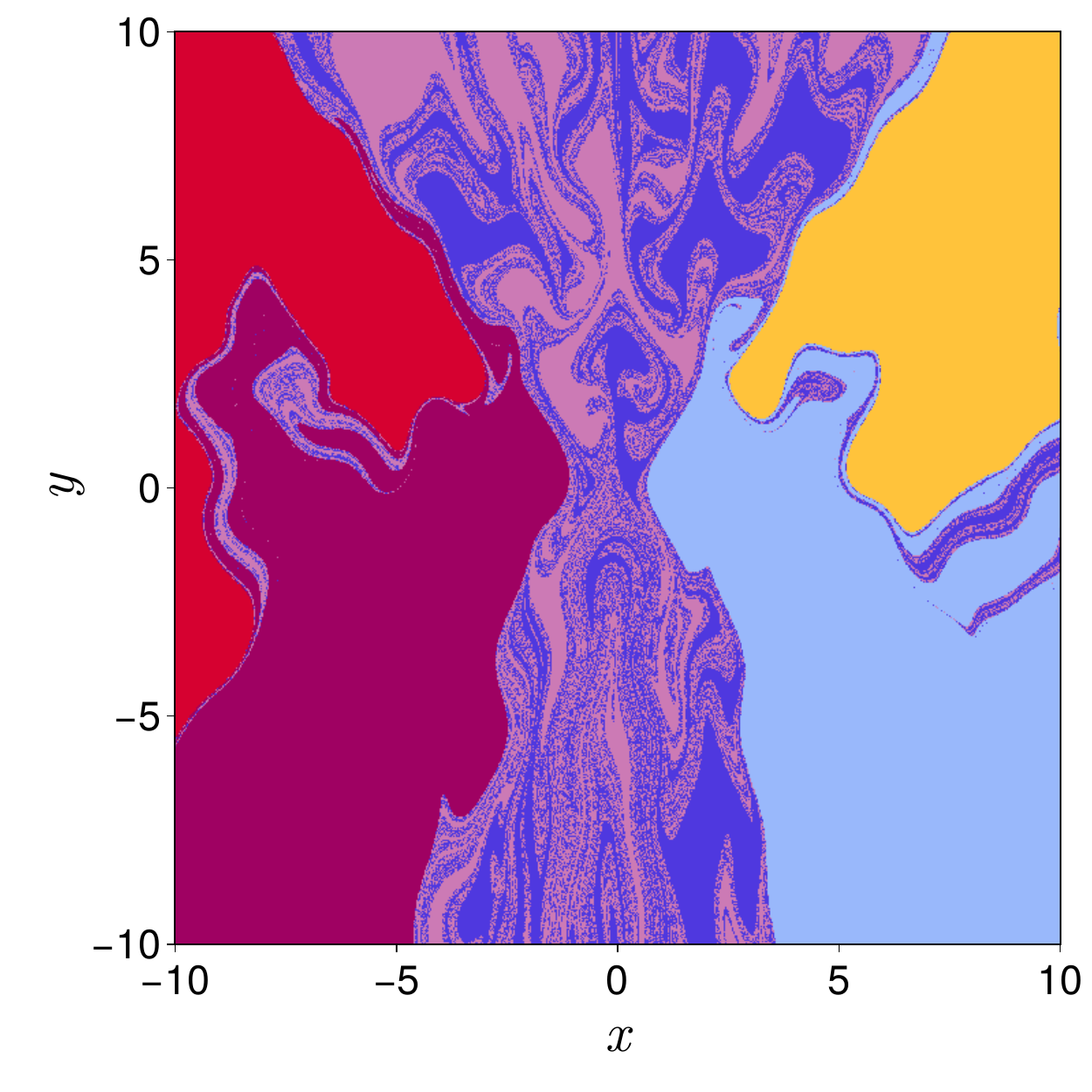}
\end{center}
\caption{\label{fig:synapse}Basins of attraction of~\ref{eq:synapse}. The attractors are firing patterns of a synapse-based neuron with an intricate boundary in the center.}
\end{figure}

Ref.~\cite{bao2022adaptive} presents an adaptive synapse-based neuron (ASN) model with a sine activation function. The model generates complex coexisting firing patterns, resulting in riddled basins of attraction. The ASN model is excited with time-varying externally applied current. Key methods include bifurcation diagrams, phase portraits, Lyapunov exponent spectra, and basin of attraction plots. The results highlight that the model can exhibit up to 12 coexisting heterogeneous attractors, accompanied by riddled basins of attraction.

The equation of the adaptive synapse-based neuron (ASN) model presented in the article is given as follows:
\begin{align}\label{eq:synapse}
\frac{dx}{dt} &= -x + H(x)H(y) + I(t), \\
\frac{dy}{dt} &= -cy + cH^2(x),
\end{align}
where $x$ represents the membrane potential of the neuron, $y$ denotes the synapse variable, $H(x)$ and $H(y)$ are the activation functions for the neuron and synapse, respectively, and $I(t) = \sin(2\pi t)$ is the externally applied current. The activation function $H$ is:
\begin{equation}
H(x) = B \sin(g~x),
\end{equation}
where $B$ is the dynamic amplitude and $g$ is the activation gradient. Fig.~\ref{fig:synapse} represents the basins for the parameters $c = 1.8$, $B = 2$, and $g = 1.7$. The boundary is interesting, featuring this fractal pattern in the center and the large area of basins on the sides.

\subsection{Biorhythm}
\begin{keywrds}SMB, ODE\end{keywrds}
\begin{figure}
\begin{center}
\includegraphics[width=\columnwidth]{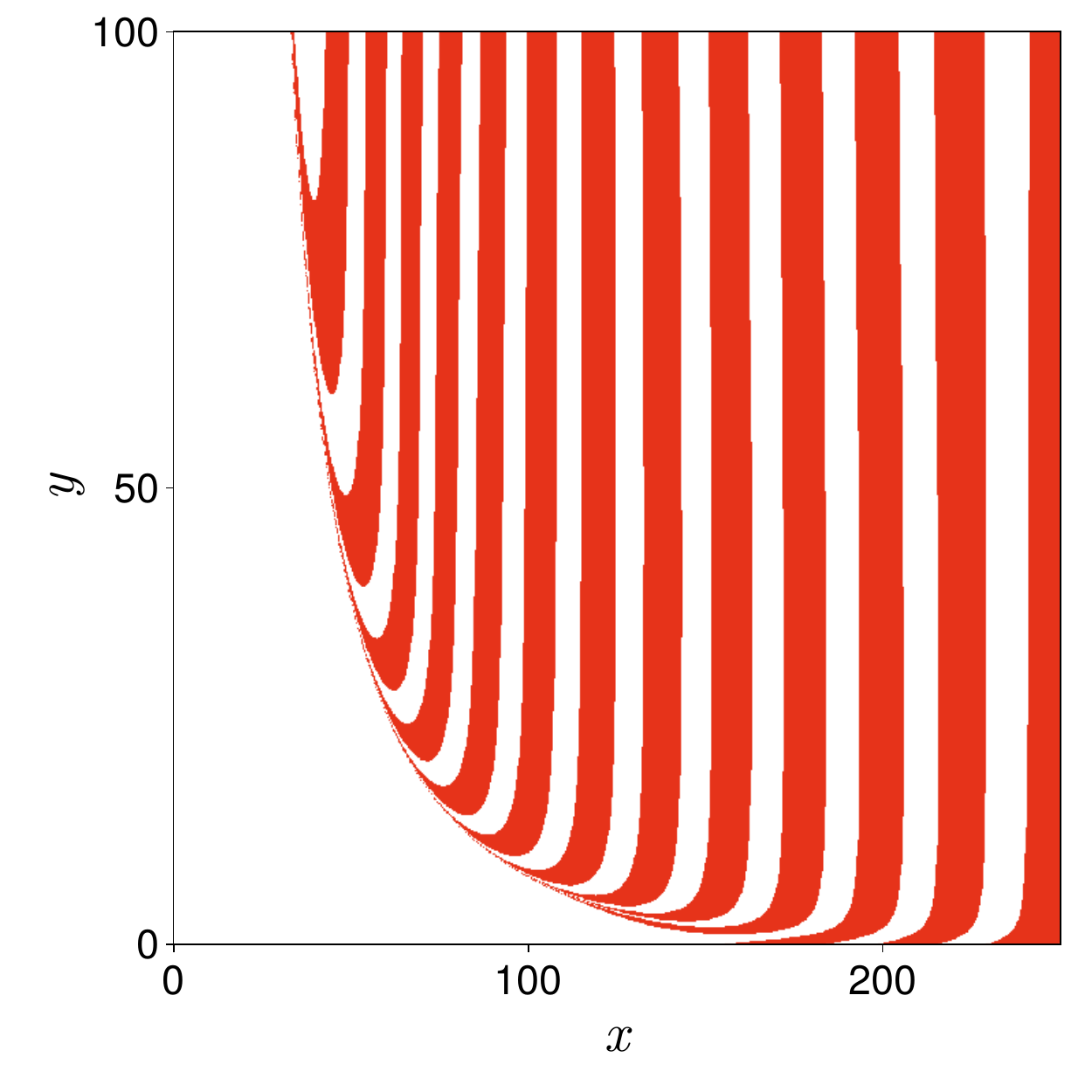}
\end{center}
\caption{\label{fig:biorhythm}Basins of attraction of~\ref{eq:biorhythm}. The attractors of the basins represent different oscillating patterns of the biochemical reactions involving three interacting enzymes. The parameters are specified in the text.}
\end{figure}

Ref.~\cite{decroly1982birhythmicity} presents a mathematical model of two allosteric enzymes activated by their respective reaction products and coupled in series. The model equations involve Michaelis-Menten kinetics and incorporate feedback loops to study the dynamic behaviors under varying conditions. The study demonstrates a range of dynamical behaviors, including simple periodic oscillations, the coexistence of two stable periodic regimes (birhythmicity), and chaos.

The interaction between two instability-generating mechanisms in a biochemical system greatly enhances the diversity of dynamic behaviors. Chaos and birhythmicity are relatively uncommon phenomena compared to regular periodic behaviors in biochemical systems. 

The mathematical model in the article involves a series of enzymatic reactions governed by ordinary differential equations that reflect the dynamics of substrates and products in the biochemical system. The core equations used in the model are as follows:
\begin{align}\label{eq:biorhythm}
   \frac{d\alpha}{dt} &= \frac{v}{K_{m1}} - \sigma_1  \cdot \Phi(\alpha, \beta)\\
   \frac{d\beta}{dt} &= q_1 \sigma_1 \Phi(\alpha, \beta) - \sigma_2 \cdot \eta(\beta, \gamma)\\
   \frac{d\gamma}{dt} &= q_2 \sigma_2 \eta(\beta, \gamma) - k_s \cdot \gamma
\end{align}

\begin{align*}
   \Phi(\alpha, \beta) &= \frac{\alpha (1 + \alpha)(1 + \beta)^2}{L_1 + (1 + \alpha)^2(1 + \beta)^2}\\
   \eta(\beta, \gamma) &= \frac{\beta (1 + d \beta)(1 + \gamma)^2}{L_2 + (1 + d \beta)^2(1 + \gamma)^2}
\end{align*}
$\alpha$, $\beta$, and $\gamma$ correspond to the concentrations of the substrates/products normalized by their respective constants. The basins presented in Fig.~\ref{fig:biorhythm} have been computed on the $\alpha$-$\beta$ plane with $\gamma = 1.0$ for initial conditions. A smooth basin boundary between the basin of a limit cycle and a chaotic attractor appears with a periodic stripe pattern. These basins have been obtained for a parameter $k_s$ slightly before a bifurcation. In the original publications, the basins do not appear; this constitutes an original contribution. Parameters of the model are set to $\sigma_1 = 10$, $\sigma_2 = 10$, $L_1 = 5\cdot 10^8$, $L_2 = 100$, $q_1 = 50$, $q_2 = 0.02$, $d = 10^{-6}$, $k_s = 1.99$, and $v/K_{m1} = 0.45$. 

\subsection{Multistability in Gene Regulatory Networks}
\begin{keywrds} SMB, ODE\end{keywrds}
\begin{figure}
\begin{center}
\includegraphics[width=\columnwidth]{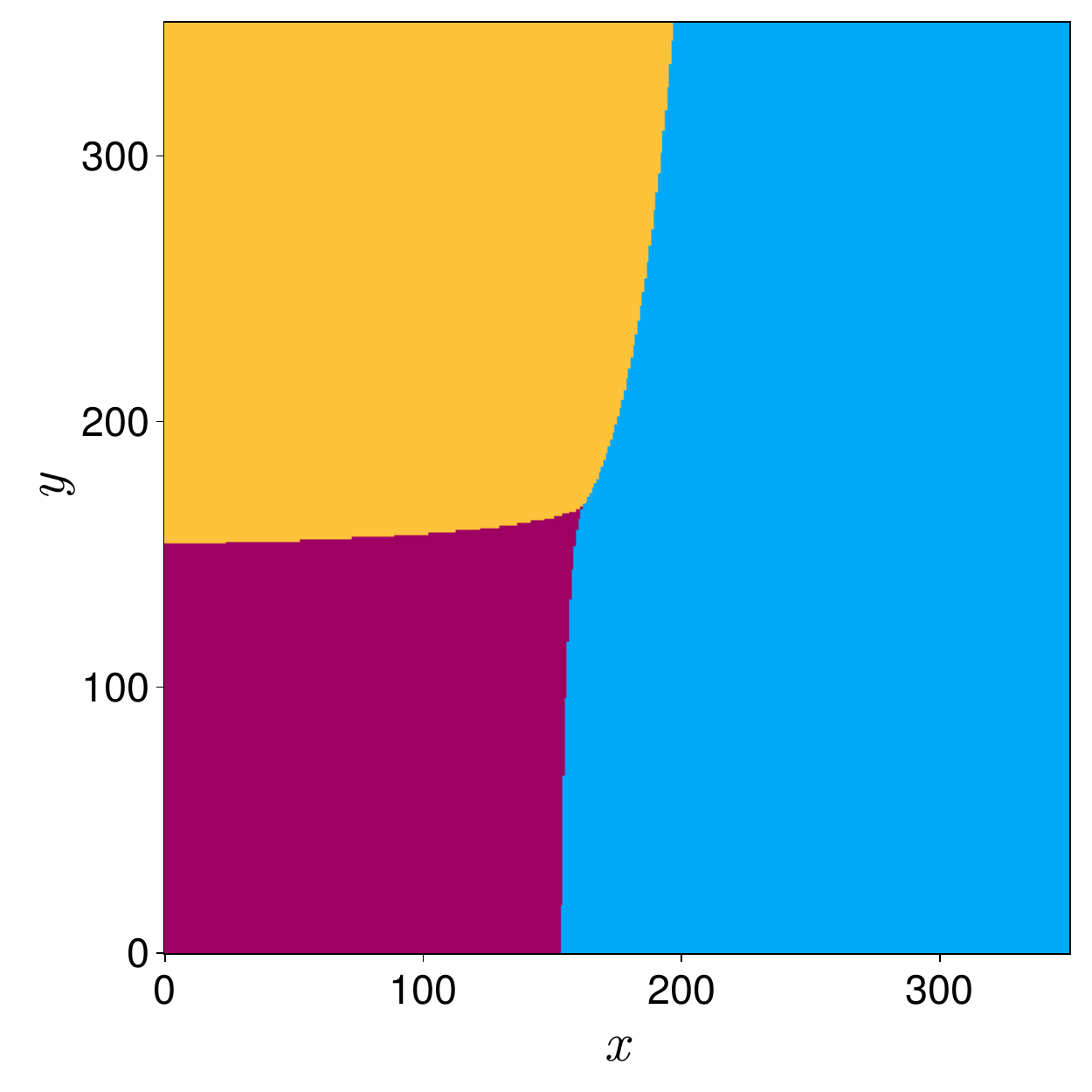}
\end{center}
\caption{\label{fig:ludp}Basins of attraction of the genetic regulatory network represented in (b) with the corresponding model in Fig.~\ref{eq:ludp}.}
\end{figure}

The design and study of genetic switches capable of multistability is a challenge in synthetic biology \cite{gardner2000construction}. Genetic switches are critical for epigenetic memory, cellular decision-making, and the construction of synthetic biological circuits.

Leon et al. \cite{leon2016computational} developed a computational tool called StabilityFinder to explore the parameter space of genetic switches and identify conditions leading to multistable behavior. Using Monte Carlo methods, the algorithm sample from parameter distributions and clusters steady-state trajectories in phase space to determine the number of stable states present in a given genetic model.  The framework provides insights into parameter regions capable of producing multistable behaviors in both deterministic and stochastic models. 

One of the model in this study is a specific genetic switch model that incorporates double positive autoregulation to enhance its multistability:
\begin{align}\label{eq:ludp}
\begin{split}
    \frac{dx}{dt} &=  g_x \left(\frac{1-0.5}{1 + (\dfrac{y}{500})^2} + 0.5\right) \left(\frac{1-10}{1 + (\dfrac{x}{250})^4} + 10\right) - k_x x,\\
    \frac{dy}{dt} &=  g_y \left(\frac{1-0.1}{1 + (\dfrac{x}{500})^2} + 0.1\right) \left(\frac{1-10}{1 + (\dfrac{y}{250})^4} + 10\right) - k_y y,
\end{split}
\end{align}
where $x$ and $y$ represent the concentrations of two transcription factors,  $g_x = 40$ and $g_y = 40$ are their respective production rates, and $k_x = 0.55$ and $k_y =0.55$ represent their degradation rates. The sigmoidal interaction functions with fixed parameters represent the positive feedback interaction between species. As the concentratio of species $x$ and $y$ increase, the production rates reach a limiting value. In the study, all the parameters have a range of variability in order to test the robustness of the structures of the phase space. The basins in Fig.~\ref{fig:ludp} show three stable regions with a smooth boundary between basins. The article shows that these basins are structurally stable over a large range of parameters.

\section{Examples in Physics}

At the core of the work of a physicist is the construction of models based on the object of study. The models are examined through different lenses: numerical, logical, and analytical. The examples presented in this section proceed from the study of different established models from the perspective of nonlinear dynamics.

\subsection{RF-driven Josephson Junction}

\begin{keywrds} FB, ODE\end{keywrds}
\begin{figure}
\begin{center}
\includegraphics[width=\columnwidth]{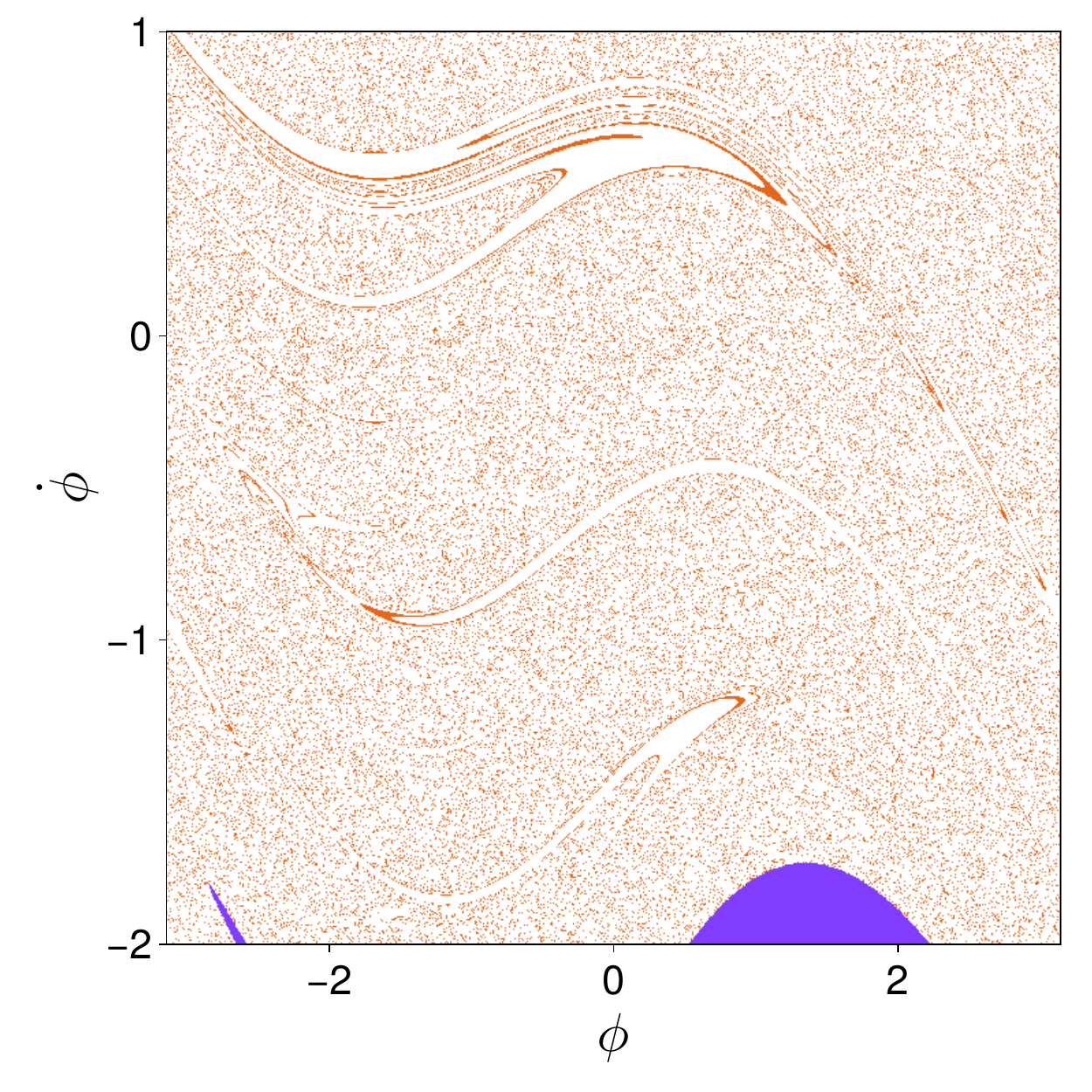}
\end{center}
\caption{\label{fig:josephson}Basins of attraction of the RF-driven Josephson junction, as described in equation \ref{eq:josephson}, for the parameters $\beta = 25$, $i_{dc} = 1.878$, $i_{rf} = 10.198$ and $\Omega = 1$.}
\end{figure}

Josephson junctions are formed by two superconducting electrodes separated by a thin insulator. Under the right conditions, an electric current can flow between the electrodes driven by the tunneling effect. In~\cite{Davidson_1987}, the model of a driven junction was proposed:
\begin{equation}\label{eq:josephson}
\ddot \phi + \beta^{-1/2}\dot \phi + \sin \phi = i_{dc} + i_{rf} \sin \Omega t 
\end{equation}
where $\beta$ is a hysteresis parameter, and $i_{dc}$ and $i_{rf}$ are the dc and rf bias currents. In the publication, the author studies the basins of this system using the cell-mapping technique~\cite{sun2018cell}. This technique allows for the recovery of the basins and some invariants of the phase space, such as the boundary. However, this comes at the cost of mapping the entire dynamical system and transforming it into a discretized system. The memory requirements and computational overhead can be prohibitive.

In Fig.~\ref{fig:josephson}, we reproduce a basin included in~\cite{Davidson_1987} with improved resolution using our numerical tools. There are three attractors present in the phase space. The computation is straightforward, and a stroboscopic map can be employed, but the computation still takes about an hour on an average laptop. The plot has been generated for the parameters $\beta = 25$, $i_{dc} = 1.878$, $i_{rf} = 10.198$ and $\Omega = 1$.

\subsection{\label{sec:rikitake}The Rikitake Oscillator: Earth's Magnetic Field Pattern}
\begin{keywrds}FB,  ODE\end{keywrds}
\begin{figure}
\begin{center}
\includegraphics[width=\columnwidth]{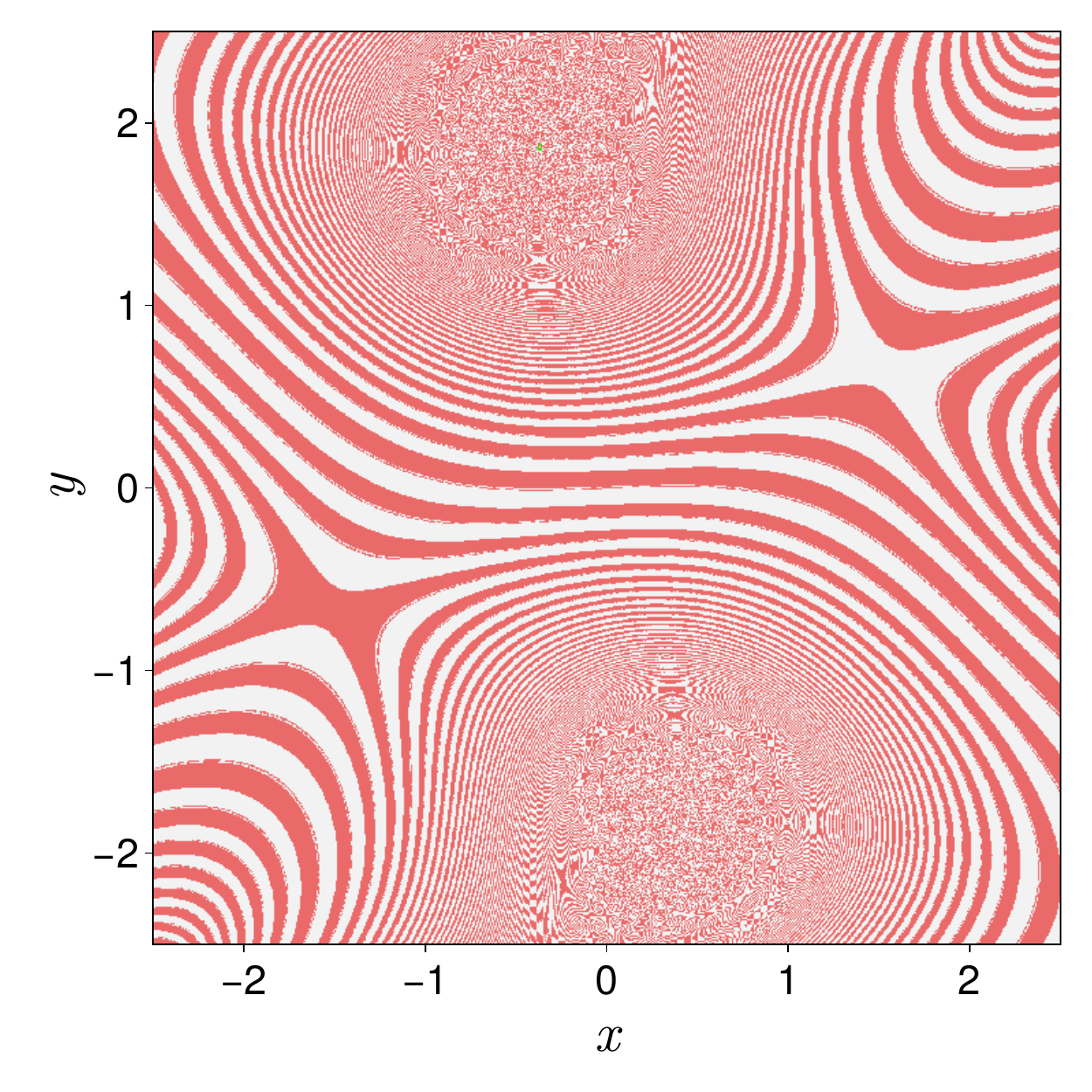}
\end{center}
    \caption{\label{fig:rikitake}Basins of attraction of the Rikitake oscillator in Eqs.~\ref{eq:rikitake}. The Rikitake model represents the magnetic field reversal of the Earth. There are two symmetric attractors with an intricate boundary. The parameters are $\alpha = 1.0$ and $\mu = 0.5$.}
\end{figure}

The Rikitake dynamo~\cite{rikitake1958oscillations} is a system modeling the magnetic reversal events of the Earth through a double-disk dynamo system. Each rotating disk generates a magnetic field. The simplicity of the model, a three-dimensional differential equation, is its main strength. The processes involved in magnetic reversal are overly complex, but this simple model has attracted considerable qualitative research on the Earth's magnetic field~\cite{munoz2023planetary}. The equations of the dynamical system are: 
\begin{align}\label{eq:rikitake}
\begin{split}
\dot{x} &= -\mu x + yz \\
\dot{y} &= -\mu y + x(z - \alpha) \\
\dot{z} &= 1 - xz
\end{split}
\end{align} 
where $x$ and $y$ represent the angular velocities of the two disks, $z$ the produced magnetic field, and $\mu$ is a damping coefficient. The basins of Eqs.~\ref{eq:rikitake} have not been studied in detail yet. For the parameters $\mu = 0.5$ and $\alpha = 1.0$, two symmetric attractors can be found with also symmetric basins. The basins form a cut in the $x-y$ plane for the initial condition $z=0$.

\subsection{Geomagnetic Field Reversal Model}
\begin{keywrds}FB, ODE\end{keywrds}
\begin{figure}
\begin{center}
\includegraphics[width=\columnwidth]{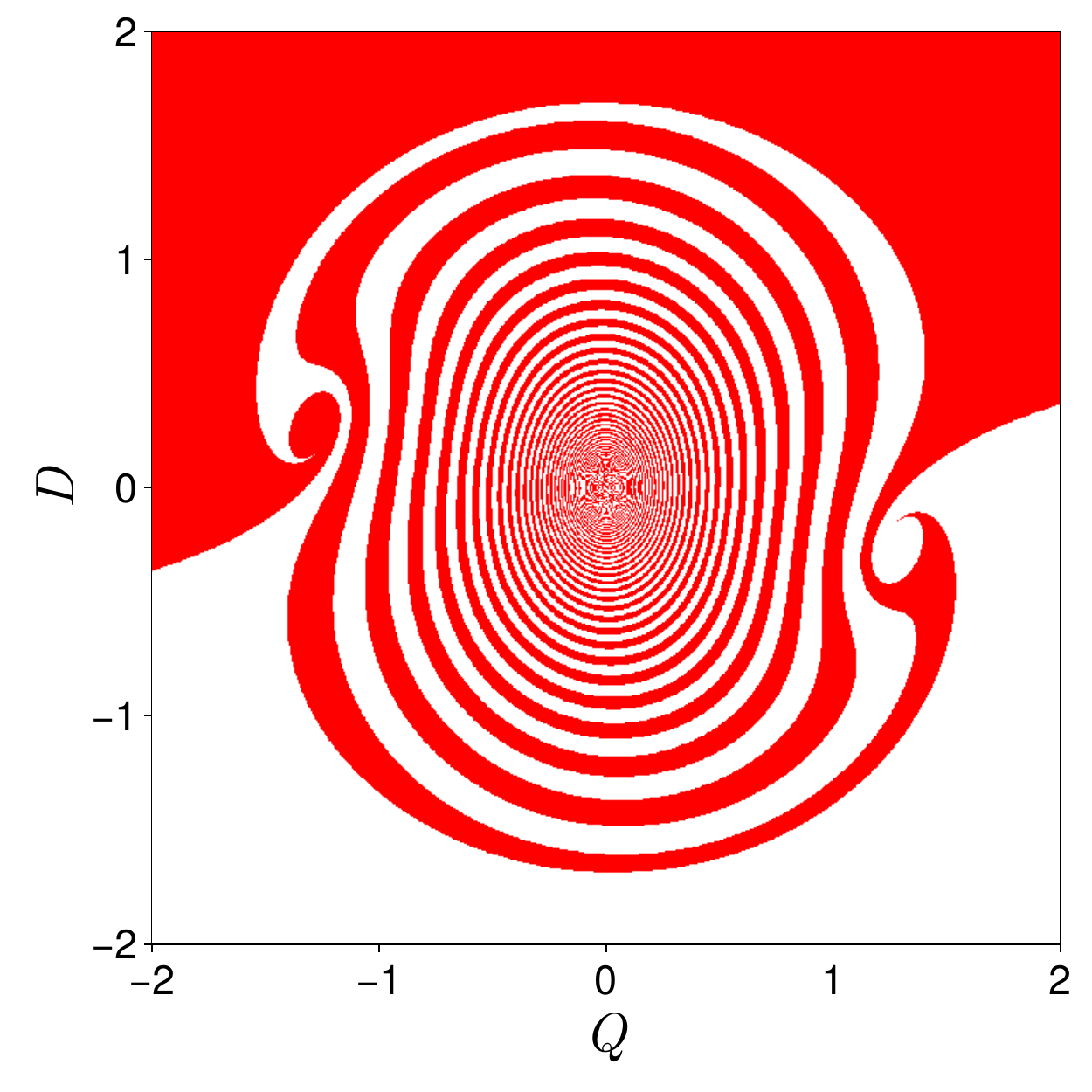}
\end{center}
\caption{\label{fig:gissinger_model}Basins of the geomagnetic field model in Eqs.~\ref{eq:gissinger_model} for the parameters $\mu = 0.1193$, $\nu = 0.1$, $\gamma = 0.9$.}
\end{figure}
The paper~\cite{gissinger2012new} presents a model of three coupled differential equations to study reversals occurring in fluid mechanics when turbulent fluctuations are important. The application is set in the context of a turbulent magnetic dynamo, as an interpretation of the reversals of the Earth's magnetic field. Similar models in the literature, such as the Rikitake model (see Sec.~\ref{sec:rikitake}) or the Nozière model, have been derived from truncations of the magnetohydrodynamic equations.

The study shows robust reversals generated through a boundary crisis between two symmetric chaotic attractors. The authors claim that the model is in good agreement with paleomagnetic observations of reversals. The model for the reversal is: 
\begin{align}\label{eq:gissinger_model}
\begin{split}
\dot{Q} &= -\mu Q + VD, \\
\dot{D} &= -\nu D + VQ,\\
\dot{V} &= \Gamma -V + QD.
\end{split}
\end{align} 
Variables $D$ and $Q$ represent the dipolar and quadripolar components of the magnetic field while $V$ is the amplitude. The basins in Fig.~\ref{fig:gissinger_model} have been obtained for the parameters $\gamma = 0.9$, $\mu = 0.1$ and $\nu = 0.1193$ and the initial condition $V(0) = -1$. There are two interesting symmetric basins spinning around the origin.

\subsection{The Lorenz-84 Atmospheric Circulation Model}
\begin{keywrds}FB, ODE\end{keywrds}
\begin{figure}
\begin{center}
\includegraphics[width=\columnwidth]{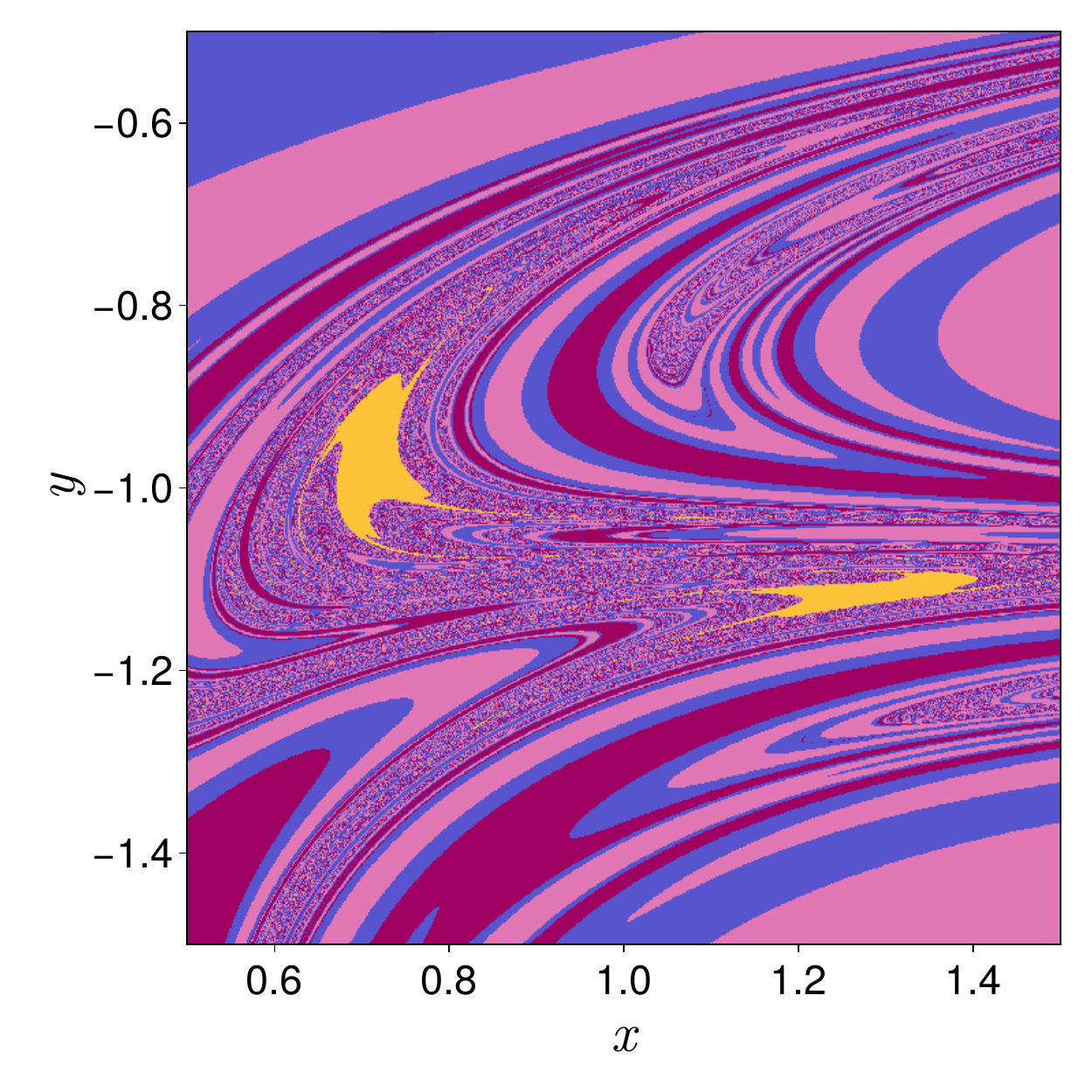}
\end{center}
\caption{\label{fig:lorenz84}Basins of the Lorenz-84 model in Eqs.~\ref{eq:lorenz84} for the parameters $F=6.846$, $G=1.287$, $a=0.25$, and $b=4.0$.}
\end{figure}

The Lorenz-84 model is a simplified atmospheric circulation model introduced by Edward N. Lorenz in 1984. This model is designed to capture some essential features of global atmospheric dynamics, particularly the interactions between different scales of motion in the atmosphere. In~\cite{Freire_2008}, the model studied has the following expression: 
\begin{align}\label{eq:lorenz84}
\begin{split}
\frac{dx}{dt} &= -y^2 - z^2 - ax + aF,\\
\frac{dy}{dt} &= xy - y - bxz + G,\\
\frac{dz}{dt} &= bxy + xz - z.
\end{split}
\end{align}
Here, $x$, $y$, and $z$ are the state variables representing different modes of atmospheric circulation. The parameters $a$, $b$, $F$, and $G$ control the behavior of the system. The parameter $F$ is often interpreted as a measure of external forcing, such as the thermal forcing due to the temperature difference between the equator and the poles. The model captures essential features of global atmospheric circulation, including the interactions between zonal, meridional, and eddy flows. 

In Fig.~\ref{fig:lorenz84}, the basins of four attractors are represented in the plane $xy$, with the initial condition $z=0$ and for the parameters $F=6.846$, $G=1.287$, $a=0.25$, and $b=4.0$. There are three periodic orbits close to one another and one fixed point. The technique used in~\cite{Freire_2008} consists of computing the Lyapunov exponents of each orbit and clustering the obtained metrics to identify the attractors. This technique limits the type of orbits that can be detected; for example, fixed points have the same Lyapunov spectrum and cannot be differentiated. The recurrence method works for this system; however, the parameters responsible for the detection of the attractors must be increased significantly due to the small distance between them.

\subsection{\label{sec:cold_atoms}Cold Atom Scattering}
\begin{keywrds}WD, OH\end{keywrds}
\begin{figure}
\begin{center}
\includegraphics[width=\columnwidth]{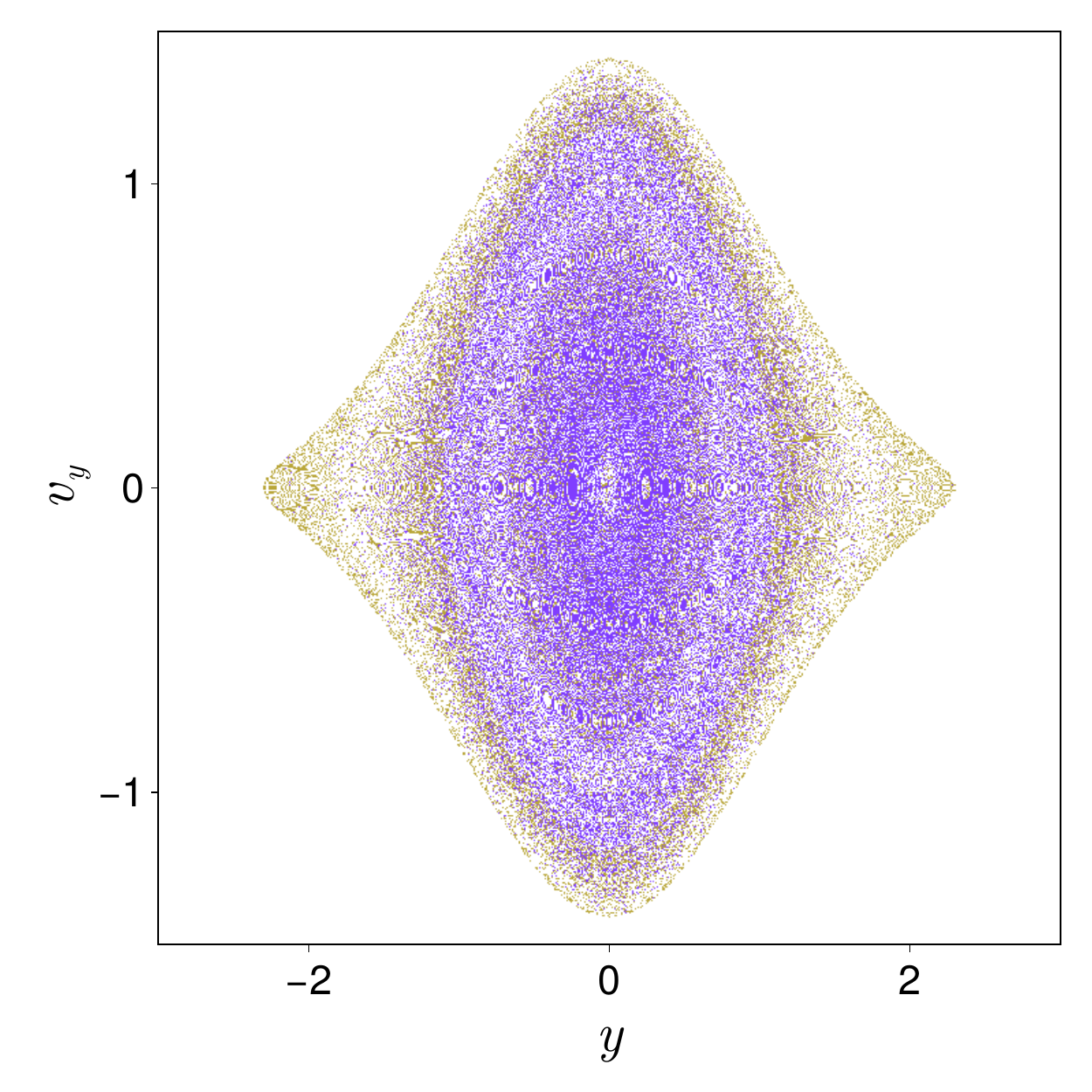}
\end{center}
\caption{\label{fig:cold_atoms}Escape basins of the Hamiltonian system in Eq.~\ref{eq:cold_atoms}. Parameters are $\alpha_1 = \alpha_2 = \beta_1 = \beta_2 = 1$, $\theta = \pi/4$.} 
\end{figure}

Ref.~\cite{Daza_2017} describes a system with two crossing laser beams guiding cold atoms. The atoms are launched toward the crossing region, and they eventually emerge through one of four possible paths defined by the lasers after some chaotic transients. These laser beams define the exit basins from which the atoms escape.

The model used in the paper involves the dynamics of atoms in the presence of the two crossed laser beams acting as waveguides. The potential energy of the system incorporates the effects of the two Gaussian dipole laser beams in the Hamiltonian:
\begin{equation}
H=\frac{1}{2} \left( \dot{x}^2+\dot{y}^2\right) -\alpha_1 e^{-\beta_1 y^2}-\alpha_2 e^{-\beta_2 (x\sin\theta+y\cos\theta)^2}.
\label{eq:cold_atoms}
\end{equation}
The features of each laser are condensed into two characteristic parameters: $\alpha_i$, which relates to the depth of the potential, and $\beta_i$,  a parameter related to the laser waist. The parameter $\theta$ is the angle formed by the two beams.

The equations of motion of the atoms are:
\begin{align}
\begin{split}
\dot x &= \frac{\partial H}{\partial p_x} = p_x, \\
\dot y &= \frac{\partial H}{\partial p_y} = p_y, \\
\dot p_x &= -\frac{\partial H}{\partial x},\\ 
\dot p_y &= -\frac{\partial H}{\partial y}.
\end{split}
\end{align}
The initial conditions of the basins pictured in Fig.~\ref{fig:cold_atoms} are $y$ and $v_y$, while $x = -500$ and $v_x = 0.1$ for unit mass atoms. The atoms are fired from very far away from the scattering region with a small velocity along the $x$-axis. The atoms oscillate in the potential defined by the laser until they reach the scattering region. Depending on the parameters, fractal basins are formed, limiting the predictability of the system. The scattering effect is notable for small velocities. For $v_y > 2$, the scattering disappears, and all atoms go through the same exit.

\subsection{Photonic Couplers}
\begin{keywrds}SMB, ODE\end{keywrds}
\begin{figure}
\begin{center}
\includegraphics[width=\columnwidth]{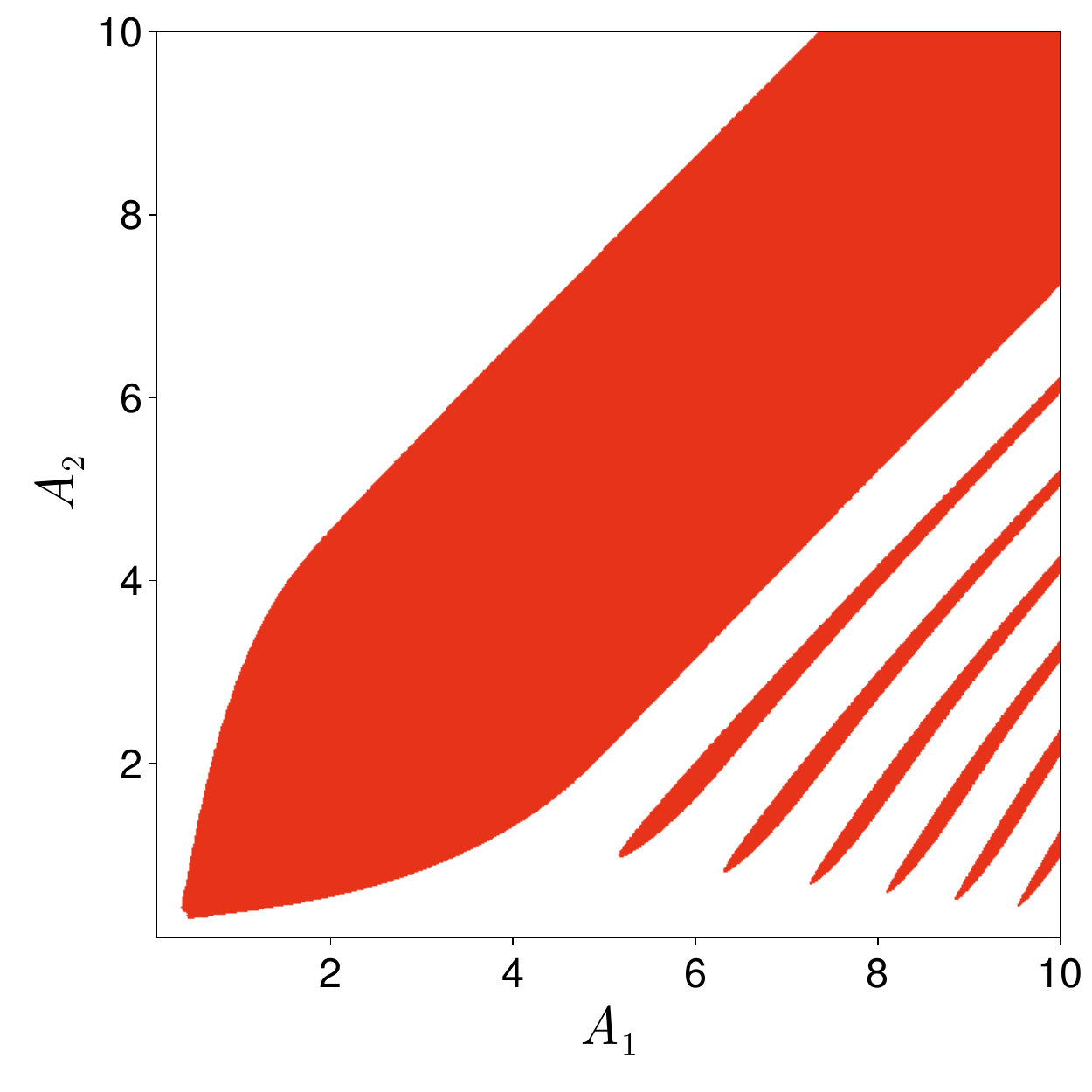}
\end{center}
\caption{\label{fig:photonic_couplers}Basins of attraction for the photonic coupler described in Eqs.~\ref{eq:photonic_couplers}. The parameters for this figure are $\alpha = 2$, $\beta = 1.5$, $\varepsilon = 0.5$, $k = 5$.} 
\end{figure}

The study~\cite{zhiyenbayev2019enhanced} explores an asymmetric active photonic coupler with saturable gain. The coupled-mode equations describe the dynamics, including the effects of asymmetry and saturable gain. This gain enhances bistability in specific parametric regions. The study details the conditions for the existence of stable nonlinear modes and the computation of their basins of attraction. The dynamics of the coupler are governed by the following coupled-mode equations:
\begin{align}\label{eq:photonic_couplers}
\begin{split}
   \dot{A_1} &= -\alpha_1 A_1 - \frac{k}{2} A_2 \sin(\phi),\\
   \dot{A_2} &= -\frac{\alpha_2}{1 + \epsilon A_2^2} A_2 + \frac{k}{2} A_1 \sin(\phi),\\
   \dot{\phi} &= (\beta_2 - \beta_1) + \gamma\left(A_2^2 - A_1^2\right) + \frac{k}{2}(A_1 / A_2 - A_2 / A_1) \cos(\phi),
\end{split}
\end{align}
The variables $A_1$ and $A_2$ are the amplitudes of the electric field of the two coupled waveguides, and $\phi$ is the phase difference between the two waves. The parameters for Fig.~\ref{fig:photonic_couplers} are $\alpha_1 = 1$, $\alpha_2 = -\alpha \alpha_1$, $\beta_1 = 1$, $\beta_2 = \beta \beta_1$, $\varepsilon = 0.5$, $k = 5\beta_1$, $\alpha = 2$, $\beta = 1.5$. To compute the basins correctly, a logarithmically spaced grid along the $x$-axis is recommended. There is one stable fixed point (mod $2\pi$ for the third variable) whose basin is represented by slanted stripes. This is a stable mode with constant amplitude in the coupler. The other basin represents diverging trajectories.

\subsection{Dynamics of a CO2 Laser}
\begin{keywrds}SMB, ODE\end{keywrds}
\begin{figure}
\begin{center}
\includegraphics[width=\columnwidth]{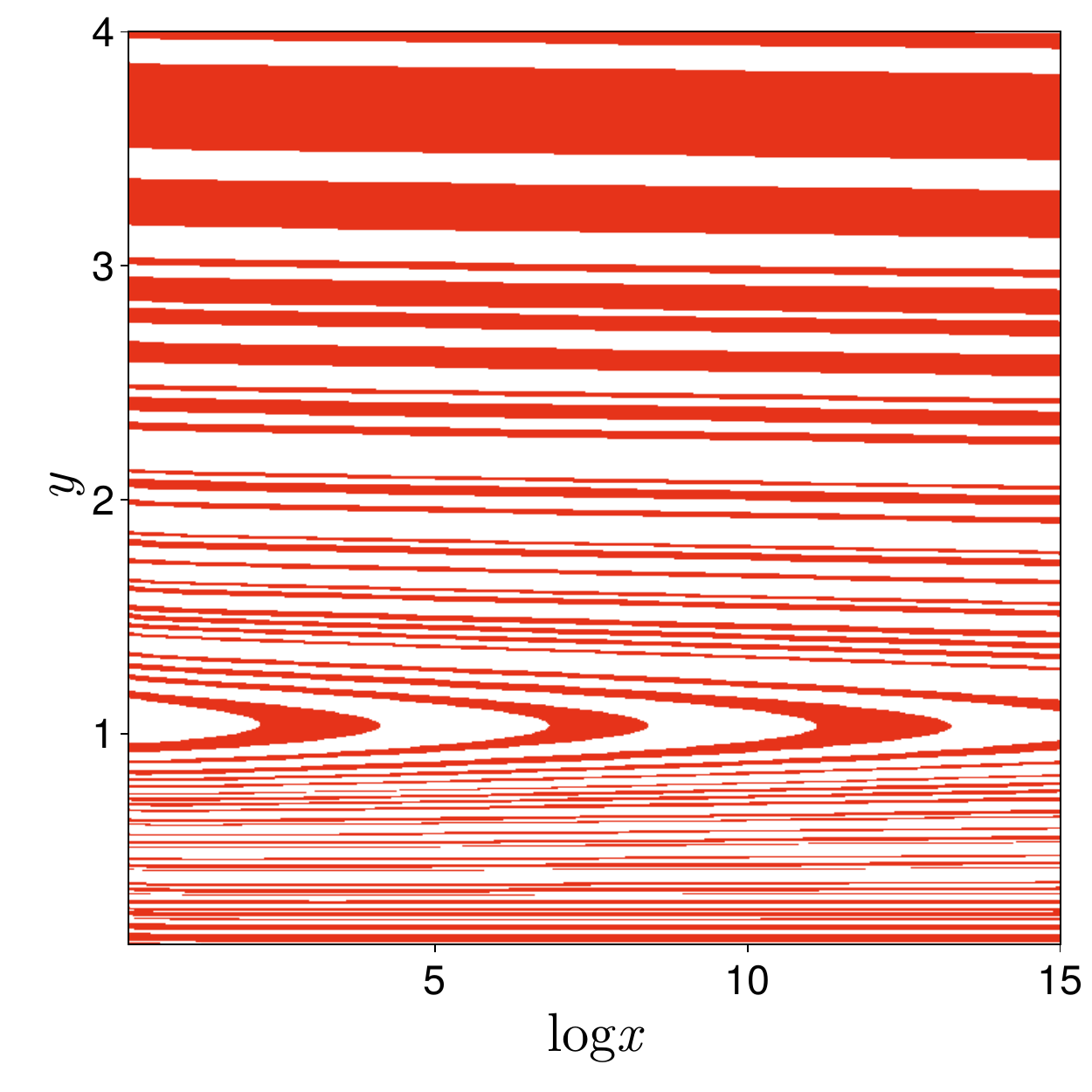}
\end{center}
\caption{\label{fig:co2_lasers}Basins of attraction of a CO2 modulated laser modeled in Eqs.~\ref{eq:co2_lasers}. Parameters for this figure are $k = 12$, $B_0 = 0.05$, $\gamma = 0.0025$, $\alpha = 0.002$, $p_0 = 1.252$, $f_{\text{mod}} = 0.005$, $m = 0.02$.}
\end{figure}

Lasers operate in a region full of nonlinearities. There are saturations, nonlinear elements, and so on. It is not surprising to find multistable modes of emission in these devices. This is the case in the study~\cite{Meucci_2022}, where a laser model with cavity loss modulation exhibits bistable modes. A secondary sinusoidal perturbation can effectively eliminate bistability when a suitably chosen phase difference is applied, a technique known as phase control.

The two-level laser model is described by the equations:
\begin{align}
\begin{split}
   \dot{x} &= - x \left( 1+k (B_0 + m\sin(2\pi f_\text{mod}t))^2 - y \right),\\
\dot{y} &= -\gamma y - \alpha xy + \gamma p_0.
\end{split}
\end{align}
However, this model has numerical issues, and a change of variable is suitable: $u = \log(x)$. The equations become: 
\begin{align}\label{eq:co2_lasers}
\begin{split}
\dot{u} &= -\left( 1+k (B_0 + m\sin(2\pi f_\text{mod}t))^2 - y \right),\\
\dot{y} &= -\gamma y - \alpha y e^u + \gamma p_0.
\end{split}
\end{align}
The basins in Fig.~\ref{fig:co2_lasers} have been computed with the modified model and the parameters $k = 12$, $B_0 = 0.05$, $\gamma = 0.0025$, $\alpha = 0.002$, $p_0 = 1.252$, $f_{\text{mod}} = 0.005$, and $m = 0.02$. The basins correspond to those published in~\cite{Meucci_2022} in a logarithmic scale on the $x$ axis.

\subsection{Pump Modulated Erbium-Doped Fiber Laser}
\begin{keywrds}FB, ODE\end{keywrds}
\begin{figure}
\begin{center}
\includegraphics[width=\columnwidth]{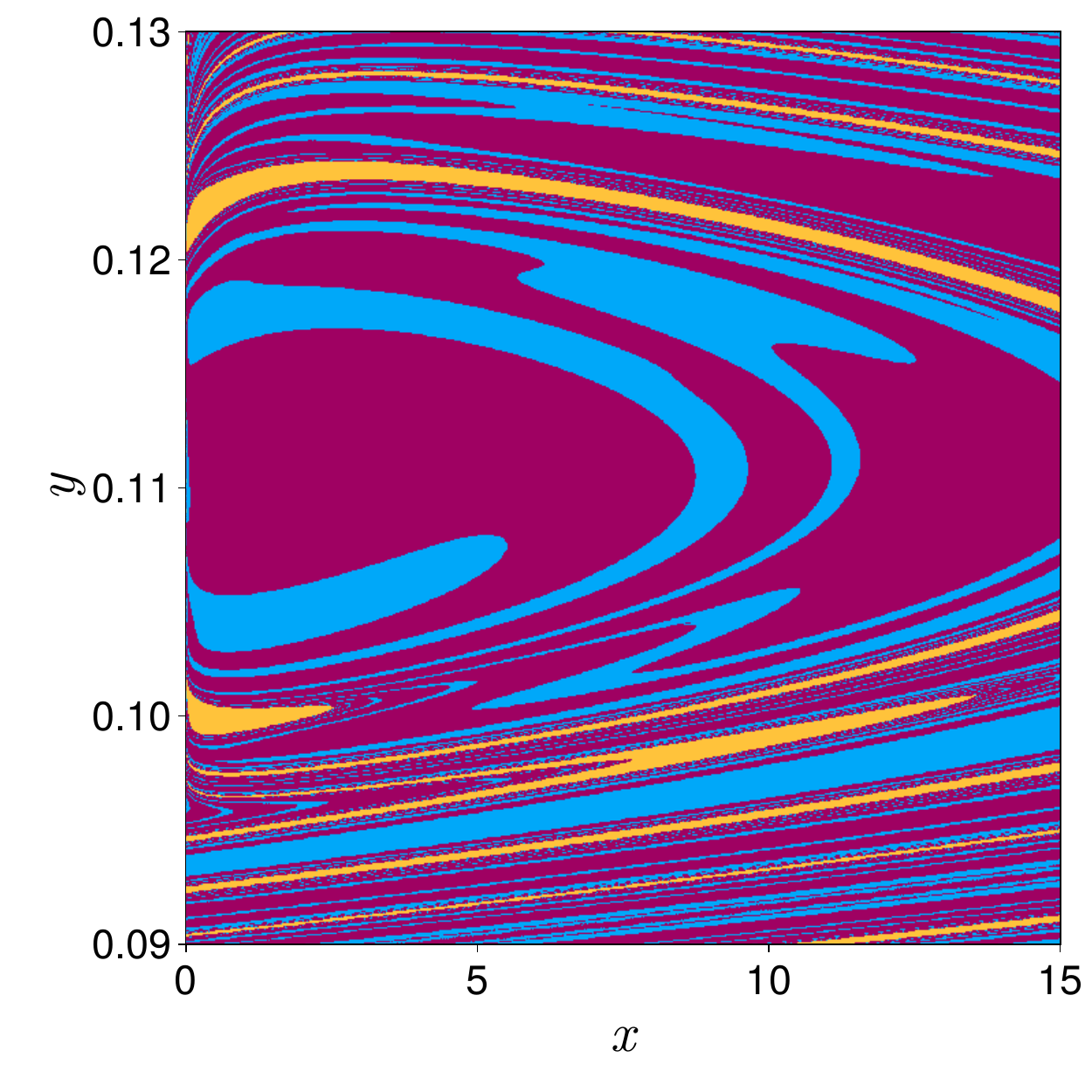}
\end{center}
\caption{\label{fig:erbium_laser}Basins of attraction of the pump-modulated erbium-doped fiber laser described in Eqs.~\ref{eq:erbium_laser}. The parameters are $a = 6.6207 \cdot 10^7$, $b = 7.4151 \cdot 10^6$, $c = 0.0163$, $d = 4.0763 \cdot 10^3$, and $\rho = 0.3075$.}
\end{figure}

The article~\cite{pisarchik2009control} investigates a multistable erbium-doped fiber laser controlled through harmonic modulation or stochastic noise applied to the pump parameter.

The study employs numerical simulations using a three-level laser model, analyzing both harmonic and stochastic modulations impacting the control parameter of the laser. The simulations are validated against experimental outcomes. Key equations governing the dynamics of the fiber laser are derived from the power-balance approach and rate equations that incorporate parameters such as pump power and population inversion.

The influence of noise on the volumes of basins of attraction is also highlighted, indicating a noise-dependent probability of reaching specific attractors. The results align closely with experimental data previously recorded on attractor behaviors in fiber lasers.

The fundamental equations governing the system dynamics are as follows:
\begin{align}\label{eq:erbium_laser}
\begin{split}
    p_m(t) &= 506 (1 + m \sin(2 \pi f t))\\
    \frac{dx}{dt} &= a x y - b x + c (y + \rho)\\
    \frac{dy}{dt} &= -d x y -(y + \rho) + p_m(t) (1 - e^{-18 (1 - (y + \rho)/0.615)}) 
\end{split}
\end{align}
The parameters for the computed basins in Fig.~\ref{fig:erbium_laser} are $a = 6.6207 \cdot 10^7$, $b = 7.4151 \cdot 10^6$, $c = 0.0163$, $d = 4.0763 \cdot 10^3$, and $\rho = 0.3075$. To compute the basins correctly, an irregular grid is necessary since the trajectories of the attractors accumulate close to the vertical axis. In these cases, the recurrence algorithm cannot differentiate between attractors if they intersect the same cell grid. A logarithmically spaced grid allows differentiation between these cases. Once the grid has been set correctly, the recurrence algorithm operates without any problems.

\subsection{Alfvén Basin Boundary}
\begin{keywrds}FB, ODE\end{keywrds}
\begin{figure}
\begin{center}
\includegraphics[width=\columnwidth]{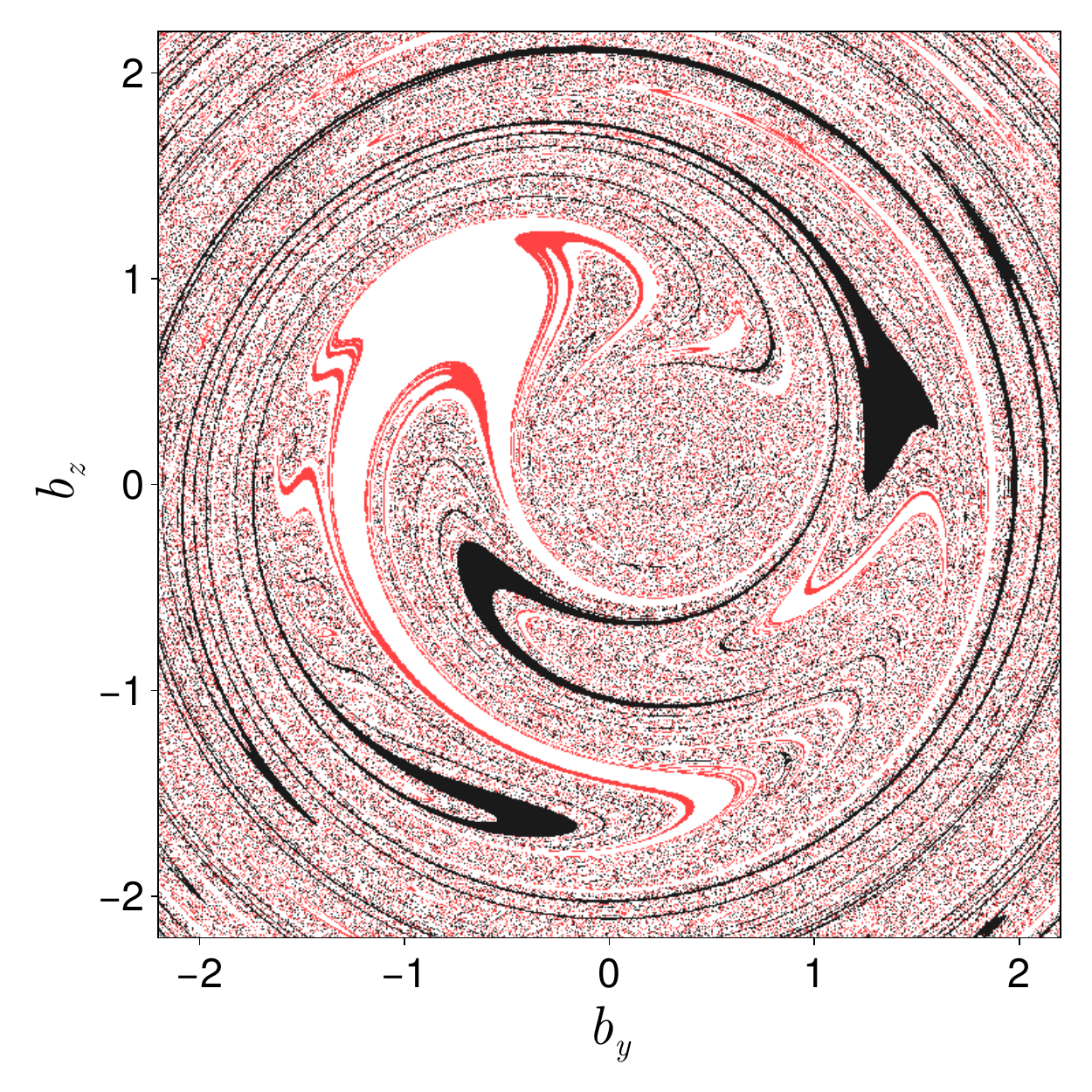}
\end{center}
\caption{\label{fig:alfven}Basins of attraction of large-amplitude Alfvén waves traveling along an ambient magnetic field, described by Eqs.~\ref{eq:alfven}.}
\end{figure}

In a particular complex plasma region, the transition mechanism to Alfvén waves can occur via boundary crises. The paper~\cite{chian2002alfven} explores a model based on the driven-dissipative derivative nonlinear Schrödinger equation for plasma dynamics. 

The study starts with the nonlinear Schrödinger equation modeling the large-amplitude Alfvén wave traveling along an ambient magnetic field in the $x$ direction. Several assumptions and modifications reduce the equations to a set of ordinary differential equations expressing the evolution of the transverse magnetic field. 

The study identifies a complex plasma region with multiple attractors and shows that double boundary crises can lead to the sudden disappearance or appearance of chaotic attractors due to homoclinic tangencies involving the same unstable periodic orbit.  

The equations of motion are: 
\begin{align}\label{eq:alfven}
\begin{split}
\dot{b}_y - \nu \dot{b}_z &= \frac{\partial H}{\partial b_z} + a \cos \theta\\
\dot{b}_z + \nu \dot{b}_y &= -\frac{\partial H}{\partial b_y} + a \sin \theta\\
\dot{\theta} &= \Omega
\end{split}
\end{align}
where $ H $ is the Hamiltonian-like function with the partial derivatives:
\begin{align*}
    \frac{\partial H}{\partial b_y} &= (b_z^2 + b_y^2 -1) b_y - \lambda (b_y -1 )\\ 
    \frac{\partial H}{\partial b_z} &= (b_z^2 + b_y^2 -1) b_z - \lambda b_z,
\end{align*}
with $ b = (b_y, b_z) $ being the amplitude of the transverse magnetic field. Additional parameters are defined as follows: $ \nu = 0.01747$, $a = 0.1$, $\omega = -1$, $\lambda = 1/4$. Since the first two differential equations are driven by periodic forcing, we can set up a stroboscopic map for the detection of the attractors. The basins in Fig.~\ref{fig:alfven} are interesting, featuring a twisted rotation deforming the plot. There are four attractors appearing just before the occurrence of a boundary crisis affecting an attractor.

\subsection{9D Shear Flow Model}
\begin{keywrds}FB, ODE\end{keywrds}
\begin{figure}
\begin{center}
\includegraphics[width=\columnwidth]{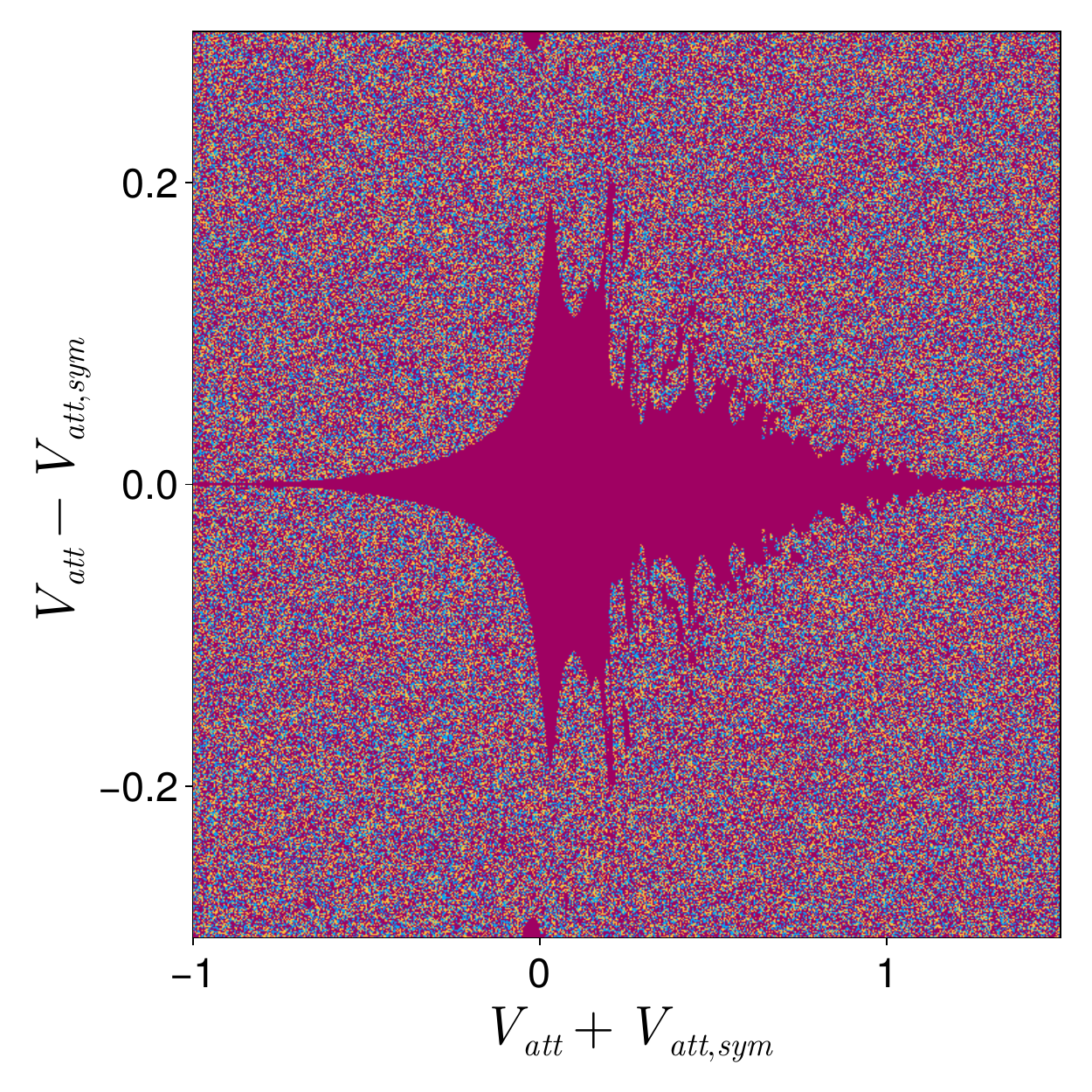}
\end{center}
    \caption{\label{fig:9d_shearflow}Basins of attraction of~\ref{eq:9d_shearflow}. The basins correspond to two non-trivial attracting orbits and the laminar attractor (in red) for $Re = 425$. The process of choosing the initial conditions and the parameters is described in the text.}
\end{figure}

Computational fluid dynamics is usually a domain where partial differential equations are the gold standard for inquiry. Nevertheless, there are successful approaches involving transformations of fluid flow into a set of ordinary differential equations facilitating tractable numerical computation. The technique, known as Galerkin projection, is applied only on a finite domain of the fluid.

Ref.~\cite{Joglekar_2015} investigates the geometry of the edge of chaos in a nine-dimensional sinusoidal shear flow model, focusing on how the edge changes with varying Reynolds numbers and the minimum perturbation required to transition from the laminar state to a turbulent one. The authors employed numerical simulations of a nine-dimensional model representing sinusoidal shear flow to analyze the edge of chaos. They computed the scaling of minimum perturbations needed to destabilize the laminar attractor and examined the lifetimes of turbulent trajectories. They followed the edge trajectories and calculated distances from the laminar attractor to the edge.

The distance of the edge of chaos from the laminar attractor scales with the Reynolds number as approximately $Re^{-2}$. As the Reynolds number increases, smaller perturbations are sufficient to induce turbulence. The average lifetime for transient chaotic behavior scales as $Re^{4.51}$. The study provides insights into the geometric structure of the edge of chaos in the context of turbulent transitions in low-dimensional systems. 

The model is described in detail in~\cite{Moehlis_2004}. We reproduce here only the numerical coefficients for our example. The flow is approximated as a linear combination of modes:
   \[
   \mathbf{u}(x, t) = \sum_{m} a_m(t) \mathbf{u}_m(x),
   \]
   where $a_m(t)$ are time-dependent amplitudes of modes $\mathbf{u}_m(x)$. The resulting ODEs for the amplitudes of the nine modes obtained through Galerkin projection are given by
\begin{align}\label{eq:9d_shearflow}
\begin{split}
    \frac{da_1}{dt} &= -\text{a}_1  \text{k}_1 / \text{Re} + \sigma_1  \text{a}_6  \text{a}_8 + \sigma_2  \text{a}_2  \text{a}_3 + \text{k}_1 / \text{Re}; \\
    \frac{da_2}{dt} &= -\text{a}_2  \text{k}_2 / \text{Re} + \sigma_3  \text{a}_4  \text{a}_6 + \sigma_4  \text{a}_5  \text{a}_7 + \sigma_5  \text{a}_5  \text{a}_8 + \sigma_6  \text{a}_1  \text{a}_3 + \sigma_7  \text{a}_3  \text{a}_9; \\
    \frac{da_3}{dt} &= -\text{a}_3  \text{k}_3 / \text{Re} + \sigma_8  (\text{a}_4  \text{a}_7 + \text{a}_5  \text{a}_6) + \sigma_9  \text{a}_4  \text{a}_8; \\
    \frac{da_4}{dt} &= -\text{a}_4  \text{k}_4 / \text{Re} + \sigma_{10}  \text{a}_1  \text{a}_5 + \sigma_{11}  \text{a}_2  \text{a}_6 + \sigma_{12}  \text{a}_3  \text{a}_7 + \sigma_{13}  \text{a}_3  \text{a}_8 + \sigma_{14}  \text{a}_5  \text{a}_9; \\
    \frac{da_5}{dt} &= -\text{a}_5  \text{k}_5 / \text{Re} + \sigma_{15}  \text{a}_1  \text{a}_4 + \sigma_{16}  \text{a}_2  \text{a}_7 + \sigma_{17}  \text{a}_2  \text{a}_8 + \sigma_{18}  \text{a}_4  \text{a}_9 + \sigma_{19}  \text{a}_3  \text{a}_6; \\
    \frac{da_6}{dt} &= -\text{a}_6  \text{k}_6 / \text{Re} + \sigma_{20}  \text{a}_1  \text{a}_7 + \sigma_{21}  \text{a}_1  \text{a}_8 + \sigma_{22}  \text{a}_2  \text{a}_4 + \sigma_{23}  \text{a}_3  \text{a}_5 + \sigma_{24}  \text{a}_7  \text{a}_9 + \sigma_{25}  \text{a}_8  \text{a}_9; \\
    \frac{da_7}{dt} &= -\text{a}_7  \text{k}_7 / \text{Re} + \sigma_{26}  (\text{a}_1  \text{a}_6 + \text{a}_6  \text{a}_9) + \sigma_{27}  \text{a}_2  \text{a}_5 + \sigma_{28}  \text{a}_3  \text{a}_4; \\
    \frac{da_8}{dt} &= -\text{a}_8  \text{k}_8 / \text{Re} + \sigma_{29}  \text{a}_2  \text{a}_5 + \sigma_{30}  \text{a}_3  \text{a}_4; \\
    \frac{da_9}{dt} &= -\text{a}_9  \text{k}_9 / \text{Re} + \sigma_{31}  \text{a}_2  \text{a}_3 + \sigma_{32}  \text{a}_6  \text{a}_8.
\end{split}
\end{align}
where $a_i$ is the amplitude of mode $i$. For the constants defined in~\cite{Moehlis_2004}, we obtain the coefficients: 
\begin{equation*}
\begin{matrix}
\sigma_i = [-1.253& 
1.4&  1.871& -0.561& -0.236& -1.4\\ 
-1.4& 0.528& -0.603& -0.467& -0.88& -0.792\\
-0.333& -0.467& 0.467& 0.264& -0.236& 0.467\\ 
0.528& 0.467& 1.253& -0.991& -1.056& 0.467\\ 
1.253& -0.467& 0.297& 0.264& 0.472& 0.937\\ 
1.4& -1.253]& 
\end{matrix}
\end{equation*}
\begin{equation*}
\begin{matrix}
k_i = [
2.467& 6.068& 5.245& 4.596& 3.774& 7.374\\
6.551&  6.551&  22.207]
\end{matrix}
\end{equation*}
The initial conditions for the basins in Fig.~\ref{fig:9d_shearflow} are taken from the linear combination of two vectors $v_{att}$ and $v_{att,sym}$ such that:
\begin{equation*}
\begin{matrix}
v_{att} = [ 
0.129992 - 1& -0.0655929& 0.0475706\\
0.0329967& 0.0753854& -0.00325098\\ 
-0.042364& -0.019685 &-0.101453]
\end{matrix}
\end{equation*}
\begin{equation*}
\begin{matrix}
v_{att,sym} = [
0.129992 - 1& 0.0655929& -0.0475706\\ 
-0.0329967 & -0.0753854& -0.00325098\\
-0.042364& -0.019685 & -0.101453]
\end{matrix}
\end{equation*}
These vectors span a plane containing the laminar attractor and two symmetric non-trivial attractors. The initial point as a function of two scalars $x,y$ is $x(v_{att} + v_{att,sym}) + y(v_{att} - v_{att,sym})$. The basins in Fig.~\ref{fig:9d_shearflow} show a clear fractal boundary between the stable states. The laminar attractor in red has the largest connected basins.

\subsection{Split Ring Resonator}
\begin{keywrds}FB, ODE\end{keywrds}
\begin{figure}
\begin{center}
\includegraphics[width=\columnwidth]{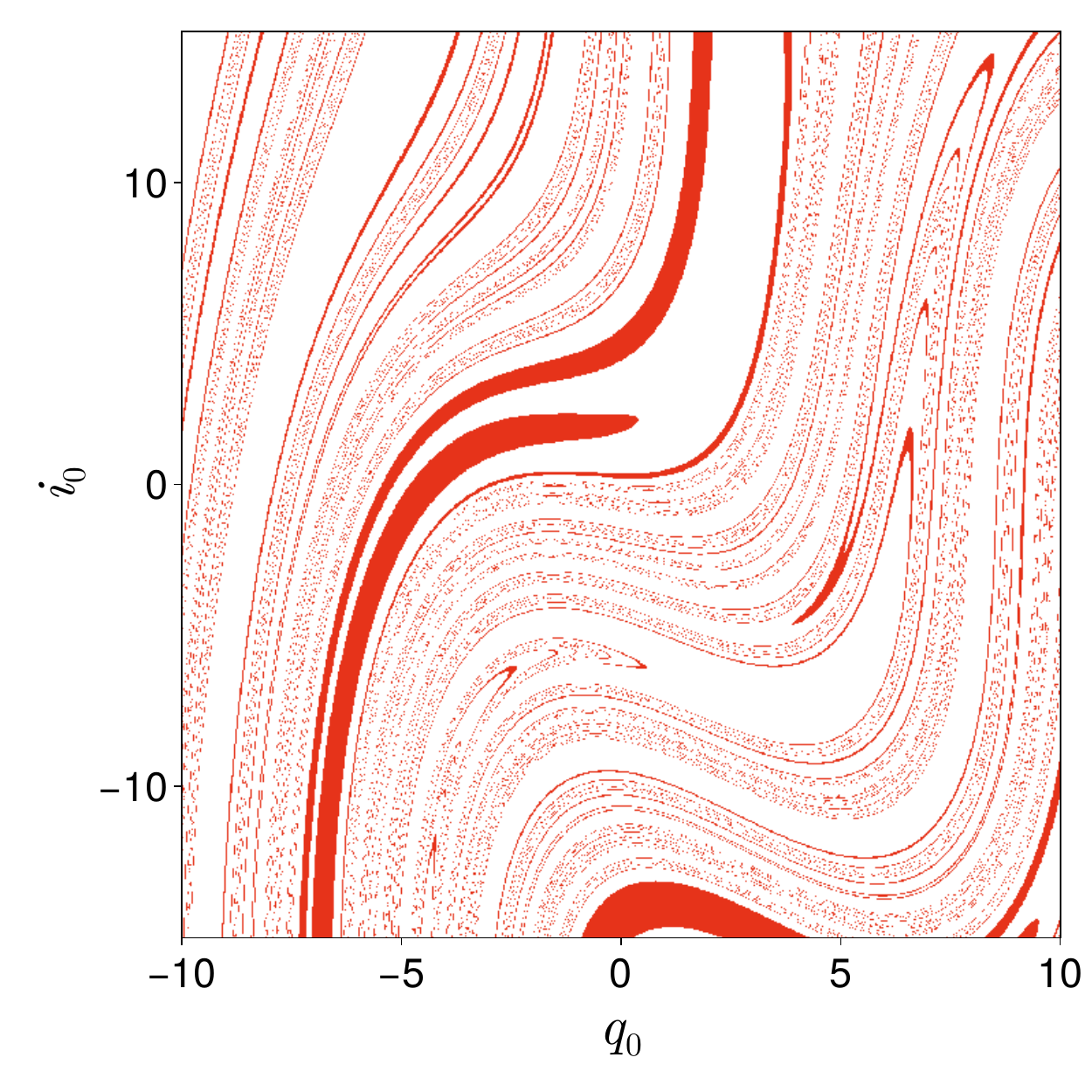}
\end{center}
\caption{\label{fig:split_ring_res}Basins of attraction of a split ring resonating at terahertz frequency, as described in Eqs.~\ref{eq:split_ring_res}. The attractors in the phase space comprise a period-6 and a chaotic attractor.}
\end{figure}

The paper~\cite{Leutcho_2023} investigates the nonlinear dynamics of a single-gap terahertz split-ring resonator under electromagnetic radiation. The study consists of the analysis of the nonlinear model using common tools of nonlinear dynamics: bifurcation diagrams, Lyapunov exponents, basins of attraction, and so on. The results reveal the presence of chaotic and periodic dynamics as the excitation amplitude and loss parameter are varied. The study also confirms the transition to chaos via period doubling and describes distinct regions in parameter space where various dynamic behaviors occur.

The normalized ordinary differential equation for the nonlinear oscillator is given by:  
\begin{equation}\label{eq:split_ring_res}
   \frac{d^2 q}{d\tau^2} + \sigma \frac{dq}{d\tau} + q - \beta q^2 + \eta q^3 = \mu \cos(\omega \tau),
\end{equation}
where $q$ is the normalized charge, $\sigma$ is the loss parameter, $\mu$ is the normalized amplitude of the applied electromagnetic force, $\beta$ and $\eta$ are constants related to the system nonlinearity, and $\omega$ is the driving frequency. The parameters for Fig.~\ref{fig:split_ring_res} are: $\omega = 1.0285$; $\mu = 35$; $\sigma = 0.38$; $\beta = 0.4$; $\eta = 0.08$.

There is a fractal boundary between the two attractors.

\subsection{6D Shear Flow Model}
\begin{keywrds}SMB, ODE\end{keywrds}
\begin{figure}
\begin{center}
\includegraphics[width=\columnwidth]{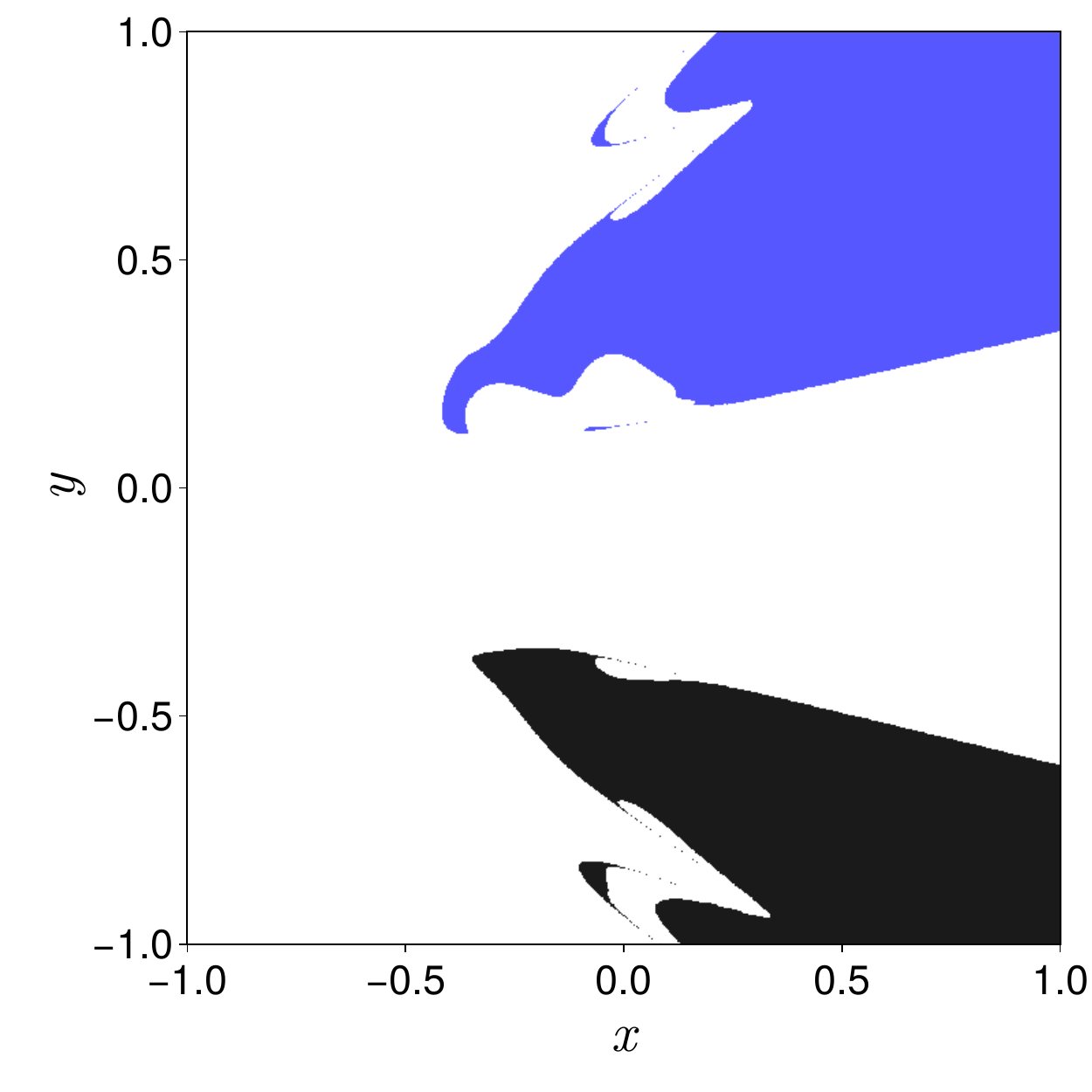}
\end{center}
\caption{\label{fig:6d_shearflow}Basins of attraction of~\ref{eq:6d_shearflow}. The parameter values of the model are: $Re = 307$, $k_1 = 2.46$, $k_2 = 15.11$, $k_3 = 12.64$, $k_4 = 11.07$, $k_5 = 6.45$, $k_6 = 13.20$, $\sigma_0 = 0.73$, $\sigma_1 = 1.39$, $\sigma_2 = 0.60$, $\sigma_3 = 0.29$, $\sigma_4 = 0.73$, $\sigma_5 = -0.066$, $\sigma_6 = 0.054$.}
\end{figure}

The paper~\cite{Lebovitz_2013} elucidates the structure of the edge, a codimension-one invariant manifold distinguishing orbits leading to a laminar state from those that do not in the context of turbulence onset in shear flows.

The authors analyze a series of low-dimensional dynamical models related to shear flows to isolate geometric features of the edge. They investigate its properties in two, three, four, and six-dimensional models to characterize the edge structure and its role in relaminarization.

The edge serves as a boundary separating trajectories in phase space; orbits 'below' the edge typically relaminarize quickly, while those 'above' the edge do so more slowly and often after circling around the edge state. The six-dimensional model described in the article involves the Navier-Stokes equations modified for a plane Poiseuille flow scenario. The model is based on a Galerkin truncation of the Navier-Stokes equations, choosing adequate modes to represent the fluid dynamics in a limited spatiotemporal domain. The mathematical representation of the system can be expressed as follows:
\begin{equation}\label{eq:6d_shearflow}
\frac{dx}{dt} = Ax + b(x)
\end{equation}
where $x$ is the state vector in $\mathbb{R}^6$ representing the different modes. The matrix $A$ is given by:
\begin{displaymath}
A =
\begin{pmatrix}
    -\dfrac{k_1}{Re} & 0 & 0 & 0 & 0 & 0 \\
    0 & -\dfrac{k_2}{Re} & \sigma_0 & 0 & 0 & 0 \\
    0 & 0 & -\dfrac{k_3}{Re} & 0 & 0 & 0 \\
    0 & 0 & 0 & -\dfrac{k_4}{Re} & 0 & -\sigma_3 \\
    0 & 0 & 0 & 0 & -\dfrac{k_5}{Re} & 0 \\
    0 & 0 & 0 & \sigma_3 & 0 & -\dfrac{k_6}{Re}
\end{pmatrix}
\end{displaymath}
Here, $k_i$ are all positive constants related to wavenumbers in the fluid flow. The nonlinear term $b(x)$ is represented as:
\begin{displaymath}
b =
\begin{pmatrix}
-\sigma_0 x_2 x_3 \\
\sigma_0 x_1 x_3 - \sigma_1 x_4 x_5 \\
-(\sigma_4 + \sigma_5)x_5 x_6\\
\sigma_2 x_2 x_5 - \sigma_3 x_1 x_6 \\
(\sigma_1 - \sigma_2)x_2 x_4 + (\sigma_4 - \sigma_6)x_3 x_6 \\
(\sigma_5 + \sigma_6)x_3 x_5 + \sigma_3 x_1 x_4
\end{pmatrix}
\end{displaymath}
The terms $\sigma_i$ are also constants determined based on the Galerkin procedure applied to the Navier-Stokes equations. In the figure caption of Fig.~\ref{fig:6d_shearflow}, we provide the values of these parameters. The basins are a projection of the $x_1$-$x_5$ phase space with the other variables being $x_2 = -0.0511$, $x_3 = -0.0391$, $x_4 = 0.0016$, $x_6 = 0.126$ for all other initial conditions. The results show the boundary of the edge in this phase space with three basins of attraction: the laminar state in white and the basins of two periodic attractors. The transients leading to the stable states can be very long, depending on the distance to the unstable manifold originating on the boundary.

\subsection{\label{sec:binary_BH}Binary Black Hole System}
\begin{keywrds}WD, OH\end{keywrds}
\begin{figure}
\begin{center}
\includegraphics[width=\columnwidth]{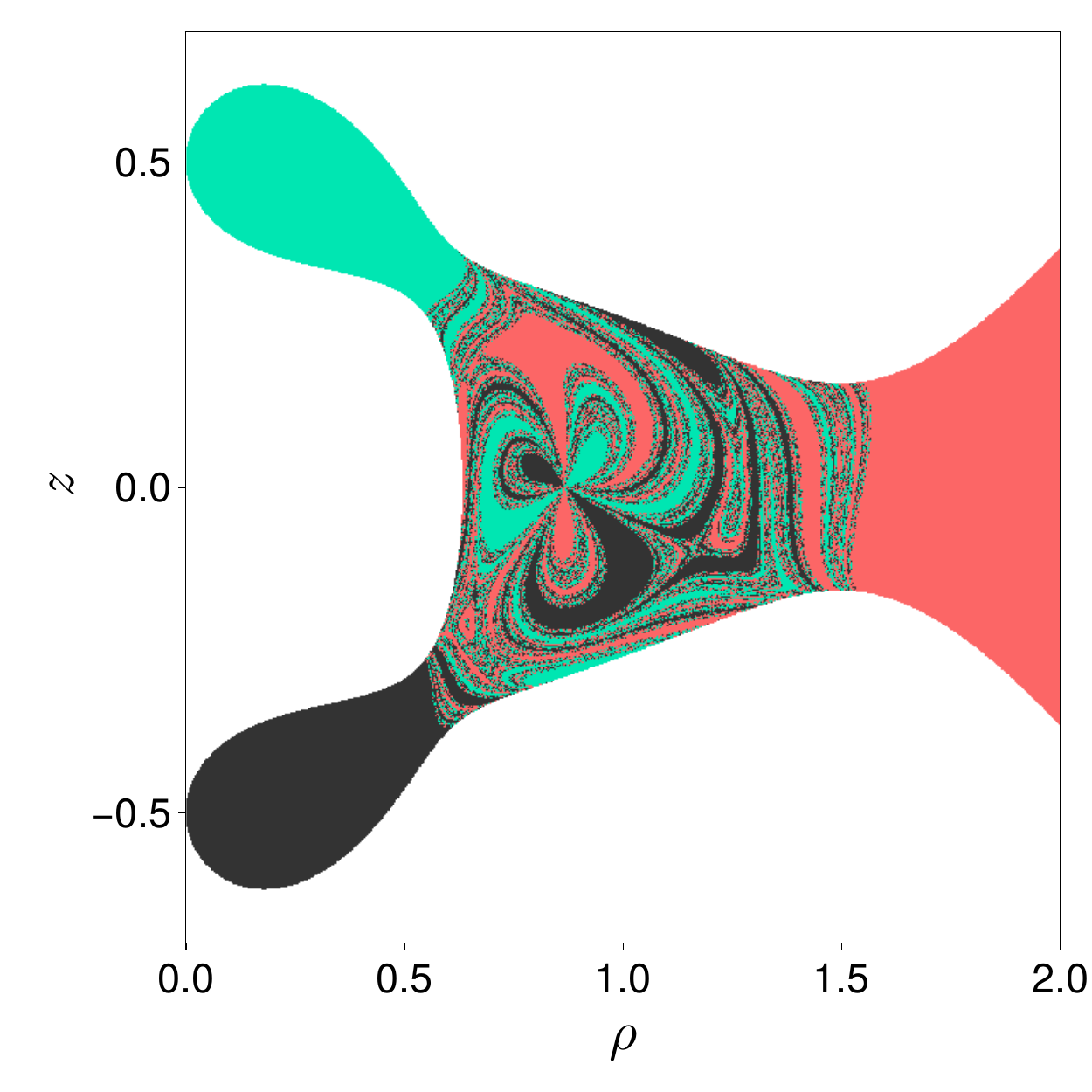}
\end{center}
\caption{\label{fig:binary_BH}Escape basins of the system described in Eqs.~\ref{eq:binary_BH}. We have set the angular momentum $p_\varphi - p_\varphi^* = 0.03$ for this simulation.}
\end{figure}

A binary black hole is a system in which two black holes orbit around each other. The dynamics of their interaction are governed by the laws of gravity and famously produce gravitational waves that have recently been detected with very sensitive instruments. A photon approaching the system can encounter three outcomes: it can ripple away and leave the system, fall into the event horizon of black hole one, or into the event horizon of black hole two.

The article~\cite{Daza_2018} investigates the fractal structure formed by the binary black hole system using techniques from nonlinear dynamics. Exit basins in the phase spaces of the Majumdar-Papapetrou binary black hole model represent the fate of photons approaching the system. Not only do some regions of the binary black hole shadows exhibit fractal characteristics, but they also display the Wada property. A qualitative transition in the Wada property is identified as the separation distance between the black holes changes. The model involves integrating the trajectories of photons interacting with a simplified Hamiltonian potential:
\begin{equation}\label{eq:binary_BH}
   H =  \frac{1}{2} (p_{\rho}^2 + p_{z}^2) + V(\rho, z) = 0,
\end{equation}
where $V$ is the effective potential, which is determined by the geometry of the black hole spacetime and can be expressed as:
\begin{equation}
   V(\rho, z) = -\frac{1}{2\rho^2} (h - p_{\phi})(h + p_{\phi}),
\end{equation}
where $h(\rho, z) = \rho U^2$ and $U$ is a function related to the spacetime metric: 
\begin{equation}
    U(\rho, z) = 1 + \frac{M}{\sqrt{\rho^2 + (z- z_1)^2}} + \frac{M}{\sqrt{\rho^2 + (z - z_2)^2}}  
\end{equation}
$M$ is the mass of one black hole, and $z_i$ are their relative positions with the following values: $M = 1$, $z_1 = 0.5$, $z_2 = -0.5$. The equations of motion are derived from Hamilton's equations:
\begin{align*}
\begin{split}
\dot x &= \frac{\partial H}{\partial p_x}, \quad \dot y = \frac{\partial H}{\partial p_y}, \\
\dot p_x &= -\frac{\partial H}{\partial x},\\
\dot p_y &= -\frac{\partial H}{\partial y}. 
\end{split}
\end{align*}
The initial conditions of the escape basins in Fig.~\ref{fig:binary_BH} involve the computation of the momentum $p_z$ and $p_\rho$. First, we set the initial angular momentum $p_\varphi$ below the critical threshold $p_\varphi < p_\varphi^* = \dfrac{1}{2} 5^{5/4}\varphi^{3/2}$, where $\varphi = \dfrac{1}{2}(1+ \sqrt{5})$. For the initial conditions, we set the coordinates $\rho$ and $z$ first and choose the initial three-momentum to be tangential to the circle of radius $\sqrt{(\rho - \rho_{max})^2 + (z - z_{max})^2}$, centered on $(\rho_{max}, z_{max})= (\sqrt{3}/2, 0)$. We provide the expressions for completeness: 
\begin{align*}
\begin{split}
p_{total} &= \sqrt{(U(\rho,z)^4-p_\varphi^2/\rho^2)},\\
p_\rho &= \frac{p_{total}}{\sqrt{(1+(\rho-\sqrt{3}/2)^2/z^2)}},\\
p_z &= \sqrt{(p_{total}^2-p_\rho^2)}.
\end{split}
\end{align*}
Once the initial conditions have been set up, the solver integrates the trajectory until it detects that a photon has escaped or has reached the shadow of one of the two black holes. In this case, the position must satisfy $|(z- z_i)| + |\rho| < \varepsilon$, where $z_i$ are the positions of the black holes and $\varepsilon = 1/50$. The basin boundary in Fig.~\ref{fig:binary_BH} is clearly fractal and has the Wada property for the parameters mentioned above.

\subsection{\label{sec:Intermingled_sommerer}Intermingled Basins of a Forced Particle}
\begin{keywrds} IB, ODE \end{keywrds}
\begin{figure}
\begin{center}
\includegraphics[width=\columnwidth]{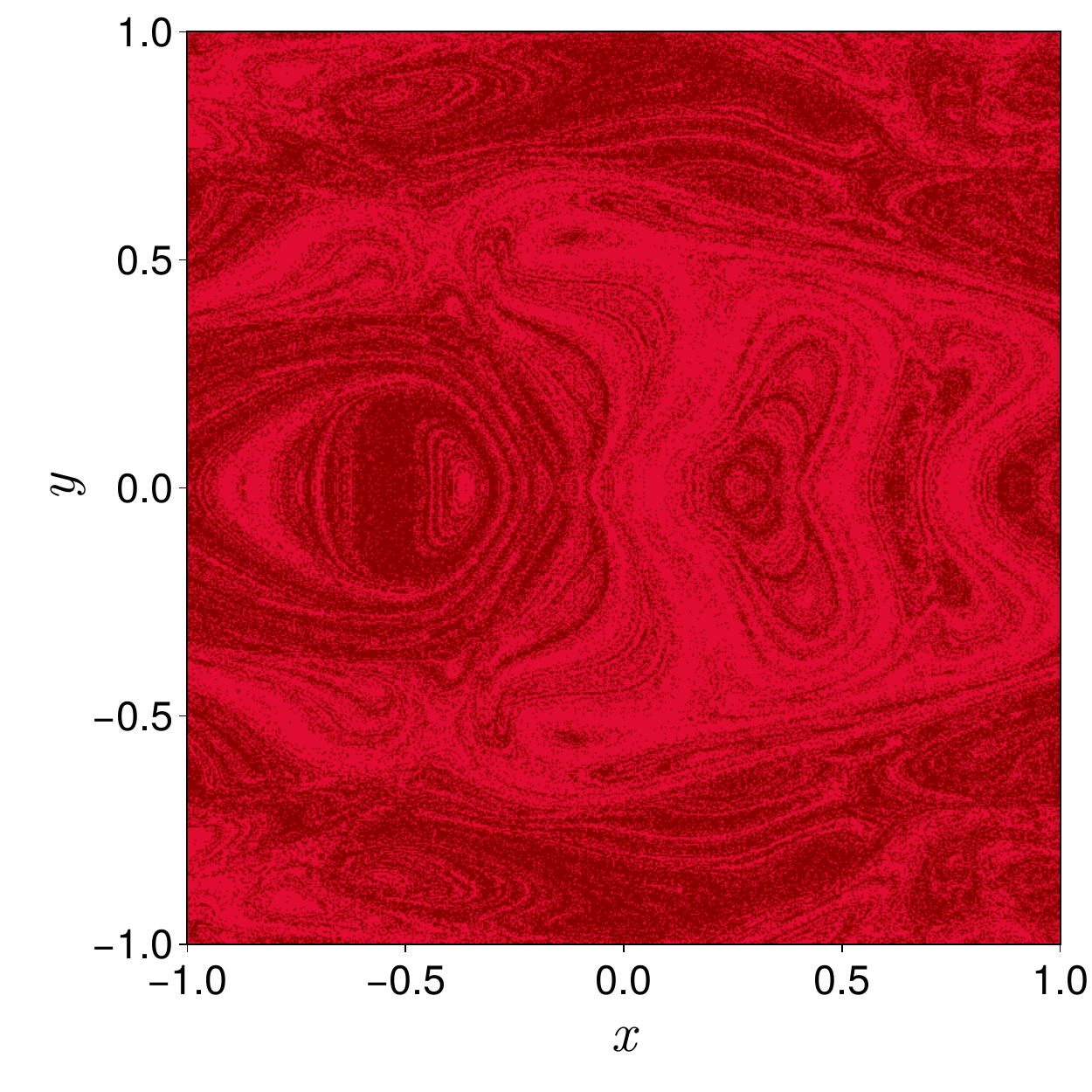}
\end{center}
\caption{\label{fig:Intermingled_sommerer}Basins of attraction of periodically driven particle in a two dimensional potential of Eq.~\ref{eq:Intermingled_sommerer}. The two basins are intermingled for the parameters: $\gamma = 0.632$, $f_0 = 1.0688$, $\omega = 2.2136$, $s = 20$, $p = 0.098$, $k = 10$. The initial conditions for the system are $dx/dt = 0$ and $dy/dt = 0$.}
\end{figure}

Ref.~\cite{sommerer1996intermingled} explores the concept of intermingled basins of attraction with a concrete example of a physical system,  a simple mechanical model of a particle subject to friction and sinusoidal forcing within a two-dimensional potential field. This model is defined by a set of ordinary differential equations that describe the particle motion. The basins of attraction exhibit intermingling and the study reveals that two chaotic attractors exist in the system. The presence of a single symmetry leading to an invariant manifold is sufficient to create intermingled basins, making such phenomena more likely to occur in common practical situations encountered in nature and engineering. 

The model are two coupled differential equation of second order:
\begin{align}\label{eq:Intermingled_sommerer}
\begin{split}
\frac{d^2x}{dt^2} & + \gamma \frac{dx}{dt} - 4x(1-x^2) + 2 s x y^2 = f_0 \sin(\omega t) \\
\frac{d^2y}{dt^2} & + \gamma \frac{dy}{dt} + 2ys(x^2 - p) + 4ky^3 = 0
\end{split}
\end{align}
where $x$ and $y$ are the coordinates of the particle. The corresponding nonlinear potential is: $V(x, y) = (1 - x^2)^2  + s y^2 (x^2 - p) + k y^4$. The basins in Fig.~\ref{fig:Intermingled_sommerer} have the intermingled property for the parameters: $\gamma = 0.632$, $f_0 = 1.0688$, $\omega = 2.2136$, $s = 20$, $p = 0.098$, $k = 10$. 

\subsection{The Single Kicked Rotor} 
\begin{keywrds}FB, MAP\end{keywrds}
\begin{figure}
\begin{center}
\includegraphics[width=\columnwidth]{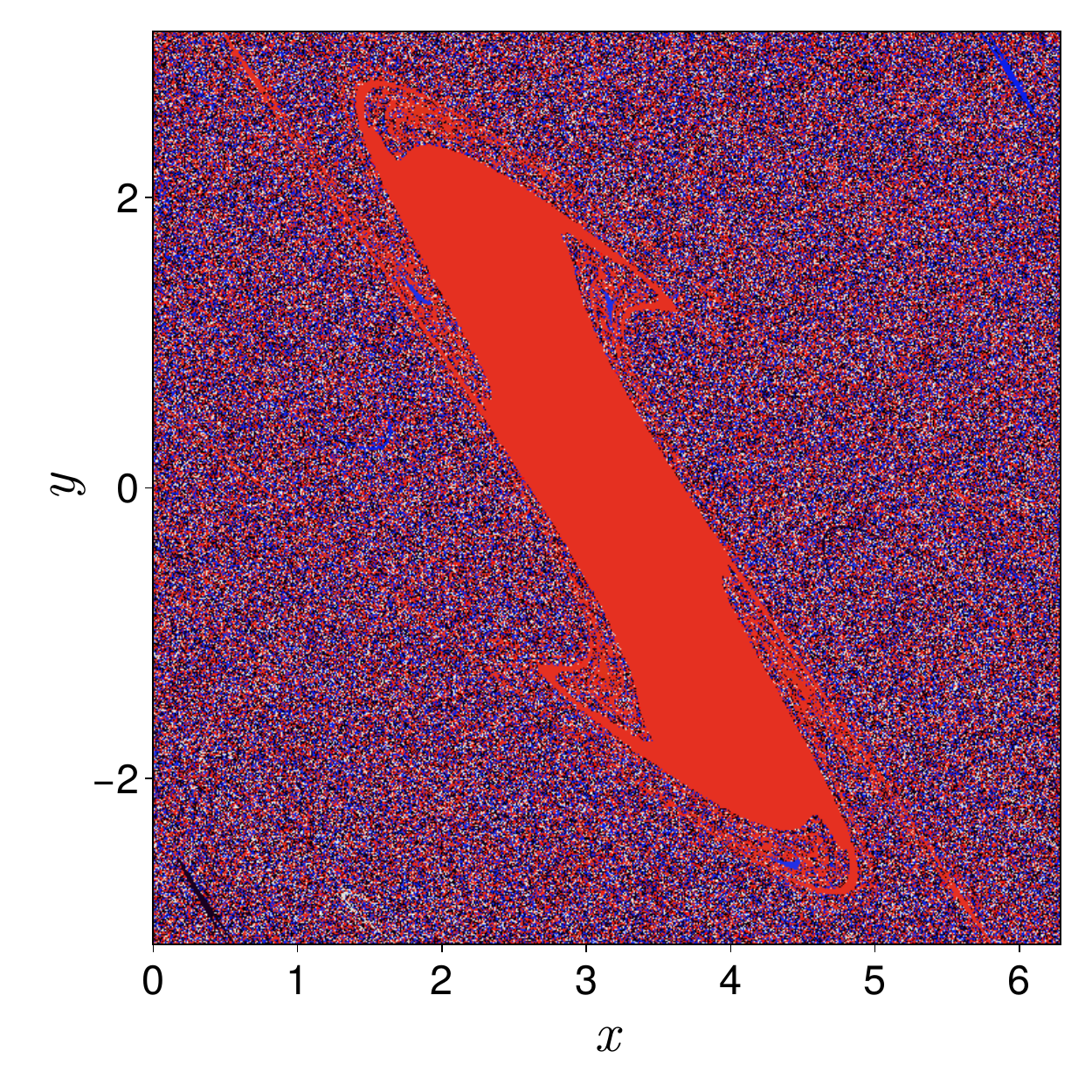}
\end{center}
\caption{\label{fig:2d_kicked_rotor}Basins of attraction of the kicked rotor for the model in Eqs.~\ref{eq:2d_kicked_rotor} with the parameters $f_0 = 4$ and $\nu = 0.02$. The algorithm has identified 64 basins in this region of phase space.}
\end{figure}

The single kicked rotor describes the time evolution of a mechanical pendulum subjected to periodic kicks of constant force. The kicks are modeled as delta functions, allowing the evolution of the pendulum to be reduced to a discrete map at the moments of the kicks. The equations for the phase $x$ and angular velocity $y$ are given by~\cite{Feudel_1996}:  
\begin{align}\label{eq:2d_kicked_rotor}
\begin{split}
x_{n+1} &= x_n + y_n \mod (2\pi)\\
y_{n+1} &= (1-\nu)y_n + f_0 \sin(x_n + y_n) 
\end{split}
\end{align} 
The significance of this system lies in the presence of many coexisting attractors for a low dissipation parameter $\nu$. For $\nu=0$, the system is conservative and there is no attractor. However, for small $\nu$, many orbits remain stable in phase space, explaining the plethora of coexisting stable states. In Fig.~\ref{fig:2d_kicked_rotor}, the figure replicates the basins published in~\cite{Feudel_1996}. In this region, 64 basins have been identified for this resolution. 

Due to the large number of possible attractors, the parameters of the mapper algorithm must be adjusted with a higher number of recurrences. In other cases, transient trajectories may pass near an existing attractor and be misinterpreted by the algorithm.

\section{Examples in Engineering}

Nonlinear dynamical systems appear pop up in all fields of engineering naturally; semiconductors, mechanical devices, materials, chemical processes, bridges: it happens everywhere. Linear systems analysis is part of the basic engineering toolbox. However it sometimes fails to grasp all the behaviors of a system. The examples in the following show multistable systems arising due to impact or nonlinearities intrinsic to the models. 

\subsection{Impact Dynamical System}
\begin{keywrds} SMB, ODE\end{keywrds}
\begin{figure}
\begin{center}
\includegraphics[width=\columnwidth]{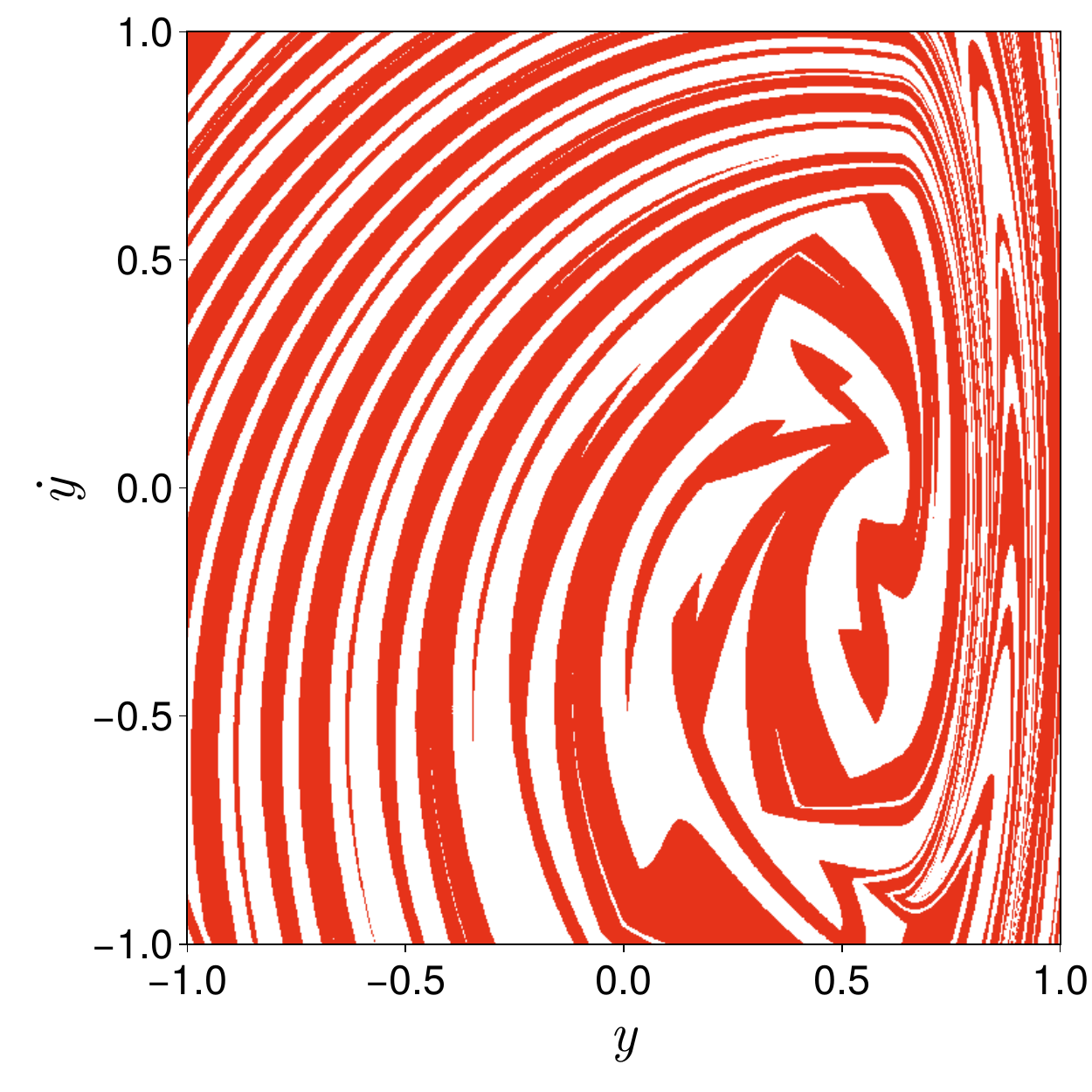}
\end{center}
\caption{\label{fig:impact}Basins of attraction of the system in Eq.~\ref{eq:impact} simulating a beam impacting due to an external periodic forcing. The two attractors in the phase space are a fixed point and a chaotic attractor.}
\end{figure}

Modeling and including the effects of impacts in dynamical systems is crucial in engineering. Discontinuities in ordinary differential equations can be handled with common numerical tools. Here we present an example of a simple cantilever beam impacting when a condition is met. The article~\cite{souza2008suppressing} investigates the ability of different stiffnesses to suppress chaotic behavior near grazing transitions.

Two different types of nonlinear stiffness are studied using different piecewise functions. The piecewise linear system exhibits a vast region of chaotic behavior upon grazing bifurcation, while the piecewise nonlinear system displays periodic attractors. This suggests that structural nonlinearity effectively suppresses chaos.  

The equation of motion for the cantilever beam impacting system in a non-dimensional form is:
\begin{equation}\label{eq:impact}
    \ddot{y} + 2\xi \dot{y} + f(y) + \alpha H(y-g)(y - g) = \beta \cos(\Omega t)
\end{equation}
$H(x)$ is the Heaviside step function, and $f$ is a function describing the restoring force. For the example at hand, we set the linear function $f(y) = y$. The basins of Fig.~\ref{fig:impact} have been computed for the parameters $\xi = 0.02$, $\beta = 0.5$, $\omega = 0.417893$, $\alpha = 20$, and $g = 0.63$. There are two attractors: a fixed point and a chaotic attractor. Besides the impact on the cantilever, some initial conditions lead to a stable fixed point. The basins of this attractor change with the nonlinearity of the stiffness $f$.

\subsection{Uncertainty in the British Power Grid}
\begin{keywrds} FB, ODE\end{keywrds}
\begin{figure}
\begin{center}
\includegraphics[width=\columnwidth]{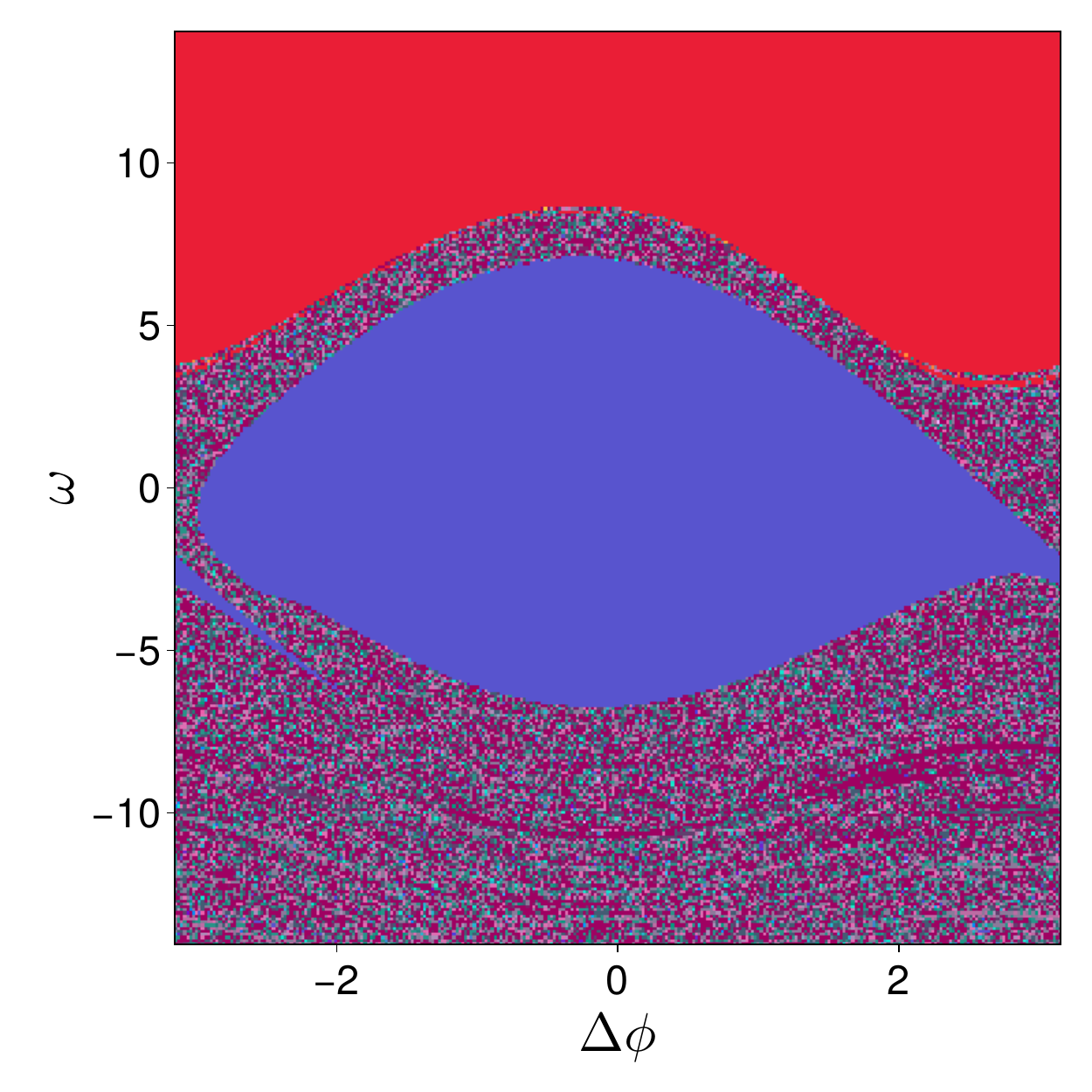}
\end{center}
    \caption{\label{fig:kuramoto_power}Local basins landscape of a perturbation of a node within the British power grid. The coupled system of ODEs is described in Eqs.~\ref{eq:kuramoto_power}. The incidence matrix and the vector $P_i$ have been retrieved from the supplementary material of the original paper~\cite{halekotte2021transient}. Additional parameters are $\alpha = 0.1$, $N=120$, and $K=5$.}
\end{figure}

Electric power distribution grids have been the subject of many studies in the dynamical systems community. They present an intriguing theoretical and practical matter at the crossroads of science and engineering. The grid can be modeled as a network of interacting agents, where each node is a dynamical system. The dynamical system represents the oscillatory dynamics of the synchronous generators producing electricity for the grid.

In~\cite{halekotte2021transient}, the authors focus on how multistability and the complexity of basin landscapes contribute to the uncertainty in the outcome of a perturbation to the synchronous state of the grid. The nodes of the grid are modeled with Kuramoto oscillators with inertia, where a second-order differential equation represents the dynamics of a synchronous generator. The basins are analyzed when the frequency and the initial phase of a node are perturbed. The final state of the network is examined after a long transient to study the influence of the perturbation.

The model of the node oscillators is as follows:
\begin{align}\label{eq:kuramoto_power}
\begin{split}
\dot \phi_n &=  \omega_n\\
\dot \omega_n &= P_n - \alpha \omega_n - K \sum_j A_{ij}\sin(\phi_i - \phi_j),
\end{split}
\end{align}
where $\phi_n$ and $\omega_n$ are the phase and frequency of the oscillator $n$. The adjacency matrix $A_{ij}$ contains all the information about the coupling of the system and is taken from field data of the British power grid. $K$ is the coupling coefficient between oscillators and is common to all nodes. $P_n$ is the net power input/output of oscillator $n$, with $P_n = P_0$ for a producer and $P_n = -P_0$ for a consumer node, where $P_0 = 1$. The signs are chosen randomly, assuming half of the nodes are consumers and half are producers. Figure~\ref{fig:kuramoto_power} illustrates the basins generated by the perturbation on one of the nodes of the power grid. In this plot, 58 basins have been detected for 300 by 300 initial conditions. The algorithm takes a long time to complete since we are looking for recurrences in a space of dimension 240 (with 120 oscillators). The detection of the attractors may take considerable time. For this kind of problem, a projection of the states onto a lower-dimensional space, a technique known as featurizing, can be a better strategy \cite{datseris2023framework}. Nevertheless, the algorithm manages to identify a variety of stable states for this local region. Many open problems related to this system have been suggested as follow-up research in~\cite{halekotte2021transient}, and there are still many unresolved questions.

\subsection{Gear Rattle Model}
\begin{keywrds} SMB, ODE \end{keywrds}
\begin{figure}
\begin{center}
\includegraphics[width=\columnwidth]{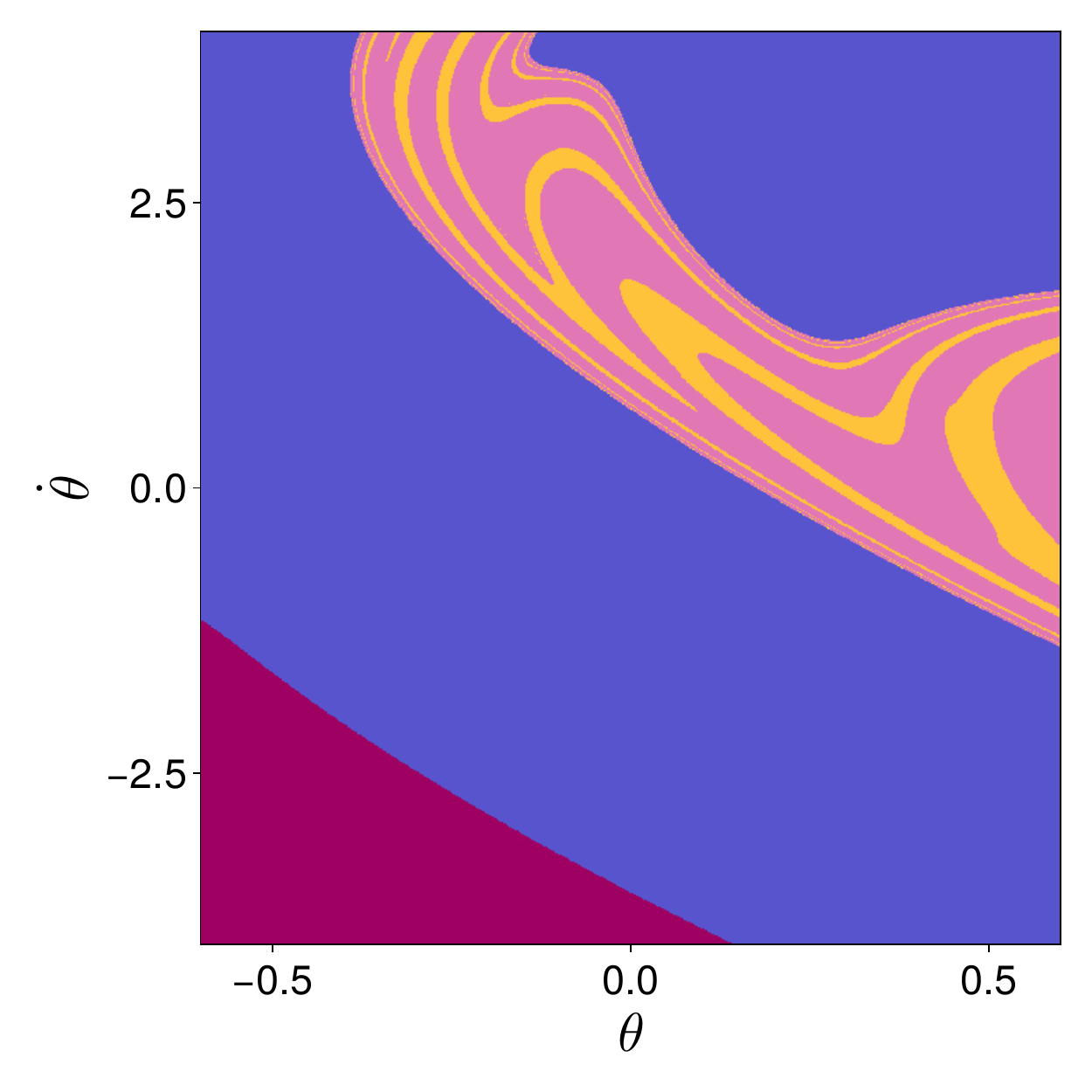}
\end{center}
\caption{\label{fig:gear_rattle}Basins of attraction of~\ref{eq:gear_rattle} with four quasiperiodic attractors. The system simulates an oscillator with backlash.}
\end{figure}

The paper~\cite{mason2009basins} computes and analyzes the basins of attraction of a backlash oscillator. The analysis is motivated by the modeling of vibrations in geared systems with impacts. The authors consider both a piecewise-linear stiffness model and an infinite stiffness impacting limit to understand the transitions in terms of smooth and discontinuous bifurcations.

The article demonstrates the importance of basin of attraction computations in understanding the relative dominance of competing solutions in the long-term dynamics of gear systems. The analysis of basins provides insights into the coexistence of different types of rattling solutions. The ordinary differential equation of the system is:
\begin{equation}\label{eq:gear_rattle}
 \Phi'' + \delta \Phi' + 2 \kappa B(\Phi) = 4 \pi \delta - 4 \pi^2 \varepsilon \cos(2 \pi t) - 2 \pi \delta \varepsilon \sin(2 \pi t),
\end{equation}
with the piecewise function $B$ representing the gear rattle:
\begin{equation*}
B(\Phi) = \begin{cases}
\Phi - \beta, & \Phi \geq +\beta \\
0, & |\Phi| < \beta \\
\Phi + \beta, & \Phi \leq -\beta
\end{cases}. 
\end{equation*}
The basins in Fig.~\ref{fig:gear_rattle} have been computed for the parameters $\beta = 0.6$, $\delta = 0.6$, $\varepsilon = 0.1$, and $\kappa = 100$. The boundaries between the basins are smooth, and the attractors detected are quasiperiodic. The basins in the original publication have been computed using a transformation of the system into a discrete map. The phase space is different, and the correspondence is not obvious, but the number of attractors is the same, and the shapes of the basins share some similarity.

\subsection{Another Gear Rattle Model}
\begin{keywrds} FB, ODE \end{keywrds}
\begin{figure}
\begin{center}
\includegraphics[width=\columnwidth]{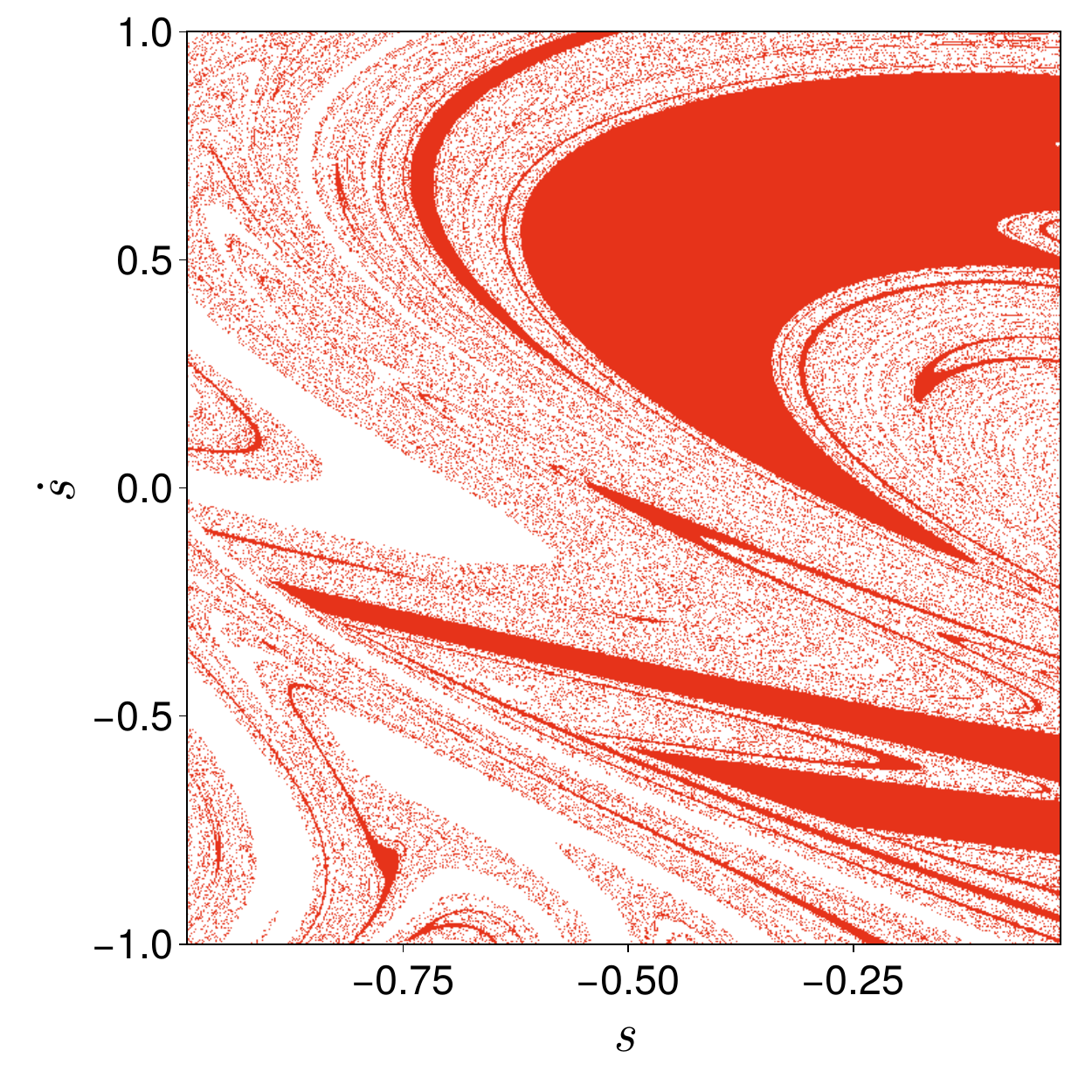}
\end{center}
\caption{\label{fig:gear_rattle2}Basins of attraction of~\ref{eq:gear_rattle2}. There is a fractal boundary between the basins of two periodic attractors of the gear rattle system.}
\end{figure}

In the article~\cite{Souza2001basins}, you can find another example of a study on gear rattle dynamics. This model involves a system with two coupled gears, characterized by impacts described by differential equations with impact conditions. The simulations revealed the presence of multiple periodic and chaotic attractors, whose basins have fractal boundaries. 

The system is:
\begin{equation}\label{eq:gear_rattle2}
    \ddot s + \beta \dot s = \gamma  -\alpha\beta\omega\cos(\omega t) + \alpha \omega^2 \sin(\omega t)
\end{equation}

The impact conditions for the gear are $s=-1$ and $s=0$. After an impact has been detected, a special procedure in the numerical solver sets the following conditions on the variables: $\dot s_0 = -r \dot s$ and $s_0 = s$. The direction of the oscillator is reversed, and there is a dissipation factor $r = 0.9$, so that there is a loss of energy during the impact. The basins in Fig.~\ref{fig:gear_rattle2} have a fractal boundary for the two periodic attractors detected. The basins in the original publication are represented using the variables of a discretized system, but we can find the same characteristics: two attractors and a fractal boundary. This example is included in the collection since it is a good example of a simple impact system with a fractal boundary. Parameters for these basins are: $\gamma = 0.1$,  $\alpha = 0.48$,  $\beta = 0.1$,  $\omega = 1$.

\subsection{Bell-Yoke System}
\begin{keywrds} SMB, ODE \end{keywrds}
\begin{figure}
\begin{center}
\includegraphics[width=\columnwidth]{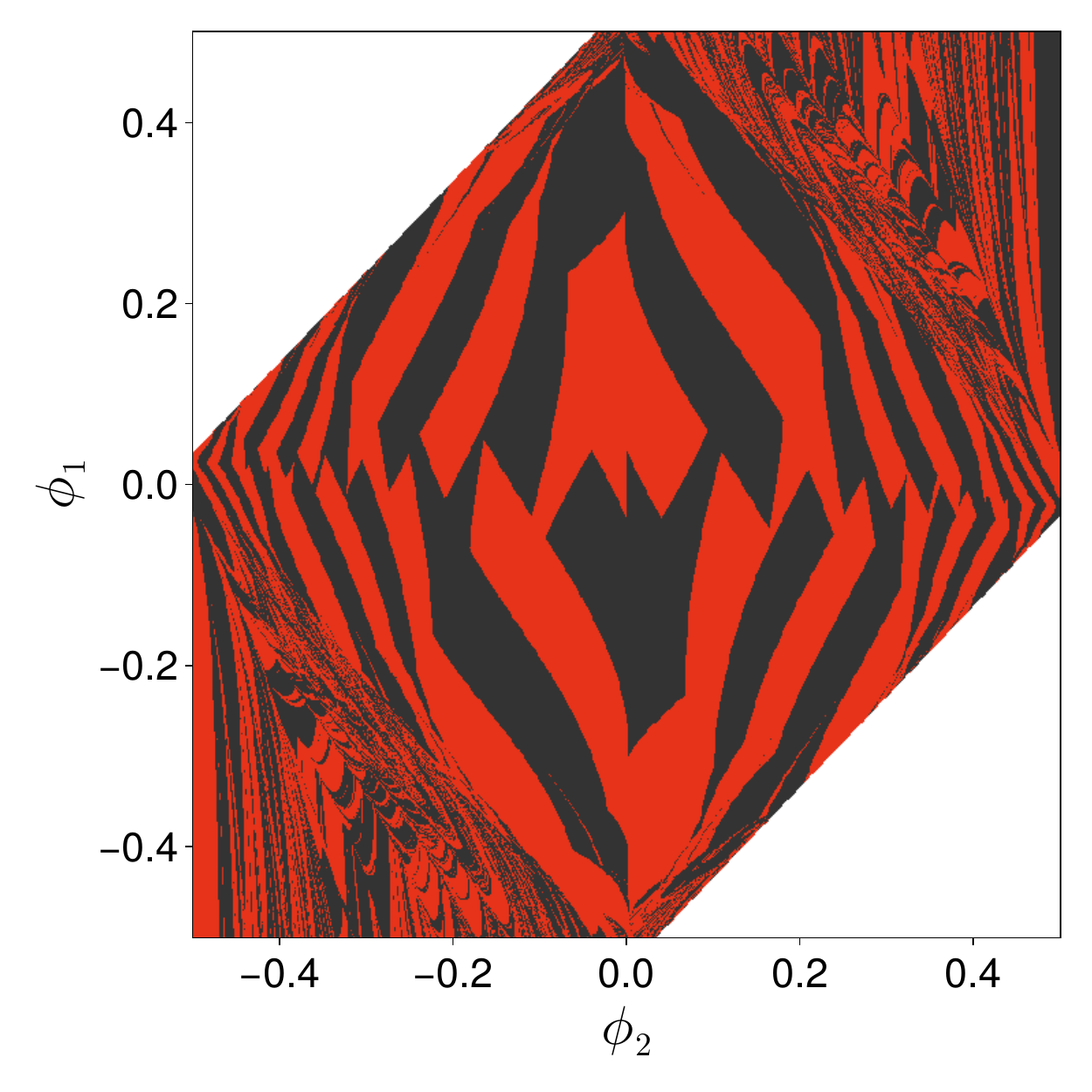}
\end{center}
\caption{\label{fig:bell_yoke}Basins of attraction of~\ref{eq:bell_yoke} for a swinging bell with a clapper impacting the bell. The diamond shape of the basins is related to the impact condition between the bell and the clapper. The parameters for this figure are: $l_r = -0.03$, $T_{max} = 150$, $M = 2633$, $m = 57.4$, $Bb = 1375$, $B_c = 45.15$, $L = 0.236$, $l = 0.739$, $l_c = -0.1$, $\alpha = 0.5349$, $D_b = 26.68$, $D_c = 4.539$, $g = 9.8$, $L_r = L - l_r$, $l_{cr} = l_c - l_r$, $Bbr = (Bb - M L^2) + M L_r^2$, $A = 15$, $\omega = 7.5$, $\beta = (M L_r + m l_{cr}) g$.}
\end{figure}

In a series of articles in mechanical engineering, a group from the University of Łódź studied the dynamics of swinging bells with a clapper in detail, both numerically and experimentally. The paper~\cite{brzeski2015analysis} is part of this series and characterizes the dynamics of church bells.

The authors classify their most common working regimes and analyze how different yoke designs and propulsion mechanisms influence these dynamics. There are also practical considerations regarding the suitable regimes for smooth operation.

The model is a hybrid dynamical system of the yoke-bell-clapper system, based on detailed measurements from the largest bell in the Cathedral Basilica of St. Stanislaus Kostka, Łódź, Poland. The two important parameters are the yoke design and propulsion mechanism. The identified working regimes of bells include symmetric and asymmetric falling clappers, symmetric and asymmetric flying clappers, double kiss, sticking clapper, and scenarios with no impacts.

Lagrange’s principle provides the equations of motion for the yoke-bell-clapper system. The main equations describing the dynamics of the system are represented as coupled second-order ordinary differential equations. The equations are as follows:
\begin{equation} \label{eq:bell_yoke}
    \begin{bmatrix}
    (B_{br} + m l_{cr}^2) &   m l_{cr} l \cos(\phi_2 - \phi_1)\\
     m l_{cr} l \cos(\phi_2-\phi_1) &  B_c
    \end{bmatrix}
\begin{bmatrix}
\ddot{\phi_1}\\
\ddot{\phi_2}
\end{bmatrix}
=
\begin{bmatrix}
     m l_{cr} l \dot \phi_2^2 \sin(\phi_1-\phi_2) - \beta\sin(\phi_1) - D_b \dot \phi_1 + D_c (\dot \phi_2-\dot \phi_1) + M_t\\
     - m l_{cr} \dot \phi_1^2 \sin(\phi_2 - \phi_1) - m g \sin(\phi_2) - D_c (\dot \phi_2 - \dot \phi_1) 
\end{bmatrix}
\end{equation}
$\phi_1$ and $\phi_2$ are the dynamical variables for the bell and the clapper, $B_b$ and $B_c$ are the moments of inertia of the bell and clapper, and $M$ and $m$ are their respective masses. $D_b$ and $D_c$ are damping coefficients, while $l$ and $l_c$ represent the distances between the rotation axis and the center of gravity of the bell and the clapper, respectively. $M_t(\phi_1, \dot \phi_1)$ is the generalized momentum generated by the motor, defined as:
\begin{equation}
   M_t(\phi_1, \dot \phi_1) =
   \begin{cases}
   T_{max} \text{sgn}(\dot{\phi_1}) \cos(\omega \phi_1), & \text{if } |\phi_1| \leq \frac{\pi}{A} \\
   0, & \text{if } |\phi_1| > \frac{\pi}{A}
   \end{cases}
\end{equation}
The parameter values of the model are included in the figure caption. A particularity of this model is the impact occurring between the clapper and the bell. The condition $|\phi_1 - \phi_2| = \alpha$ defines the occurrence of an elastic impact between the two masses. When this event is detected, the angular velocities are updated using a restitution coefficient $k = 0.05$ and the conservation of angular momentum. The following linear system must be solved: 
\begin{equation}
\begin{bmatrix}
a_1   & a_2 \\
1 & -1 
\end{bmatrix}
\begin{bmatrix} 
\dot \phi_{1,AI}\\
\dot \phi_{2,AI}
\end{bmatrix} 
= 
\begin{bmatrix} 
 a_1 & a_2 \\
 -\sqrt{k} & \sqrt{k}
\end{bmatrix} 
\begin{bmatrix} 
\dot \phi_{1,BI}\\
\dot \phi_{2,BI}
\end{bmatrix} 
\end{equation}
Coefficients $a_1$ and $a_2$ are defined as follows: $a_1 = (B_{br} + m l_{cr}^2 + m l_{cr} l \cos(\phi_2 - \phi_1))$, 
$a_2 = (B_c + m l_{cr} l \cos(\phi_2 - \phi_1))$. 
The angular velocities with index AI denote the quantities after the impact, while BI denotes the known quantities before the impact. Each time the impact event occurs, we update the system with the new velocities following the impact. The basins of attraction in Fig.~\ref{fig:bell_yoke} have two possible symmetric periodic attractors for these parameters. This plot is not included in the publication and represents an original contribution. Further investigation may explore other regimes of parameters for this interesting mechanical system.

\subsection{Chaos in a DC/DC Buck Converter}
\begin{keywrds}FB, ODE\end{keywrds}
\begin{figure}
\begin{center}
\includegraphics[width=\columnwidth]{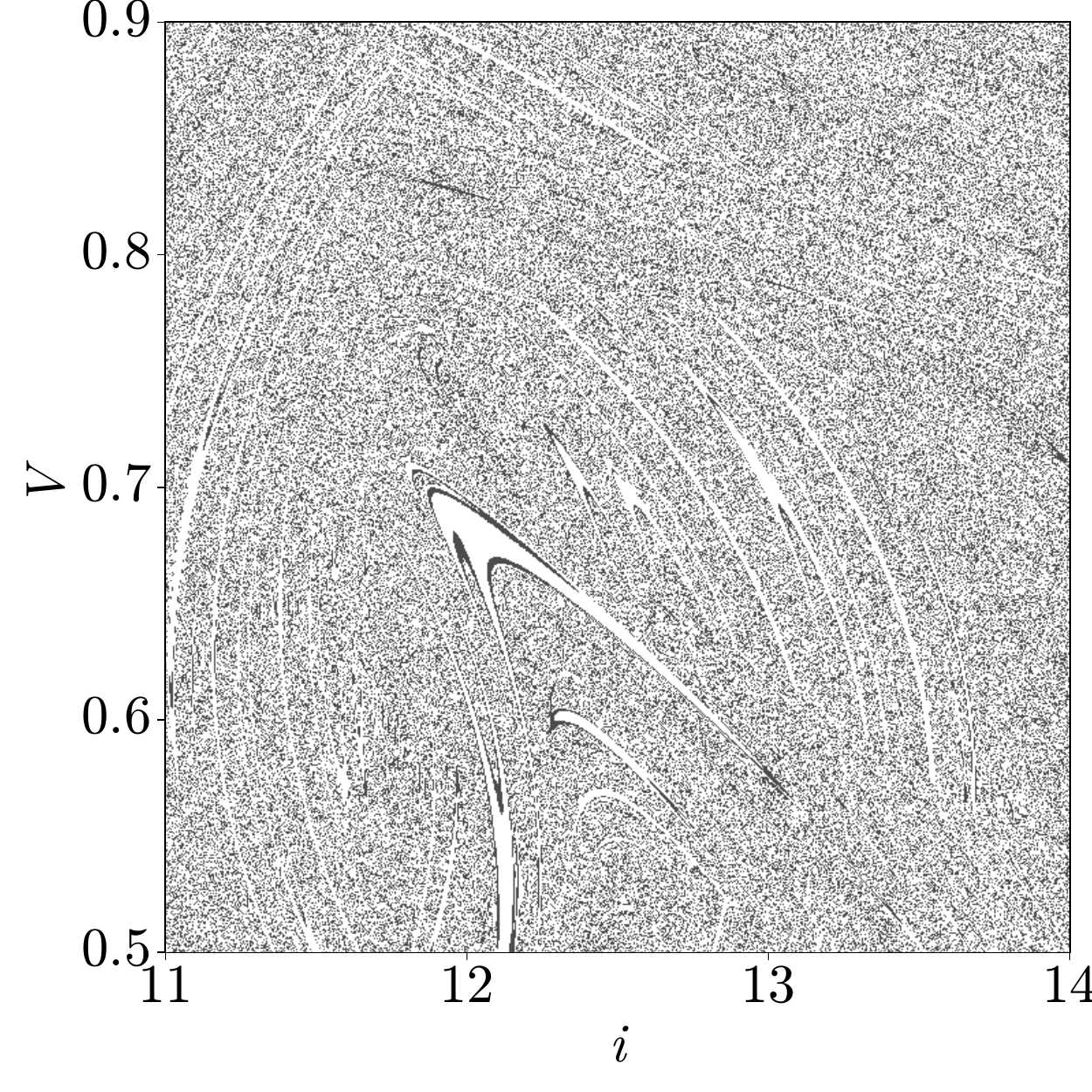}
\end{center}
\caption{\label{fig:buck_converter}Basins of attraction of a DC/DC Buck converter described in Eqs.~\ref{eq:buck_converter} with parameters $R = 22\,\Omega$, $C = 47 \cdot 10^{-6}\,\text{F}$,  $L = 20 \cdot 10^{-3}\,\text{H}$, $\gamma = 11.75$, $\eta = 1309.52$, $T = 400 \cdot 10^{-6}\,\text{s}$,  $V_{in} = 30.1\,\text{V}$.}
\end{figure}

Nonlinear electronic circuits have attracted a great deal of attention in the nonlinear dynamics community. In particular, with the popularity of switching electronic circuits, studies on the appearance of chaos in circuits have become very common. Most efforts have been dedicated to the demonstration of chaos, while less attention has been drawn to the multistable aspect. Ref.~\cite{di1997secondary} is an example of a multistable voltage-controlled DC/DC buck converter. This device is designed to transform a direct current supply from one value to another with a periodic switching of a transistor. The article details bifurcation routes, chaos, and the multistable system by changing the parameters of the model.

The state variables of the converter are the output voltage $v(t)$ across the capacitor and the inductor current $i(t)$. The dynamics of the circuit can be described by the following linear differential equations:
\begin{align}\label{eq:buck_converter}
\begin{split}
   \frac{d}{dt} \begin{pmatrix} V \\ i \end{pmatrix} = 
    \begin{pmatrix} -\frac{1}{RC} & \frac{1}{C} \\ 
           -\frac{1}{L} & 0 \end{pmatrix} 
    \begin{pmatrix} V \\ i \end{pmatrix} + 
    \begin{pmatrix} 0 \\ \frac{V_{in}}{L} \end{pmatrix} u(t)
\end{split}
\end{align}
$u(t)$ is a control signal that is 0 when $V(t) > V_r(t)$ (converter OFF) and 1 when $V(t) < V_r(t)$ (converter ON). The ramp signal controlling the switching is $V_r(t) = \gamma + \eta(t \mod T)$. $\gamma$ and $\eta$ are constants, and $T$ is the period of the ramp signal. Figure~\ref{fig:buck_converter} shows the phase space of the system, $V-i$, where period-2 and period-12 attractors exist. The interesting dynamics are caused by the switching condition of the transistor. In an engineering context, chaos and multistability are undesired behaviors. Everything has to be linear and predictable. However, semiconductor devices are far from being linear, at best over a limited range of operation. Engineering usually consists of thinking on a linear basis, but the article~\cite{di1997secondary} points out the limits of this mindset.

\subsection{Multistable Chaotic Gyrostat System}
\begin{keywrds}SMB, ODE\end{keywrds}

\begin{figure}
\begin{center}
\includegraphics[width=\columnwidth]{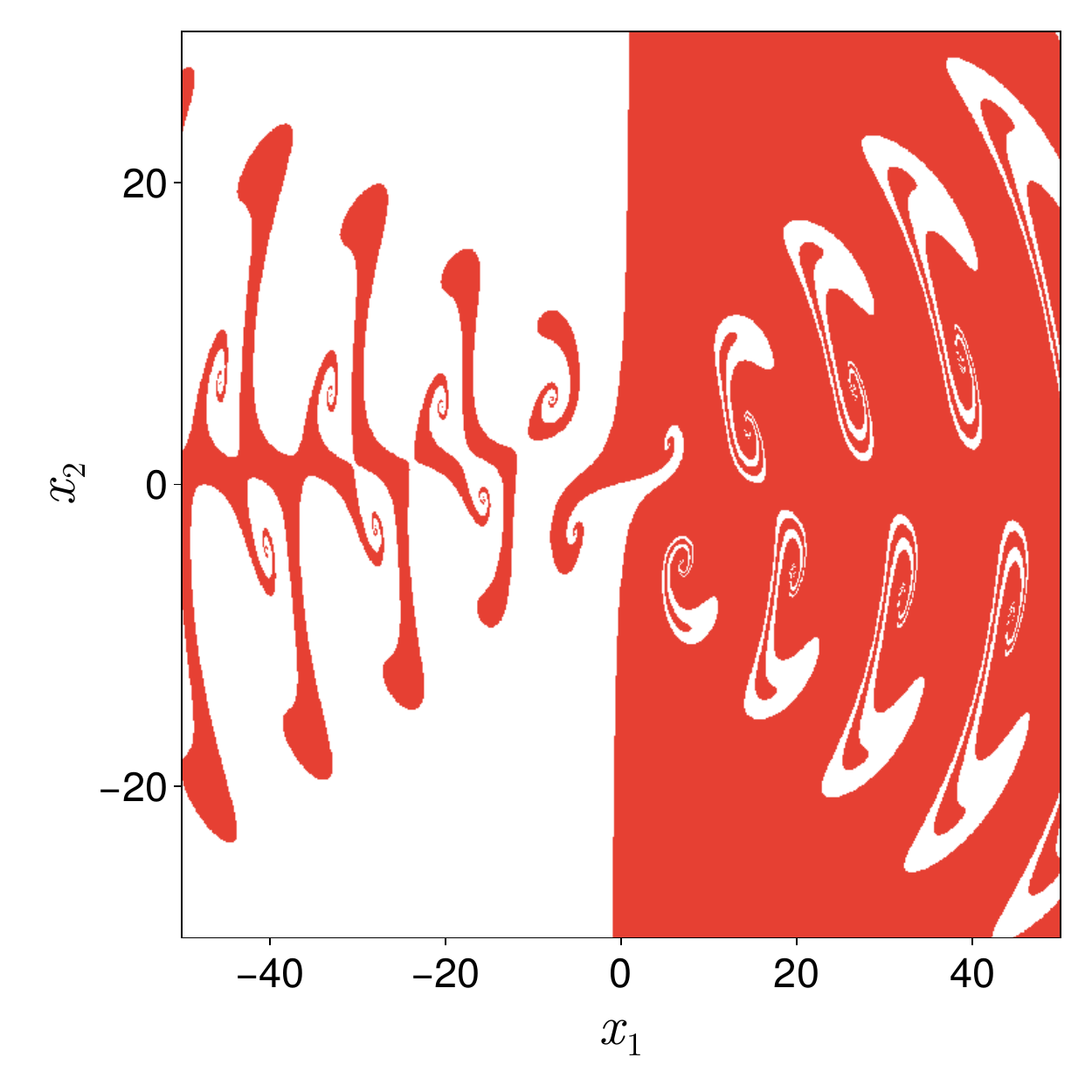}
\end{center}
\caption{\label{fig:gyrostat}Basins of attraction of a chaotic gyrostat described in Eqs.~\ref{eq:gyrostat}. Parameters are described in the text.}
\end{figure}

The study \cite{marwan2022coexisting} investigates the presence of coexisting attractors in a gyrostat chaotic system through the analysis of basins of attraction. The researchers first identified multistability, then switched on the design of a chaotic controller to synchronize the gyrostat with a aerial vehicle system. 

A gyrostat is a type of mechanical device or system that utilizes the principles of gyroscopic motion to maintain stability or orientation in space. Essentially, it is a rigid body that has one or more rotors  mounted to it, which can generate gyroscopic effects. The simplified equations of motions on the three axis are: 
\begin{align}\label{eq:gyrostat}
\begin{split}
\dot{x}_1 & = -b_{11} x_1 - b_{12} x_2 + b_{13} x_3 + F_{1m} x_2 x_3 + L_{1m} \\
\dot{x}_2 & = b_{21} x_1 + b_{22} x_2 + F_{2m} x_1 x_3 + L_{2m} \\
\dot{x}_3 & = -b_{31} x_1 - b_{33} x_3 + F_{3m} x_1 x_2 + L_{3m}
\end{split}
\end{align}
The simulation of the system on the $x_1 x_2$ plane with the initial conditions $x_3 =0 $ showed a periodic and a chaotic attractors with their basins represented in Fig.~\ref{fig:gyrostat}. The parameters used for the basins are $b_{11} = 2$,  $b_{12} = 0.7933$,  $b_{13}= 0.1914$, $b_{21} = 1.19$, $b_{22} = 3.48$, $b_{31} = 0.5742$, $b_{33} = 5.8$, $F_{1m} = 1/3$, $F_{2m} = -1$, $F_{3m} = 1$,  $L_{1m} =0$, $L_{2m} = 0$, $L_{3m} = 22.8$.

\section{Systems with Delay}

This section has few examples for two reasons. First, there are very few studies of multistable systems in delayed differential equations. The second reason is the problem of the infinite-dimensional phase space. The initial conditions of the delayed differential equations also include a history for booting up the system. The examples presented here project the basins in a two-parameter differential space. This is a subject worth investigating, but traditional tools have to be adapted. Since the recurrence algorithm only requires a grid in the phase space and the state of the system, we can still detect attractors. The following two examples have been computed this way.

\subsection{Wada in Systems with Delay}
\begin{keywrds} WD, DDE\end{keywrds}
\begin{figure}
\begin{center}
\includegraphics[width=\columnwidth]{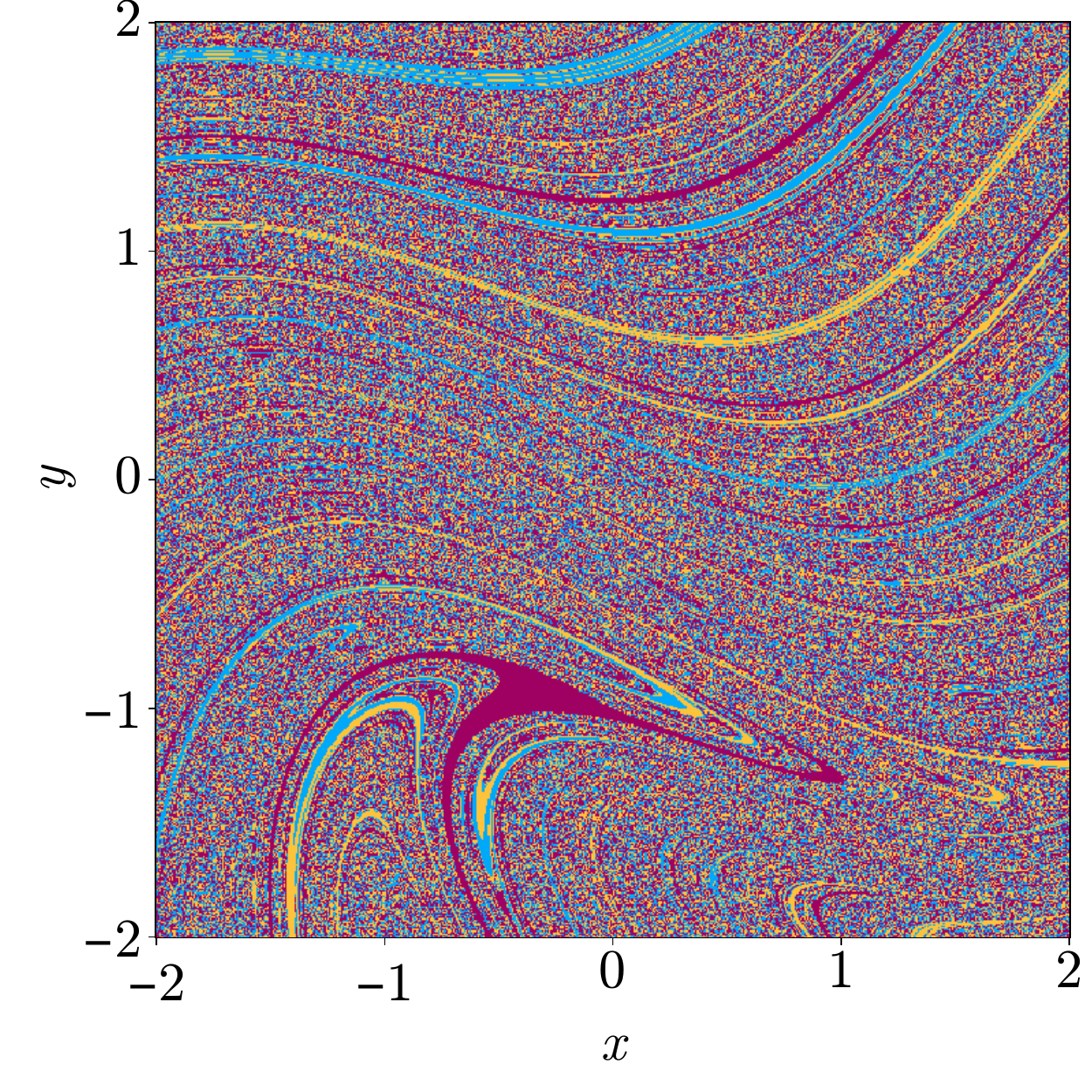}
\end{center}
    \caption{\label{fig:wada_delay}Basins of attraction of the lagged oscillator with nonlinear feedback in Eqs.~\ref{eq:wada_delay}. Parameters of the simulations are $\alpha = -0.925$, $F = 0.525$, $\omega = 1.0$, and $\tau = 1.065$. Additionally, we set the initial conditions at $t=0$: $x = 1.0$ and $\dot{x} = 1.0$.
}
\end{figure}

The delay differential equation in~\cite{daza2017wada} presents the unpredictability of the systems in phase space. The simple oscillator, provided with nonlinear delayed feedback, can exhibit transient chaos and Wada basins. 

There are no initial conditions for delayed differential equations since the history functions define an infinite-dimensional space. The sensitivity to initial history must be restricted to a small subspace of functions for study with the available tools. The authors chose a simple two-dimensional projection for the initial history of the following nonlinear delayed feedback system:
\begin{equation}\label{eq:wada_delay}
    \dot{x} + x((1 + \alpha)x^2 - 1) - \alpha x(t - \tau) = F \sin(\omega t)
\end{equation}
This is a periodically forced system with nonlinear delayed feedback. The numerical integration of these equations requires a special algorithm to account for the previous history of the system. The authors chose the initial history $x(t) = x_1 + x_2 t$ for $-\tau \leq t < 0$ to define a two-dimensional projection of the initial history. The result of the numerical integration is shown in Fig.~\ref{fig:wada_delay}, where three periodic attractors with a Wada basin boundary appear.

\subsection{Extreme Events and Riddled Basins in Neuron Networks}
\begin{keywrds}RB, DDE\end{keywrds}
\begin{figure}
\begin{center}
\includegraphics[width=\columnwidth]{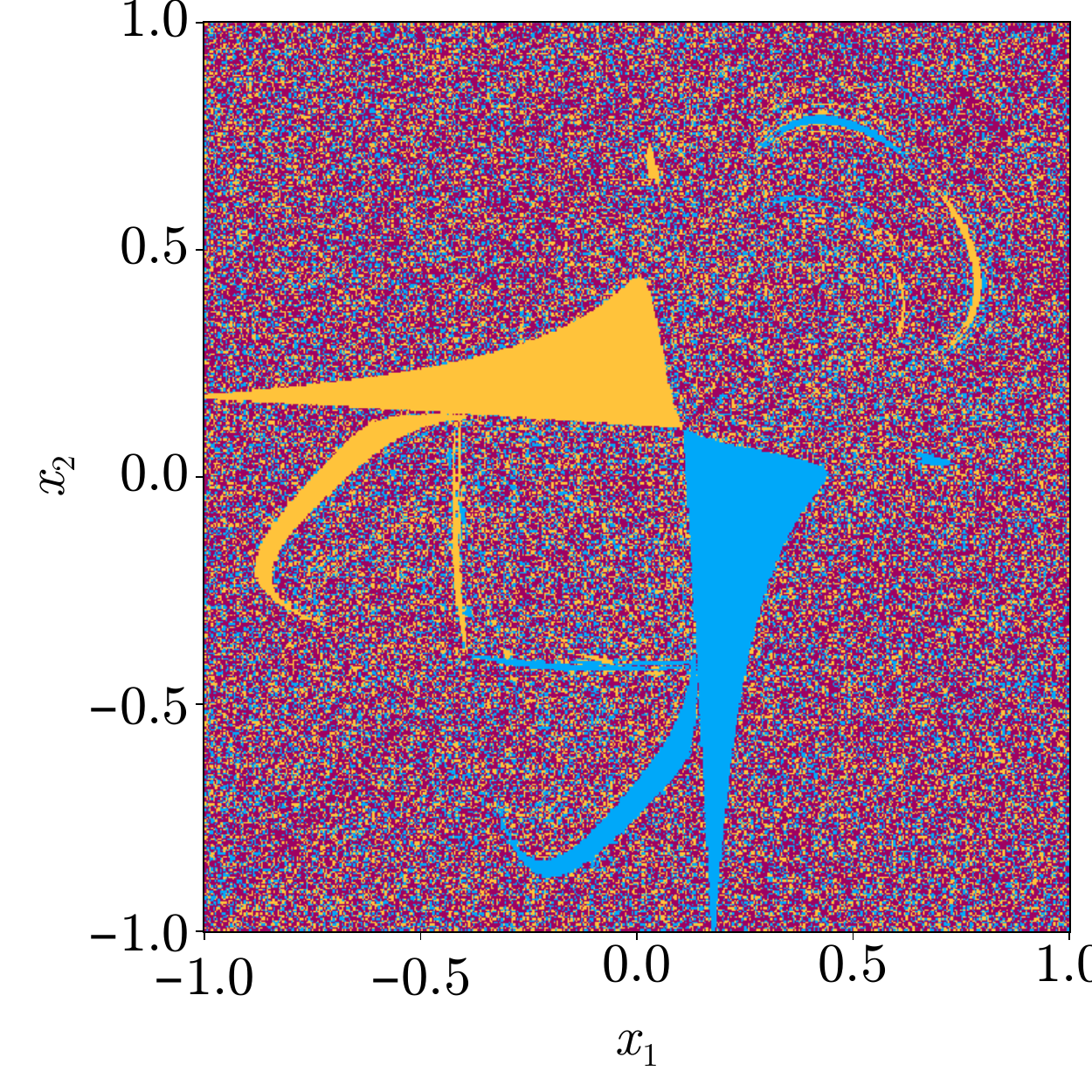}
\end{center}
\caption{\label{fig:neuron_delay}Basins of attraction of two delayed coupled FitzHugh-Nagumo neuron models described in Eqs.~\ref{eq:neuron_delay}. Additional parameters for this model are $M_1 = 0.01$ and $M_2 = 0.026$.}
\end{figure}

In ~\cite{saha2018riddled}, the authors explore the appearance and implications of riddled basins of attraction in delay-coupled systems. They use a system of two FitzHugh-Nagumo units coupled with delays to demonstrate the formation of riddled basins of attraction.

The equations governing the dynamics of the model with delay coupling are given as follows:
\begin{align}\label{eq:neuron_delay}
\begin{split}
\dot{x}_i &= x_i (a - x_i)(x_i - 1) - y_i + \sum_{k=1}^{2} M_k \left( x_j(t - \tau_k) - x_i \right) \\
\dot{y}_i &= b x_i - c y_i + \sum_{k=1}^{2} M_k \left( y_j(t - \tau_k) - y_i \right)
\end{split}
\end{align}
Here, $x_i$ and $y_i$ are the state variables of the $i$-th FHN unit (with $i = 1, 2$). The parameters are $a = -0.025$, $b = 0.00652$, $c = 0.02$, and $M_k$ represents the coupling strengths, with $k = 1, 2$. There are also two different delay feedbacks: $\tau_1 = 80$ and $\tau_2 = 65$. The basins represented in Fig.~\ref{fig:neuron_delay} have been computed with a special solver adapted for delay differential equations. However, the recurrence algorithm can still be used with the state variables $x_i$ and $y_i$.

The initial conditions in the figures represent a projection of the infinite-dimensional phase space where the initial conditions and history of the variables $y_i$ are set to $y_i = 0.01$. The history of the time-delayed variables is chosen so that $x_i(t) = x_0$ for $t \leq 0$. This is a constant history equal to the value at $t = 0$.

The trajectories can exhibit very long transient behavior before settling to a fixed point. We must take care to set a long integration time before running the recurrence algorithm; otherwise, the transient can be mistaken for oscillatory dynamics.

\section{Conclusions}
This closing section is not really necessary for this work. I suppose most of the interested people will pick and choose the system they are interested in. Therefore, I should comment a little on the motivations behind this work.

In the process of researching the material, I have been fascinated (and obsessed) by the variety of shapes produced by these mathematical equations. Many of these images are at the crossroads of science and art, and there is a vibrant community of design artists publishing fractal pictures. One of the goals of this article is to share my enthusiasm for these objects and to provide the tools to other researchers so they can also appreciate the aesthetics of basins and pursue the investigations of these objects. The zoo is closing its doors.

\section*{Acknowledgments}
I would like to thank A. Daza, A. García, G. Datseris and K. Rossi for their constructive comments and suggestions. I also thank Pr. Sanjuán for his mentoring. This research was supported by the European Regional Development Fund (ERDF, EU) under Project No. PID2023-148160NB-I00 (MCIN/AEI/10.13039/501100011033).

\bibliographystyle{ws-ijbc}
\bibliography{refs.bib} 
\end{document}